\documentclass[encoding=utf8,english,noloadhyperref]{tumphthesis}
\usepackage{amsmath}
\usepackage{listings} 
\usepackage{cite}
\usepackage{appendix}
\usepackage{adjustbox}
\usepackage{placeins}
\usepackage{cleveref}

\crefname{table}{Tab.}{Tables}
\definecolor{mygreen}{rgb}{0,0.6,0}
\definecolor{mygray}{rgb}{0.5,0.5,0.5}
\definecolor{mymauve}{rgb}{0.58,0,0.82}

\subject{Abschlussarbeit im Masterstudiengang Physik von}
\title{Verbesserte Tuningmethoden von Monte Carlo Generatoren}
\subtitle{\foreignlanguage{british}{Improved tuning methods for Monte Carlo generators}}
\author{Fabian Klimpel}
\date{06.~September 2017}
\cooperators{Max-Planck-Institut f\"ur Physik}

\lowertitleback{Erstgutachter (Themensteller): PD Dr. Stefan Kluth}

\begin{document}

\frontmatter
\maketitle
\tableofcontents


\mainmatter
\chapter{Introduction}

\begin{figure}[htbp]
	\centering
	\includegraphics[width=0.7\textwidth]{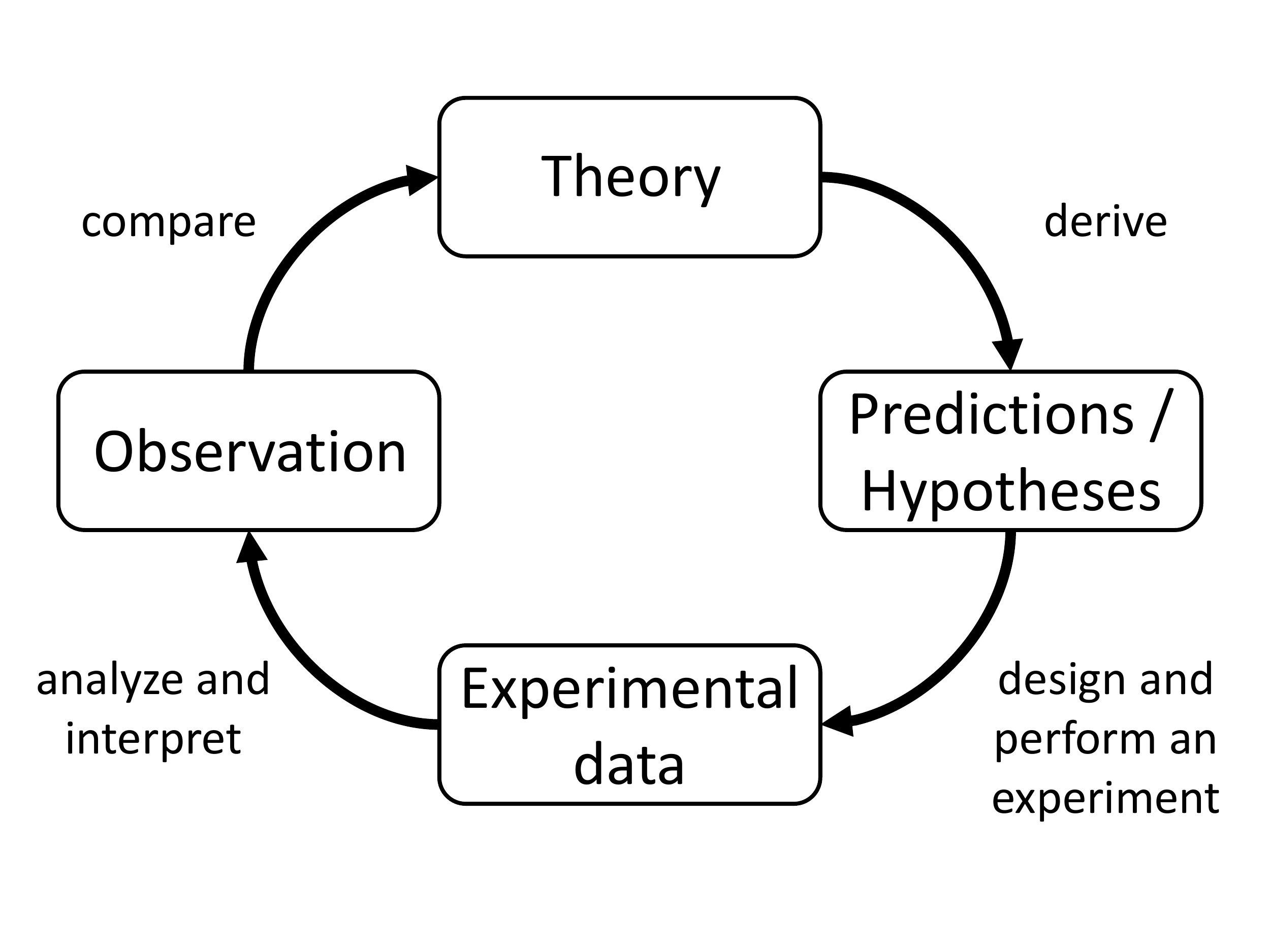}
	\caption{Illustration of the interaction between theory and experiment. This graphic is based on a graph from Ref.~\cite{ref:boris_grube}.}
	\label{fig:cycle_theory_experiment}
\end{figure}

The general purpose of particle physics is the identification of elementary particles, the measurement of their properties and the interactions between them. This led to the construction of the standard model of particle physics (SM)\cite[p.~46ff]{ref:griffiths}. This model is a theoretical construction that was created in agreement with experimental data and generated predictions for further measurements. The general interaction between theory and experiment is displayed in Fig.~\ref{fig:cycle_theory_experiment}.

The data analyzed in high-energy physics is mainly produced in collider experiments. In such a setup, it is possible to collect data from a controllable environment that allows precise measurements. Also, these experiments are (theoretically) repeatable. This is important due to the fact that the interactions in particle physics are not deterministic and the measurements cannot be performed with infinite accuracy. If a process under study has a large production cross-section and/or the collider has a large interaction rate, the statistical uncertainty on the measurements can be strongly reduced which allows to derive meaningful results.

Another problem in collider experiments arises from the setup itself. The initial state is the particle beam itself. Its properties like energy distribution or the spatial particle distribution in a bunch can be measured before the actual series of measurements start. The precision and accuracy of a detector can also be determined beforehand. The detector allows a collection of final state information. But both of these measurements carry a certain uncertainty or can only be modeled. This underlines the importance of repeatable measurements. Additionally, the processes that happen between the particle collision and the signal in the detector remain at most indirectly measured. The collection of all those information of a single particle collision is called an ``event''.\\

In order to describe the properties and interactions of elementary particles, a theoretical description is needed. These predictions are e.g. indirect limits on the masses based on experimental data. Also data were interpreted afterwards as the discovery of new elementary particles like the strange quark\cite[p.~28ff]{ref:griffiths}.

The theoretical description which is used today is the SM. Although the SM is a very successful theory for data description, several extensions exist which predict new particles (e.g. the ``Minimal Supersymmetric Standard Model'' (MSSM)\cite{ref:mssm}). The particles in the SM as well as their interactions will be described in the first part of Chapter~\ref{ch:theorie}.

Theoretical models are used to describe the transition between the initial collision and the final state measurement in a particle collider. The probabilistic models used for that description will be presented in the second part of Chapter~\ref{ch:theorie}.\\

These models are part of Monte Carlo (MC) simulation frameworks like \mbox{\textsc{Pythia~8}}\cite{ref:pythia6,ref:pythia8}. The goal of such a simulation is to reproduce the actual measurements and serve therefore as the link between theory and experiment. For that purpose, model parameters need to be set in order to give a good data description. A procedure for the model parameter estimation will be described and performed in Chapter \ref{ch:alter_ansatz}. A detailed investigation of several parts that will be mentioned there will be discussed and analyzed in the Chapters \ref{ch:bayes} and \ref{ch:adaptive}. The investigation of the tuning procedure will be summarized in the Chapter \ref{ch:summary}.\\

\chapter{The Standard Model of Particle Physics}
\label{ch:theorie}

In this Chapter, the elementary particles in the SM and the interactions between them will be introduced. At first, the leptons and quarks will be introduced together with their most important properties. Secondly, the interaction between the particles of the SM will be presented. In the second part of this Chapter, the working principles of Monte Carlo (MC) generators will be discussed.

\section{Elementary particles}
This Section is meant take a closer look at the question which elementary particles in the standard model of particle physics (SM) exist. After a short overview over the fermions in the first part of Subsection \ref{subsec:thestandardmodelofparticlephysics}, the possible interactions will be presented in the Subsection \ref{subsec:interactions}. The content of the short overview of particle properties is mainly extracted from Ref.~\cite{ref:griffiths}.

\subsection{The standard model of particle physics}
\label{subsec:thestandardmodelofparticlephysics}
The SM is meant to describe the constituents of all matter. Additionally its goal is to describe the interaction between those constituents. The elementary particles exist either with a half-integer or a integer spin. Half-integer spins are called fermions and exist in the SM as leptons and quarks. Interactions are mediated by particles with integer spin, called bosons. Up to date, six leptons and six quarks are known. These leptons are split in three generations. Each of those generations contains two quarks, one charged and one neutral lepton. Those parts will be explained further in the following.\\

For leptons, the first generation contains the electron $e$ and the electron neutrino $\nu_e$. Beside their different electric charge $q$, their masses are also significantly different. To distinguish the particles of this generation from the other two, beside the differences in their mass three further characterization numbers were introduced: electron number $L_e$, muon number $L_\mu$ and tau number $L_\tau$. Furthermore, in every generation there is always a charged particle (electron $e$, muon $\mu$, tau $\tau$) and a neutral particle (electron neutrino $\nu_e$, muon neutrino $\nu_\mu$, tau neutrino $\nu_\tau$). Their characteristic numbers are shown in Table \ref{tab:lepton_numbers}. Additionally, every lepton has a \textit{weak isospin} $T$. For left-handed particles (cf. \cite[p.~46ff]{ref:qft}), a lepton generation forms a doublet state. The charged particle always has a third component of $T_z = -1/2$ and the respective neutrino has the component $T_z = 1/2$\cite[p.~299]{ref:qft}. For the case of right-handed particles, the weak isospin forms a singlet state with a zero third component for the charged lepton, since neutrinos cannot be right-handed.\\

\begin{table}[htbp]
  \centering
  \begin{tabular}{ c | c | c | c | c c c }
    \hline
    generation & lepton & $q~[e]$ & $m~[MeV]$ & $L_e$ & $L_\mu$ & $L_\tau$ \\ \hline
    1 & $e$ & -1 & 0.5109989461 $\pm$ 0.0000000031 & 1 & 0 & 0\\
    1 & $\nu_e$ & 0 & < 0.460 $\cdot~10^{-3}$ (CL = 95\%) & 1 & 0 & 0 \\ \hline
    2 & $\mu$ & -1 & 105.6583745 $\pm$ 0.0000024 &  0 & 1 & 0 \\
    2 & $\nu_\mu$ & 0 & < 0.19 (CL = 90\%) & 0 & 1 & 0 \\ \hline
    3 & $\tau$ & -1 & 1776.86 $\pm$ 0.12 &  0 & 0 & 1 \\
    3 & $\nu_\tau$ & 0 & < 18.2 (CL = 95\%) & 0 & 0 & 1 \\
    \hline
  \end{tabular}
  \caption{Overview of the leptons, their generations, charge $q$, mass $m$, electron number $L_e$, muon number $L_\mu$ and tau number $L_\tau$. The masses are taken from Refs.~\cite{ref:pdg_leptons, ref:pdg_neutrinos}.}
  \label{tab:lepton_numbers}
\end{table}

Similar to the leptons, the quarks can also be characterized by their charge and mass inside a generation. Here, the difference to the lepton numbers is the introduction of six \textit{flavors}. Those are called downness $D$, upness $U$, strangeness $S$, charmness $C$, bottomness $B$ and topness $T$. Their properties are shown in Table \ref{tab:quark_numbers}. Quarks carry additionally a quantum number called color charge that is either red, green or blue. This charge is necessary in order to create an additional degree of freedom to describe the existence of hadrons that contain three quarks of the same flavor. Otherwise, this would be forbidden by the \textit{Pauli-principle}.\\


\begin{table}[htbp]
\makebox[\textwidth][c]{
  \begin{tabular}{ c | c | c | c | c c c c c c }
    \hline
    Generation & Quark & $q~[e]$ & $m~[MeV]$ & $D$ & $U$ & $S$ & $C$ & $B$ & $T$ \\ \hline
    1 & $d$ & -1/3 & $4.8^{+0.5}_{-0.3}$ & -1 & 0 & 0 & 0 & 0 & 0 \\
    1 & $u$ &  2/3 & $2.3^{+0.7}_{-0.5}$ & 0 & 1 & 0 & 0 & 0 & 0 \\ \hline
    2 & $s$ & -1/3 & $95 \pm 5$ & 0 & 0 & -1 & 0 & 0 & 0 \\
    2 & $c$ &  2/3 & $1275 \pm 25$ & 0 & 0 & 0 & 1 & 0 & 0 \\ \hline
    3 & $b$ & -1/3 & $4180 \pm 30$ & 0 & 0 & 0 & 0 & -1 & 0 \rule{0pt}{2.6ex} \\
    3 & $t$ &  2/3 & $(173.21\pm 0.51 \pm 0.71)\cdot 10^3$ & 0 & 0 & 0 & 0 & 0 & 1 \\
    \hline
  \end{tabular}
}
  \caption{Overview over the quark, their generations, charge $q$, mass $m$ and flavor. The masses are taken from Ref.~\cite{ref:pdg_quarks}.}
  \label{tab:quark_numbers}
\end{table}

For the case of leptons and quarks, these were only half of the constituents. The other half consists of their antiparticles. In that case, every characterization number introduced above but the mass is inverted. For quarks, the color is also inverted and referred to as anti-red etc. With respect to the antiparticles, the lepton numbers, flavors, colors are as the charge conserved in interactions.\\

\subsection{Interactions}
\label{subsec:interactions}
Now that the fermions are introduced, the goals will lie on the interaction between them in this Subsection. Interaction in particle physics means the exchange of particles in order to transfer energy, momentum etc. In the SM, one can distinguish four different forces of interaction. The first one is the electromagnetic force. Its exchange particle is the photon $\gamma$. This force affects every electrically charged particle.

The second force, the weak force, occurs due the electric neutral $Z^0$, the positively charged $W^+$ and the negatively charged $W^-$ boson. This force affects every fermion mentioned before. Also, interaction between those three bosons is possible.

\begin{figure}
	\centering
	\includegraphics[width=0.7\textwidth]{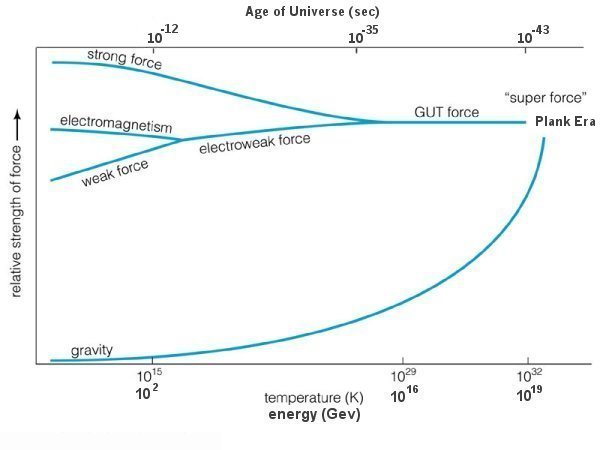}
	\caption{Relative strength of the forces at certain energies/temperatures/ages of the universe. The graph is taken from Ref.~\cite{ref:strength_forces}.}
	\label{fig:strength_forces}
\end{figure}

The third force, the strong force, is driven by gluons $g$. This kind of interaction is exclusively for quarks. In other words, the gluon only interacts with particles that carry a color charge. The gluon carries itself two color charges: a color and an anti-color. This construct leads to an overall of eight different gluons according to their color configuration. Since gluons carry a color charge themselves, they can interact with each other.\\
The last force is the gravitation, driven by the (hypothetical) graviton, but since this force is too weak, it is not considered in the SM yet. However it is a part in physics beyond the SM. Therefore, the gravity will not be mentioned further. The relative strengths of these forces are shown in Fig.~\ref{fig:strength_forces}. In the following those forces will be described in more detail.

\subsubsection{Electromagnetic interaction}
\begin{figure}
	\centering
	\includegraphics[width=0.4\textwidth]{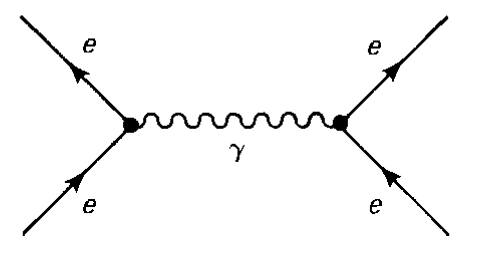}
	\includegraphics[width=0.7\textwidth]{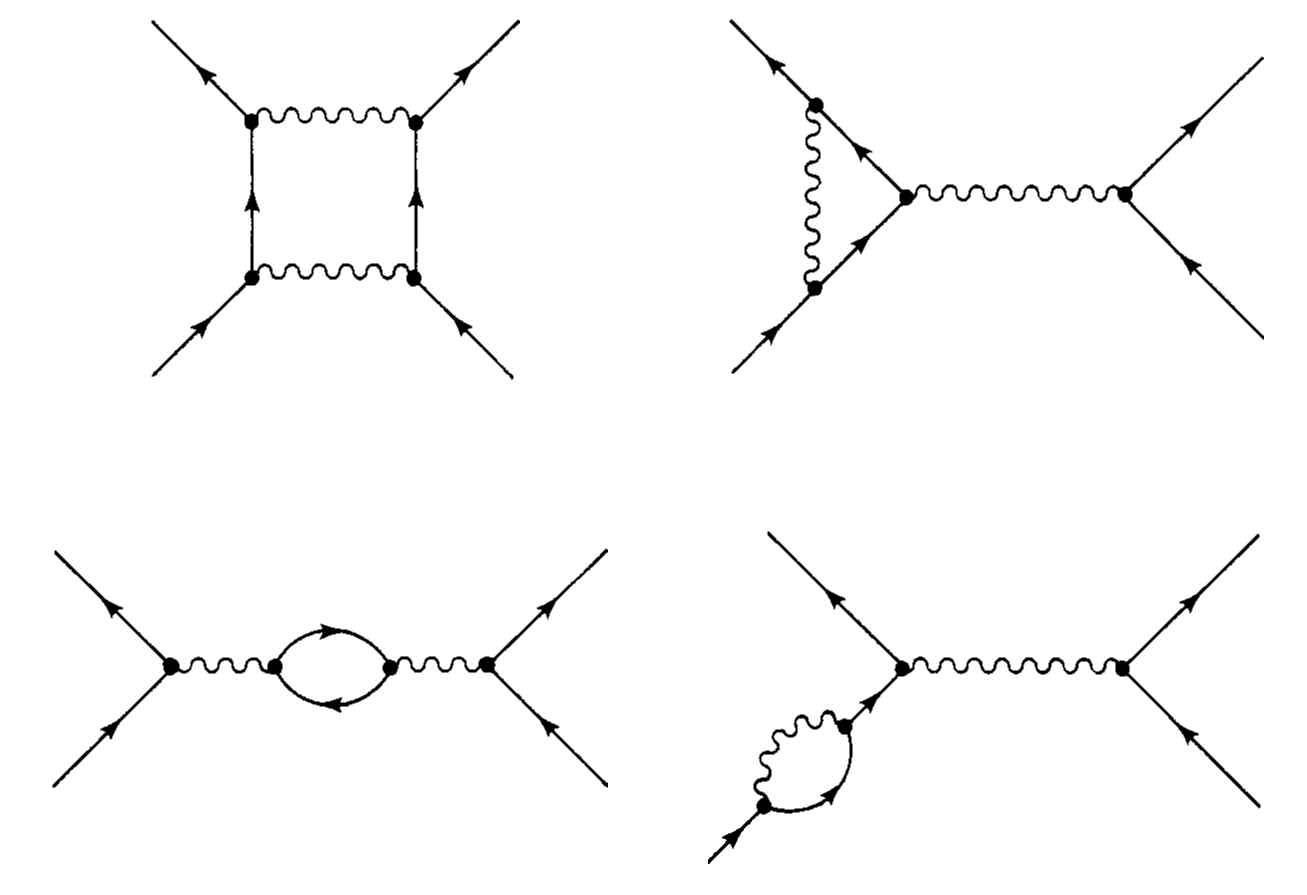}
	\caption{Feynman diagrams of electrodynamic interactions. The top one represents a simple interaction between two leptons. The four diagrams below show possible perturbations in the process.}
	\label{fig:ed_interactions}
\end{figure}

The electromagnetic interaction is the oldest theory and formally described by the theory of quantum electrodynamics (QED). This interaction only affects electrically charged particles. In such an interaction, particles exchange photons $\gamma$. A graphical representation of the exchange are Feynman diagrams (see Fig.~\ref{fig:ed_interactions}). Important is the feature, that an interaction can become arbitrary complicated due to the fact that only the ingoing and outgoing particles underlie the particles boundary like the mass. In addition the ingoing and outgoing system needs to match in conservation of quantum numbers such as the flavor, lepton number etc. Inside the interaction process, there can be many exchanges, influencing the predicted results. Some examples are shown in the lower row of Feynman diagrams in Fig.~\ref{fig:ed_interactions}. The regulation for the number of vertices is given by the coupling constant $\alpha_{EM}=\frac{e^2}{\hbar c}\approx \frac{1}{137}$. Since $\alpha_{EM} < 1$, the probability for a higher number of interactions $n$ is suppressed by $\mathcal{O}(\alpha_{EM}^n)$.

\subsubsection{Weak interaction}
\begin{figure}[htbp]
	\centering
	\includegraphics[width=0.4\textwidth]{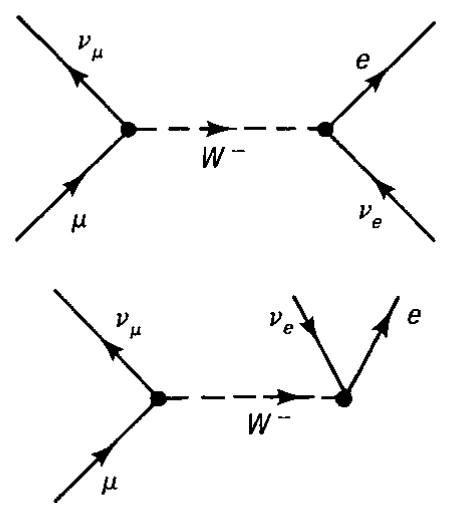}
	\caption{Feynman diagrams of charged weak interactions of the processes \mbox{$\mu\nu_\mu\rightarrow\nu_e e$} (top) and $\mu\rightarrow\nu_\mu\nu_e e$ (bottom). The additional charged current allows more possible in the weak interaction than the in electrodynamic interaction.}
	\label{fig:w_flavorchange}
\end{figure}

The weak force interacts with every quark and lepton as well as the mediators of the weak force. For a leptonic case, the process is only observed if the lepton numbers mentioned in the previous part are conserved additionally to the regular conservation laws like the energy, charge etc. For a neutral current interaction via the exchange of a $Z^0$ boson, the participating ingoing and outgoing particles with respect to their conservation numbers underlie the same constraints like in the electromagnetic case. With the charged current interaction mediated by the $W^+$ and $W^-$ bosons, the weak force allows further interactions constrained to a charge of $\pm 1e$. This leads to a larger number of possible interactions in the weak force than in the electrodynamic case.
Such processes in the leptonic cases are illustrated in Fig.~\ref{fig:w_flavorchange}.\\

The comparisons of the weak force to the electromagnetic were not chosen accidentally. As the Fig.~\ref{fig:strength_forces} already implies, both forces can be formulated combined in the \textit{electroweak force} (cf. Ref.~\cite{ref:pich}).

For the case of a charged weak interaction including quarks, the conservation of flavors is not directly fulfilled due to possible transformations like $d\rightarrow u+W^-$. Furthermore, the description of quarks in the context of weak interaction leads to an understanding that every quark can be expressed as a superposition of quarks of the same charges (including color). A numerical treatment of the probabilities that a quark flavor couples to another quark flavor is given by the Cabibbo-Kobayashi-Maskawa-matrix (CKM-matrix):\\

\begin{equation}
\begin{pmatrix}
V_{ud} & V_{us} & V_{ub} \\
V_{cd} & V_{cs} & V_{cb} \\
V_{td} & V_{ts} & V_{tb}
\end{pmatrix}
=
\begin{pmatrix}
0.9705~to~0.9770 & 0.21~to~0.24 & 0.~to~0.014 \\
0.21~to~0.24 & 0.971~to~0.973 & 0.036~to~0.070 \\
0.~to~0.024 & 0.036~to~0.069 & 0.997~to~0.999
\end{pmatrix}
\end{equation}

If the CKM-matrix would be a unitary matrix, then this would allow a conservation in the sum of the flavors of a generation, e.g. upness + downness. This construction would be comparable to the leptonic numbers. But since this is not the case, decays as $\Lambda\rightarrow p^+ + \pi^-$ or $\Omega^-\rightarrow \Lambda + K^-$ are possible. Beside pure partonic (between color-charged particles) or leptonic interaction, in a weak interaction, an exchange boson e.g. emitted by a quark can interact with a lepton as in the electromagnetic case. Even interactions between weak bosons are possible.

\subsubsection{Strong interaction}
\begin{figure}[htbp]
	\centering
	\includegraphics[width=0.5\textwidth]{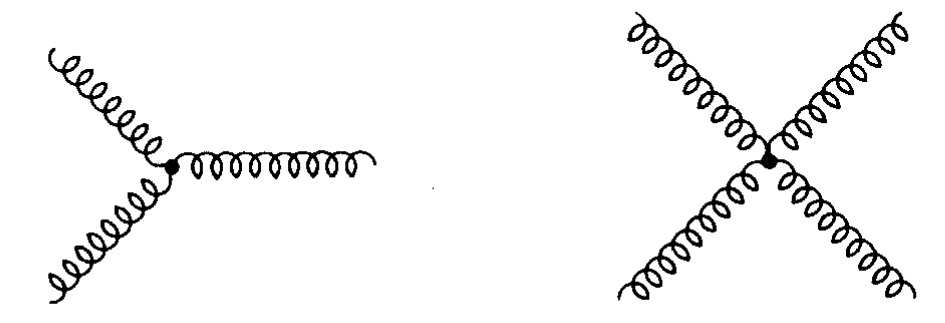}
	\caption{Feynman diagrams of vertices with three and four gluons.}
	\label{fig:loops}
\end{figure}

The strong force is described by quantum chromodynamics (QCD). The exchange particle is the gluon $g$. This force only occurs between colored objects. Gluons carry a color and an anti-color charge and are therefore able to change the color of a quark but not its flavor. Also, in strong interaction, the color is conserved. Due to the fact that gluons carry those colors they are able to couple with other gluons. This results in vertices with three and four gluons (cf. Fig.~\ref{fig:loops}).

\begin{figure}[htbp]
	\centering
	\includegraphics[width=0.8\textwidth]{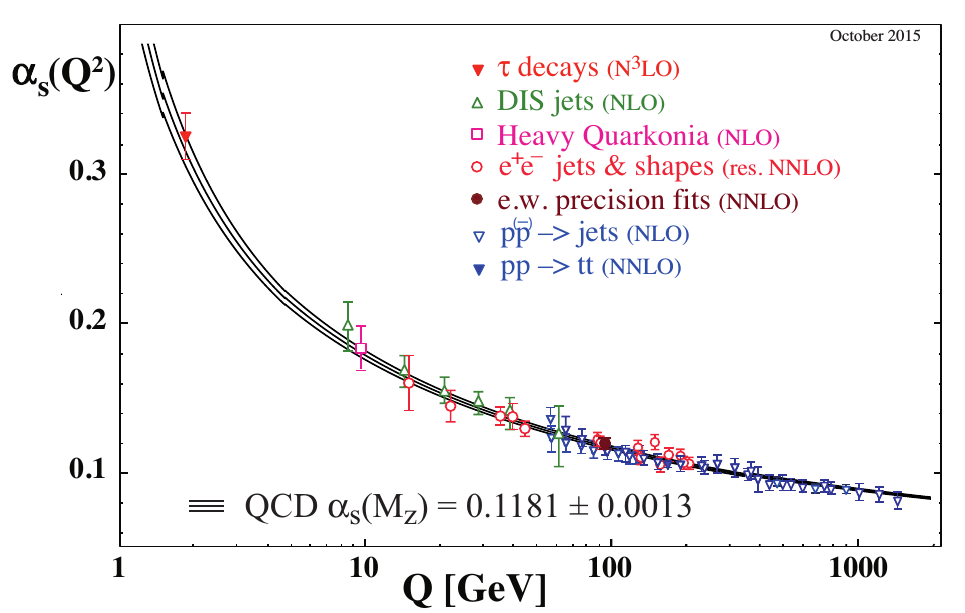}
	\caption{Dependence of the strong coupling constant $\alpha_s$ of the energy scale $Q$. At higher $Q$ values, the coupling strength decreases further, so that in the limit a colored particle is \textit{asymptotically free}. The picture is taken from Ref.~\cite{ref:alphas_energy}.}
	\label{fig:alphas_energy}
\end{figure}

An important difference between the electroweak and the strong force is the coupling constant. While $\alpha_{EM}$ can be treated as almost a constant, the strong coupling $\alpha_s$ is not as shown in Fig.~\ref{fig:strength_forces}. 
Known as the \textit{asymptotic freedom}, the coupling between the strong interacting particles strongly depends on their energy scale as shown in Fig.~\ref{fig:alphas_energy}. An alternative graph could be drawn that shows the dependence of the coupling strength as a function of the distance between the quarks. The relation between energy scale $Q$ and distance $r$ is given by $Q(r)\propto r^{-1}$. In that case, the asymptotic freedom would be reached at smaller distances, the coupling would be stronger at bigger distances. At large distances between particles, the asymptotic freedom leads under the confinement to the creation of a quark-anti-quark-pair ($q\bar{q}$). The confinement states that colored particles cannot be isolated. Following that rule, hadrons are always color-neutral. This property needs to be fulfilled by the combination of all quarks + gluons (collectively called \textit{partons}) in a hadron.

\subsubsection{Summary of the SM}
As a summary, the main properties of the mediators are presented in Tab.~\ref{tab:mediator_numbers}.

\begin{table}[htbp]
  \centering
  \begin{tabular}{ c | c | c | c | c}
    \hline
    mediator & $m~[eV]$ & $q~[e]$ & spin & color charge \\ \hline
    $\gamma$ & $<10^{-18}$ & $<10^{-35}$ & 1 & X \rule{0pt}{2.6ex} \\
    $Z^0$ & $(91.1876\pm0.0021)\cdot 10^9$ & 0 & 1 & X \\
    $W^+$ & $(80.385\pm0.015)\cdot 10^9$ & 1 & 1 & X \\
    $W^-$ & $(80.385\pm0.015)\cdot 10^9$ & -1 & 1 & X \\
    $g$ & <$\mathcal{O}(MeV)$ & 0 & 1 & $\surd$ \\
    \hline
  \end{tabular}
  \caption{Overview over masses $m$, the electrical charges $q$, the spin and the color charge of the mediators in the SM. The information is taken from Ref.~\cite{ref:pdg_mediators}.}
  \label{tab:mediator_numbers}
\end{table}

\begin{figure}
	\centering
	\includegraphics[width=0.7\textwidth]{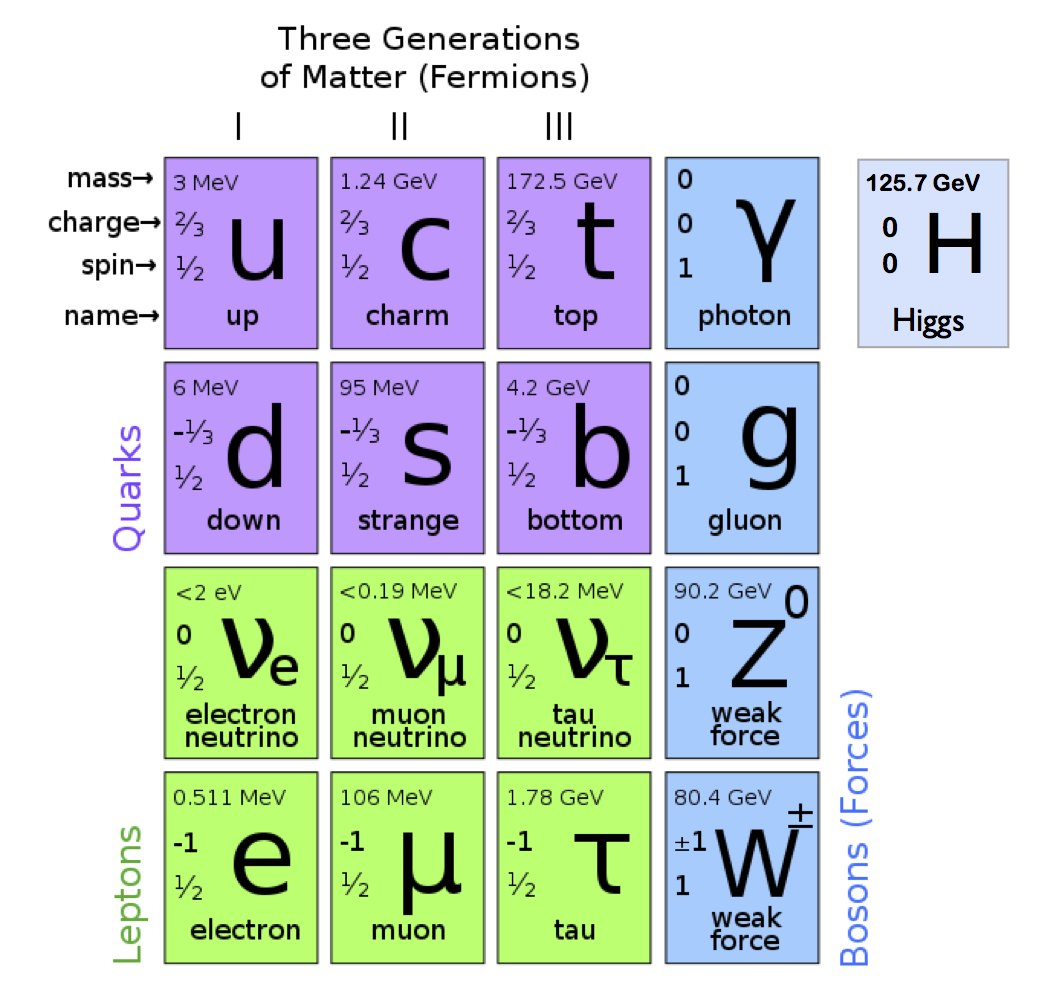}
	\caption{Summary of the particles of the SM without anti-particles. The quarks are shown in violet, the leptons in green, the bosons in blue and the Higgs boson in gray. Quarks and leptons are divided in generations column wise. Furthermore, the most important properties are displayed for every particle. The picture is taken from Ref.~\cite{ref:sm_bild}.}
	\label{fig:sm}
\end{figure}

\begin{figure}
	\centering
	\includegraphics[width=0.7\textwidth]{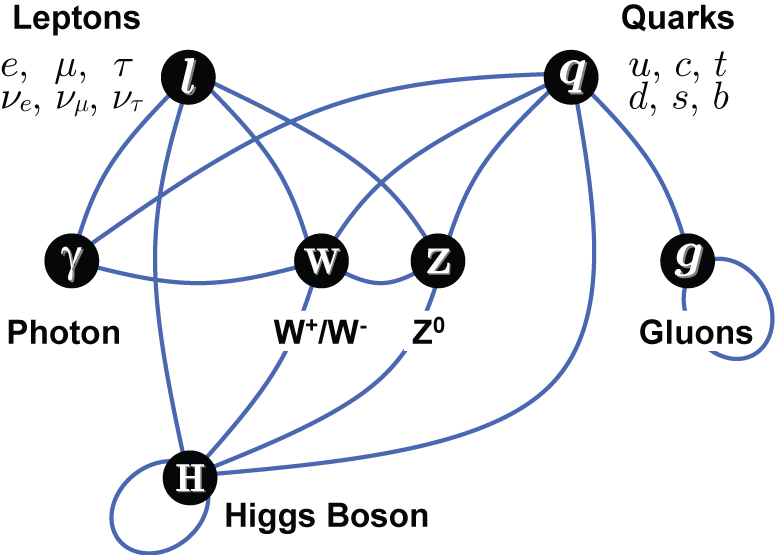}
	\caption{Overview over the particles in the SM as the black dots. The blue lines represent the possible interactions between the constituents. The graph is taken from Ref.~\cite{ref:sm_interactions}.}
	\label{fig:sm_interaction}
\end{figure}

Summing up all particles, this results in a total number of 12 (6 leptons + 6 anti-leptons) + 36 ((6 quark + 6 anti-quarks) $\cdot$ 3 color charges) + 12 (1 electromagnetic + 3 weak + 8 strong mediators) + 1 (Higgs boson) = 61 particles that form the SM. These are shown in the summary Fig.~\ref{fig:sm}. Their interactions are summarized in Fig.~\ref{fig:sm_interaction}.

\section{The evolution from initial state to final state hadrons}
\label{sec:shower_evolution}
In the previous Section, the constituent particles of the SM were presented. In this Section, high energy physics (HEP) will be considered. The interactions will remain the same as before, but there are more particles and more possible processes to consider due to the high energy.

\begin{figure}
	\centering
	\includegraphics[width=0.8\textwidth]{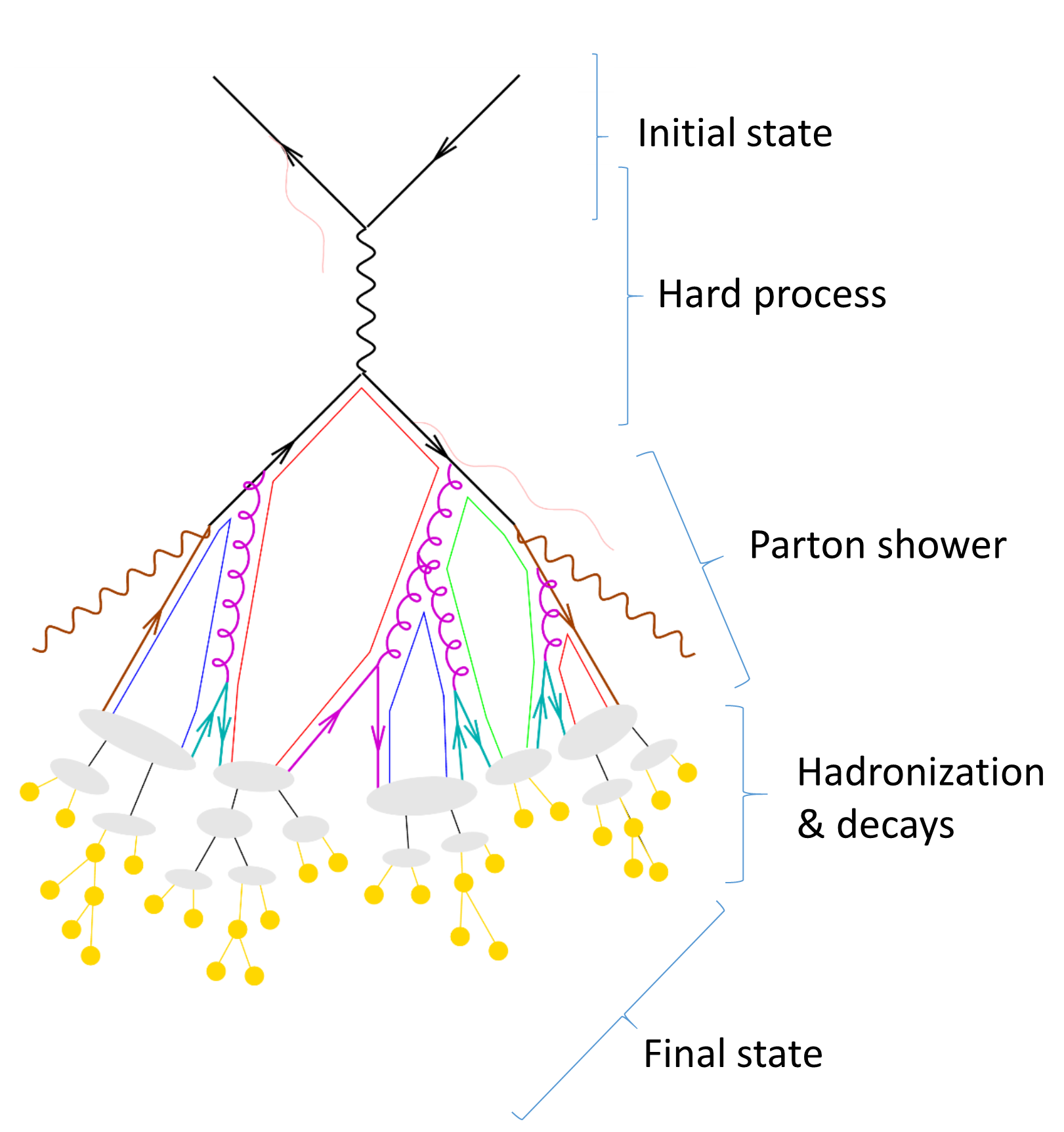}
	\caption{Illustration of the evolution of an initial state before a particle collision to a final state. The picture is taken from Ref.~\cite{ref:zeppenfeld}.}
	\label{fig:initial_to_final_state}
\end{figure}

In order to handle those more complex systems, the evolution from an initial state, e.g. in $e^+e^-$-collisions with fixed center-of-mass energy $\sqrt{s}$, to observable hadrons, leptons and photons is divided in three parts. The first one is the hard process. Here, the particle(s) of interest can be created together with additional partons e.g. in the process $e^+e^-\rightarrow Z^0/\gamma^* \rightarrow q\bar{q}$. In the second step, the outgoing particles radiate further particles and thus create a so-called parton shower. During this period, their energy is reduced until the third stage is reached. This is the phase of hadronization. Since the average energy of the partons gets reduced due to the radiation, this phase treats the confinement of the partons in contrast to the previous phase. At that point, hadrons are created, fragmented and decayed until they are registered in the detector. An overview over the evolution is shown in Fig.~\ref{fig:initial_to_final_state}.\\

In reality, these phases are not distinguished but it is a simplification so that it can be treated better in simulation programs, the so-called MC generators. Since it is the working principle of the generators, in the following the three phases will be described in chronological order in more detail. The content of this Section, if not stated otherwise, is extracted from Ref.~\cite{ref:pskands}.

\subsection{Hard processes}
The first step is the calculation of the matrix element for the transition of the initial state to the final state that will be handed over to the following shower phase. This calculation can be done at fixed order in $\alpha_s$. But since effects like self coupling gluons and $q\bar{q}$-loops are possible inside e.g. a $pp$-collision, the inner structure of a proton would be needed to be known in order to be able to calculate a cross-section. At that point the so called \textit{factorization theorem} becomes important. This states for a parton $i$ with the momentum $\vec{p}_i$ of a hadron with a momentum $\vec{p}_h$ and the relation $\vec{p}_i = x_i \vec{p}_h$ the separation of the parton density function (PDF) $f$ from the actual cross-section $\sigma$ of the process in a calculation. For a lepton-hadron-collision, this delivers the form

\begin{equation}
\sigma_{lh}=\sum_i \int_0^1 dx_i \int d\Phi_f f_{i,h}(x_i,\mu_F^2)\frac{d \hat{\sigma}_{li\rightarrow f}(x_i, \Phi_f, \mu_F^2)}{dx_i d\Phi_f}
\label{eq:factorization_theorem_lh}
\end{equation}

for a cross-section $\sigma_{lh}$. The variable $i$ runs over all incoming partons, $f$ states the number of final state partons.$\Phi_f$ represents the Lorentz-invariant phase space, the PDF is represented by $f_{i,h}$, $\hat{\sigma}_{li\rightarrow f}$ is the partonic cross-section for the actual process of an interaction of a lepton $l$ and the parton $i$ to form a final state with $f$ partons. $\mu_F$ is the \textit{factorization scale} and serves as dividing term between the PDF and the partonic cross-section term. While the PDF needs to be measured, $d\hat{\sigma}$ can be calculated within perturbation theory. For hadron-hadron-collisions, this would deliver a similar form (cf. Ref.~\cite[p.20]{ref:pskands}).\\

Since measurements and particle properties are related to observables, the cross-section for a certain observable will be considered. The according formula for the matrix element is given by

\begin{equation}
\frac{d\sigma_F}{d\mathcal{O}} \bigg\rvert_{\text{matrix element}} = \underbrace{\sum_{k=0}^\infty \int d\Phi_{F+k}}_{\sum legs} \bigg\lvert \underbrace{\sum_{l=0}^\infty M_{F+k}^{(l)}}_{\sum loops}\bigg\rvert^2 \delta\left(\mathcal{O}-\mathcal{O}\left(\Phi_{F+k}\right)\right)
\label{eq:matrixelement}
\end{equation}

for the differential cross-section $\frac{d\sigma_F}{d\mathcal{O}}$, a final state $F$ and an observable $\mathcal{O}$. The amplitude for the production of $F$ together with $k$ additional partons, so called \textit{legs}, and $l$ additional loops is denoted as $M^{(l)}_{F+k}$. The integration is performed over the phase space $\Phi_{F+k}$. The $\delta$-function leads to the evaluation of $\mathcal{O}$ on the momentum configuration $\Phi_{F+k}$. This is denoted $\mathcal{O}(\Phi_{F+k})$\cite[p.~24]{ref:pskands}.

The variables $k$ and $l$ can be used for order classification as shown in Tab.~\ref{tab:order_jets}. Limiting the nested sums in Eq.~\ref{eq:matrixelement} leads to a fixed-order truncated calculation in the perturbative QCD (pQCD). Applying pQCD is a legitimate tool for this problem, since $\alpha_s$ is considered small (e.g. $\alpha_s(M_Z) = 0.1181(11)$\cite{ref:alphas_pdg}) in the hard process.

\begin{table}
  \centering
  \begin{tabular}{ | c c | c |  }
    \hline
    $k$ & $l$ & perturbative order and amount of jets \\ \hline
    0 & 0 & LO for F production \\
    n & 0 & LO for F+ n jets\\
    $k+l\leq n$ && N$^n$LO for F,N$^{n-1}$LO for F+1 jet ...\\
    \hline
  \end{tabular}
  \caption{Name giving scheme for a number of legs and loops. The former increases the number of outgoing jets, the latter the order of calculation.}
  \label{tab:order_jets}
\end{table}

The problem arising from a simple integration of Eq.~\ref{eq:matrixelement} over the phase space $d\Phi_{F+k}$, is the so called \textit{infrared divergence} due to up to $k$ collinear or soft partons. Hence, these must be regulated by cuts in the angles, energies etc. The cut is set in a way, that the relation $\sigma_{k+1}\ll \sigma_k$ holds\cite[p.~25]{ref:pskands}. This allows the truncation of Eq.~\ref{eq:matrixelement}.

\begin{figure}
	\centering
	\includegraphics[width=\textwidth]{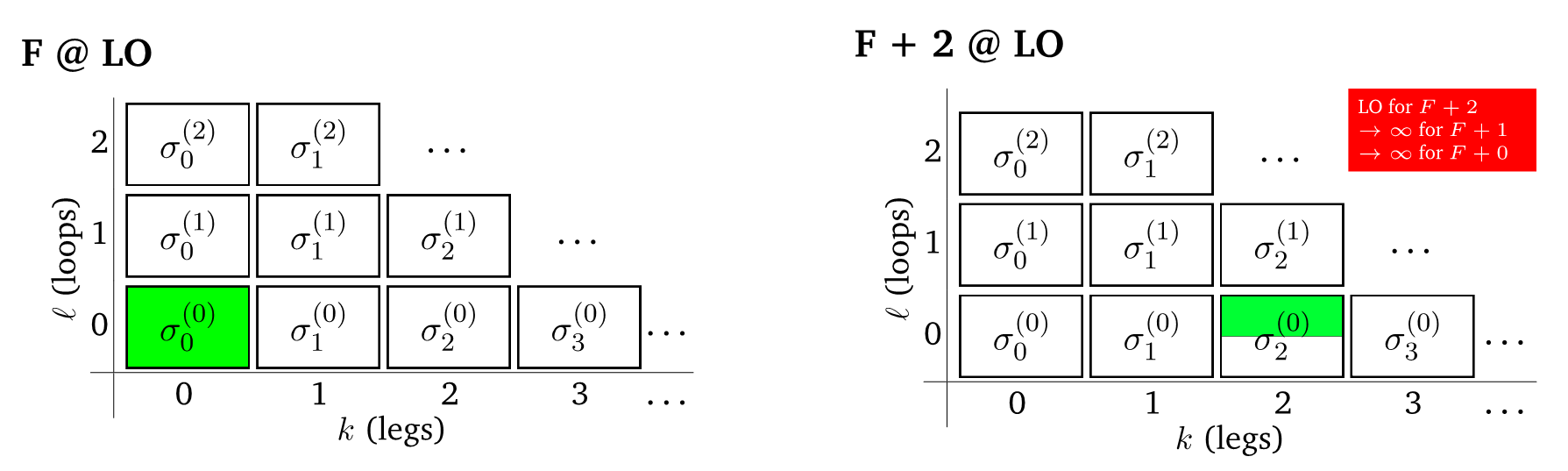}
	\caption{Coefficients used for a perturbative calculation in LO. The left picture shows the production of $F$ final state partons, the right $F+2$ partons. The half shaded box implies the restriction, that exactly two jets are resolved. In other words, this implies the need for a phase-space cut.}
	\label{fig:lo}
\end{figure}

In practice, the divergences are restrained by the KLN theorem\cite{ref:kinoshita, ref:klnrest} using multiple matrix elements that cancel divergent parts out. Those parts can be used in the pQCD regime. While in leading order ($l=0$), the case $n=0$, that represents the Born-level cross-section, can be handled without further problems, more legs need a phase space cut to remain finite. This cut is meant to the phase space with exactly $k$ resolved jets. This is illustrated in Fig.~\ref{fig:lo}.\\ 

\begin{figure}
	\centering
	\includegraphics[width=\textwidth]{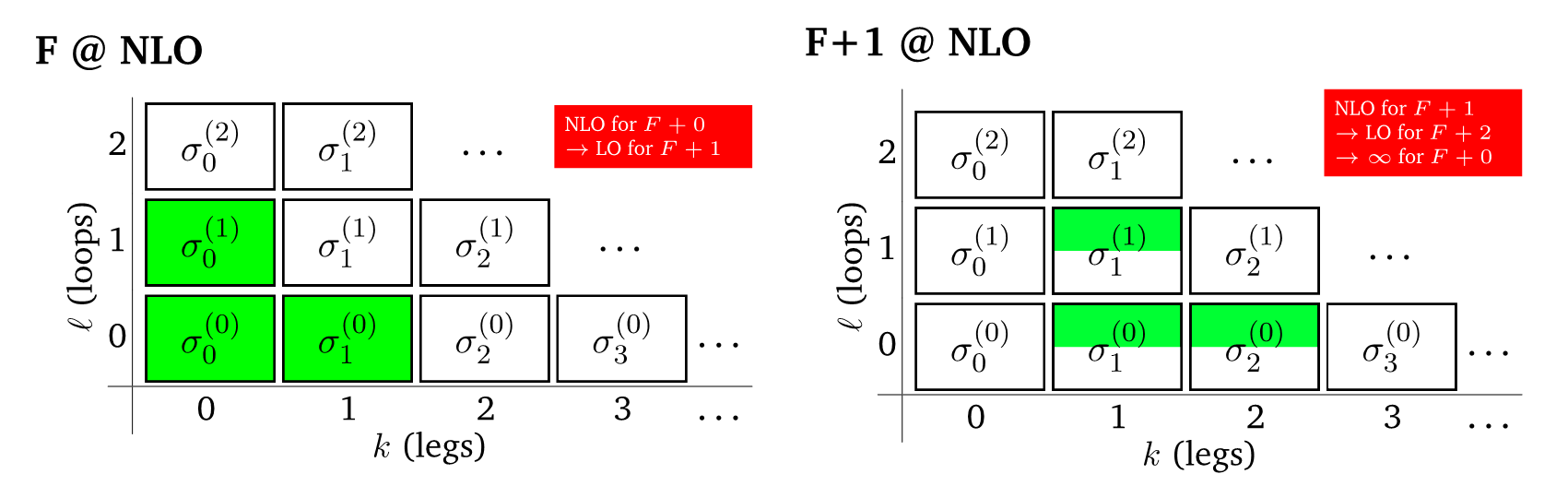}
	\caption{Terms used for a perturbative calculation in NLO. The left picture shows the production of $F$ final state partons, the right $F+1$ partons. The half shaded box implies the restriction, that exactly one jets are resolved. In other words, this implies the need for a phase-space cut.}
	\label{fig:nlo}
\end{figure}

Going to the next-to-leading order (NLO, $l=1$), the respective element is not finite anymore. In order to handle this, more matrix elements are used (see Fig.~\ref{fig:nlo}). This leads to the formula

\begin{align}
\sigma_0^{NLO} & = \int d\Phi_0 \lvert M_0^{(0)} \rvert^2 && + \int d\Phi_1 \lvert M_1^{(0)} \rvert^2 && + \int d\Phi_0 2Re\left[ M_0^{(1)} M_0^{(0)*} \right] \\
\label{eq:sigmanlo}
& = \sigma_0^{(0)} && + \sigma_1^{(0)} && + \sigma_0^{(1)}
\end{align}

with $\sigma^{(l)}_{k}$. The divergence of the second and third term will cancel out, while the Born-level cross-section is finite. If $k\ne 0$, the cut mentioned above concerning the exact number of $k$ resolved jets still needs to be performed. Going further to next-to-next-to-leading order (NNLO), further matrix elements need to be taken into account in a cascade manner down to LO matrix elements. Since the relation $\sigma_{k+1}\ll \sigma_k$ needs to be fulfilled by the phase-space cuts, the parts with $k+1$ are only corrections in pQCD.

\subsection{Parton showers}
In this part, the evolution of the particles from the high energy final state of the hard process to a hadronic state will be presented. For the evaluation of a shower evolution one still needs to keep the restriction to small $\alpha_s$ values and to the relation $\sigma_{k+1}\ll \sigma_k$. This needs to be used in order to control soft/collinear divergences in the shower evolution. Furthermore, the shower leads to a reduction of the energy of the particles while radiating further particles. So the algorithm needs to hold from a high energy regime down to the hadronic scale, where increasing $\alpha_s$ values lead to breakdown of pQCD. In order to handle this problem, one can start with reconsidering at Eq.~\ref{eq:factorization_theorem_lh}. The PDF $f_{i,h}$ ``includes so-called resummations of perturbative corrections to all orders from the initial scale of order of the mass of the proton, up to the factorization scale, $\mu_F$''\cite[p.~35]{ref:pskands}. In order to calculate the differential cross-section as in Eq.~\ref{eq:matrixelement} one needs to take a look at the fixed-order calculations in $d\hat{\sigma}_{ij\rightarrow f}$. These calculations need to be concerned from the QCD up to the factorization scale. This will be performed in two steps. First of all an infinite amount of legs will be treated. In a second part, an infinite amount of loops will be considered. Using a combination will lead to a possibility to calculate the cross-section for $F+k$ jets.

\subsubsection{Infinite number of legs}
In order to calculate the cross-section for an infinite amount of legs, a possible approach relies on resummation techniques. Such an approach can start with considering two color-connected partons $J$ and $K$ in the state $F$. The squared amplitude of the emission of a parton is given by

\begin{equation}
\lvert M_{F+1}\rvert^2=\underbrace{g_s^2 N_C \left( \frac{2s_{ik}}{s_{ij}s_{jk}} + \text{collinear terms}\right)}_{\equiv Antenna Function}\lvert M_F\rvert^2
\label{eq:recursion}
\end{equation} 

with $g_s^2=4\pi\alpha_s$, the color factor $N_C$ and $s_{ij}$ the invariant between parton $i$ and $j$. The indices $i$ and $k$ represent the partons after the emission of $j$. With respect to the parameter $s_{ij}$, the possible infrared (IR) divergences in the soft or collinear limit can be extracted. The definition of this parameter delivers for massless partons the relation

\begin{align}
s_{ij} &\equiv 2p_ip_j \\
&=(p_i + p_j)^2 - m_i^2 - m_j^2 \\
&=2|\vec{p}_i||\vec{p_j}|(1-\cos(\theta))\\
&=2E_iE_j(1-\cos(\theta))
\label{eq:sij}
\end{align}

with the angle $\theta$ between $i$ and $j$. In the soft limit (|$\vec{p}_j|\rightarrow 0$) $s_{ij}\rightarrow 0$. The same result would be in the case of $\theta\rightarrow 0$. This formulation becomes more visible in another calculation. The parameters $s_{ij}$ and $s_{jk}$ are directly linked to the phase-space by the four-vectors, such that a connection between those can be formulated. This proceeds by calculating the differential of Eq.~\ref{eq:sij}:

\begin{equation}
\frac{ds_{ij}}{s_{ij}}\frac{ds_{jk}}{s_{jk}} \propto\frac{dE_j}{E_j}\frac{d\theta_{ij}}{\theta_{ij}}+\frac{dE_j}{E_j}\frac{d\theta_{jk}}{\theta_{jk}}
\label{eq:differential}
\end{equation}

\begin{figure}
	\centering
	\includegraphics[width=0.5\textwidth]{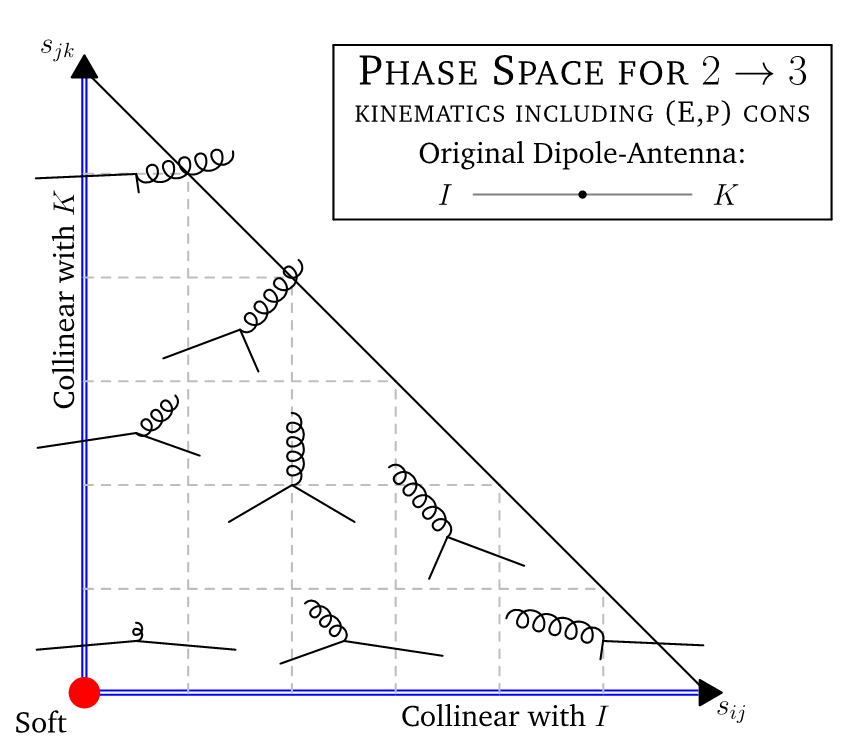}
	\caption{Visualization of the branching $q\bar{q}\rightarrow qg\bar{q}$ for different values of $s_{ij}$ and $s_{jk}$.}
	\label{fig:antenna}
\end{figure}

Both cases would produce a singularity in the amplitude due to $1/E_j$, $1/\theta_{ij}$ or $1/\theta_{jk}$ term. Besides the better visibility of the singularities, Eq.~\ref{eq:differential} shows the reason for the names \textit{soft} and \textit{collinear}. This is graphically shown in Fig.~\ref{fig:antenna}. Therefore this must be constrained in order to use Eq.~\ref{eq:recursion} as the amplitude for the calculation of Eq.~\ref{eq:matrixelement} since the integration is performed over the whole phase-space. For that purpose, a cutoff parameter $\mu_{IR}^2$ will be used as a minimum scale. This parameter is also called \textit{minimum perturbative cutoff scale}.\\

\begin{figure}
	\centering
	\includegraphics[width=\textwidth]{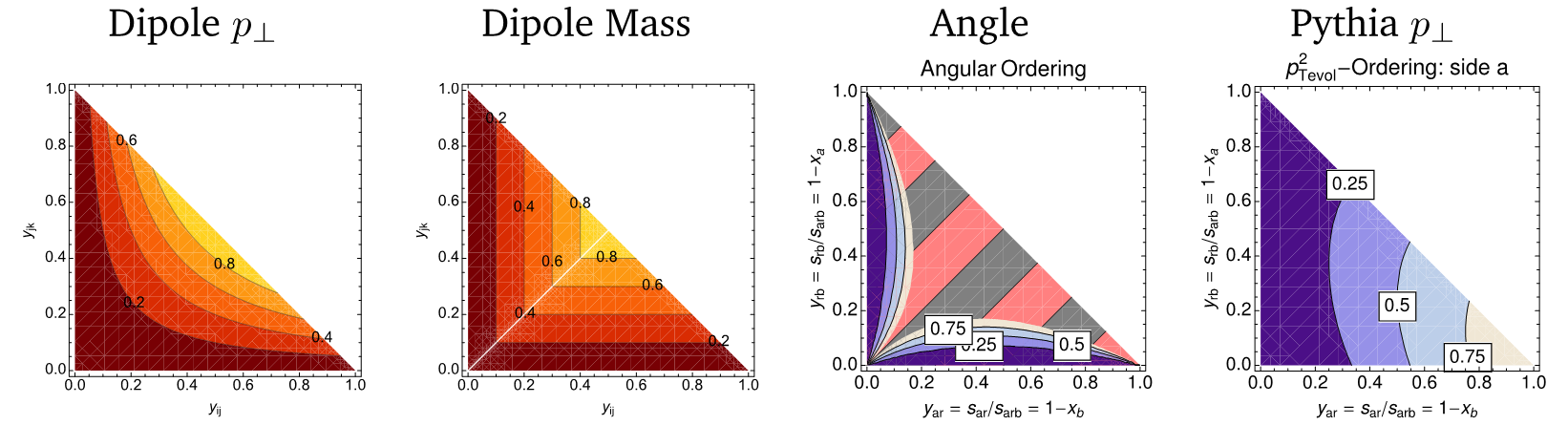}
	\caption{An overview over the most common shower evolution parameters used in MC generators. Variants of the dipole approach are implemented in \textsc{ARIADNE}, \textsc{SHERPA} and \textsc{VINCIA}. Angular ordering is used by \textsc{HERWIG 7}. The $p_T$ ordering is implemented in \textsc{PYTHIA 6} and \textsc{8}. The rightmost represents the evolution of only one parent.}
	\label{fig:ordering_variables}
\end{figure}

The Eq.~\ref{eq:differential} additionallly allows an ordering of the shower. A collection of typically used ordering parameters are shown in Fig.~\ref{fig:ordering_variables}. Which ordering is used depends on the MC generator. During the shower evolution, the algorithm will integrate over the whole ordering parameter over and over again. The choice of the ordering parameter empowers or weakens certain radiations. This affects the virtual corrections via the so-called \textit{Sudakov factor}\cite{ref:sudakov}. In the end, Eq.~\ref{eq:recursion} delivers a powerful recursion formula that is able to calculate a parton shower with arbitrarily many particles until the hadronization. Note that all those calculations are only performed in leading order.\\

The recursion formula of the amplitude $|M_{F+n}|^2$ needs to be integrated over the (cut) phase-space in order to get the respective cross-section as shown in Eq.~\ref{eq:sigmanlo}. Performing this calculation delivers a result with the algebraic structure

\begin{equation}
\sigma_{F+n}^{0}=\alpha_s^n(\ln^{2n}+\ln^{2n-1}+\ln^{2n-2}+...+\ln+\mathcal{F})
\end{equation}

with the form $\ln^\lambda$ that denotes a \textit{transcendentality} $\lambda$. The function $\mathcal{F}$ denotes a rational function with $\lambda=0$. This series is usually cut at a certain term in order to speed up the computation. The easiest cut includes only the $\ln^{2n}$ term and is called the \textit{double logarithmic approximation} (DLA). Including additionally the $\ln^{2n-1}$ term is called the \textit{leading-logarithmic} (LL). In order to improve the calculations with the underlying series cut, further calculations are performed ``such as explicit momentum conservation, gluon polarization and other spin-correlation effects, higher-order coherence effects, renormalization scale choices, finite-width effects, etc''\cite[p.~39]{ref:pskands}. Using more terms are called next-to-leading-log (NLL) etc. One needs to keep in mind that all those are approximations of the real solution and that a scale cut at $\mu_{IR}$ still needs to be performed. In order to get rid of the scale cut, a similar approach as done in this Subsection will be performed but for infinite loops in the next Subsection.

\subsubsection{Infinite number of loops}
The underlying idea of this Subsection is the same as in the KLN theorem for the NLO calculation. With the goal of an integration over the full phase-space, the terms with $l>0$ will be taken into account such that the singularities will cancel out. For a $l=1$ correction, this results in a contribution in the form of

\begin{equation}
2Re\left[M_F^{(0)}M_F^{(1)*}\right] \supset -g_s^2N_C\lvert M_F^{(0)}\rvert^2\int \frac{ds_{ij}ds_{jk}}{16\pi^2s_{ijk}}\left(\frac{2s_{ik}}{s_{ij}s_{jk}}+\text{less singular terms}\right)\; .
\label{eq:infiniteloops_recursion}
\end{equation}

\begin{figure}
	\centering
	\includegraphics[width=\textwidth]{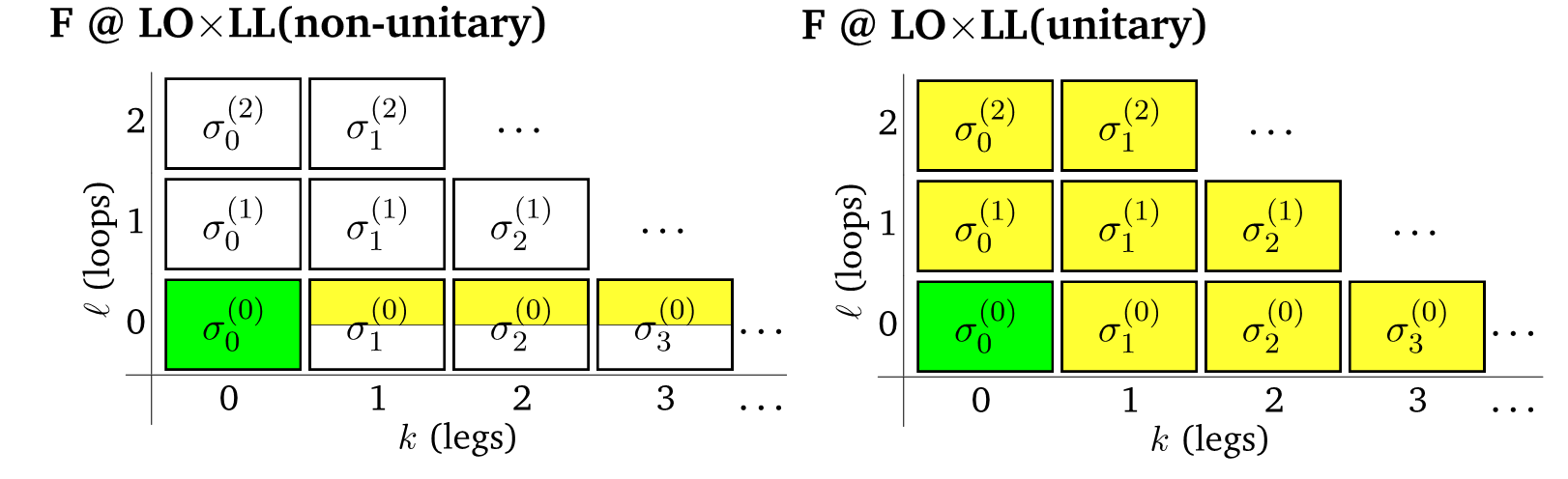}
	\caption{Coefficients used for LO+LL approximation in the parton shower. The green box indicates that no approximation is needed for the calculation. In contrast, the yellow boxes indicate a LL approximation. The filled boxes show an integration over the whole phase-space without the need of scale cuts / IR divergencies. In the half filled boxes, the cuts are needed. The left picture shows the infinite legs used in LO only, the right shows a KLN theorem like approach for the calculation of a $F+n$ final state including infinite loops.}
	\label{fig:shower_pqcd}
\end{figure}

The right-hand side is equivalent to the term of $\sigma_0^{(1)}$ in Eq.~\ref{eq:sigmanlo}. This expression is a part of the term $2Re\left[M_F^{(0)}M_F^{(1)*}\right]$. Eq.~\ref{eq:infiniteloops_recursion} cancels the singularity from Eq.~\ref{eq:recursion}. But since the calculation was performed for a series of leading-order terms for a $F+n$ parton shower and each term contains the singularities that need to be cut, that type of correction needs to be performed for every term. This is the actual \textit{resummation} calculation. In a graphical illustration corresponding to Fig.~\ref{fig:lo}, this can be shown as in Fig.~\ref{fig:shower_pqcd}. While the recursion of Eq.~\ref{eq:recursion} implies a horizontal walk over a fixed $l$ value, the Eq.~\ref{eq:infiniteloops_recursion} represents a diagonal movement along a fixed $n=k+l$.

\subsection{Hadronization and fragmentation}
During the parton shower, the energy of the partons is reduced due to further radiated partons. Therefore, the next phase treats the confinement of the partons. During that phase, the partons will build color-neutral hadrons. Those hadrons will further decay until a collection of stable hadrons is reached. In MC generators, the fragmentation of hadrons is described by a model.

A famous model for hadron fragmentation is called \textit{Lund-string-model}. This model is implemented in the \textsc{Pythia} MC generator. In the following section, the main parts of the model will be described.\\
\medskip

The Lund-string-model describes the hadronization and fragmentation. This model simplifies the hadronization by connecting partons in the same phase space via ``strings''. In order to describe the fragmentation, these strings are allowed to ``break''. A breaking string creates a new $q\bar{q}$-pair.

\begin{figure}
	\centering
	\includegraphics[width=0.7\textwidth]{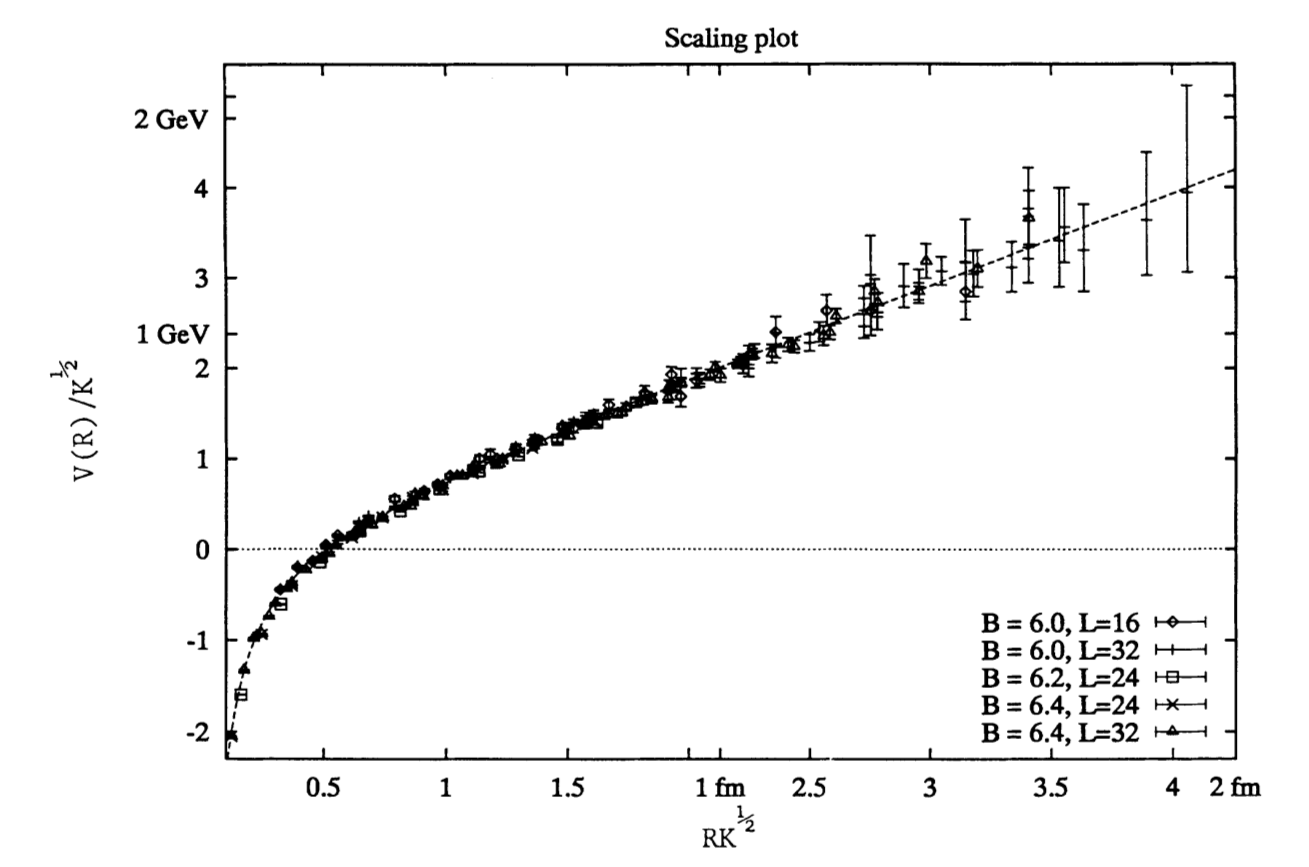}
	\caption{Potential of a $q\bar{q}$-pair as function of the distance between the quarks. The picture is taken from \cite{ref:superposition_confinement}.}
	\label{fig:qqbar_potential}
\end{figure}

Firstly, two partons will be considered that are connected by a string. If both partons will be moved further apart, the string tension will rise. Analogously to classical mechanics, a potential $V$ can be formulated as

\begin{equation}
V(R)=\kappa R
\end{equation}

with the distance $R$ between the partons and the string tension $\kappa$. As the value of $\kappa$, one can use values around $0.9~GeV/fm$. If on the other hand the partons will be moved closer together, one needs to take Coulomb interactions with a $1/R$ dependency into account. The superposition of these two potentials is shown in Fig.~\ref{fig:qqbar_potential} for a $q\bar{q}$-pair. In fact, the Coulomb part is neglected in the model itself.\\

\begin{figure}
	\centering
	\includegraphics[width=\textwidth]{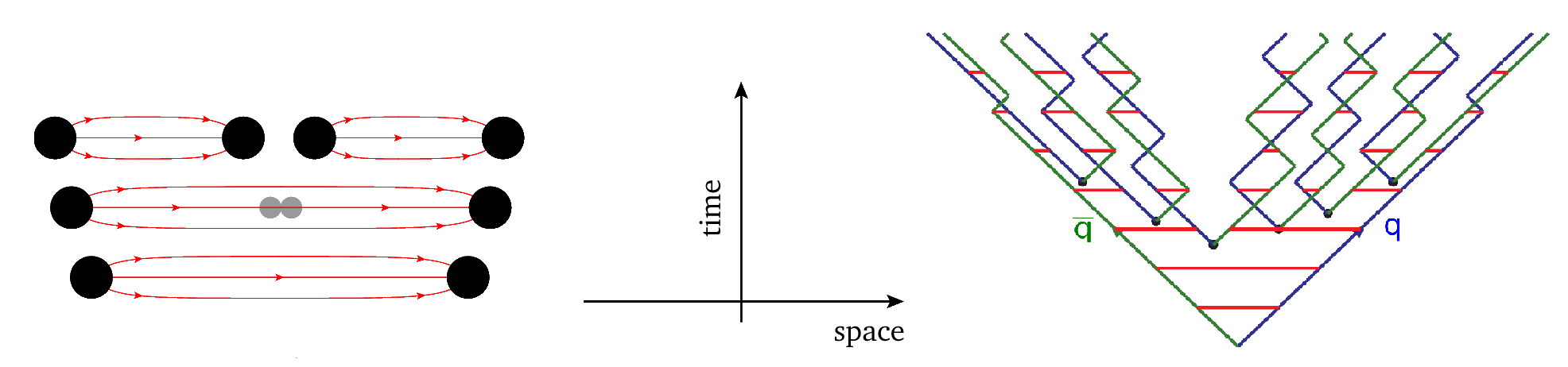}
	\caption{Time dependent evolution of a $q\bar{q}$ production during a string breaking (left). The right graph shows the evolution over a longer time period including further fragmentations. The right picture is taken from Ref.~\cite{ref:durham}.}
	\label{fig:string_breaking}
\end{figure}

As the potential rises for bigger distances between the partons, the string \textit{breaks} at a certain point. In the string breaking process, a new $q\bar{q}$-pair will be created as shown on the left-hand side in Fig.~\ref{fig:string_breaking}. If the distance between both partons of a newly produced hadron will also increase, the breaking will be repeated. This is shown on the right-hand side in Fig.~\ref{fig:string_breaking}.\\

Since the model does not provide an individual handling of the string breaking from first principles, the model explains the phenomenological treatment by a quantum tunneling. The creation of a $q\bar{q}$-pair is connected with a certain transverse momentum $p_T$ for both partons. Here, the transverse direction is defined as the direction of the original string. The probability density for the actual $p_T$-value of the partons is in a gaussian shape of the form

\begin{equation}
\mathcal{P}(m_q^2,p_{Tq}^2)\propto exp\left(-\frac{\pi m_q^2}{\kappa}\right)\left(-\frac{\pi p_{Tq}^2}{\kappa}\right)
\end{equation}

with the parton mass $m_q$. Note, that this is the transverse momentum for a single parton created in the string breaking. The second parton will get the negative value of the first one's transverse momentum. Furthermore, the actual value of the $p_T$ is flavor independent because of the factorization of the mass and $p_T$ and allows therefore a universal treatment of the string breaking in this model. The masses of the resulting hadrons after a string break are given by Breit-Wigner distributions.

An additional parameter that can be derived from this equation is the width of the gaussian functions given by

\begin{align}
Var[p_{Tq}] &= E[p_{Tq}^2] - E[p_{Tq}]^2 \\
&= E[p_{Tq}^2] \\
&= \kappa / \pi \approx (240~MeV)^2
\end{align}

with the mean $E$ of a distribution and the respective variance $Var$, later referred as $\sigma^2$.

Another important aspect of the Lund-string-model is the fragmentation function parametrization, called the \textit{Lund symmetric fragmentation function},

\begin{equation}
f(z)\propto \frac{1}{z}(1-z)^a exp\left(-\frac{b(m_h^2 + p_{Th}^2)}{z}\right)
\end{equation}

with the free parameters $a$ and $b$ and the fragmentation variable $z$, e.g. \mbox{$z=(E+p_z) / (E+p_z)_{total}$}\cite[p.~6]{ref:durham}. The mass $m_h$ and the transverse momentum $p_{Th}$ refer to the hadrons produced in the fragmenting string.

One important aspect of the Lund-string-model is the handling of baryons. Until now, only strings between two partons were considered. In the case of three partons, the model will be expanded by a \textit{diquark}, an object that is made out of two quarks. In that case, the string connects a quark with a diquark. Using this construction, the formalism mentioned before almost stays the same. In the case of the fragmentation function, the parameter $a$ will be modified by the addition of the parameter $a_{\textrm ExtraDiquark}$ in order to handle differences between mesons and baryons\cite{ref:pythia_manual}.

\chapter{Parameter-based tuning approach using the \textsc{Professor} framework}
\label{ch:alter_ansatz}
\setcounter{page}{23}

For measurements and searches in high energy physics experiments, good simulations of the underlying physics processes are of utmost importance. For that purpose, so-called Monte Carlo generators are used. These MC generators are used to make predictions of physics processes. The calculations performed by the MC generators are model based. However, the models depend on parameters which cannot be derived from first principles.

\begin{figure}[htbp]
\makebox[\textwidth][c]{
	\begin{adjustbox}{max width=1.\textwidth}
	\includegraphics[width=\textwidth]{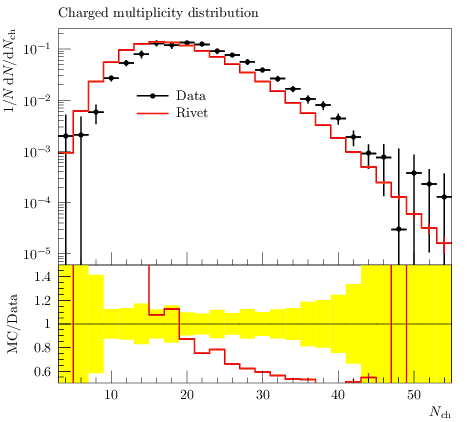}
	\includegraphics[width=\textwidth]{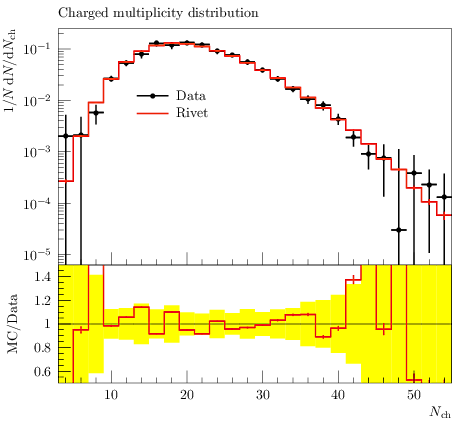} 
	\end{adjustbox}
}
	\caption{Illustration of the charged multiplicity distribution of a hadronic $Z^0$ decay in an $e^+e^-$ collision with a center of mass energy of $\sqrt{s}=m(Z^0)=91.2~GeV$. The black dots are the measurements performed by the ALEPH collaboration\cite{ref:aleph_1996} and provided by the \textsc{Rivet} framework\cite{ref:rivet}, the red line is simulated with \textsc{Pythia 8}\cite{ref:pythia6,ref:pythia8} based on two different sets of input parameters (cf. Tab.~\ref{tab:parameter_values_example}).}
	\label{fig:example_param_variation}
\end{figure}

To illustrate the dependence of MC generator predictions on the parameters, an example is shown in Fig. \ref{fig:example_param_variation}. This Figure shows a comparison of the charged particle multiplicity distribution of the process $e^+e^-\rightarrow Z^0\rightarrow q\bar{q}$ at a center-of-mass energy $\sqrt{s} = 91.2~GeV$ between the measurements performed by the ALEPH experiment\cite{ref:rivet_analysis_link} and a MC simulation using two different parameter sets. By visual inspection the graph on Fig.~\ref{fig:example_param_variation} (right) shows a better agreement of the MC prediction with the data compared to the graph on Fig.~\ref{fig:example_param_variation} (left).
The goal is to estimate the parameter configuration for the simulation such that the MC generator prediction fits the data best. The search procedure for those parameter values is called \textit{tuning} and has to be performed in a well defined way.

In this chapter, a tuning procedure will be described. This method is implemented in the \textsc{Professor} framework\cite{ref:professor}. Beside explaining the tuning method in Sec.~\ref{sec:tuning_approach}, the explicit implementation of such a tuning using \textsc{Professor} will be presented. For that purpose, the reproduction of a previously performed tuning from Ref.~\cite{ref:nadine_fischer} will be described in Sec.~\ref{sec:applying_tune}. As a last step, the tuning will be re-performed with a modified setup in Sec.~\ref{sec:double_peak}.


\section{Tuning approach}
\label{sec:tuning_approach}
Several different tuning approaches are available and used in HEP. In general, these approaches can be separated into three different types\cite{ref:professor}. The first type mentioned here is the \textit{manual tuning}. This approach is in general done by hand and requires an appropriate expertise. The tuning can become complex for a more comprehensive parameter set.

The second type is called \textit{brute force tuning}. Approaches in this category are constructed as a direct search. A direct search can be defined as a Markov Chain (MC), meaning the evaluation of the $n$-th step determines the $(n+1)$-th step through the parameter space. Such a point lies around the current point. An algorithm will then test the goodness of the proposal point and decide whether to change to that point or to stay at the current position. The benefit of such an approach is a minimum of needed assumptions in order to find a best parameter set in order to describe the measurements. On the other hand in this approach the MC generator needs to run for every new step resulting possibly in larger run-times. This can become especially complicated for more complex problems.

The third tuning type is the \textit{parametrization-based tuning}. This approach is the one that is used in this thesis. The basic idea of the model parameter estimation is the calculation of a function which describes the MC generator response depending on the model parameters that need to be tuned. Using this function, the parameters are chosen such that the description of the data by the MC predictions is optimal. For the purpose of calculating the needed function, many different configurations of model parameters are used and a function of the associated MC generator response is calculated. Since many configurations and therefore many simulations are needed, the calculation of such a response function is computing intensive. These calculations were complicated to realize formerly but are easier/possible nowadays (cf. Moore's law\cite{ref:moores_law}).
\medskip

\begin{figure}
	\centering
	\includegraphics[width=\textwidth]{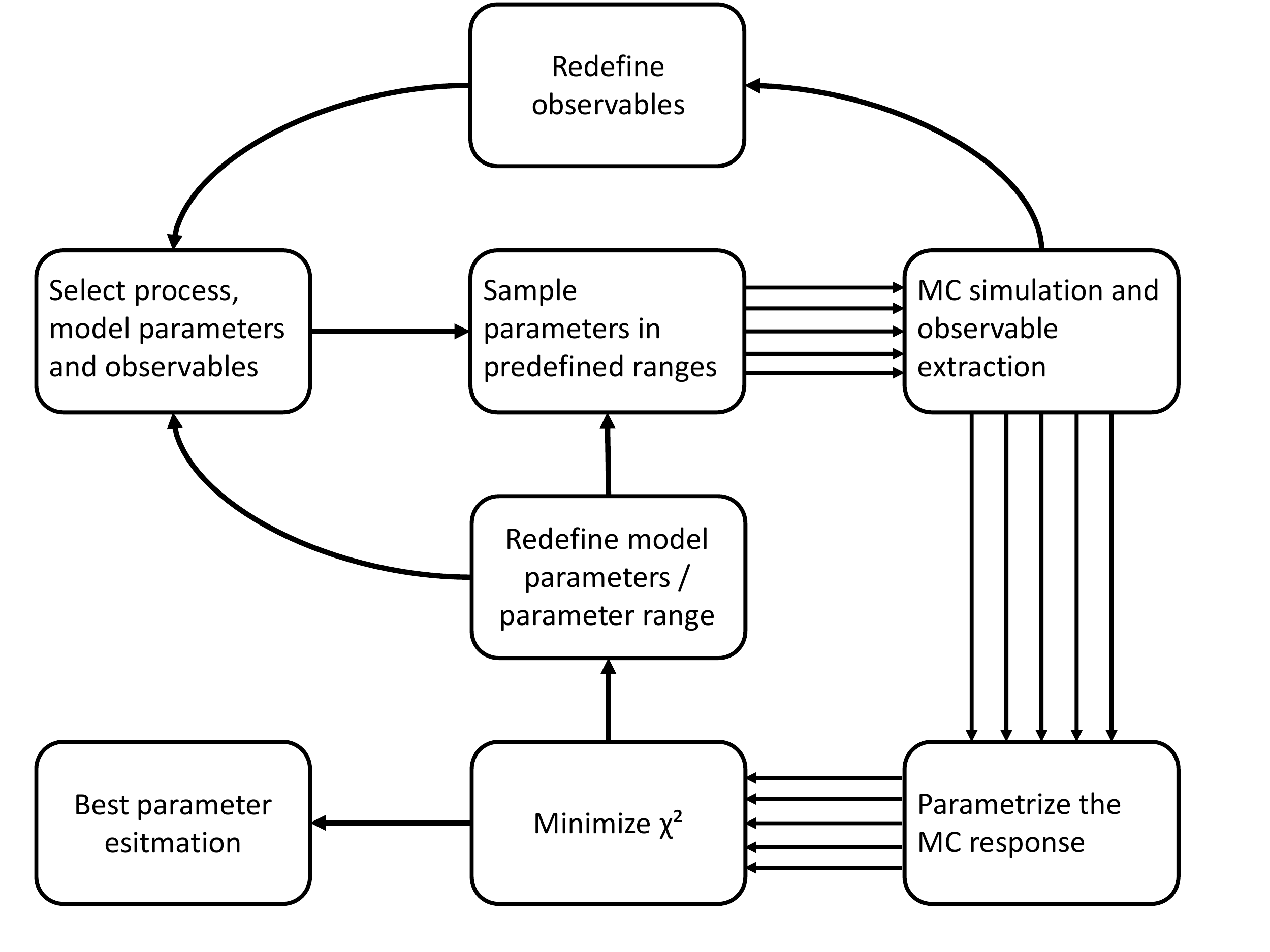}
	\caption{Procedure of a parametrization based MC generator tuning.}
	\label{fig:tuning_steps}
\end{figure}

An overview over the needed steps for this tuning approach is shown in Fig.~\ref{fig:tuning_steps}. Before a tune can be performed, the $n$ parameters of the MC generator that need to be tuned need to be specified. Secondly, each parameter needs a specific range. Those are restricted either by the meaningfulness in the context of physics or at least limited by the developers of the MC generator framework itself.

In order to tune those parameters, observables and corresponding reference data are needed. The choice of the observables depends upon the parameters chosen. The most important property in order to select them is the \textit{sensitivity}. This term describes the dependency of the value of an observable on the parameter value. A further selection of the observables depends on the MC generator and the prediction possibilities, e.g. whether it works in LO or NLO etc. After the selection is performed, the reference data are provided by the \textsc{Rivet}\cite{ref:rivet} framework.

The first tuning step consists of sampling randomized values for each parameter in its range. Thus, the results are $m$ parameter vectors $\vec{p_i}$ with $i\in [1,m]$ and $dim(\vec{p_i}) = n$. The set of all those vectors will be called $S = \{\vec{p_i}\}_{i=1,..,m}$.

In the second tuning step, those vectors are used as the configuration of the MC generator. Every single configuration produces an output file containing all processes and properties of the particles in every simulated event. Such an output file can have a size of $\mathcal{O}$(GB) per 100,000 events. The large amount of storage can limit the number of events calculated in a single configuration or the number of configurations $m$.

In the third tuning step, a collection of observables from one or several experiments can be extracted for every configuration out of the simulated events. The calculated observables from the simulations allow a direct comparison with the reference data. Using only the sampled parameter sets for a comparison would provide the best sampled parameter vector but this vector does not have to be the best possible parameter vector.

\begin{figure}
	\centering
	\includegraphics[width=\textwidth]{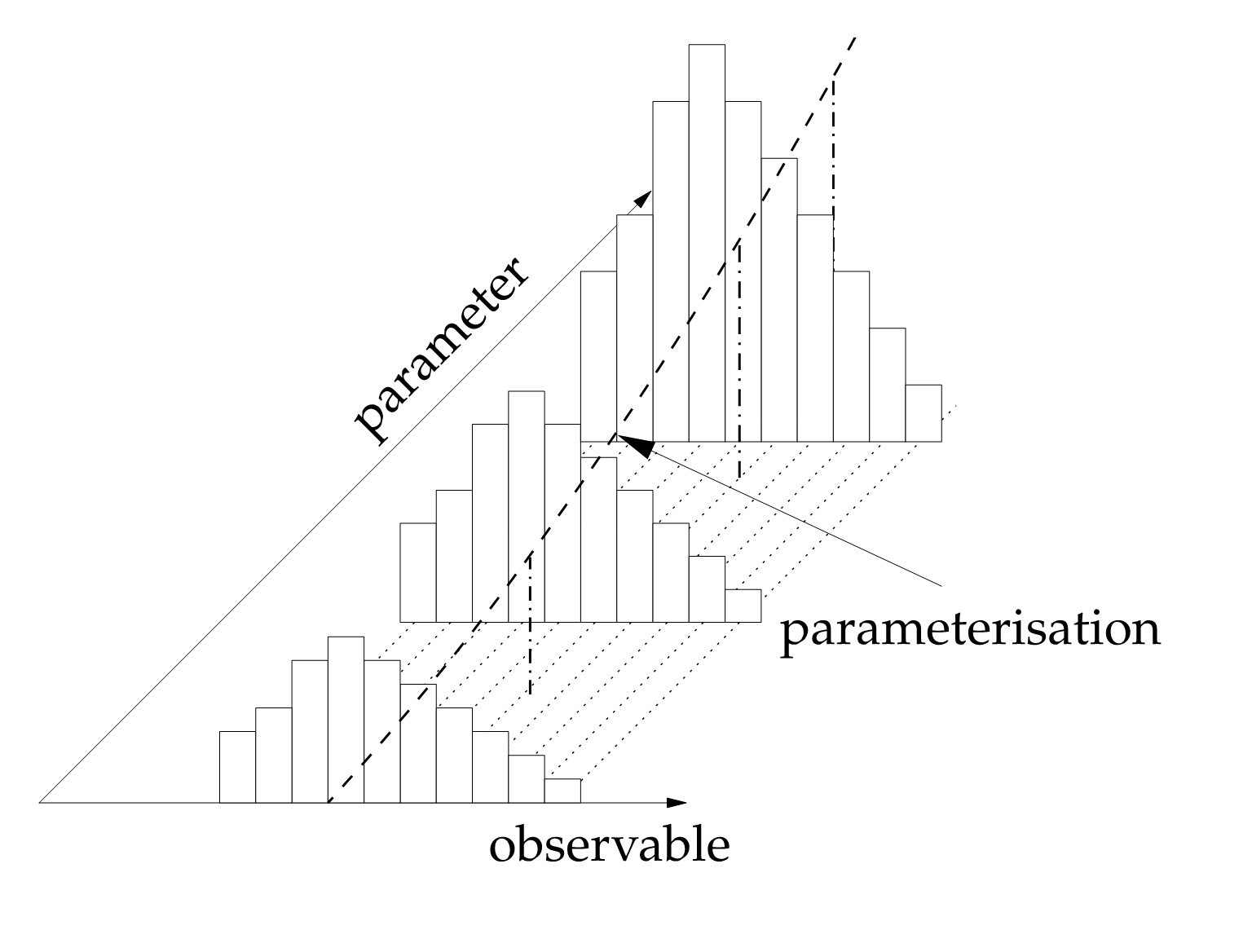}
	\caption{Illustration of the impact of parameter values on an observable. This leads to a parametrization of the data points that only depends on the parameter values themselves. The sketch is taken from Ref.~\cite{ref:holger_schulz}.}
	\label{fig:parameter_changing}
\end{figure}

To avoid this, the MC response will be parametrized in the fourth step. The assumption that the variation per data point is sufficiently smooth while changing the parameter values (shown in Fig.~\ref{fig:parameter_changing}) leads to the possibility of an interpolation approach between all sampled parameter sets with a smooth function. Furthermore, for the approach used in this thesis, all data points will be fitted individually. In this way, the total number of obtained interpolation functions is equal to the total number of data points.

In order to get a good description by a function of the MC response inside the parameter space, the sample-density needs to be sufficiently high. Under the assumption such a sufficient amount of sample points $m$ in one dimension and within a given parameter range exists, a $n$-dimensional tuning process would need $\mathcal{O}(m^n)$ samples. Therefore the number of samples with respect to the memory consumption and computing power needed in order to receive such an parametrization of the MC response is limited in terms of $n$. This leads to a limitation of the model parameters that can be optimized in this tuning process.

In the last step, all interpolation functions should be optimized simultaneously in order to find the smallest difference between the functions and the measurements and thus the best $\vec{p}$ for reproduction of the measurements using the MC simulation.\\

For the purpose of sampling the anchor points $\vec{p_i}$ as well as for parametrization and minimization, the \textsc{Professor 2.1.4} framework will be used. The working principle of the framework will be described in the Subsection \ref{subsec:professor}. The MC simulation is provided by the \textsc{Pythia 8}\cite{ref:pythia6,ref:pythia8} framework. The output file of \textsc{Pythia 8} is a complete record of all events. This output is used as the input for the \textsc{Rivet~2.4.2} framework. This framework extracts physical observables out of the output files from MC generators. Furthermore, \textsc{Rivet} provides the measured data for comparison with the MC generator response. Those two frameworks will be presented in the Subsection \ref{subsec:simulating}.

\subsection{The Professor 2.1.4 framework}
\label{subsec:professor}
The Professor framework in general is used for systematic parametrization-based tuning. For that purpose, the framework provides \textsc{Python}-scripts and \textsc{C++}-classes. In the following, the most important scripts for the intended tuning will be presented. Those will be described based on the chronological order as displayed in Fig.~\ref{fig:tuning_steps}. This will be only a brief overview over the possibilities that the framework provides, but since the content in this thesis needs comparability, some configurations need to remain unchanged and are therefore neither used in this thesis nor presented here. Note that this description considers \textsc{Professor} in the version 2.1.4. If not mentioned otherwise, the following explanations are based on Ref.~\cite{ref:professor} and the source code of \textsc{Professor 2.1.4}\cite{ref:professor_download} itself.\\

\subsubsection{\textsc{prof2-sample}}
This script serves as a generator of random vectors $\vec{p}_i$. The sampling is based on the \textsc{Python} standard library \textsc{random}\cite{ref:python_random}. The pseudo-random sampling algorithm is based on Ref.~\cite{ref:wichmann_hill} and it samples uniformly in a predefined range. A benefit of using \textsc{Professor} in the version 2.x is, that the prof2-sample script directly writes the sampled vectors into the configuration files for a MC generator. So, after sampling, the MC generator can be run directly with these configuration files.

\subsubsection{\textsc{prof2-ipol}}
This script is the heart of the Professor framework. The script calculates a fit function in order to parametrize the MC response for the given set $S$. Since it is assumed that the parameter changes yield a sufficiently slow and smooth change in the MC response $MC_b$ in a data point $b$, the Professor framework calculates for that purpose a polynomial fitting function $f^{(b)}(\vec{p})$. The calculation will be performed for every data point $b$ independently but with in the same fixed and predefined order. In the case of a polynomial function of second-order, this looks like

\begin{equation}
MC_b(\vec{p})\approx f^{(b)}(\vec{p})=\alpha^{(b)}_0 + \sum_i \beta^{(b)}_i p'_i + \sum_{i\leq j} \gamma^{(b)}_{ij} p'_ip'^j
\label{eq:prof_fit_func}
\end{equation}

with the fit parameters $\alpha^{(b)}_0$, $\beta^{(b)}_i$ and $\gamma^{(b)}_{ij}$. The prime of the MC model parameters indicates the mapping

\begin{equation}
p'_i=\frac{p_i - p_{i,min}}{p_{i,max} - p_{i,min}}
\label{eq:prof_mapping}
\end{equation}

with the minimum (maximum) value $p_{i,min}$ ($p_{i,max}$) in the dimension $i$ of all sampled parameters in $S$. Hence, the parameters are mapped onto a [0,1]-interval. This is performed for numerical stability, since now the fit parameters are more likely to be of the same order. Furthermore, since this transformation is a bijective shift, the mapping mentioned in Eq.~\ref{eq:prof_mapping} is possible without changing the function behavior.

The number of fit parameters needed for the interpolation depends on the order of the polynomial function $n$ and $dim(\vec{p}) = P$. A general formula for the number of parameters $N^{(P)}_n$ is given by\cite[p.~3]{ref:professor}

\begin{equation}
N^{(P)}_n = 1 + \sum^n_{i=1}\frac{1}{i!}\prod^{i-1}_{j=0}(P+j)\; .
\end{equation}

In order to calculate the fit parameters, at least 

\begin{equation}
(|S| = N^{(P)}_n)\wedge(\vec{p}_i~\neq~\vec{p}_j \forall~i,j~\in~[1,N^{(P)}_n])
\end{equation}
 
needs to be fulfilled. Further anchor points provide additional information about the dependency on the parameters of the MC generator prediction. The authors of the framework recommend an oversampling of $|S| \geq 2 N^{(P)}_n$.

For retrieving the fit parameters, Eq.~\ref{eq:prof_fit_func} can be generalized to arbitrary dimension and order, leading to

\begin{equation}
MC_b(\vec{p})\approx f^{(b)}(\vec{p})=\sum^{N^{(P)}_{n}}_{i=1} c^{(b)}_i \tilde{p}_i 
\label{eq:prof_fit_func_rewritten}
\end{equation}
with the fit coefficients $c^{(b)}_i$. This vector $\vec{c}^{(b)}$ contains a dimension-wise combination of all fit parameters like $\alpha^{(b)}_0$, $\beta^{(b)}_i$ and $\gamma^{(b)}_{ij}$ from Eq.~\ref{eq:prof_fit_func}. The vector $\vec{\tilde{p}}$ of the $\tilde{p}_i$ is called \textit{extended parameter vector}. This vector contains every combination of every $p_i$ corresponding to the fit parameter $c^{(b)}_i$.

For a given set $S$ and the corresponding set $\{MC^{(b)}\}$ the fitting problem from Eq.~\ref{eq:prof_fit_func_rewritten} can be formulated as

\begin{equation}
\vec{v}^{(b)} = \vec{\tilde{P}}\vec{c}^{(b)}
\label{eq:prof_mat_vec}
\end{equation}

with the MC response vector $\vec{v}^{(b)}$, the matrix $\vec{\tilde{P}}$ that row-wise consists of $\vec{\tilde{p}}$ and the extended parameter vector $\vec{c}^{(b)}$. Explicitly, this can be written for a two dimensional polynomial function of second order as

\begin{equation}
\underbrace{\begin{pmatrix}
v_1 \\ v_2 \\ \vdots \\ v_{|S|}
\end{pmatrix}}_{\vec{v} \text{(values)}} = \underbrace{\begin{pmatrix}
1 & p'_1 & p'_2 & p'^2_1 & p'_1p'_2 & p'^2_2 \\
1 & p'_1 & p'_2 & p'^2_1 & p'_1p'_2 & p'^2_2 \\
&&&\vdots&&\\
1 & p'_1 & p'_2 & p'^2_1 & p'_1p'_2 & p'^2_2
\end{pmatrix}}_{\vec{\tilde{P}} \text{(sampled parameter sets)}} \underbrace{\begin{pmatrix}
\alpha_0 \\ \beta_{p'_1} \\ \beta_{p'_2} \\ \gamma_{p'_1p'_1} \\ \gamma_{p'_1p'_2} \\ \gamma_{p'_2p'_2}
\end{pmatrix}}_{\vec{c} \text{(coeff.)}}\; .
\label{eq:eqsys}
\end{equation}

With $|S| > N^{(P)}_n$, this equation is overdetermined. The solution of the linear equation system delivers the fit parameters for the polynomial function for a given data point $b$. The calculation of the solution is based on a \textit{singular value decomposition} (SVD). This algorithm decomposes the matrix $\vec{\tilde{P}}$ of size $M\times N$ into $\vec{\tilde{P}} = U\cdot W \cdot V^T$\cite{ref:cp1,ref:svd}. The matrix $V$ is an orthonormal matrix of size $N\times N$. $U$ is a column-orthogonal matrix of size $M\times N$ and $W$ is a diagonal matrix with $W_{ii} = w_i$ and size $N\times N$. Inverting this decomposition with respect to the problem in Eq.~\ref{eq:prof_mat_vec} yields

\begin{equation}
\vec{c}^{(b)} = \tilde{\mathcal{I}}\left[\vec{\tilde{P}}\right]\vec{v}^{(b)} = V\cdot W^{-1}\cdot U^T \vec{v}^{(b)}
\end{equation}

with the inverse operator $\tilde{\mathcal{I}}\left[\cdot\right]$ and the inverse $W_{ii}=1/w_i$. This method delivers the shortest $\vec{c}^{(b)}$. If $\vec{\tilde{P}}$ is not invertible, the \textit{Roger-Penrose-Inverse}\cite{ref:penrose} can be calculated. That way, this method delivers in the invertible and non-invertible case the shortest length of $\vec{\tilde{P}}\vec{c}^{(b)}-\vec{v}^{(b)}$\cite{ref:cp1}.\\
\medskip

In order to calculate the propagation of the uncertainty for the fit parameters, which has its origin in the statistical uncertainty of the MC generated events due to the limited number of events, the framework provides several different options. The option that is used in this thesis is called \textit{symm} inside the \textsc{Professor} framework. This option is the closest to a fit-parameter-by-fit-parameter uncertainty estimation. Here, the uncertainty calculation follows the same construction as in the fit parameter calculation, but the MC response values are exchanged by their corresponding uncertainties. Doing so, the correlation between the fit parameters are neglected and the uncertainties of the fit parameters will be overestimated.

\subsubsection{\textsc{prof2-tune}}
The third \textsc{Python}-script that will be used seeks the minimum of a goodness-of-fit function $\chi^2$. In the case of Professor, this function is given by

\begin{equation}
\chi^2(\vec{p}) = \sum_\mathcal{O}\sum_{b\in \mathcal{O}}w_b\frac{(f^{(b)}(\vec{p})-\mathcal{R}_b)^2}{\Delta^2_b}
\label{eq:chi2_prof}
\end{equation}

with the weights $w_b$ for each data point $b$ of every observable $\mathcal{O}$ and the corresponding measurement $\mathcal{R}_b$. The squared uncertainty \mbox{$\Delta^2_b = \Delta^2_f + \Delta^2_\mathcal{R}$} is a combination of the uncertainty $\Delta^2_f$ of the fit function $f^{(b)}(\vec{p})$ and the uncertainty $\Delta^2_\mathcal{R}$ of the measurement $\mathcal{R}_b$. The weights can be chosen manually for each data point independently in order to provide a better MC response for some data points/observables, because the impact of a data point/observable becomes more important for the overall $\chi^2$-value.

This function needs to be minimized which is done using the \textsc{Minuit}\cite{ref:minuit} package. This package provides a gradient based method, called \textsc{Migrad}, ``which copes with high dimensional problems better than the SciPy \textit{Nelder-Mead simplex} minimizer.''\cite[p.~7]{ref:professor}. In addition, the parameters can be calculated using parameter limits in order to stay in physical meaningful regions. Also, the resulting correlation matrix $\rho_{ij} = C_{ij} / \sqrt{C_{ii}C_{jj}}$ of the MC model parameters with the corresponding covariance matrix $C$ is written out.

\subsubsection{Run-combinations}
A last recommendation of the authors of the Professor framework involves the usage of so called \textit{run-combinations}. This term means that the tuning is run multiple times using different parameters sets $S_j$ with the condition $S_j \subseteq S$. That way, the tuning will deliver a set of tuned parameter sets. In the case of $\left(S_i \cap S_j \neq 0\right) \wedge \left(S_j \neq S\right) \forall i,j$ 
the result is a variation of the information given for the interpolation that appears overall similar, but does not necessary result in exactly the same function. With respect to the distribution of the tuned model parameter values, a best fitting parameter set can be extracted.

\subsection{Process simulation and extraction of observables}
\label{subsec:simulating}
This subsection is meant to describe the MC simulation and the corresponding observable extraction. Since those two parts can be used as pure black-boxes, they will be presented only briefly here.

The first part will be the MC generator. In this thesis, only \textsc{Pythia 8} will be used. Its physical working principle was presented in Sec.~\ref{sec:shower_evolution}. Since only a subset of model parameters will be tuned in the following sections and chapters, many model parameters remain in a default or a predefined setting. \textsc{Pythia 8} provides predefined tunes for e.g. $pp$ and $e^+e^-$ collisions. The chosen set of default parameters is the \textsc{Pythia 8} \textit{eeTune} = 5 or 6. The parameters that will be kept constant remain for both predefined tunes the same. A list of the parameter values is shown in App.~\ref{asec:pythia_configuration}. Besides the constant model parameters, one needs to specify the initial and final state of the physics process as well as the center-of-mass energy $\sqrt{s}$.

The MC generator will produce an output file in the \textsc{HepMC}\cite{ref:hepmc} format. This file contains all simulated events, including all particles and their properties. Those events are experiment- and therefore detector-independent. Since the parametrization-based tuning will be performed on mostly histograms of observables, those need to be extracted out of those \textsc{HepMC}-files. The \textsc{Rivet 2.4.2} framework extracts the distributions of the specified observables from the \textsc{HepMC}-files. The \textsc{Rivet} framework also provides access to measured data.

Since the simulated events do not have detector effects while actual measurements on the other hand do, this difference needs to be corrected in order to produce comparable observables. For that purpose, the data measured by collider experiments are unfolded\cite{ref:unfold} to stable particle level\footnote{In a MC generator, particles decay with a lifetime shorter than a certain threshold. The term ``stable'' denotes that a particle does not decay within this time.} such that detector effects are removed. The reference data of an experiment and the associated observable extraction out of the \textsc{HepMC}-files are stored in an ``analysis'' in the \textsc{Rivet} framework. This allows the comparison between simulated and reference data.

\section{Applying a parameter tuning based on a previously performed tune}
\label{sec:applying_tune}
The tuning approach that will be used in this thesis is directly linked to the master thesis of Nadine Fischer\cite{ref:nadine_fischer}. Although the method itself has already a longer history and was used multiple times before\cite{ref:holger_schulz, ref:abreu, ref:hamacher_weierstall}, the configuration and observables mentioned in Ref.~\cite{ref:nadine_fischer} will be used as a starting point for an investigation of the tuning method itself.

The thesis of Nadine Fischer used tunes from different MC generator setups. Also, the observables were used with different weight sets. The idea of Nadine Fischer's project was to compare the predictions made by the selection of tuned MC generators with actual measurements. The goal of this thesis is not using different MC generators and weights in order to control the prediction power for new observables, but an investigation of the tuning method itself. So, only a single MC generator will be used. In order to give an example implementation of the method described in Sec.~\ref{sec:tuning_approach} and to receive a reference for future modification of this method, a pure reconstruction of the tuning is presented in this section.

For that purpose, in the Subsection \ref{subsec:experiments}, the experiments that provided the measurements for the analyzes in Ref.~\cite{ref:nadine_fischer} will be described. In a second step the setup for the tuning will be presented in Subsection \ref{subsec:tuning_setup}. In the last part of this section, the result of the reproduction of the tuning will be presented in Subsection \ref{subsec:nadine_tuning_results}.\\

\subsection{Experiments}
\label{subsec:experiments}
The experiments for the thesis in Ref.~\cite{ref:nadine_fischer} were mostly located at the Large Electron-Position collider (LEP)\cite{ref:lep_tdr, ref:lep} at CERN. The data taking program of LEP was divided into two runs. The first run operated from 1989 to 1995 at a center-of-mass energy of $\sqrt{s} = 91~GeV = m(Z^0)$, the second at higher energies up to $209~GeV$ until 2000. Since the main concern of the thesis was based on the process \mbox{$e^+e^- \rightarrow Z^0 \rightarrow q\bar{q}$}, the data that will be used is mainly based on the first run of LEP.

The data acquisition at LEP was performed by four detectors: ALEPH\cite{ref:aleph_tdr, ref:aleph}, DELPHI\cite{ref:delphi_tdr, ref:delphi}, OPAL\cite{ref:opal_tdr, ref:opal} and L3\cite{ref:l3_tdr}. The data measured from the first three will be used here. All detectors were constructed in a onion-layered structure around the beam pipe with the collision point in the center of the detector. Those layers are designed to measure different particles and their properties. In the following, a short example of the detector working principle will be shown. Since the basic detection principles are very similar for the detectors, only the ALEPH detector will be presented. The ALEPH detector has as most inner part a vertex detector. The next layer is a drift chamber for particle tracking and a time projection chamber. Both are used to measure charged particles. In order to measure the energy of electrons and photons, the next layer is an electromagnetic calorimeter. The next outer layer is a hadronic calorimeter that measures the energy of hadrons. All those parts mentioned so far are encompassed in a magnetic field, which allows to measure the charges and momenta of the particles. Muons are the only charged particles traversing the detector without being absorbed. Therefore they can be identified best in the outermost layer where their tracks can be measured. This allows ALEPH to detect particles and measure their properties.

\begin{figure}
	\centering
	\includegraphics[width=\textwidth]{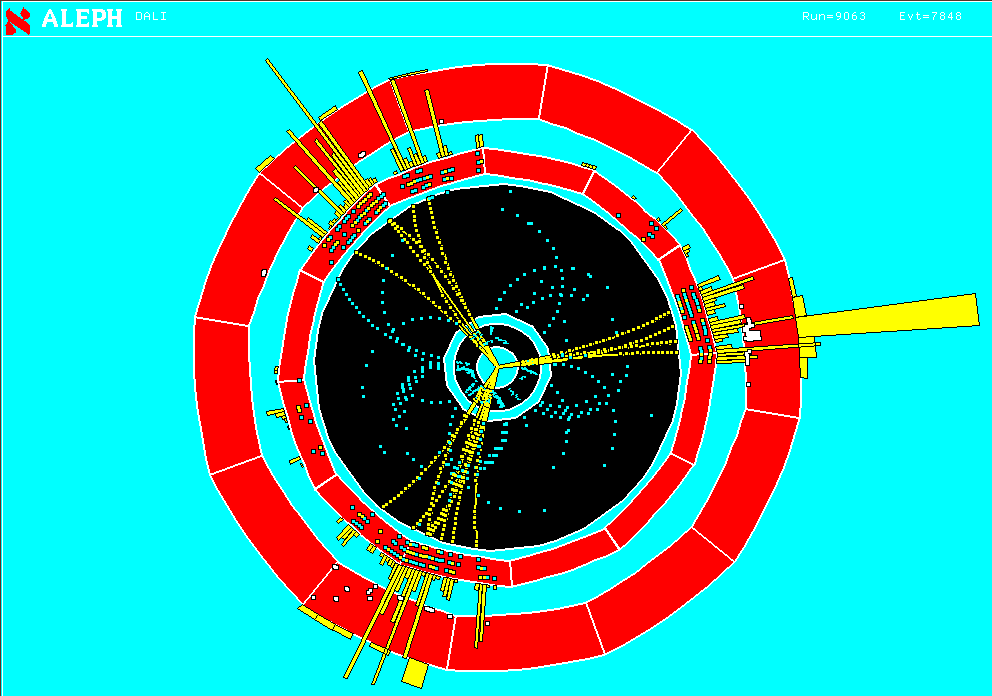}
	\caption{Measurement of a hadronic $Z^0$ decay performed by ALEPH\cite{ref:aleph_zdecay}. The different colors represent the different layers of the detector. Every dot in the colored areas represents an electric signal. The heights of the bars represent the deposited energy.}
	\label{fig:aleph_zdecay}
\end{figure}

As an example of the layer based detection, an event-display recorded by the ALEPH detector is shown in Fig.~\ref{fig:aleph_zdecay}. This picture shows the measurement of a possible three-jet event from the hadronic decay of a $Z^0$ boson. Various measurements were performed with ALEPH and published in \textsc{Rivet}. These measurements of event shapes, momentum distributions of several hadrons and constraints upon $b$-quarks will be used in the later tuning processes. In addition, more detailed jet shape measurements were provided by the DELPHI experiment and jet rate measurements were provided by OPAL.

The last data source is provided by the particle data group (PDG). From this group several hadron multiplicities were provided as further constraints to the MC tuning. The data of these multiplicities have multiple sources from several experiments.

\subsection{Tuning setup}
\label{subsec:tuning_setup}
The \textsc{Pythia 8} simulations utilized in Ref.~\cite{ref:nadine_fischer} were performed for the process $e^+e^- \rightarrow Z^0 \rightarrow q\bar{q}$ with the quark flavor $q$ at $\sqrt{s}=91.2~GeV=m(Z^0)$. In order to receive a sufficiently small statistical uncertainty, 500,000 events are generated for every single parameter set $\vec{p}_i$. The hard process of every simulation is performed at LO, the parton shower in LL. Overall, $m = 650$ different parameter sets are sampled in the ranges mentioned in Tab.~\ref{tab:parameter_ranges_nadine}. An explanation of the meaning of the model parameters can be found in Sec.~\ref{sec:shower_evolution}. Out of the simulated events, several histograms of different observables are provided using the \textsc{Rivet} framework. Those histograms refer to the data mentioned in the previous Subsection. A detailed list is shown in the \cref{tab:particle_spectrum,tab:pdg_multiplicities1,tab:pdg_multiplicities2,tab:b_fragmentation,tab:event_shape,tab:diff_jet_rate} in App.~\ref{asec:observablesandweights}. A translation table between the \textsc{Rivet} internal name-giving and the actual observable that it represents is shown in App.~\ref{asec:rivet_analyzes}.\\

\begin{table}[htbp]
  \centering
  \begin{tabular}{ l l l }
    \hline
    Parameter & Minimum & Maximum \\ \hline
    $a_{\textrm Lund}$ & 0.2 & 0.7 \\
    $b_{\textrm Lund}~[GeV^{-2}]$ & 0.5 & 1.5 \\
    $a_{\textrm ExtraDiquark}$ & 0.5 & 1.0 \\
    $\sigma~[GeV]$ & 0.2 & 0.4 \\
    $\alpha_s(M_Z)$ & 0.12 & 0.139 \\
    $p_{T,min}~[GeV]$ & 0.4 & 1.0 \\
    \hline
  \end{tabular}
  \caption{Parameter ranges of the tuning parameters as defined in Ref.~\cite{ref:nadine_fischer}.}
  \label{tab:parameter_ranges_nadine}
\end{table}

The results are a total of 1115 data points from 103 observables. Every data point is interpolated using the \textsc{Professor} framework with a polynomial function of fourth order. A minimization is performed to receive the tuned parameter values using \textsc{Professor}. In order to study the stability of the result, a run-combination is performed using 300 times a subset of 500 samples out of the full set of 650 samples. For every single subset out of these 300, the interpolation and minimization process was repeated.

The used analyzes as provided by the \textsc{Rivet} framework contain more than the mentioned 103 observables but those are the only ones that are actually in use, since the other observables carry redundant information in connection to other analyzes. Furthermore, the chosen observables are weighted, in order to increase or decrease the importance of observables and therefore optimize the resulting tuned parameters for several aspects like particle multiplicities, event shapes etc. In addition to the weighting sets in Ref.~\cite{ref:nadine_fischer}, no (neutral) weighting is used. An overview over the weights is given in \cref{tab:particle_spectrum,tab:pdg_multiplicities1,tab:pdg_multiplicities2,tab:b_fragmentation,tab:event_shape,tab:diff_jet_rate} in App.~\ref{asec:observablesandweights}.

The weights are not chosen based on physical processes but to improve a tuning on certain observables. Therefore, they can be chosen arbitrarily. The value itself results mostly from the fact that in Eq.~\ref{eq:chi2_prof} the $\chi^2$-value gets calculated for every single data point. Since some observables contain for example a single data point while other observables contain 20 data points, the last one would contribute more to the overall $\chi^2$-value on the left hand side of Eq.~\ref{eq:chi2_prof}, if the contribution of all data points would be almost equal. The implicit statement would result in a more important observable that contains more data points than the one with less data points. In order to encounter that numerical problem, several observables need to be treated with an artificially higher importance. This manifests in a change of the weights.

The tuning in Ref.~\cite{ref:nadine_fischer} contains two sets of weights. The first tuning tries to treat every observable almost as equal. The second tuning is more focused on the event shapes. So, particle multiplicities are neglected, particle spectra remain unchanged but every event shape parameter is weighted five times higher than in the first tuning.

\subsection{Tuning results}
\label{subsec:nadine_tuning_results}

The tuning in Ref.~\cite{ref:nadine_fischer} is based on two different weight-sets as mentioned in \cref{tab:particle_spectrum,tab:pdg_multiplicities1,tab:pdg_multiplicities2,tab:b_fragmentation,tab:event_shape,tab:diff_jet_rate} in App.~\ref{asec:observablesandweights}. The run-combinations performed in this thesis delivered as tuned parameters in the \textsc{Pythia 8} framework the values in Tab.~\ref{tab:nadine_tune}. This table also contains the tuned parameter values obtained in this thesis. The re-performing is additionally performed with no weighting, i.e. weighted every observable with weight unity.\\

\begin{table}[htbp]
  \centering
  \begin{tabular}{ l | l l | l l l}
    \hline
    Parameter & \multicolumn{2}{c}{Original tuning} & \multicolumn{3}{| c}{Re-performed tuning} \\
    & Tune 1 & Tune 2 & Tune 1 & Tune 2 & Tune neutral\\ \hline
    $a_{\textrm Lund}$ & 0.35 & 0.42 & 0.61 & 0.56 & 0.62 \\
    $b_{\textrm Lund}~[GeV^{-2}]$ & 0.94 & 1.05 & 1.20 & 1.17 & 1.25 \\
    $a_{\textrm ExtraDiquark}$ & 0.95 & 0.83 & 0.94 & 0.87 & 0.93 \\
    $\sigma~[GeV]$ & 0.284 & 0.285 & 0.292 & 0.291 & 0.292 \\
    $\alpha_s(M_Z)$ & 0.139 & 0.139 & 0.139 & 0.139 & 0.139 \\
    $p_{T,min}~[GeV]$ & 0.41 & 0.40 & 0.42 & 0.41 & 0.43\\
    \hline
  \end{tabular}
  \caption{Original tuning: Tuning result with two different weight-sets as it is presented in Ref.~\cite{ref:nadine_fischer}. Re-performed tuning: Tuning result from the reproduction of the tuning.}
  \label{tab:nadine_tune}
\end{table}

As Tab.~\ref{tab:nadine_tune} shows, the resulting tuned parameter values of tuning 1 and 2 are consistent except for $a_{\textrm Lund}$ and $b_{\textrm Lund}$. These parameters have correlation coefficients up to 0.96. A re-run of the minimization with fixed $a_{\textrm Lund}$ values is shown in Tab.~\ref{tab:fixed_alund}. As values for $a_{\textrm Lund}$ the tuned values from the original tuning in Tab.~\ref{tab:nadine_tune} are used. Since $a_{\textrm Lund}$, $b_{\textrm Lund}$ and $a_{\textrm ExtraDiquark}$ are correlated the results also differ from the tuning results in \cite{ref:nadine_fischer}. A possible reason for the differences shown in Tab.~\ref{tab:nadine_tune} is based on different \textsc{Professor}, \textsc{Pythia} and \textsc{Rivet} versions that were released since the creation of the thesis in Ref.~\cite{ref:nadine_fischer}. Especially the \textsc{Rivet} analyzes changed partly since the thesis was released. Additionally, the result is based on MC procedure and therefore, the results can vary by its statistical construction. For the further investigations, the reproduced parameter set will be used as reference values, which allows for a direct comparison.

The values of the reproduced tuning are extracted from the mean of the distribution of tuned parameter values after the calculation of the run-combinations. Any correlation between the parameter values is neglected in that statement. A complete overview over the distributions can be found in App.~\ref{asec:distribution_reproduction}.

\section{Minimization without limits of parameter values}
\label{sec:double_peak}
\begin{figure}[htbp]
\makebox[\textwidth][c]{
	\begin{adjustbox}{max width=1.\textwidth}
	\includegraphics[width=\textwidth]{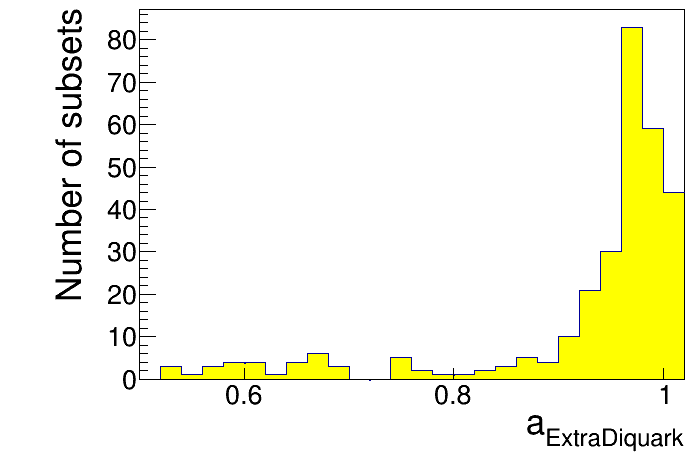}
	\includegraphics[width=\textwidth]{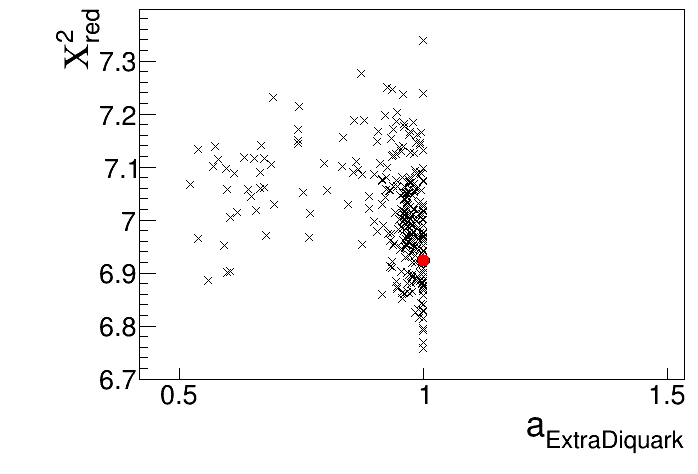} 
	\end{adjustbox}
}
	\caption{Distribution of the run-combinations using the neutral weights from \cref{tab:particle_spectrum,tab:pdg_multiplicities1,tab:pdg_multiplicities2,tab:b_fragmentation,tab:event_shape,tab:diff_jet_rate} in App.~\ref{asec:observablesandweights}. The left hand side graph shows the histogram with number of tuned parameter values of $a_{\textrm ExtraDiquark}$. The right hand side shows the corresponding values of $\chi^2_{red}=\chi^2/n_{dof}$ for the tuned parameter values of $a_{\textrm ExtraDiquark}$ with $n_{dof}=1109$. The red dot represents the result of the tuning using all 650 samples.}
	\label{fig:aed_prof}
\end{figure}

If using such a method, the result needs to be trustworthy, since it is almost impossible to check if the resulting parameter values are the \textit{best} tuning parameters. This is especially important, if the parameter space is higher dimensional. In order to investigate the trustworthiness of the result, at first a selected graph from App.~\ref{asec:distribution_reproduction} is shown in Fig.~\ref{fig:aed_prof}.\\


The number of degrees of freedom $n_{dof}=1109$ results from the number of data points (1115) minus the number of model parameters that need to be tuned (6). The interesting result is that many tunings are at the upper limit of the parameter range of $a_{\textrm ExtraDiquark}$. This is especially visible in the $\chi^2_{red}$-distribution (see the right hand side of Fig.~\ref{fig:aed_prof}). Overall, the $\chi^2_{red}$-distribution indicates a usual spread along the ordinate. Also, the tunings of the model parameters $\alpha_s$ and $p_{T,min}$ are at their limit.

\begin{figure}[htbp]
\makebox[\textwidth][c]{
	\begin{adjustbox}{max width=1.\textwidth}
	\includegraphics[width=\textwidth]{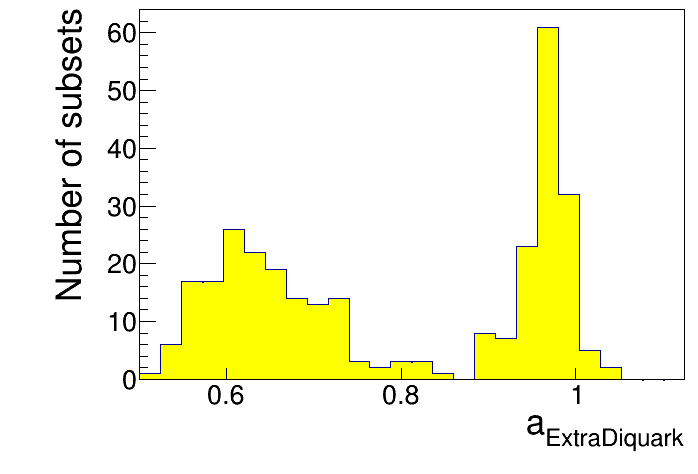}
	\includegraphics[width=\textwidth]{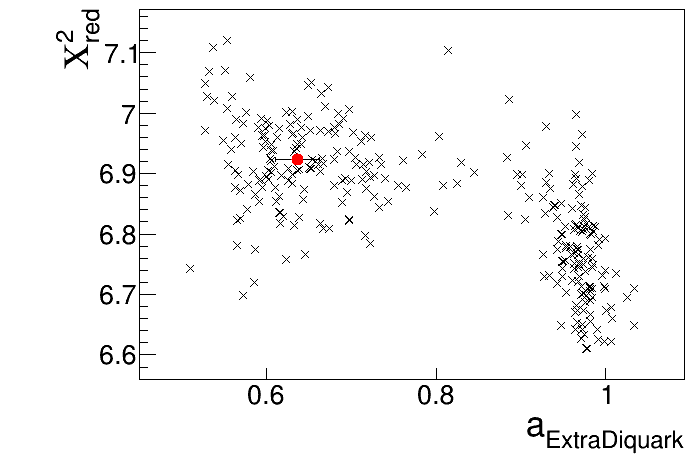} 
	\end{adjustbox}
}
	\caption{Distribution of the run-combinations using the neutral weights from \cref{tab:particle_spectrum,tab:pdg_multiplicities1,tab:pdg_multiplicities2,tab:b_fragmentation,tab:event_shape,tab:diff_jet_rate} in App.~\ref{asec:observablesandweights} without parameter limits. The left hand side graph shows the histogram with number of tuned parameter values of $a_{\textrm ExtraDiquark}$. The right hand side shows the corresponding $\chi^2_{red}$-value of the tuned parameter values of $a_{\textrm ExtraDiquark}$ with $n_{dof}=1109$. The red dot represents the result of the tuning using all 650 samples.}
	\label{fig:aed_prof_nolimits}
\end{figure}

Since 300 subsets were used with 500 out of 650 different parameter sets, many samples are shared in the determination of the interpolation functions. Thus the behavior of the interpolation functions should be similar and therefore the tuning results should also be similar. Since the tuned parameter values are at the upper limit, this cannot be shown in this graph, if it is assumed that such a distribution should look like a normal distribution. For this purpose the parameter limits will be dropped in the next step. The interpolation functions will remain the same as before but the minimizer will be allowed to extrapolate the interpolation functions in regions without any samples in order to minimize the $\chi^2$-value. Since no simulated information exists in a region of extrapolation, such a result is not usable as final tuning result, but it will give an insight into the function behavior. Due to the usage of the same reference data and the same functions, the result should be a distribution of tuned parameter values for $a_{\textrm ExtraDiquark}$ around a value. A total overview over the results is shown in App.~\ref{asec:distribution_reproduction_nolimits}. In order to compare to the Fig.~\ref{fig:aed_prof}, Fig.~\ref{fig:aed_prof_nolimits} shows the results of a tuning without parameter limits.


An interesting result of a tuning without limits for parameter values is the existence of two attractors for the parameter values. 
The double peak in Fig.~\ref{fig:aed_prof_nolimits} (left) indicates that each result depends on the subset used for the individual tuning. The $\chi^2_{red}$-values do not allow clear separation of the attractors along the abscissa. Indeed, the $\chi^2_{red}$-values around a $a_{\textrm ExtraDiquark}$-value of 1 are on average a little bit smaller than the values around 0.6, but the overlap of the $\chi^2_{red}$-values is very big. Additionally, the tuning using all 650 samples provides a tuned parameter values around 0.6. The fact that the resulting tuning parameters for $a_{\textrm ExtraDiquark}$ are still almost inside the original parameter space is just a coincidence. The other tuned parameter values of that run-combinations can also lie outside the original parameter range (cf. Fig.~\ref{fig:tune_nadine_weights0_nolimits}) and thus affect the value of $a_{\textrm ExtraDiquark}$. This happens due to the correlations between those parameters.\\
\medskip

In order to obtain this result, no further changes than the removing of the parameter limits were made. The results were assumed to be centered around a single value, such that a tuned parameter could be extracted by mean, mode, median etc. If the run-combinations provide multiple attractors for the model parameters, reliable parameter values cannot be extracted. 
To study this problem, the tuning will be modified step by step in order to identify the problem. This will be the main content of the following chapters.

The structure for such an investigation is to modify the single steps in Fig.~\ref{fig:tuning_steps}. Since the sampling is assumed to not cause the problem, this will not be changed at all. The simulation and observable extraction is assumed to be independent of the subset chosen for interpolation and tuning in the run-combinations step. Therefore, the interpolation and minimization algorithm will be investigated. As a first step, the $\chi^2$-minimization will be replaced by another algorithm. This will be described and tested in Chapter \ref{ch:bayes}. The Chapter \ref{ch:adaptive} will concern the interpolation algorithm used by \textsc{Professor}.

\chapter{Bayesian Analysis Toolkit}
\label{ch:bayes}
\setcounter{page}{39}

As shown in Sec.~\ref{sec:double_peak}, the run-combinations performed by the \textsc{Professor} framework can produce more than one peak in the distribution of tuned parameter values. A first explanation approach is based on the last step in the tuning process displayed in Fig.~\ref{fig:tuning_steps}. The $\chi^2$-minimization algorithm used in the framework uses the \textsc{Minuit} package. A possible behavior could be that the gradient based algorithm (\textsc{Migrad}) does not necessarily converge to the global minimum of the $\chi^2$-function. If so, the existence of the double-peak could be fixed by exchanging that package by another minimizer. For that purpose, the $\chi^2$-function minimization will be tested in this Chapter using another framework, called \textsc{Bayesian Analysis Toolkit} (BAT)\cite{ref:batdl}.

Since the basic working principle of BAT is based on Bayesian analysis, the underlying theory will be explained in Sec.~\ref{sec:bayesian_analysis}. Based on that model, the working principle of the toolkit will be described in Sec.~\ref{sec:working_principle}. Since BAT is designed to serve for several purposes, only the most important parts for the intent of this thesis will be presented. A corresponding implementation for the purpose of model parameter estimation of MC generators in BAT is described in Sec.~\ref{sec:implementation}. In the last part of this Chapter, the obtained results of this approach will be presented and compared with the minimization from the \textsc{Professor} framework using the \textsc{Minuit} package.

\section{Bayesian analysis}
\label{sec:bayesian_analysis}
In this Section, the background of Bayesian analysis will be explained and the most important properties derived. If not stated otherwise, the information in this Section is based on Ref.~\cite{ref:caldwell}.

As a motivation it will be assumed, that a given probability function $P(\vec{x}|M)$ for the data $\vec{x}$ with a given underlying model $M$ is available. As an example, the binomial distribution can be such a model $M$. Binomial processes have two outcomes, \textit{hit} or \textit{miss} for each event. The corresponding model parameters of that distribution are the number of events and the probability for a \textit{hit}. These model parameters will later be collected in a vector $\vec{\lambda}$. The corresponding data are the number of hits that are measured. This calculation of the probability assumes already that $\vec{\lambda}$ is given but usually the question is about these parameters themselves, so that the probability distribution function $P(\vec{\lambda}|\vec{x})$ is needed. Exactly this problem is addressed by the Bayesian analysis.\\

As first step in the derivation of Bayes theorem, the probability will be considered for two events $A$ and $B$ happening. The corresponding probability is

\begin{equation}
P(A\cap B) = P(A|B) \cdot P(B).
\label{eq:bayes1}
\end{equation}

This equation is derivable from Kolmogorov's axioms\cite{ref:boris_grube}. It states the combined probability that two events happen. First of all the event $B$ with its corresponding probability $P(B)$ needs to happen. If that event occurred, the event $A$ needs to occur, too. Therefore, $P(A|B)$ states the probability that $A$ can happen, but that needs to be under the condition that $B$ already happened. This can be graphically represented in Euler diagrams. A special case of Eq.~\ref{eq:bayes1} would be the independence of both events. This leads to $P(A|B) = P(A)$ and therefore to $P(A\cap B) = P(A)\cdot P(B)$. Another special case would be $A = B$ and therefore $P(A\cap B) = P(A) = P(B)$.\\


The next step in the derivation is given by the symmetry property \mbox{$A\cap B = B\cap A$} (in an Euler diagram these sets would be represented by the exact same area). This symmetry can be used in combination with the Eq.~\ref{eq:bayes1} to formulate Bayes theorem under the condition $P(B)\neq0$:

\begin{align}
P(A\cap B) &= P(B\cap A) \\
\leftrightarrow P(A|B) \cdot P(B) &= P(B|A) \cdot P(A) \\
\leftrightarrow \qquad \quad P(A|B) &= \frac{P(B|A)\cdot P(A)}{P(B)}
\label{eq:bayes2}
\end{align}

The factor $P(B|A)$ in the final Eq.~\ref{eq:bayes2} is called \textit{likelihood}, the term $P(A)$ \textit{prior} and the term $P(B)$ \textit{evidence}. The left-hand side is called \textit{posterior}.

As Eq.~\ref{eq:bayes2} implies, Bayes theorem connects $P(A|B)$ to $P(B|A)$. According to the example of a binomial distribution at the beginning of this Section, $B$ corresponds to the data and $A$ to the model parameters. This theorem therefore allows to switch the dependency of a conditional probability. Due to the fact that the data is usually measured, a probability(-density) that depends on the model parameters can be calculated with the given measured data. In order to formulate the conditional probability in terms of model parameters and data, $A$ can be replaced by the model parameters $\vec{\lambda}$ and $B$ by the data $\vec{x}$.\\

As shown in Eq.~\ref{eq:bayes2}, the evidence $P(B)$ needs to be given in order to switch the dependency of the conditional probability. This means that the probability for the data itself needs to be known. In order to handle this problem, the law of total probability can be used, that states

\begin{equation}
P(B) = \sum_i P(B|A_i)\cdot P(A_i).
\end{equation}
 
\begin{figure}
	\centering
	\includegraphics[width=0.7\textwidth]{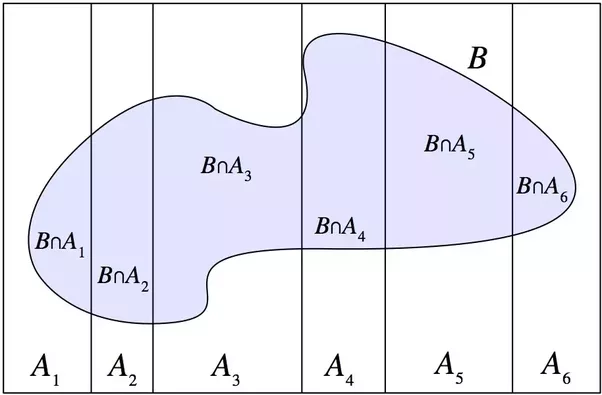}
	\caption{Illustration of the space splitting in the law of total probability. In every region $A_i$, the content of $B$ will be determined. That splitting can be performed with arbitrary small $A_i$. The image is taken from Ref.~\cite{ref:lawoftotalprob}.}
	\label{fig:lawoftotalprob}
\end{figure}

This relation can be understood as a splitting of the whole sample space in parts $A_i$. The splitting can be performed such that for every $A_i$ the corresponding probability $P(A_i)$ can be obtained. Then one needs to take a look at the probability of $B\cap A_i$. This will give $P(B|A_i)$. In an Euler diagram it would look like Fig.~\ref{fig:lawoftotalprob}. Using the law of total probability in combination with Eq.~\ref{eq:bayes2}, the Bayes-Laplace-theorem can be formulated as

\begin{equation}
P(A|B) = \frac{P(B|A) \cdot P(A)}{\sum_i P(B|A_i)\cdot P(A_i)}.
\end{equation}

This equation is already the final theorem in the case of a probability distribution that depends on discrete parameters.\\

Since many model parameters are not discrete like in the case of the mean or the width of a normal distribution, this theorem needs to be expanded for the case of continuous parameters. In this case, a likelihood becomes a probability density (pdf) instead of a probability. Its relation to a probability is given by $P(A)dA$ and so the evidence becomes $P(B) = \int P(B|A)dA$. This leads to the continuous case of the Bayes-Laplace-theorem:

\begin{equation}
P(A|B) = \frac{P(B|A) \cdot P(A)}{\int P(B|A)dA}.
\label{eq:bayes3}
\end{equation}

In the context of data and model parameters, the Eq.~\ref{eq:bayes3} allows the calculation of the posterior pdf for the model parameters. From such a posterior pdf, the model parameters can be extracted. Another measurement (and therefore new data) does not require a whole new calculation but the posterior pdf of the previous measurement can be used as prior for the next one. Altogether, this theorem states a learning rule for parameter values based on data. This learning rule underlies assumptions like the selected model so that the posterior does not intend to be a \textit{true} distribution but is led by the data itself.

Besides the iterative usage of posterior pdfs as priors, it can be shown that the same result can be obtained by using the product of the probability functions as likelihood and therefore combine all data in a single likelihood. This can only be done, as mentioned before, if the measurements are independent. Otherwise, the correlation between the data needs to be treated additionally.

The approach of the usage of the posterior pdf as prior pdf for the next calculation leads to the question about the first prior before any data is collected. Due to the fact that there is no information about the parameters from the data, the prior needs to be defined by the data analyst himself/herself. This prior can be used to represent his/her prior belief in the parameter values but keeps a subjective fingerprint in the procedure. The impact of that fingerprint can be reduced due to many or \mbox{strong/precise data/measurements}. By using this theorem, a possibly wrong belief in the parameter value can be corrected in the posterior pdf and the posterior pdf is able to represent the data better than the prior. For the explicit choice of the prior itself, no rule exists.

\subsection{Likelihood}
In this subsection, the focus lies on the likelihood and prior. Since the evidence is independent of the model parameters, it just remains as a normalization for the posterior in order to handle it as a pdf. Hence, the evidence is a scaling factor, that can be ignored as long as no normalization is needed.

Therefore, the likelihood $\mathcal{L}(\vec{\lambda}) \equiv P(\vec{x}|\vec{\lambda})$ can now be defined. A flat prior can be treated as another scaling factor and so Eq.~\ref{eq:bayes3} becomes the relation $P(\vec{\lambda}|\vec{x}) \propto \mathcal{L}(\vec{\lambda})$. That means that the most probable values for the model parameters with respect to the data correspond to a maximum of the likelihood.

If the prior is not flat but a function of the model parameters, the likelihood can be defined in a modified way as $\mathcal{L}(\vec{\lambda}) \equiv P(\vec{x}|\vec{\lambda})\cdot P(\vec{\lambda})$. This modification inherits the property, that a maximum likelihood still refers to the most probable parameter values with respect to the prior information.\\

In case of $i$ independent measurements $\vec{x}_i$, the likelihood can be expressed as

\begin{equation}
\mathcal{L}(\vec{\lambda}) = \prod_i \mathcal{L}_i(\vec{\lambda}) = \prod_i P(\vec{x}_i|\vec{\lambda})~[\cdot~ P(\vec{\lambda})].
\end{equation}

The square brackets are meant to indicate a possible expansion of the likelihood function by the prior in the case that it is not constant.

Maximizing this likelihood is possible but often not easy to calculate (if at least analytically possible). Often it is more simple to use the natural logarithm of the likelihood (or log-likelihood in short) $\ln(\mathcal{L}(\vec{\lambda}))$. This is possible due to the fact, that the natural logarithm is a bijective function on $(0,\infty)$ and so it is a simple rescaling without changing the properties of the function. Even quantiles remain unchanged after the transformation. The likelihood function becomes

\begin{equation}
\ln(\mathcal{L}(\vec{\lambda})) = \sum_i \ln(P(\vec{x}_i|\vec{\lambda}))~[ +~ \ln(P(\vec{\lambda}))]
\end{equation}

and so the maximum of the function can be calculated by using a sum instead of a product. The rescaling also helps to obtain numerically more stable calculations.

As an example a measurement $x$ with a known uncertainty $\sigma$ is distributed according to a normal distribution with mean $\mu$ is considered. The log-likelihood is

\begin{equation}
\ln(\mathcal{L}(\mu)) = -\ln(\sqrt{2\pi}\sigma) - \frac{1}{2} \frac{(x-\mu)^2}{\sigma^2}.
\label{eq:log-likelihood}
\end{equation}

If $\sigma$ is independent of the mean, the first term can be neglected, since it is constant. The result contains neither the natural logarithm nor an exponential function. With respect to a finite machine precision in numerical approaches, such a log-likelihood is a more stable function than using the likelihood itself.

The usage of a normal distribution in the log-likelihood is often applied in data science in order to receive the best model parameters. To derive the $\chi^2$-function as shown in Eq.~\ref{eq:chi2_prof}, and its relation to that log-likelihood, one needs to consider Eq.~\ref{eq:log-likelihood} with $\sigma$ being independent of $\mu$ and a flat prior. This leads to the relation

\begin{equation}
-2\ln(\mathcal{L}(\mu)) = \frac{(x-\mu)^2}{\sigma^2} = \chi^2.
\label{eq:loggauschi2}
\end{equation}

Therefore, the $\chi^2$-function is nothing else than a special case of a normal distributed likelihood that is used for many different parameter estimation problems. The underlying reason why the normal distribution is so important and often used for this kind of problems will be explained in the following Subsection.

\subsection{Central limit theorem}

The central limit theorem (CLT) is one of the most powerful theorems in data analysis. Since it will also be used in this thesis, it will be derived and its application will be shown in this Subsection.

To derive the theorem, as a first step the \textit{characteristic function} $\phi(k)$ of a pdf $p(x)$ will be introduced as a Fourier transform of $p(x)$

\begin{equation}
\phi(k) = \int_{-\infty}^\infty e^{ikx}p(x)dx
\label{eq:characteristicfunc}
\end{equation}

or for a discrete case with a probability function $P(x)$ as $\phi(k)=\sum_j P(x_j)e^{ix_jk}$. The discrete case is given here for completeness, but will not matter in this thesis.

The inverse transformation in the continuous case is given by the inverse Fourier transformation

\begin{equation}
p(x) = \frac{1}{2\pi}\int_{-\pi}^{\pi}\phi(k)e^{-ixk}dk.
\label{eq:inversecharacertisticfunction}
\end{equation}

The benefit of the characteristic function is given by its properties. For the purpose of the derivation of the CLT, the most important property relies on the combination $z$ of two independent variables $x$ and $y$, such that $z=x + y$. The pdf of $x$ will be given by $p_x(x)$, the pdf of $y$ will be $p_y(y)$. The requirement of independence of both variables gives the combined pdf $p(x,y)=p_x(x)p_y(y)$.

The assumption that a characteristic function $\phi_z(k)$ of the variable $z$ exists leads to

\begin{align*}
\phi_z(k) &= \int e^{ikz}p_z(z)dz \\
&=\int e^{ikz} \left[\int \delta(z-x-y)p(x,y)dxdy\right]dz \\
&=\int e^{ikz} \left[\int \delta(z-x-y)p_x(x)p_y(y)dxdy\right]dz \\
&=\int\left[\int e^{ikz} \delta(z-x-y)dz\right] p_x(x)p_y(y)dxdy \\
&= \int e^{ikx}p_x(x)dx\int e^{iky}p_y(y)dy \\
&=\phi_x(k)\phi_y(k)
\end{align*}

with the characteristic functions $\phi_x(k)$ and $\phi_y(k)$ of $p_x(x)$ and $p_y(y)$ respectively. The result shows that the sum of two independent variables with their corresponding pdfs can be formulated as a product of characteristic functions.

Using this feature, the CLT can be derived. Assuming $n$ independent measurements of a quantity $x$ that follows a pdf $p(x)$, a finite mean $\bar{x}=\frac{1}{n}\sum_{i=1}^n x_i$ must exist. For such a quantity, the expectation value $E\left[\bar{x}\right] = \mu$ can be calculated. In order to use the properties of the characteristic function, the quantity $z=\bar{x}-\mu$ will be used. As shown above, the characteristic function of $z$ can be written as

\begin{equation}
\phi_z(k) = \prod_{i=1}^n \phi_{\frac{x_i-\mu}{n}}(k).
\label{eq:characteristicfunctions}
\end{equation}

Every characteristic function on the right-hand side depends by its definition in Eq.~\ref{eq:characteristicfunc} on an $e$-function, that can be expanded in a \textit{Taylor series}. Doing so, the re-ordering  and the evaluation of the integral leads to

\begin{equation}
\phi_{\frac{x-\mu}{n}}(k) = 1 + \frac{ik}{n}E\left[x-\mu\right] - \frac{k^2}{2n^2}E\left[(x-\mu)^2\right]+\mathcal{O}\left(\frac{E\left[(x-\mu)^3\right]}{n^3} \right)+...
\label{eq:taylor}
\end{equation}

Under the assumption that higher moments of $p(x)$ are finite leads to the possibility to drop higher order terms in the limit $n\rightarrow \infty$. Doing so transforms the Eq.~\ref{eq:characteristicfunctions} into

\begin{align}
\lim_{n\rightarrow\infty} \phi_z(k) &= \lim_{n\rightarrow\infty} \prod_{i=1}^n \phi_{\frac{x_i-\mu}{n}}(k) \\
&\approx \lim_{n\rightarrow\infty} (1-\frac{k^2}{2n^2}\sigma_x^2)^n \\
&= e^{-\frac{k^2}{2n}\sigma_x^2}.
\end{align}

Here, the properties $E\left[x-\mu\right] = E\left[x\right]-E\left[\mu\right] = 0$ and $E\left[(x-\mu)^2\right] = \sigma_x^2$ were used in Eq.~\ref{eq:taylor}. Calculating the inverse transformation given by Eq.~\ref{eq:inversecharacertisticfunction} gives the pdf of $z$ as

\begin{equation}
p(z) = \frac{1}{\sqrt{2\pi}\sigma_z}e^{-\frac{1}{2}\frac{z^2}{\sigma_z^2}}
\end{equation}

with $\sigma_z = \sigma_x/\sqrt{n}$. An analogous derivation could be performed using $z=\bar{x}$ or $z=\sum_{i=1}^n x_i$ but with a slightly different final function.\\
\medskip

The property that the combination of many samples in terms of a mean becomes a normal distribution is the essence of the CLT. The power of the theorem is based on the few assumptions needed in order to derive it and to become applicable. The implementation of this theorem into this thesis is given by the assumption, that the value in each data point is given by a pdf that fulfills the demands of the CLT. Since many thousands of samples are generated by the MC generator or measurements are performed by an experiment, the mean of the statistical fluctuation can be assumed to be similar enough to a normal distribution. The estimator of the mean of the simulated data, represented by the interpolation function, can be assumed to be normal distributed for a certain parameter vector. As a result of this approximation, the uncertainty of the interpolation function can be represented by a normal distribution, too.

Thus, the approximation of a normal distributed process and therefore the usage of a corresponding likelihood or a $\chi^2$-function can be justified. Based upon the derivations performed in this Subsection, normal distributed processes will be used as likelihood-function in the implementation in BAT.

\section{Working principle of BAT}
\label{sec:working_principle}
BAT is designed to handle Bayesian analysis for pdfs with multiple variables. If the underlying pdf is complicated, a static and analytic treatment of the maximization of the likelihood-function in parameter space can be problematic. If discrete sampling points, e.g.\ on a grid, would be used in order to get information about the distribution, the approach can cause two different problems. The first one would be an unfortunate choice of the distance of the grid points (step size), such that the parameter vectors lie between the points a region of interest and therefore skip those regions. If on the other hand the step size would be small enough, e.g.\ in the order of the machine precision as the minimal step size, the other problem would be a very large run-time due to function evaluations and large sampling size if the dimension of the parameter vectors is high or the parameter space is large. Furthermore, the generated information density in a region of interest is the same as in the rest of the parameter space. Therefore, there would be much useless run-time and memory consumption without gaining any knowledge.

These problems can be handled by the numerical methods used by BAT as explained in Sec.~\ref{sec:numericalmethods}.

Using a numerical method needs to be setup carefully in order to optimize the work of the algorithm. The implemented approach is addressed in \ref{sec:prerun}.\\
\subsection{Numerical methods}
\label{sec:numericalmethods}

The information gathering and integration, e.g. for a normalization in order to treat a likelihood as pdf, can be performed by Markov Chains. Such a chain is a type of random walk. A random walk means that points will be selected at random inside the parameter space and all calculations like likelihood evaluation will be performed using the sampled parameter vector. This will be denoted as $\vec{p}^{(i)}$ for a sampled parameter vector $\vec{p}$ for the $i$-th step. In the case of MC generator tuning, these are the model parameter values in the parameter space. The key feature of a Markov Chain is that the next point $\vec{p}^{(i+1)}$ in this chain only depends on current point $\vec{p}^{(i)}$.

The construction of such chains requires two steps. The first one is the selection of a starting point $\vec{p}^{(0)}$. The other step is obtaining the transition probability $p_{jk}$ for $\vec{p}^{(i + 1)} = \vec{k}$, if $\vec{p}^{(i)} = \vec{j}$. The probabilities can be extracted from the probability distribution/pdf.\\

The method implemented in BAT for the transition between continuous parameters is the Metropolis-Hastings-algorithm\cite{ref:metropolis,ref:hastings}. The algorithm is given by

\begin{lstlisting}[mathescape=true]
Given $\vec{p}^{(0)}$
for(int i=0;i<n;i++){ 
	generate $\vec{y} \sim q(\vec{y}|\vec{p}^{(i)})$
	if(rnd(0,1) < $\rho(\vec{p}^{(i)}, \vec{y})) \{$ 
		$\vec{p}^{(i + 1)} = \vec{y}$
	} else { 
		$\vec{p}^{(i + 1)} = \vec{p}^{(i)}$
	} 
} 
\end{lstlisting}

with the sampling pdf $q(\vec{y}|\vec{p}^{(i)})$ for the next point in the parameter space and $\rho(\vec{p}, \vec{y}) = min \left\{\frac{p(\vec{y})}{p(\vec{p})}\frac{q(\vec{p}|\vec{y})}{q(\vec{y}|\vec{p})},1\right\}$ with the pdf of the Bayesian problem $p(\vec{p})$. $rnd(0,1)$ describes a function that samples a random number in the interval [0,1]. $n$ is the number of iterations.\\

\begin{figure}
	\centering
	\includegraphics[width=0.7\textwidth]{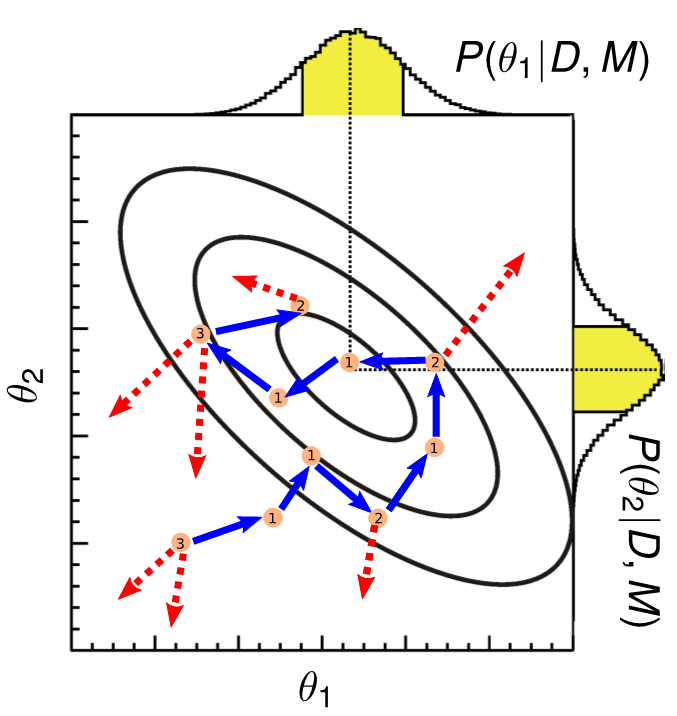}
	\caption{Illustration of the random walk of a Markov Chain using the Metropolis-Hastings algorithm through a parameter space of the parameters $\theta_1$ and $\theta_2$. It is assumed that a model $M$ and data $D$ exist. The blue lines indicate accepted proposal points, the red lines indicate declined ones. The numbers in each step represent the number of steps that the chain remained at a certain position in the parameter space. On the right-hand side and on the top the marginalized posterior likelihoods are shown. The image is taken from Ref.~\cite{ref:mcmc}.}
	\label{fig:mcmc}
\end{figure}

As this algorithm implies, the transition probability is not known before but only after every step evaluated. The function $q(\vec{y}|\vec{p})$ is a pdf representing a random walk, referred to as a Markov chain due to the current point $\vec{p}$ as the only dependence for the next sampling point $\vec{y}$. The function $\rho(\vec{p},\vec{y})$ is a veto function. If the new sampling point $\vec{y}$ is more probable/likely than the current point $\vec{p}$, this function is 1 and therefore the new point will be always accepted. The condition in the if-clause becomes important if it is less probable/likely. If the probability $p(\vec{p})$ is just slightly higher than $p(\vec{y})$, then it is likely that $\vec{p}^{(i+1)} = \vec{y}$. If the difference is bigger then it becomes more unlikely but not impossible. As a result of this algorithm, a Markov Chain does not get caught in a local maximum. Therefore, the algorithm is able to escape regions of (locally) high probability, gather information about the surrounding area and even find regions of a global maximum. An example for such a random walk is shown in Fig.~\ref{fig:mcmc}.\\

This algorithm generates a new point in each iteration. The result is a set $\left\{\vec{p}^{(0)},\vec{p}^{(1)},...,\vec{p}^{(n)}\right\}$. This algorithm is ergodic regarding $p(\vec{p})$. Therefore, a large $n$ would represent the likelihood function as needed by BAT. Furthermore, the algorithm is able to scan the whole parameter hypercube, if $n$ is large enough. Those two features reduce the importance on the choice of $\vec{p}^{(0)}$ for a large $n$. Otherwise, the first few sampling points can be deleted afterward. This is called \textit{warm-up} or \textit{burn-in}.

Furthermore, the sampling in BAT follows a normal distribution for $q(y|p)$ and therefore $\rho(\vec{p},\vec{y})$ simplifies since $q(p|y) = q(y|p)$ to $\rho(\vec{p}, \vec{y}) = min \left\{\frac{p(\vec{y})}{p(\vec{p})},1\right\}$.

\subsection{Pre-run}
\label{sec:prerun}
In BAT it is possible to run multiple Markov Chains in parallel in order to scan the whole parameter space in a smaller amount of time. The chains are independent from each other. In order to optimize the usage of the Markov Chain Monte Carlo (MCMC), those chains need to be adapted to the likelihood that they should investigate and maximize. Such an adaption process is performed in the so called \textit{pre-run}. The idea behind this phase is that the actual Markov Chains should work optimally in the actual MCMC. For that purpose, two features need to be adapted.

The first feature is the \textit{scaling factor}. BAT samples the proposal point in the Metropolis-Hastings-algorithm with a normal distribution with mean 0 and width 1. The obtained value will be scaled with that scaling factor and the result will be added to the value of the current point. This procedure is repeated for every parameter independently. Since the parameter space size is problem dependent and the likelihood function can be implemented by the user, the scale factor cannot be set prior by the authors of BAT. Rather, after a couple of steps, the scaling factor needs to be adjusted. This adjustment is based upon the efficiency of the certain Markov chain. The efficiency $\epsilon$ is defined as $\epsilon = \frac{\text{number of accepted proposal points}}{\text{number of proposal points}}$. If the efficiency is too low, e.g. if a Markov chain is on a maximum of the likelihood, then the scaling factor is reduced in order to find points that deliver a likelihood that is closer to the current value. On the other hand, if the efficiency is too high, the scaling factor is increased. In BAT, the default values for a good efficiency are given by 15-35\%. Inside this range, the scaling factor remains unchanged. If every Markov Chain is within this interval, the convergence of the scaling factors is assumed to be reached. Besides an overall scaling factor for the chain, BAT also provides the possibility to use a scaling factor for every parameter individually.\\

The second parameter that needs to be adapted is the \textit{R-value}\cite{ref:rvalue}. This parameter is an indicator for the convergence of the Markov chains with the demand that ``effective convergence of Markov chain simulation has been reached when inferences for quantities of interest do not depend on the starting point of the simulations''\cite[p.~435]{ref:rvalue}. In order to find a parameter representing convergence, the paper suggests to start with an estimation of the variance $\sigma^2$ of an underlying distribution. For the derivation, a random variable $\psi$ is assumed to be sampled according to an underlying distribution with mean $\mu$ and the variance $\sigma^2$. Those random variables are sampled from $m$ Markov Chains $n$ times. Out of the sampled values, an estimation of the mean $\hat{\mu}$ can be derived. The notation $\psi_{jt}$ represents the random variable from the $j$-th chain in the $t$-th iteration. A mean over one/both the indices will be presented as dot.
For the purpose of variance estimation, two parameters are introduced:

\begin{align}
B/n &= \frac{1}{m-1}\sum_{j=1}^m(\bar{\psi}_{j.} - \bar{\psi}_{..})^2 \\
W &= \frac{1}{m(n-1)}\sum_{j=1}^m\sum_{t=1}^n(\psi_{jt} - \bar{\psi}_{j.})^2
\end{align}

The parameter $B/n$ is called \textit{between-sequence variance} and handles the variance of the mean values. The parameter $W$ is called the \textit{within-sequence variance}. This parameter calculates the mixed variances of all random variables of all Markov Chains. Both parameters are meant to pool the random variables from the independent chains into two scalars.

The suggested unbiased variance based on the random variables $\hat{\sigma}_+^2$ is stated as

\begin{equation}
\hat{\sigma}_+^2 = \frac{n - 1}{n} W + \frac{B}{n}.
\end{equation}

Additionally, for ``accounting [..] the sampling variability of the estimator $\hat{\mu}$ yields a pooled posterior variance estimate of $\hat{V} = \hat{\sigma}_+^2 + B/(nm)$''\cite[p.~437]{ref:rvalue}. Using this, the estimation of the variance can be brought into a relation with the true variance $\sigma^2$ of the underlying distribution as

\begin{equation}
R = \frac{\hat{V}}{\sigma^2}
\end{equation} 

with the \textit{scale reduction factor} $R$. Since this parameter cannot be calculated due to the missing value of $\sigma^2$, $R$ is overestimated by using the (under)estimating value of $W$ instead of $\sigma^2$.\\

Since the sampling algorithm in BAT is not fixed due to the adaption of the scale factor for a single Markov Chain and different scale factors for every Markov Chain, this needs to be reflected in the calculation of the $R$-value because the variance of the sampling can change over time during the pre-run. For that purpose, the parameter $d$ will be used. This parameter is called \textit{estimated degrees of freedom} and treats the scale factor issues via the approximation $d\approx 2\hat{V}/\widehat{var}(\hat{V})$ with the estimated variance $\widehat{var}(\hat{V})$ of $\hat{V}$. The final, corrected and estimated $R$-value is stated as

\begin{equation}
\hat{R}_c = \frac{d + 3}{d + 1	\frac{\hat{V}}{W	}}.
\end{equation}

The hat above the $\hat{R}_c$ is meant to indicate that the value is only an estimation. The $c$ in the index is meant to represent the corrected by $d$ term of $\hat{R}$.

This $R$-value needs to reach a value close to 1 in order to represent convergence of the Markov chains. If the value is bigger, ``this suggests that either estimate of the variance $\hat{\sigma}^2$ can be further decreased by more simulations, or that further simulation will increase $W$, since the simulated sequences have not yet made a full tour of the target distribution''\cite[p.~437]{ref:rvalue}. The default configuration of BAT assumes a convergence for $\hat{R}_c \leq 1.1$\\
\medskip

\begin{figure}
	\centering
	\includegraphics[width=0.8\textwidth]{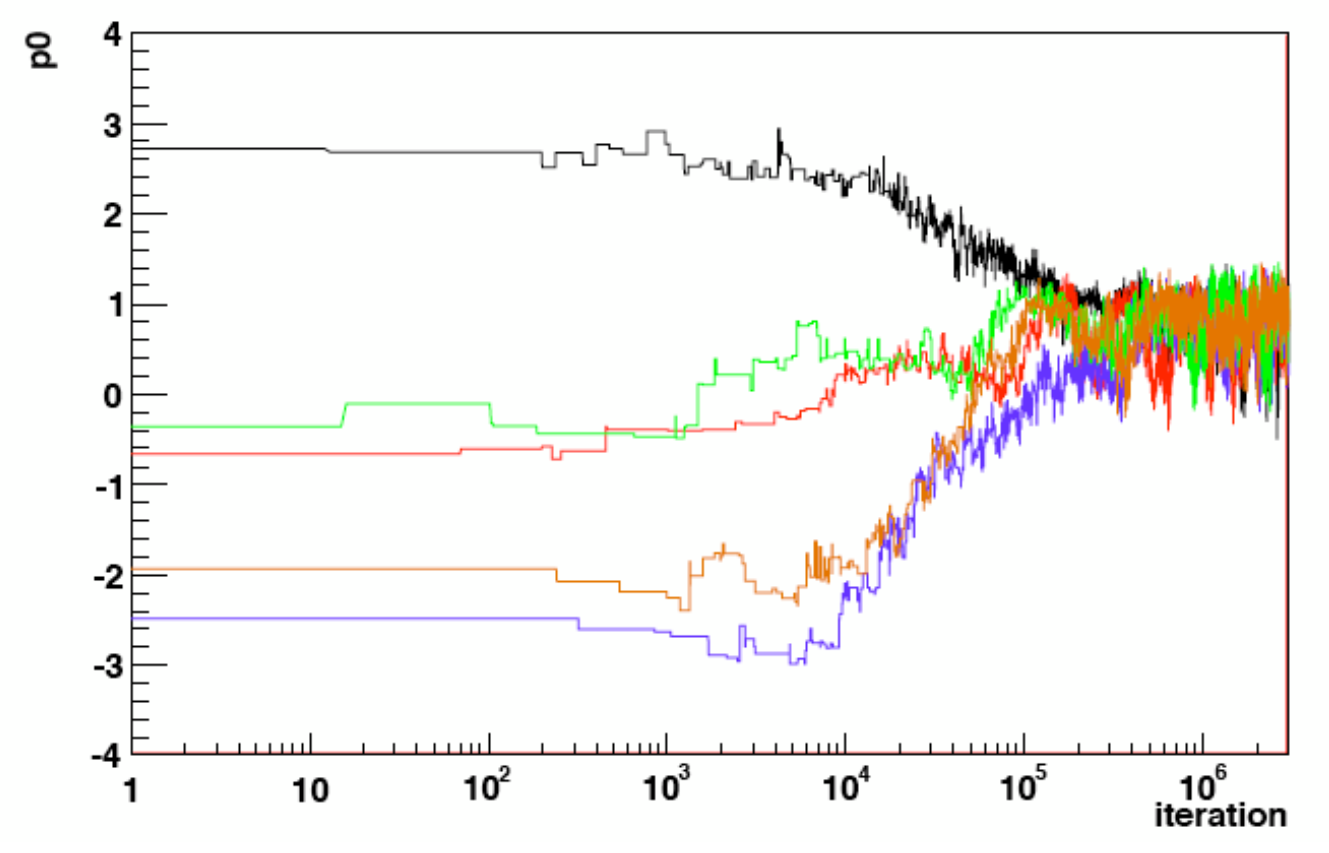}
	\caption{Example history of five Markov Chains for a parameter $p0$. The graph shows the evolution of the Markov Chains in the parameter range after every iteration. A convergence occurs after around $10^5$ iterations. This convergence is given by the observation that all Markov Chains are in the same region in the parameter space. The image is taken from Ref.~\cite{ref:mcmc}.}
	\label{fig:pretomainrun}
\end{figure}

The scale factor of each Markov chain and the $R$-value of each parameter will be adjusted during the pre-run and only during that phase. If a convergence is reached once, the \textit{main-run} starts. The evolution of multiple Markov Chains is shown in Fig.~\ref{fig:pretomainrun}. During the main-run, the pdf information of the pre-run will be ignored. The only information that remains in the main-run is the adjustment parameters. Using the likelihood-dependent configuration, the main-run will gather information of the log-likelihood function. With this information gathered after the main-run, the maxima of the posterior likelihood-function will finally be found by the \textsc{Minuit} package. The main difference between the \textsc{Migrad}- and the BAT-approach is the previous information gathering before the optimization process starts.

\section{Results}
\label{sec:result_bat}
The final tuning step as shown in Fig.~\ref{fig:tuning_steps} was repeated using BAT with an implementation similar to \textsc{Professor} except for the introduction of the \textit{Metropolis-Hastings-algorithm} before the optimization and a non-constant offset in the likelihood function, as it was shown in this Chapter. The parameters and corresponding parameter ranges were taken from Tab.~\ref{tab:parameter_ranges_nadine}. Since the approach implemented in BAT does not only work as an optimizer but also tries to gather information about the topology inside the parameter space, the run-time of this toolkit is significantly longer than in the case of a \textsc{Professor} based likelihood optimization. Hence, only a single run-combination will be optimized in BAT and compared to the result of the same run-combination optimized with \textsc{Professor}. The result cannot provide a ``proof'', but could give a hint to the source of the problem mentioned in Sec.~\ref{sec:double_peak}.

The same run-combination with all 650 samples was used as mentioned in Chapter \ref{ch:alter_ansatz} with the tuning result of \textsc{Professor} shown as the red dots in the images in App.~\ref{asec:distribution_reproduction}. The neutral weights mentioned in \cref{tab:particle_spectrum,tab:pdg_multiplicities1,tab:pdg_multiplicities2,tab:b_fragmentation,tab:event_shape,tab:diff_jet_rate} in App.~\ref{asec:observablesandweights} were used for both tunes.

The tuned parameter values obtained with \textsc{Professor} are shown in Tab.~\ref{tab:prof_tune} and the corresponding correlation matrix of the model parameters is shown in the upper part of Tab.~\ref{tab:bat_tune}.

\begin{table}[htbp]
  \centering
  \begin{tabular}{ l l l | l l }
    \hline
    & \multicolumn{2}{c |}{\textsc{Professor}} & \multicolumn{2}{c}{BAT} \\
    Parameter & Value & Uncertainty & Value & Uncertainty \\ \hline
    $a_{\textrm Lund}$ & 0.65 & 0.01 & 0.65 & 0.01 \\
    $b_{\textrm Lund}~[GeV^{-2}]$ & 1.33 & 0.02 & 1.32 & 0.02 \\
    $a_{\textrm ExtraDiquark}$ & 1.00 & 0.02 & 1.00 & 0.01 \\
    $\sigma~[GeV]$ & 0.291 & 0.001 & 0.291 & 0.001 \\
    $\alpha_s(M_Z)$ & 0.139 & 0.001 & 0.139 & 0.001 \\
    $p_{T,min}~[GeV]$ & 0.40 & 0.01 & 0.40 & 0.01 \\
    \hline
  \end{tabular}
   \caption{Tuned parameter values using neutral weights in the \textsc{Professor} framework and in BAT.}
  \label{tab:prof_tune}
\end{table}


Using the same weights and interpolation functions, the resulting parameter values and the corresponding uncertainties are shown on the right-hand side of Tab.~\ref{tab:prof_tune}. The stated values are the mode of the likelihood function and the smallest interval, as it was calculated by BAT. The correlation matrix of the parameters is shown in the lower part of Tab.~\ref{tab:bat_tune}. A complete overview over the marginalized posterior pdfs is shown in App.~\ref{asec:bat_prof}.

\begin{table}[htbp]
  \centering
  \begin{tabular}{ l  p{1.5cm} p{1.5cm} p{1.5cm} p{1.5cm} p{1.5cm} p{1.5cm}}
  	\hline
     & $a_{\textrm Lund}$ & $b_{\textrm Lund}$ & $a_{\textrm ExtraDiquark}$ & $\sigma$ & $\alpha_s(M_Z)$ & $p_{T,min}$ \\ \hline\hline
    $a_{\textrm Lund}$ & 1.00 & 0.64 & -0.03 & 0.02 & 0.00 & 0.02 \\
    $b_{\textrm Lund}$ & & 1.00 & -0.03 & -0.35 & -0.00 & -0.59 \\
    $a_{\textrm ExtraDiquark}$ & & & 1.00 & 0.00 & 0.00 & 0.02 \\
    $\sigma$ & & & & 1.00 & 0.00 & 0.19 \\
    $\alpha_s(M_Z)$ & & & & & 1.00 & -0.00 \\
    $p_{T,min}$ & & & & & & 1.00 \\
    \hline
  \end{tabular}
\newline
\vspace*{1 cm}
\newline
\centering
  \begin{tabular}{ l  p{1.5cm} p{1.5cm} p{1.5cm} p{1.5cm} p{1.5cm} p{1.5cm}}
  	\hline
     & $a_{\textrm Lund}$ & $b_{\textrm Lund}$ & $a_{\textrm ExtraDiquark}$ & $\sigma$ & $\alpha_s(M_Z)$ & $p_{T,min}$ \\ \hline\hline
    $a_{\textrm Lund}$ & 1.00 & 0.75 & -0.1 & 0.034 & -0.0053 & -0.0084 \\
    $b_{\textrm Lund}$ & & 1.00 & -0.078 & -0.3 & 0.051 & -0.42 \\
    $a_{\textrm ExtraDiquark}$ & & & 1.00 & 0.006 & -0.0051 & 0.021 \\
    $\sigma$ & & & & 1.00 & 0.06 & 0.13 \\
    $\alpha_s(M_Z)$ & & & & & 1.00 & 0.00027 \\
    $p_{T,min}$ & & & & & & 1.00 \\
    \hline
  \end{tabular}
  \caption{Upper table: Correlation matrix of the tuned model parameters using neutral weights in the \textsc{Professor} framework. Lower table: Correlation matrix of the tuned model parameters using neutral weights in BAT.}
  \label{tab:bat_tune}
\end{table}

Comparing first of all the tuned model parameter values provided by \textsc{Professor} with the modes extracted by BAT shows these are almost identical. Additionally, the correlation matrices show a similar correlation between the model parameters. Especially noticeable is the strong correlation between $a_{\textrm Lund}$ and $b_{\textrm Lund}$. Since those two model parameters belong to the same model, this could be expected. On the other hand, $a_{\textrm ExtraDiquark}$ and $\alpha_s(M_Z)$ do not show a strong correlation with any other parameter. With respect to the parameter ranges, both parameters are at the upper limit of their parameter range. Therefore, their correlation can be suppressed. The parameters $\sigma$ and $p_{T,min}$ are mostly correlated with $b_{\textrm Lund}$ and therefore with the fragmentation model. Since the first parameter is an event shape parameter, this parameter depends on the fragmentation. If more fragmentations occur then $\sigma$ is smaller in order to reproduce an event shape. This is described by the anti-correlation between $\sigma$ and $b_{\textrm Lund}$. The second parameter $p_{T,min}$ describes the end of the parton shower and the start of the hadronization and fragmentation. If the fragmentation starts at lower energies then more fragmentations need to occur in order to receive event shapes and particle multiplicities. This is described by the anti-correlation between $p_{T,min}$ and $b_{\textrm Lund}$. Additionally, the dependency of the fragmentation and the start of the hadronization is also showed by the correlation between $\sigma$ and $p_{T,min}$ in Tab.~\ref{tab:bat_tune}.\\
\medskip

\begin{figure}[htbp]
	\centering
	\includegraphics[width=0.7\textwidth]{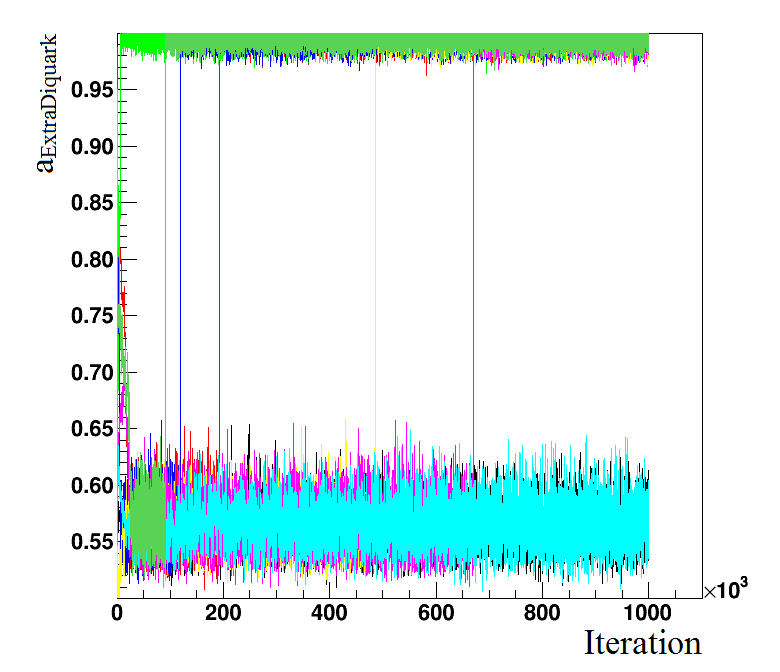}
	\caption{Overview over the history of the $a_{\textrm ExtraDiquark}$ values of all Markov Chains in the pre-run. The parameter range is given by Tab.~\ref{tab:parameter_ranges_nadine}.}
	\label{fig:aextradiquark_prof}
\end{figure}

\begin{figure}[htbp]
	\centering
	\includegraphics[width=0.7\textwidth]{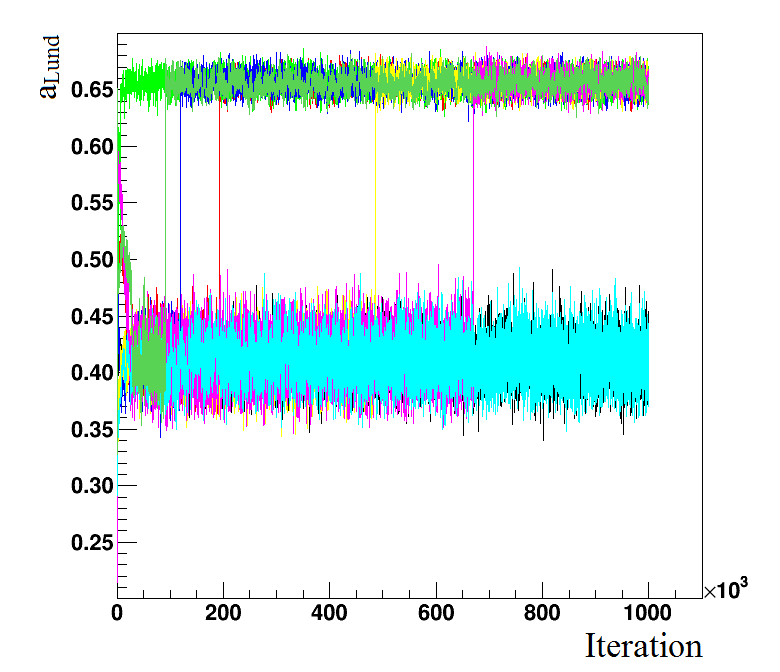}
	\includegraphics[width=0.7\textwidth]{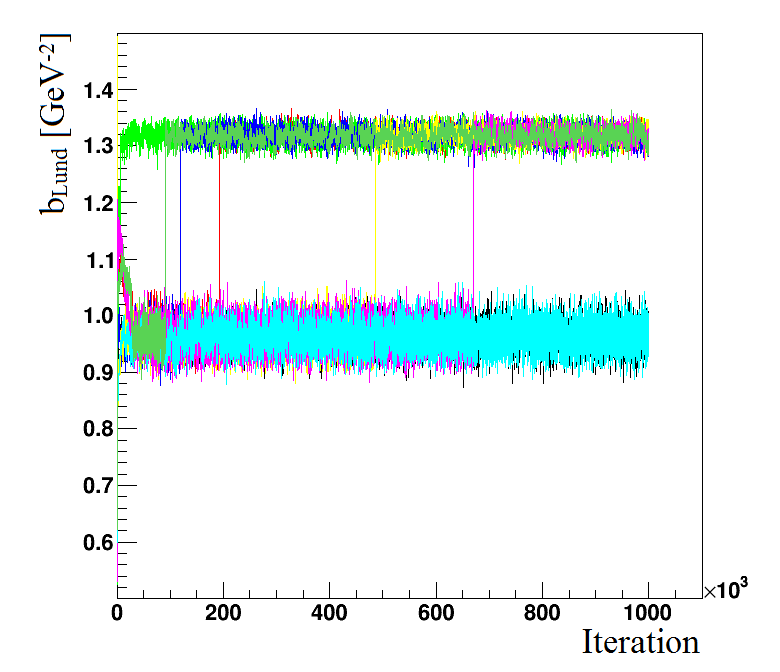}
	\caption{Top: Overview over the history of the $a_{\textrm Lund}$ values of all Markov Chains in the pre-run. Bottom: Overview over the history of the $b_{\textrm Lund}$ values in $[GeV^{-2}]$ of all Markov Chains in the pre-run. The parameter ranges are given by Tab.~\ref{tab:parameter_ranges_nadine}.}
	\label{fig:alund_prof}
\end{figure}

If the number of iterations is high enough, a MCMC approach can be used in order to minimize/maximize a function value. Without specifying how many iterations are sufficient, in the case of a limited number of steps, the result can still depend on the starting points. Therefore, a re-run is performed in order to check the correctness of the result. The configuration remains unchanged except for the number of iterations in the pre-run. Those are decreased from $10^7$ to $10^6$. The results of the pre-run for $a_{\textrm ExtraDiquark}$, $a_{\textrm Lund}$ and $b_{\textrm Lund}$ are shown in Fig.~\ref{fig:aextradiquark_prof} and \ref{fig:alund_prof}. Those histories provide a more detailed insight into the behavior of the Markov Chains. First of all, each chain shows an almost identical behavior in the three model parameters. If a chain has a lower value in $a_{\textrm ExtraDiquark}$, then the value of the same chain is also lower in $a_{\textrm Lund}$ and $b_{\textrm Lund}$ and vice versa. This behavior shows a strong correlation between these three parameters. In the correlation matrix of the re-run, these are given by coefficients between 0.95 and 0.98.

Another noticeable behavior is given by the fact that a chain only jumps from a lower attractor to a higher attractor but not in the other direction. Since those jumps should be quite symmetrical as shown in Subsection \ref{sec:numericalmethods}, this observation cannot be explained by referring to the construction of the Metropolis-Hastings algorithm. The explanation is given by the definition of the efficiency $\epsilon$ in Subsection \ref{sec:prerun}. The efficiency $\epsilon$ states whether the scale factor of the sampling should be increased, decreased or remain unchanged. The implemented calculation of this parameter considers the number of accepted sampling points. In BAT, a sampled proposal point outside the parameter range is treated as a declined point. As a result of this, if a Markov Chain is at the limit of the given range of one parameter, the symmetrical sampling leads to an efficiency of around 50\%, if the likelihood function is flat in that region. This leads to a decrease of the scaling factor as soon as a Markov Chain jumps to the limit of $a_{\textrm ExtraDiquark}$. The usage of a joint scaling factor of a single Markov Chain for every model parameter leads in the end to the behavior that the chains also remain in the upper attractor of $a_{\textrm Lund}$ and $b_{\textrm Lund}$ after they jumped to that value. Since the sampling is described by a normal distribution, a jump in the other direction is not forbidden, but unlikely due to the decreased scaling factor.

To summarize the behavior, using Markov Chains with BAT is quite problematic, if attractors are at or near the limit of the parameter range. 

A complete overview over the marginalized posterior pdf of the re-run is shown in App.~\ref{asec:rerun}. Using the higher values of the marginalized posterior pdfs with two peaks and the mode in the case of a single peak provides almost identical results compared to the \textsc{Minuit} optimization as shown in Tab.~\ref{tab:prof_tune} and \ref{tab:bat_tune}. Since the only change between both minimizations with BAT was given by the number of iterations in the pre-run, the same result could be obtained in the second execution of the toolkit if enough steps would be given for the pre-run. This argument is based on the previously described jumping behavior. The benefit of the reduction on the other hand was a better insight into the likelihood itself. All three model parameters from Fig.~\ref{fig:aextradiquark_prof} and \ref{fig:alund_prof} show the existence of at least one second attractor that seems to be only suppressed by the toolkit's algorithm.

The usage of BAT in order to control the minimization of \textsc{Professor} showed that the minimization seems not to be the problem because of the similarity of the results. On the other hand, BAT revealed that the interpolation function itself seems to have a possible second attractor that could be enhanced or suppressed by a subset in the run-combinations. The consequence of this observation is that the source of the double peaks shown in Fig.~\ref{fig:tune_nadine_weights0_nolimits} could be caused by a problem of the interpolation algorithm. This will be the content of the next chapter.

\chapter{Adaptive polynomial interpolation}
\label{ch:adaptive}
\setcounter{page}{58}
As shown in Sec.~\ref{sec:result_bat}, the problem mentioned in Sec.~\ref{sec:double_peak} could be caused by the interpolation algorithm used by \textsc{Professor}. In order to investigate this problem, another interpolation algorithm is tested. Since many different algorithms for different purposes exist, it is needed to specify what the algorithm should provide. Firstly, the algorithm should have a simple structure like the polynomials used by \textsc{Professor}. The benefit of such a function is given by a less memory consuming parametrization of the simulated data. This is especially important if the number of samples is large. Additionally, a polynomial structure allows a fast evaluation of the function value and its derivatives.

The choice of a polynomial function provides good comparison possibilities but needs further specifications about the algorithmic structure. The first criterion is based upon the fact that \textsc{Professor} allows the user to specify the order of the polynomial function but uses this order for each data point. At that point, the framework ignores if the order is suitable for the data point or not. This can lead to under- or overfitting. The first term refers to a function with too low an order in order to describe the data. The second term refers to a function with too high an order and therefore possible oscillations between data points. In order to avoid these problems, the polynomial order, or more specifically the number of terms in the polynomial function, should adapt itself to the corresponding problem. Such an iteratively performed adaption also needs a better quality criterion in order to judge if a number of terms is better or worse than another. If such a criterion would be only based upon the distance between the MC response and the function, the number of terms would always be the number of samples used for the construction of function. Therefore, this would always lead to overfitting.

In order to avoid this problem and provide a possible adaptive method for the interpolation, an approach is described in Sec.~\ref{sec:mathematicalmodel}. The explicit implementation of this approach is described in the Sec.~\ref{sec:implementationipol}. The result of a tuning using this algorithm is shown in Sec.~\ref{sec:resultsipol}. Using the new algorithm, a further consistency test of the equality of computing between BAT and \textsc{Professor} will be performed in Sec.~\ref{sec:adaptive_comparison}. In the last Section of this Chapter, an improved uncertainty calculation for the new interpolation will be presented and a more detailed insight into the tuning setup will be provided such that a tuned parameter vector can be obtained.

\section{Mathematical model}
\label{sec:mathematicalmodel}
The basic idea of the fitting algorithm which will be presented in this section is given in Ref.~\cite{ref:stablefit}. This paper addresses the problem of polynomial fitting without knowing the underlying order. For that purpose, the authors propose a procedure that adaptively determines the order without increasing the computing time significantly and still tries to remain stable without reducing the local accuracy too much.

In order to approach the problem, the initial formulation of an interpolation function $f_n(\vec{x})$ of degree $n$ with the tuple $\vec{x}=(x~y~z)$ in the case of three dimensions is given by

\begin{align}
f_n(\vec{x}) &= \sum_{0\leq i,j,k;i+j+k\leq n} a_{ijk}x^iy^jz^k \\
&= \underbrace{(1~x~...~z^n)}_{\vec{m}(\vec{x})^T}\underbrace{(a_{000}~a_{100}~...~a_{00n})^T}_{\vec{a}}
\end{align}

with the fit parameters $a_{ijk}$ that need to be determined. This is analogous to the formulation of Eq.~\ref{eq:prof_fit_func_rewritten}. The dot product can be extended to a set of tuples $\vec{x}_i$ introducing the matrix $M$. This matrix consists row-wise of the vectors $\vec{m}(\vec{x}_i)$. The matrix $M$ allows the formulation of the interpolation problem for a given set $\vec{b}$ of function values at the given $\vec{x}_i$. The relation $\|\{\vec{x}_i\}\|=dim(\vec{b})$ is fulfilled and allows an algebraic formulation as given in Ref.~\cite{ref:stablefit}:

\begin{equation}
M^T M\vec{a} = M^T \vec{b}.
\label{eq:stable_mateq}
\end{equation}

The solution of this equation system provides the fit parameters $\vec{a}$. The authors targeted a problem while solving this equation system, since ``one important reason for global instability is the collinearity of column vectors of matrix $M$, causing the matrix $M^TM$ to be nearly singular''\cite[p.~562]{ref:stablefit}. In order to handle this problem, the Eq.~\ref{eq:stable_mateq} will be modified by the \textit{ridge regression (RR) regularization}. This regularization shifts ``almost'' singular parts in a way that ``leads to losing much local accuracy while global stability is improved''\cite[p.~562]{ref:stablefit}. The explicit expression is given by

\begin{equation}
(M^T M + \kappa D)\vec{a} = M^T \vec{b}
\end{equation}

with the so called RR parameters $\kappa$ and a diagonal matrix $D$. $\kappa$ is specified to have a small positive value while the entries of $D$ are supposed to be positive. Numerically, this shift is a first step in order to regulate the interpolation parameters and shift the result towards the same order of magnitude. This is important with respect to machine precision. If the parameters would not be treated in the case of collinearity, the resulting parameters would become too large. Such a result could also become a problem with respect to a machine induced maximum value.

Since this shift, if necessary, can be included into the matrices, this extra term will not be mentioned explicitly, but remains in use.\\

In order to solve the equation system, the authors propose a QR-decomposition based on the Gram-Schmidt orthogonalization\cite{ref:qr}. The QR-decomposition \mbox{$M=Q_{NXm}R_{mXm}$} is based upon the matrix $Q$ which satisfies the property $Q^TQ=I$ with the identity matrix $I$. The matrix $R$ needs to be an upper triangle matrix. Using these properties Eq.~\ref{eq:stable_mateq} is transformed into

\begin{align}
R^TQ^TQR\vec{a} &= R^TQ^T\vec{b} \\
R\vec{a} &= Q^T \vec{b} \\
R\vec{a} &= \tilde{\vec{b}}
\label{eq:stable_qr}
\end{align}

with the modified $\tilde{\vec{b}} = Q^T\vec{b}$. Since $R$ is a triangular matrix, this system can be solved immediately. The missing part is that the matrix $M$ needs to be specified, especially if the order of the polynomial function should be adaptive to the problem. Additionally, the explicit construction of the matrices $Q$ and $R$ has not been explained yet.

These problems are addressed by the Gram-Schmidt orthogonalization. This procedure builds both matrices in an iterative way based upon a given $M$. The corresponding algorithm is given as

\begin{align}
\vec{q}_1 &= \vec{c}_1 / \|\vec{c}_1\| \\
r_{1,1} &= \|\vec{c}_1\| \\
\text{for } &1\leq i \leq \|\vec{m}(\vec{x})\|: \nonumber \\
r_{j,i+1} &= \vec{q}_j^T\vec{c}_{i+1}, \text{for } j\leq i \\
\vec{q}_{i+1} &= \vec{c}_{i+1} - \sum_{j=1}^i r_{j,i+1}\vec{q}_j \\
r_{i+1, i+1} &= \|\vec{q}_{i+1}\| \\
\vec{q}_{i+1} &= \vec{q}_{i+1} / \|\vec{q}_{i+1}\|
\label{eq:stable_qr}
\end{align}

with the $i$-th column $\vec{c}_i$ of $M$, the $i$-th column $\vec{q}_i$ of $Q$ and the $j$-th row and $i$-th column $r_{j,i}$ of the matrix $R$. Since $\|\vec{m}(\vec{x}_i)\| = \|\vec{m}(\vec{x}_j)\| ~\forall i,j$ for a given $M$, the index was neglected.

\begin{figure}[htbp]
	\centering
	\includegraphics[width=\textwidth]{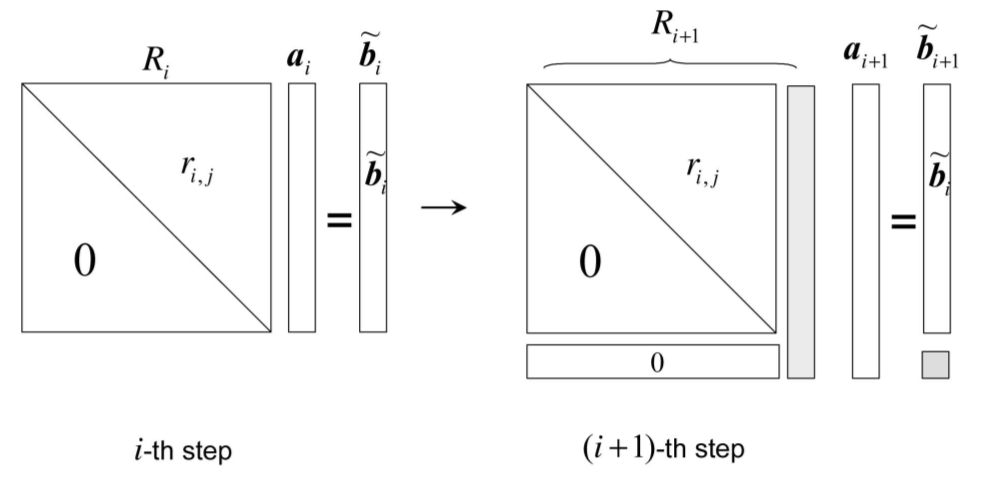}
	\caption{Left: Picture of the linear algebra problem that needs to be solved in order to obtain the fit parameters $\vec{a}_i$ after the $i$-th iteration. Right: Picture of the differences that need to be changed in order to formulate the problem after the $(i+1)$-th iteration. The gray shaded elements represent the part that actually needs to be calculated. The image is taken from Ref.~\cite{ref:stablefit}.}
	\label{fig:stableiterations}
\end{figure}

As the algorithm shows, the Gram-Schmidt orthogonalization provides the possibility to increase the matrices $Q$ and $R$ after every iteration. For both, only a new column needs to be calculated in every step until the fit reaches a certain quality. In order to modify the equation system in Eq.~\ref{eq:stable_qr}, the only further calculation needed is given by $\tilde{b}_{i+1}=\vec{q}_{i+1}^T\vec{b}$. This is summarized in Fig.~\ref{fig:stableiterations}.\\

After this equation system is set up, the solution can be calculated. In order to numerically stabilize these calculations, the RR-constraint can be used inside the decomposition to provide faster calculation. The impact of the $\kappa D$ term propagates inside this decomposition to the contributions

\begin{align}
\hat{r}_{ii} &= \sqrt{r_{ii}^2 + \kappa d_{ii}} \\
\hat{b}_i &= \frac{r_{ii}}{\sqrt{r_{ii}^2 + \kappa d_{ii}}}\vec{q}_i^T\tilde{\vec{b}} = \frac{r_{ii}}{\hat{r}_{ii}}\tilde{b}_i
\end{align}

with the diagonal element $d_{ii}$ in the $i$-th row of $D$.

In the case of a RR-constraint, these modifications lead to a regulated fit parameter

\begin{equation}
\hat{a}_i = \frac{\hat{b}_i}{\hat{r}_{ii}}=\frac{r_{ii}}{r_{ii}^2 + \kappa d_{ii}}\tilde{b}_i < \frac{\tilde{b}_i}{r_{ii}}=a_i \; ,
\label{eq:kappa}
\end{equation}

and therefore decreases the value of the fit parameter, since $\kappa D > 0$.

Another approach that was suggested by the authors to find fit parameters in the same order of magnitude is given by column-wise normalization of $M$ and $b$:

\begin{align}
M' &= \left(\frac{\vec{c}_1}{\|\vec{c}_1\| },\frac{\vec{c}_2}{\|\vec{c}_2\| },...,\frac{\vec{c}_n}{\|\vec{c}_n\| }\right) \\
\vec{b}' &= \frac{\vec{b}}{\|\vec{b}\| }
\end{align}

In the same way, $D$ can be ``normalized'' with the elements $d_{ll}' = \frac{d_{ll}}{\|\vec{c}_l\|^2 }$. The resulting, normalized QR-decomposition delivers $r_{ii}\in [0,1]$ and $Q'^T\vec{b}'\in (-1,1)$. The RR-constraint becomes therefore only important if $r_{ii}$ is too close to 0.

Since this normalization changes the fit parameter values in the interpolation, named $\vec{a}'$, the original fit parameters can be obtained via the rescaling $a_i = \frac{\|\vec{b}\|}{\|\vec{c}_i\|}a_i'$.

\begin{figure}[htbp]
	\centering
	\includegraphics[width=0.7\textwidth]{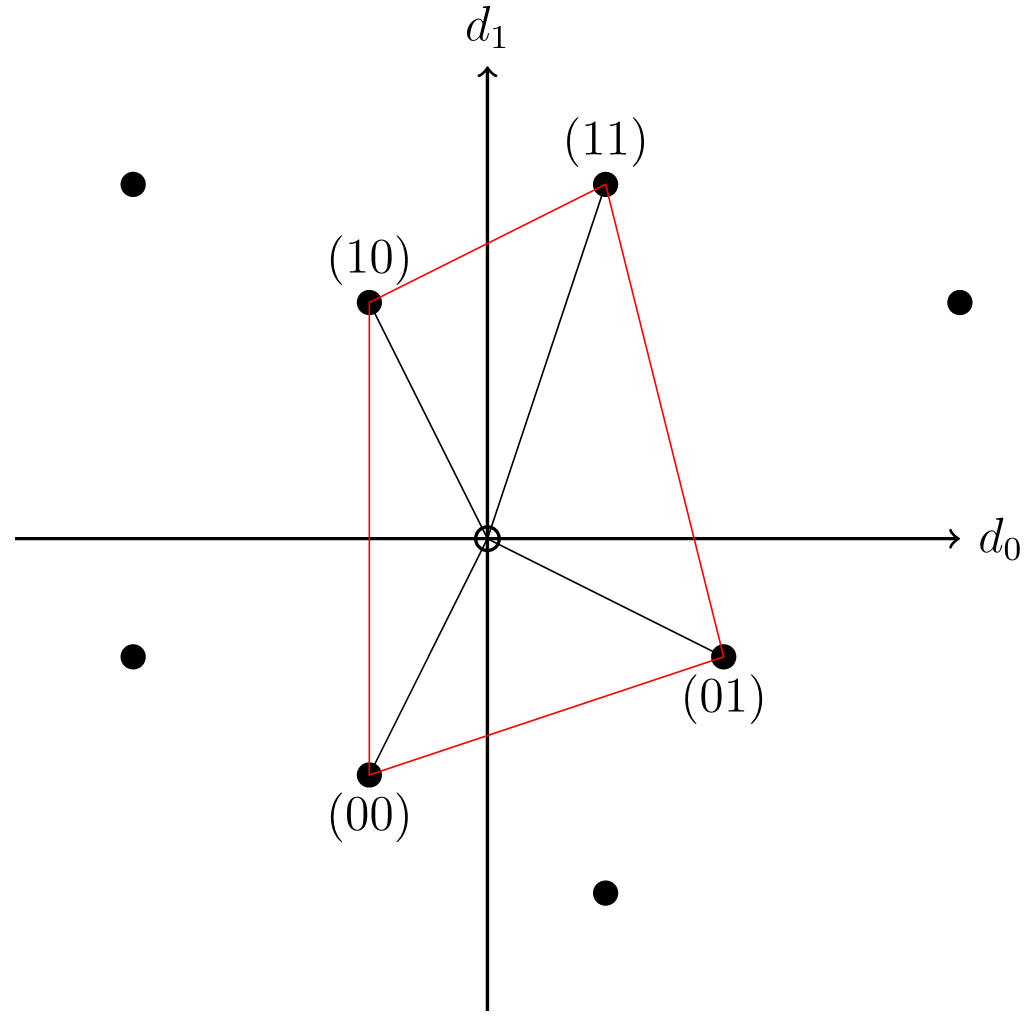}
	\caption{Selection of points surrounding a central point in the origin of the coordinate system. The selection surrounds the origin for the two-dimensional case with the parameters $d_0$ and $d_1$. The first number represents if a point has lower $d_1$-value (0) than the center or a higher one (1). The same numbering is performed for the second number with respect to the $d_0$-value. This binary representation implies that a $n$-dimensional parameter space needs $2^n$ surrounding points. The image is taken from Ref.~\cite{ref:nasa}.}
	\label{fig:nasa}
\end{figure}

The stopping criteria of this algorithm was also proposed by the authors, but this needs to be modified in order to fit better to the actual problem encountered in this thesis. Besides a single distance parameter (e.g. $\chi^2$) as a goodness-of-fit function, an additional parameter is introduced. This parameter is called $D_{smooth}$ and serves as shape judging parameter. The definition of this parameter is given as

\begin{equation}
D_{smooth} = \frac{1}{N} \sum_{i=1}^N (\vec{N}_i \cdot \vec{n}_i)
\end{equation}

with the number of points $N$, the normalized gradient vector $\vec{N}_i$ of the data points and the vector $\vec{n}_i = \frac{\nabla f(\vec{x}_i)}{\|\nabla f(\vec{x}_i)\|}$. While the latter can be calculated analytically, the former gradient vector $\vec{N}_i$ needs a special treatment. $\vec{N}_i$ should be a parameter that describes the shape of the data points without involving the interpolation function. As an easy approach to find its values, a surrounding hypercube needs to be found (cf. Fig.~\ref{fig:nasa}). Under the assumption that these points are close to the $i$-th point, a polynomial function of first order will be calculated based on all selected points and the $i$-th point. This analytic function allows to calculate the gradient vector and to normalize it.

Since the surrounding points do not need to be close to the center, they need to be re-weighted in a way that the weights decrease with increasing distance to the central point. Additionally, the central point needs to have the highest weight, since it is by construction the most important point.

Based on these two properties, a weighting scheme is chosen, that weights a point with index $j$ in contrast to all other $N-1$ points with $\left(1-\frac{dist_j}{\sum_{k=1}^N dist_k}\right)$. The variable $dist_j$ represents the Euclidean distance to the $i$-th point.\\
\medskip

In order to use a pure distance related parameter as stopping criterion, the $\chi^2$-function as defined in Eq.~\ref{eq:chi2_prof}, is used. The corresponding weights in that function are set to unity. The authors of Ref.~\cite{ref:stablefit} used the two stopping criteria independently and stopped the adaption via an \textit{AND}-connection of both quality requirements. Since first tests of the MC based problem showed many instabilities with this approach, a new parameter was build, which combines both into a single value. The problem in a mere product of $\chi^2$ and $D_{smooth}$ is, that a ``good'' fit is given for small $\chi^2$ whereas ``good'' function shapes are given by $D_{smooth}\rightarrow 1$. Therefore, the latter parameter was mapped onto a [0,$\infty$]-interval, such that smaller values refer to better fits. The resulting quality judgment parameter $\mathcal{F}$ is given by

\begin{equation}
\mathcal{F} = \chi^2 \cdot \frac{1 - D_{smooth}}{1 + D_{smooth}} \; .
\label{eq:def_f}
\end{equation}

A last modification is the explicit stopping itself. Consider that after the $i$-th iteration a small value of $\mathcal{F}$ is found and the next iteration delivers a larger $\mathcal{F}$-value. This could lead to a stopping although more iterations would yield a better $\mathcal{F}$-value than the $i$-th iteration. For this purpose, the sliding mean value over several iterations is calculated. In this thesis 100 iterations are used. If a $\mathcal{F}$-value is above this mean, then the iterations ends. Since the last iteration cannot be the best iteration with respect to the calculated participants of the mean, the best iteration that was found so far is re-calculated and used as the best result.\\
\medskip

So far, the calculation of the fit parameter estimation was described without concerning the uncertainties of the parameters. As Ref.~\cite[p.~10]{ref:blobel} shows, the $\chi^2$-function as mentioned in Eq.~\ref{eq:chi2_prof} is related to the covariance matrix as

\begin{equation}
\chi^2(\vec{a}) = (\vec{f}(\vec{a}|\vec{x}) - \vec{b})^T\Sigma^{-1}(\vec{f}(\vec{a}|\vec{x}) - \vec{b})
\label{eq:chi2covmat}
\end{equation}

with the vector $\vec{f}(\vec{a}|\vec{x})$ that contains the fit function $f(\vec{a}|\vec{x}_i)$ as the $i$-th component and $\vec{b}$ is constructed as shown in Eq.~\ref{eq:stable_mateq}. The matrix $\Sigma^{-1}$ describes the inverse covariance matrix. The vector $\vec{x}_i$ denotes the input parameter set that lead to the MC response $b_i$ at the corresponding data point. Since $\vec{x}_i$ is fixed during the calculation of the fit function, the function $f$ is only a function of the fit parameters $\vec{a}$. Therefore, the covariance matrix refers to the covariances between the fit parameters. $f(\vec{a}|\vec{x})$ is a polynomial function and therefore linearly dependent on the fit parameters. This allows to calculate derivatives from Eq.~\ref{eq:chi2covmat} such that the relation

\begin{equation}
\Sigma = \left(\frac{1}{2} \frac{\partial^2\chi^2}{\partial a_ia_j}\right)^{-1}
\label{eq:covmatchi2}
\end{equation}

is fulfilled. Performing this calculation yields the covariance matrix of the fit parameters.

In order to perform a $\chi^2$-minimization, the uncertainty of the function $f(\vec{a}|\vec{x})$ as shown in Eq.~\ref{eq:chi2_prof} needs to be calculated. For that purpose, a Taylor expansion of $f(\vec{a}|\vec{x})$ is performed around the fit parameter estimation $\hat{\vec{a}}$ \cite{ref:boris_grube}. In the case of a scalar function, this leads to

\begin{equation}
f(\vec{a}|\vec{x}) = f(\hat{\vec{a}}|\vec{x}) + \left. J\right|_{\vec{a} = \hat{\vec{a}}} (\vec{a} - \hat{\vec{a}}) + \mathcal{O}(\vec{a}^2)
\label{eq:errprop}
\end{equation}

with the Jacobian vector $J_{i} = \frac{\partial f(\vec{a}|\vec{x})}{\partial a_i}$. Using the definition of the covariance matrix for the scalar case $\sigma^2_f = E\left[(f - E[f])^2\right]$ with the expectation value operation $E[\cdot]$ and using Eq.~\ref{eq:errprop} leads to

\begin{align}
\sigma^2_f &= \sum_{i,j} \left.\frac{\partial f(\vec{a}|\vec{x})}{\partial a_i}\right|_{\vec{a} = \hat{\vec{a}}}\left.\frac{\partial f(\vec{a}|\vec{x})}{\partial a_j}\right|_{\vec{a} = \hat{\vec{a}}}E\left[(a_i - \hat{a}_i)(a_j - \hat{a}_j)\right] \\
&= \sum_{i,j} J_i \, \Sigma_{ij} \, J_j \\
&= J \, \Sigma \, J^T \; .
\label{eq:sigmaf}
\end{align}

In order to obtain this result, it was assumed that second and higher orders in Eq.~\ref{eq:errprop} could be neglected. Additionally, it was assumed, that $f(\hat{\vec{a}}|\vec{x}) \approx E[f]$.

Using $\chi^2$ allows for an uncertainty propagation of the function uncertainty with respect to linear correlations. Since the uncertainty calculation in the case of \mbox{\textsc{Professor}} neglected an explicit treatment of the correlation between the fit parameters, this will be done here, too. This allows a better comparability between both methods. The complete covariance matrix is used in Sec.~\ref{sec:covmat}.

\section{Results}
\label{sec:resultsipol}

\begin{figure}[htbp]
	\includegraphics[width=0.8\textwidth]{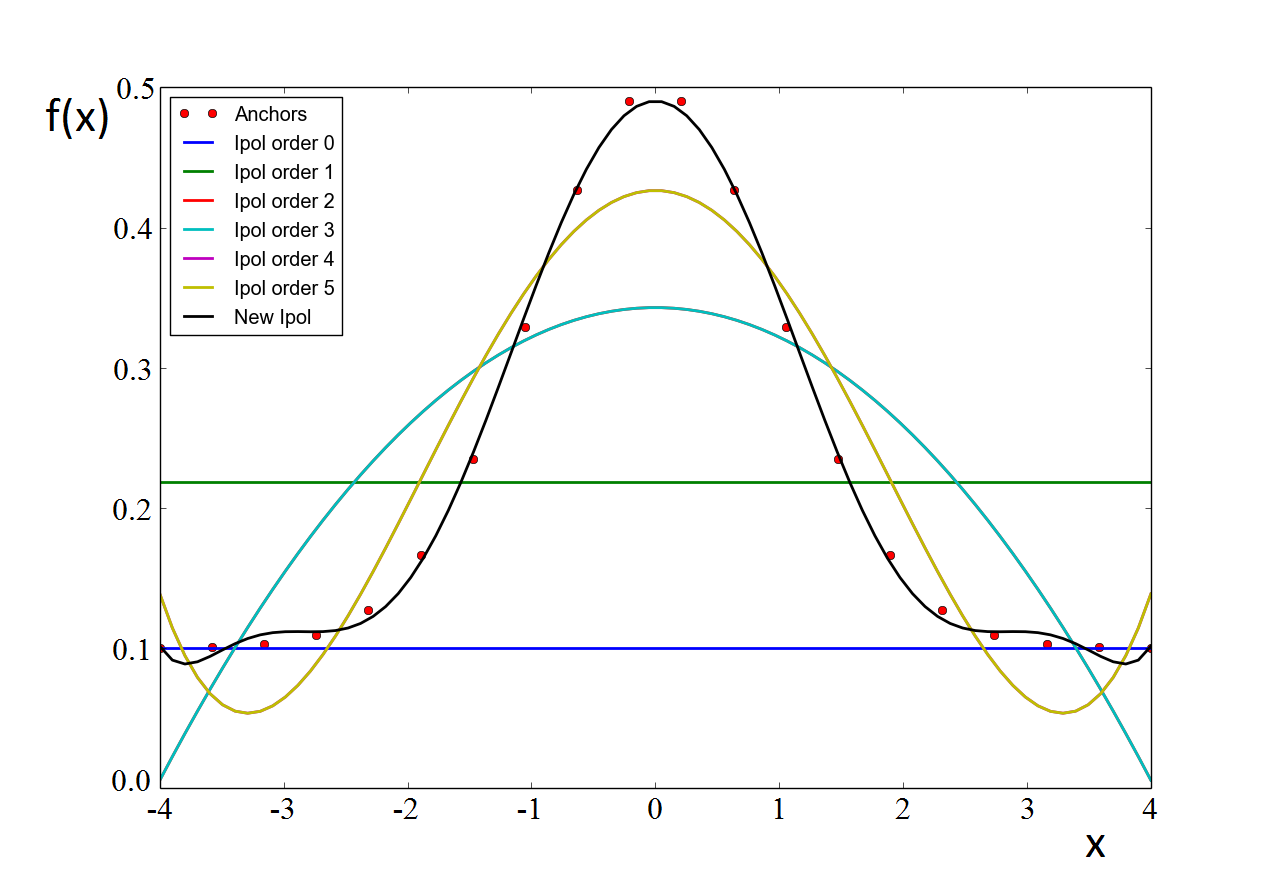}
	\includegraphics[width=0.8\textwidth]{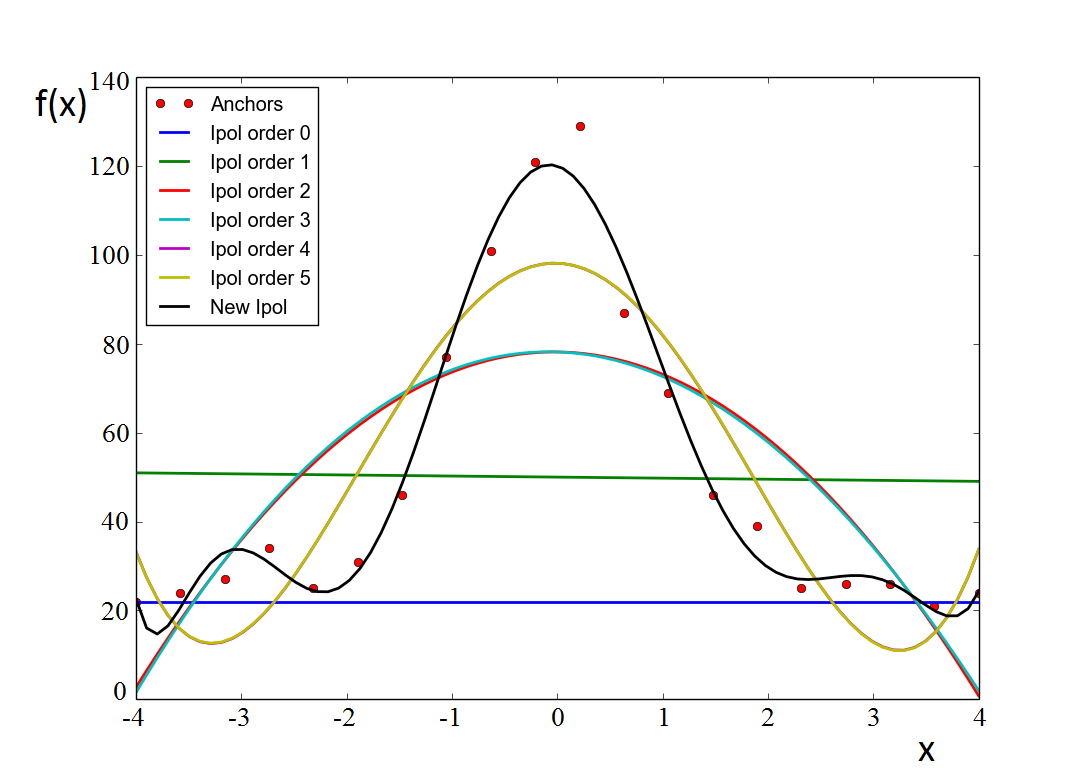} 
	\caption{Comparison between an interpolation performed by \textsc{Professor} up to the fifth order and the adaptive algorithm. Upper: Fitting of points that are normal distributed. Lower: Fitting of points that are normal distributed with additional noise.}
	\label{fig:stable_test}
\end{figure}

In order to present the result of the adaptive interpolation algorithm and to compare it with the \textsc{Professor} interpolation, first of all a toy example is used. Since the tuning problem for MC generators is high dimensional, a simpler problem depending on only one parameter is used in the beginning. This allows on the one hand a graphic representation and on the other hand the usage of a known underlying function. The example function in use is a normal distribution with a mean of zero and a width of one. Since this function cannot be reproduced by a polynomial function of finite order, the approximation of those is investigated in this example. Equidistant points are taken and the corresponding uncertainties of the points are set to the square-root of their function value. Interpolating this problem with various orders with \textsc{Professor} and the new interpolation algorithm leads to the top graph of Fig.~\ref{fig:stable_test}. The superposition of the functions shows that the new algorithm is able to adapt itself to the normal  distribution using a polynomial function of eighth order. By eye, the result fits the anchor points quite well, but shows signs of oscillations at the outer regions. A possible reason for this behavior could be a problem arising from the construction of $R$ in Eq.~\ref{eq:stable_qr} since the highest order terms are calculated first and can therefore empower those oscillations. With the interpolations from \textsc{Professor} up to the fifth order, the anchor points are not well described everywhere by the interpolation function.

Since the simulated data from MC generators are subject to statistical fluctuations, the bottom graph in Fig.~\ref{fig:stable_test} shows the same interpolation problem with an underlying noise in the normal distribution. The construction of the noise signal is based upon inverse transform sampling\cite[p.~13f]{ref:cdf}. The adaptive polynomial of ninth order follows the noise fluctuation more than the fixed order polynomials. This can be seen at the small bump around an abscissa value  of -3, that is at least roughly described by the new interpolation function. The fixed order polynomials however do not change significantly compared to the noiseless example.\\ 


In order to summarize the quality of the individual interpolation functions, Tab.~\ref{tab:stable_test_results} shows the resulting $\chi^2$ values of the interpolation functions displayed in Fig.~\ref{fig:stable_test}. The values of $\chi^2_{pred}$ represent the predictive power of the interpolation function. Those are constructed by using four times the amount of anchor points and are meant to control if the functions are able to predict points that were not used for the fitting itself. Since the underlying function is known in this case, those can be easily calculated. In this example, the parameter $\chi^2_{pred}$ is used as a quality criterion of the function shape. This parameter replaces $D_{smooth}$ in this example.\\

\begin{table}[htbp]
 	\centering
  	\begin{tabular}{ l | c c | c c}
    		\hline
    		& \multicolumn{2}{c |}{Noiseless} & \multicolumn{2}{c}{Noise} \\
    		Professor order & $\chi^2$ & $\chi^2_{pred}$ & $\chi^2$ & $\chi^2_{pred}$ \\ \hline
		0 & 1.66 & 6.90 & 393.97 & 2484.40 \\
		1 & 1.78 & 6.97 & 414.77 & 18862.16 \\
		2 & 0.77 & 2.82 & 221.87 & 15440.04 \\
		3 & 0.77 & 2.82 & 222.13 & 15437.09 \\
		4 & 0.20 & 0.73 & 91.97 & 11780.19 \\
		5 & 0.20 & 0.73 & 91.94 & 11780.65 \\ \hline
		New Ipol & $3.67\cdot 10^{-3}$ & $19.08\cdot 10^{-3}$ & 7.04 & 10699.88 \rule{0pt}{2.6ex} \\
    		\hline
  	\end{tabular}
\caption{Result of the interpolation of the Fig.~\ref{fig:stable_test}. The noiseless columns are for the upper, the noise columns for the lower graph.}
\label{tab:stable_test_results}
\end{table}

For the noiseless case, the even orders of the \textsc{Professor} interpolation improve the quality of the fit significantly. This behavior is related to the symmetric property of the normal distribution. On the other hand, the adaptive algorithm is able to fit the problem much better and reduces the $\chi^2$ by two orders of magnitude compared to the highest interpolation order used in \textsc{Professor}. A similar observation about the improvements can be made for the prediction test of the functions.

In the case of a noisy normal distribution, the \textsc{Professor} interpolations increase their goodness-of-fit continuously with increasing orders. As in the noiseless case, the new algorithm is able to provide a better $\chi^2$ value than the fixed order algorithms. A noticeable observation is that a polynomial function of zeroth order seems to have the best prediction power but this is mostly based upon the construction of the uncertainty calculation. The fact that smaller values have smaller uncertainties leads in this case to the situation that the polynomial function describes the anchor points with the smallest uncertainties best. This results in a misleading high prediction power. Again, the new algorithm is overall able to predict the underlying function best.

In this example, a normal distribution was used and the highest order of the fixed order interpolation was set to five. The general case will be that the underlying distribution is neither known nor can it be visualized. Additionally, the order of the interpolation function needs to be set a-priori. The corresponding recommendation of the \textsc{Professor} authors states ``that a second-order polynomial has so far been sufficient for almost all purposes in generator tuning''\cite[p.~7]{ref:professor}. Although that observation was made, it cannot be stated that higher orders are not needed in a tuning process. Furthermore, every a-priori guessed order cannot be justified as sufficiently high enough.\\

For the purpose of MC generator tuning, the configuration of the adaptive interpolation algorithm needs to be specified. The threshold of the RR-constraint as well as $\kappa$ is set to $10^{-10}$ in order to just evade singularities. The number of iterations used for the calculation of the sliding mean is set to 100. The inverse distance is weighted by a power of one. Since the construction of the normalized gradient vectors of the simulated data by using a polynomial function of first order needs a high point density, the number of parameter vectors is increased from 650 to 1500.

\begin{figure}[htbp]
	\centering
	\includegraphics[width=0.8\textwidth]{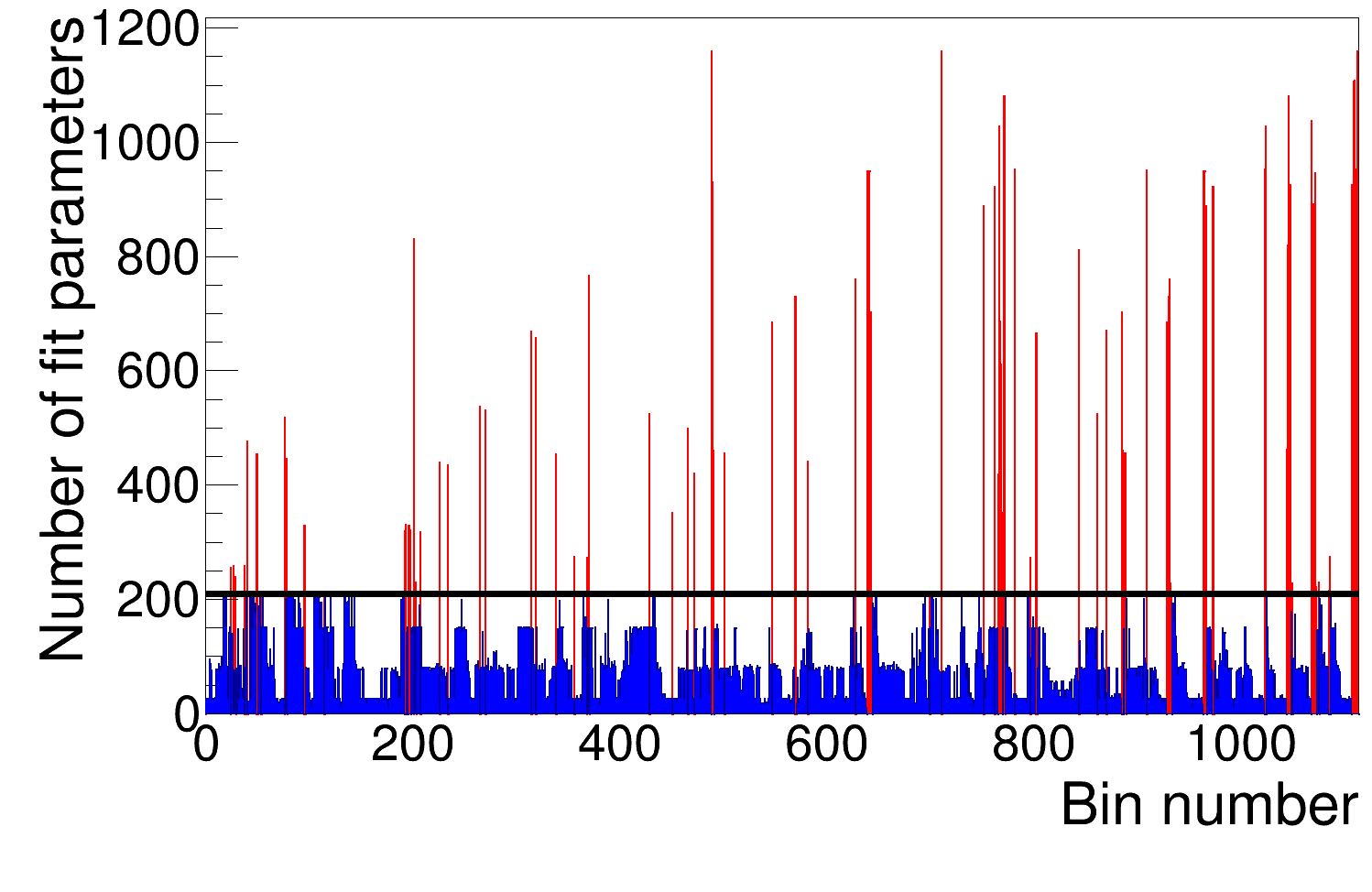}
	\caption{Number of fit parameters needed by the adaptive interpolation algorithm per data point. The data points are enumerated according to the order they were read. The black line indicates the number of terms used by \textsc{Professor} in order to interpolate the Problem using a polynomial function of fourth order. The blue parts are data points that needed less, the red parts are data points that needed more terms than \textsc{Professor}.}
	\label{fig:num_fitparams}
\end{figure}

In the case of the MC generator tuning, the adaptive algorithm uses a data point dependent number of terms. An overview over the number of terms needed for the 1115 data points that are interpolated for the tuning problem mentioned in Chapter \ref{ch:alter_ansatz} are shown in Fig.~\ref{fig:num_fitparams}. As the colors indicate, most of the data points can be interpolated with a polynomial function with a lower order than the fourth order used in Ref.~\cite{ref:nadine_fischer}, but some data points seem to need more terms in order to describe the response of the MC generator.

\begin{figure}[htbp]
	\includegraphics[width=\textwidth]{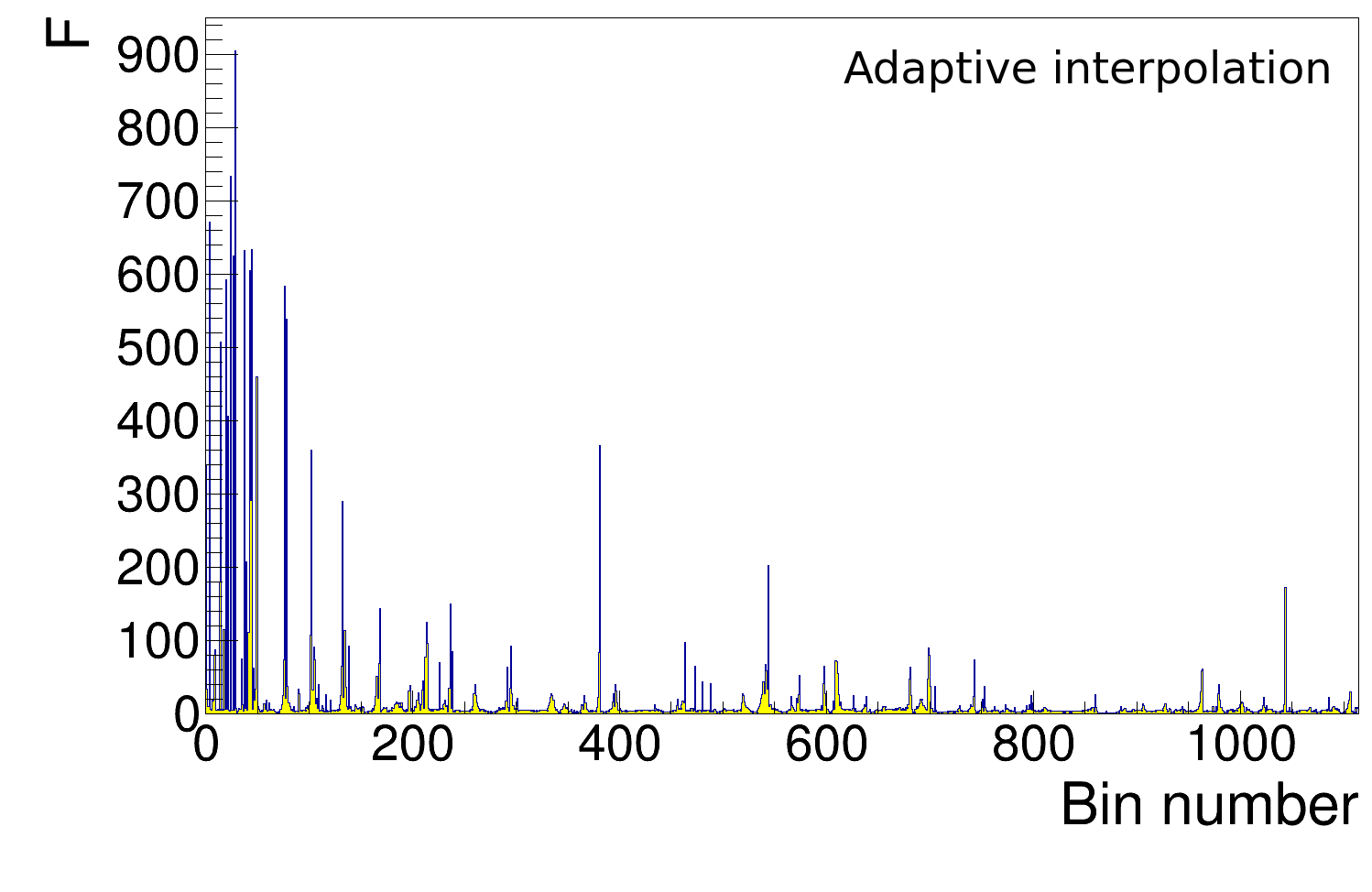}
	\includegraphics[width=\textwidth]{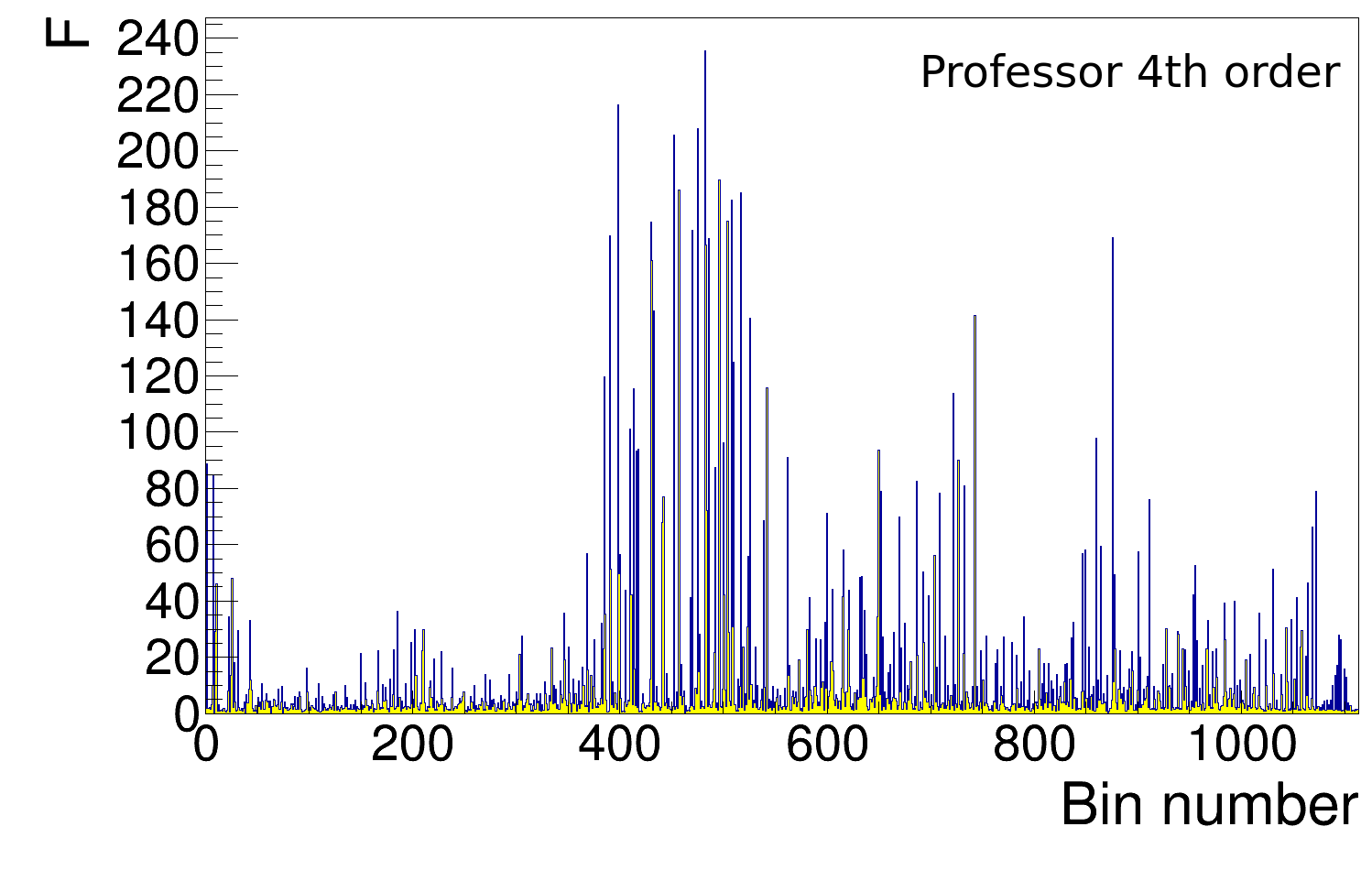} 
	\caption{Resulting $\mathcal{F}$ values for the different data points. The data points are enumerated according to the order they were read. The top graph shows the $\mathcal{F}$ values of the adaptive, the bottom one of the fixed order interpolation.}
	\label{fig:f}
\end{figure}

This leads to the question, if the MC response really has a complex structure. In order to investigate the quality of the fits and to compare it with the result of the fixed order polynomial, several parameters will be used. The first parameter is the $\mathcal{F}$ parameter (see Eq.~\ref{eq:def_f}) itself. The data point-wise values are shown in Fig.~\ref{fig:f}. The new algorithm shows for the low numbered data points high values for $\mathcal{F}$, but in the higher numbered data points almost no large values. The \textsc{Professor} algorithm shows smaller highest values but has around the data point number between 400 and 600 rather high values. Above the 600th data point, the values become lower but remain rather high.

The adaptive algorithm is built such that it optimizes the $\mathcal{F}$ value. If this is the case, the low numbered data points need to produce higher values for other number of terms. The fixed order polynomial was just used without optimizing this parameter. But the values are found to be smaller. A possible source of this problem is that the construction of the quality criterion treats the $\chi^2$ linearly, but the smoothness in a non-linear way. Such a combination could be the source of the high resulting $\mathcal{F}$ values shown here. On the other hand, the $\mathcal{F}$ value seems to be a better quality criterion for the data points in the region of data point number 400 and higher.


\begin{figure}[htbp]
	\centering
	\includegraphics[width=0.8\textwidth]{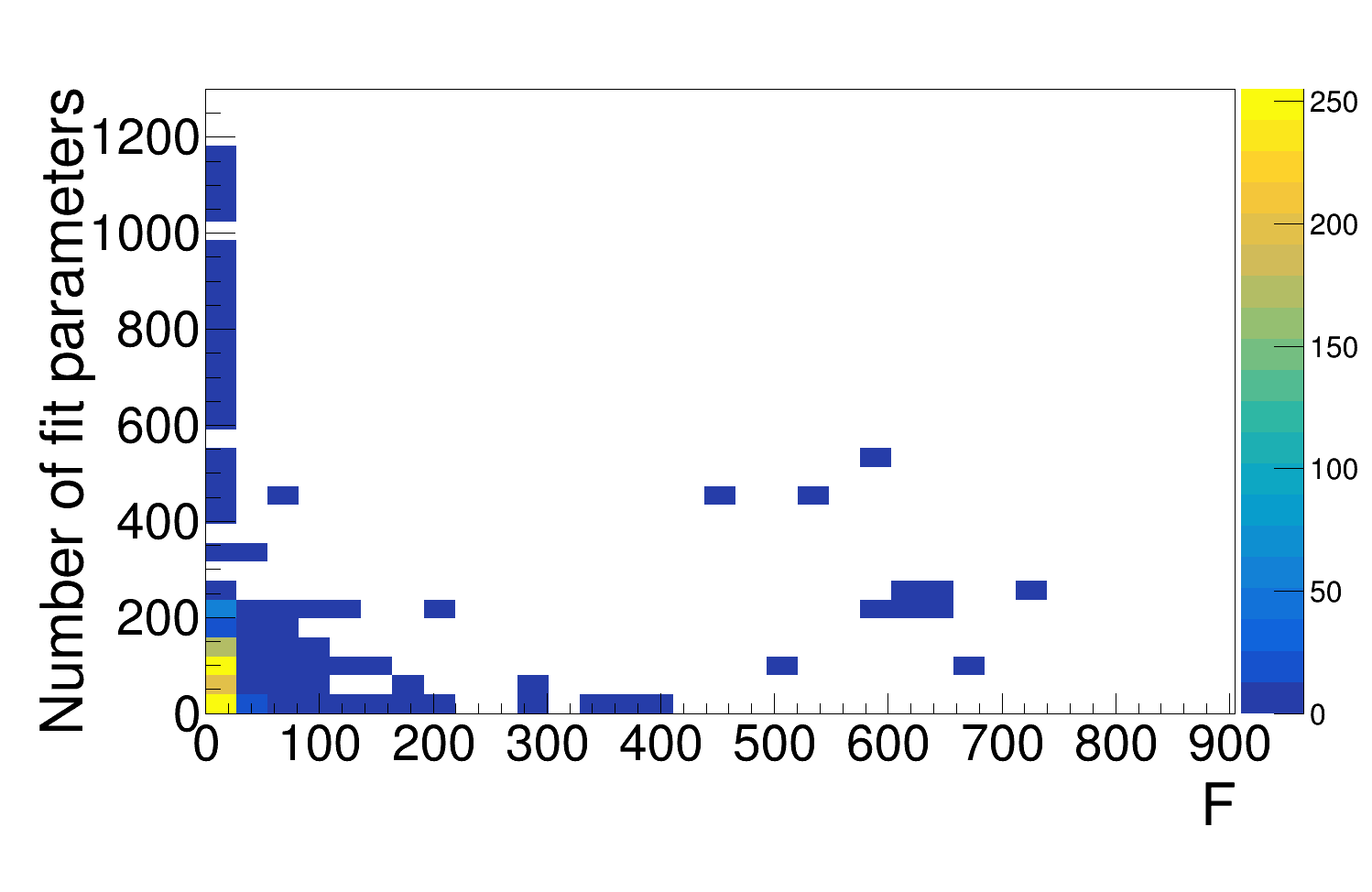}
	\caption{Number of interpolation parameters needed for every data point and the corresponding $\mathcal{F}$ values.}
	\label{fig:f_num_fitparams}
\end{figure}

Until now, the number of fit parameters and the corresponding $\mathcal{F}$ value were treated independently. As final judgment about the impact of a complicated MC response, and therefore a higher number of terms in the polynomial function upon the resulting $\mathcal{F}$ value, the Fig.~\ref{fig:f_num_fitparams} will be used. As the map shows, the highest number of terms produces very good results with respect to the $\mathcal{F}$ value. On the other side, the data points with less interpolation parameters show the largest $\mathcal{F}$ values. This observation could indicate that the MC response in such data points is possibly not sufficiently smooth as it was assumed in Sec.~\ref{sec:tuning_approach}. The result would be that such a MC response could not be well described and further terms lead to bad oscillations rather than optimizing the parametrization. Since the sliding mean is used with a given number of terms, a cutoff could be performed before a sufficiently high order was reached but this is, due to the high dimension of the problem, just an assumption.

\begin{figure}[htbp]
\makebox[\textwidth][c]{
	\begin{adjustbox}{max width=1.\textwidth}
	\includegraphics[width=\textwidth]{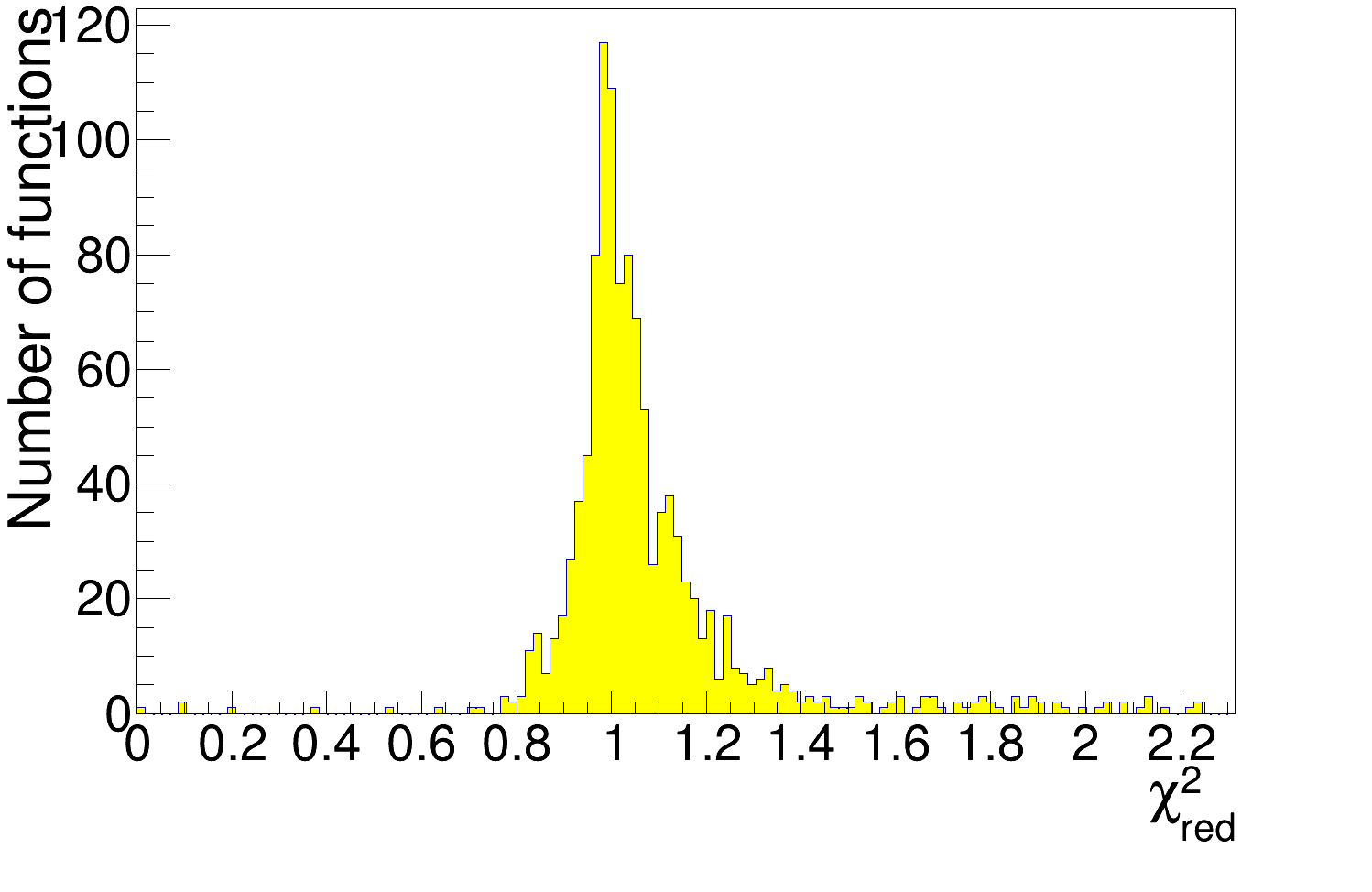}
	\includegraphics[width=\textwidth]{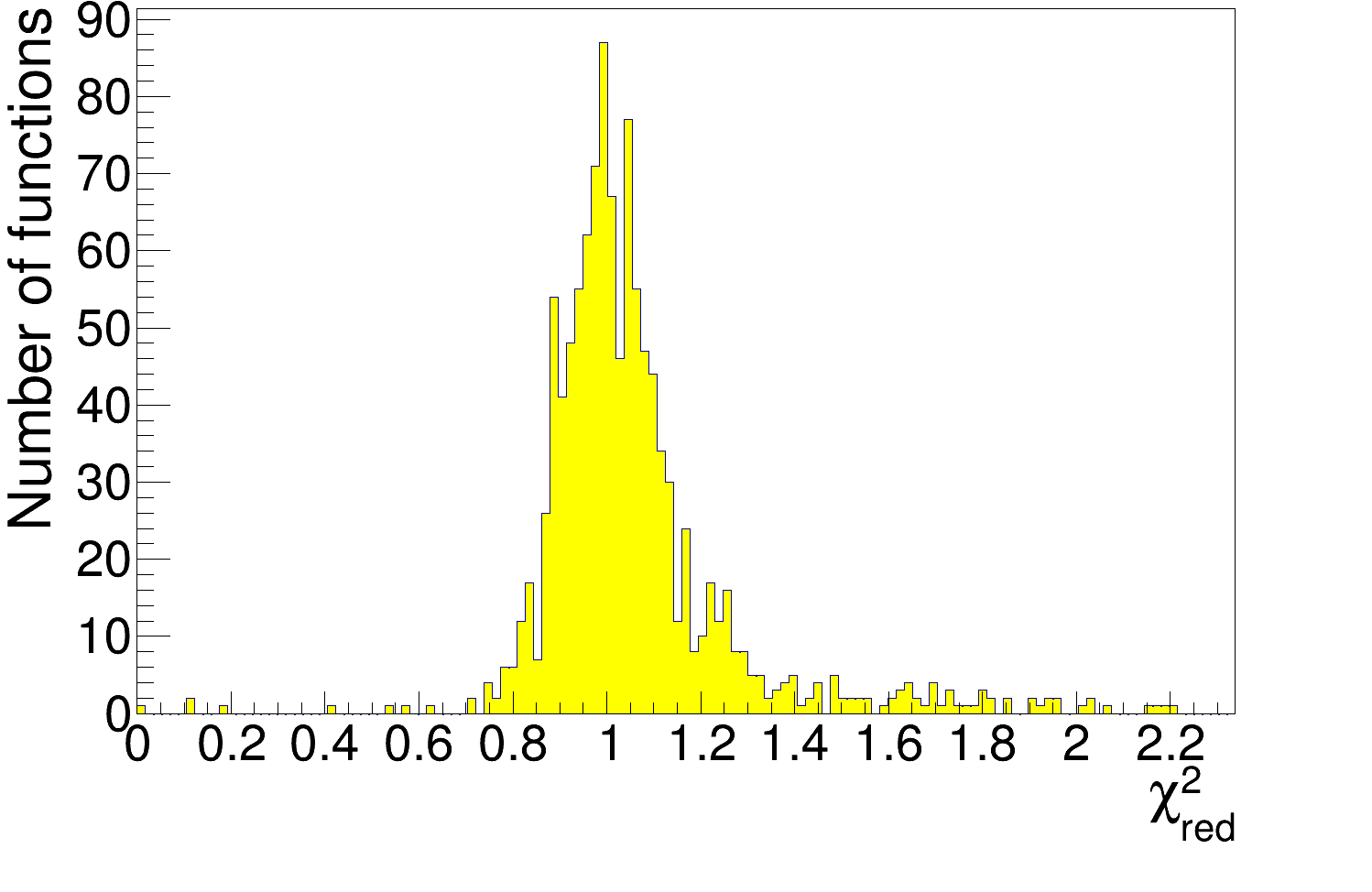}
	\end{adjustbox}
}
	\caption{Distribution of the $\chi^2_{red}$ values for every interpolated data point. The left graph shows the distribution of the adaptive, the right one the distribution of the fixed order interpolation.}
	\label{fig:chi2_distr}
\end{figure}

A further parameter that can be used in order to judge the quality of the fit is the usage of the $\chi^2$ alone. But since the number of terms used for the interpolation is different for every data point, a graph showing the $\chi^2$ distribution would be useless. Rather the usage of the $\chi^2_{red}$ distribution could be used. The corresponding distributions of both interpolation algorithms are shown in Fig.~\ref{fig:chi2_distr}. It can be shown\cite{ref:caldwell} that such a distribution is a normal distribution with a mean of one, if the used model, in this case a polynomial function, is the right model in order to describe the data. Both distributions indicate such a distribution, but with an additional tail area for $\chi^2_{red}>1$. The adaptive algorithm has a mean of $\overline{\chi^2_{red}}=1.08$ with a mean error of $\Delta\chi^2_{red}=0.22$. The fixed order approach has a mean of $\overline{\chi^2_{red}}=1.06$ with a mean error of $\Delta\chi^2_{red}=0.22$.


\begin{figure}[htbp]
	\centering
	\includegraphics[width=0.8\textwidth]{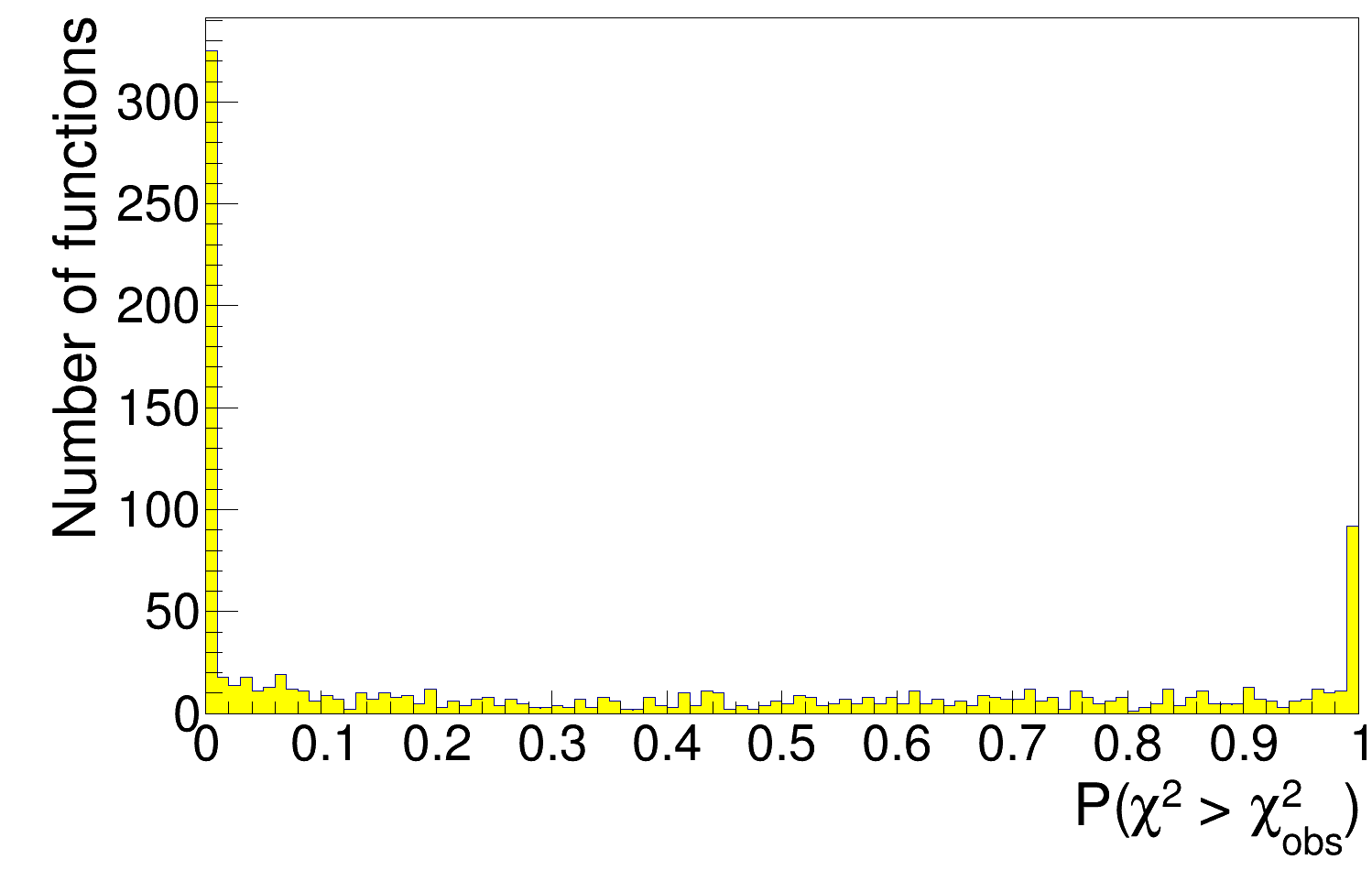}
	\includegraphics[width=0.8\textwidth]{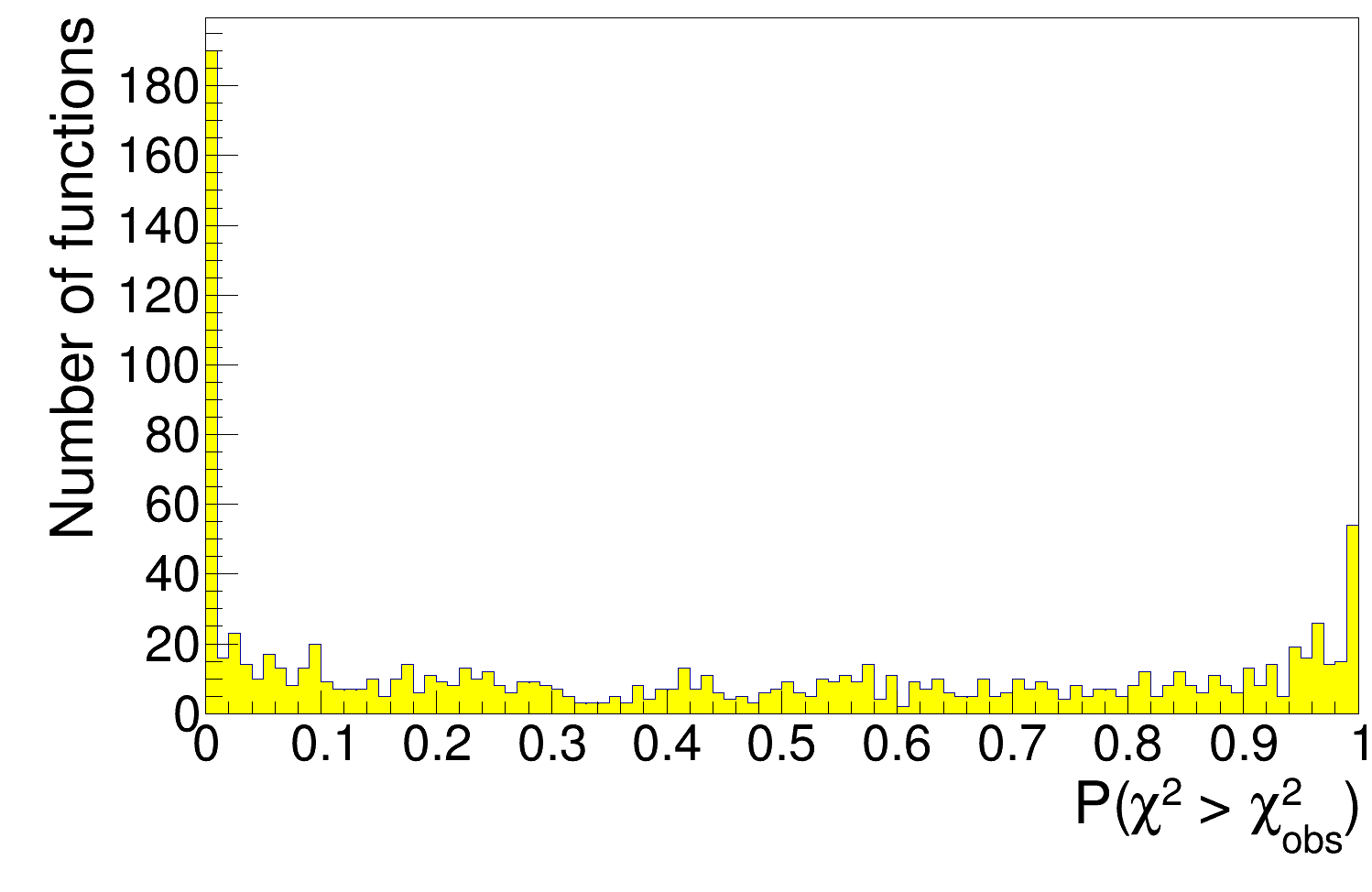}
	\caption{Distribution of the p-values of every interpolated data point using the adaptive interpolation. Upper: Distribution of the adaptive interpolation. Lower: Distribution of the fixed order polynomial.}
	\label{fig:pvalue_distr}
\end{figure}

Using the $\chi^2$ values allows the investigation of another parameter, the $p$-value distribution. This is shown for both algorithms in Fig.~\ref{fig:pvalue_distr}. If the used function would behave according to the underlying model of the data, the $p$-value distribution would be flat\cite{ref:caldwell}. As can be seen in the region of abscissa values around zero and one, that is not the case. The very small values correspond to the tail of $\chi^2_{red}>1$ in Fig.~\ref{fig:chi2_distr}, the very big to the tail of $\chi^2_{red}<1$. A statement that can be derived of that distribution is that the smaller width of the $\chi^2_{red}$ distribution is beneficial for the purpose of the description of the MC response. But this graph also shows that a polynomial function does not seem to be able to correctly describe the underlying distribution of the MC response.

\begin{figure}[htbp]
	\centering
	\includegraphics[width=0.8\textwidth]{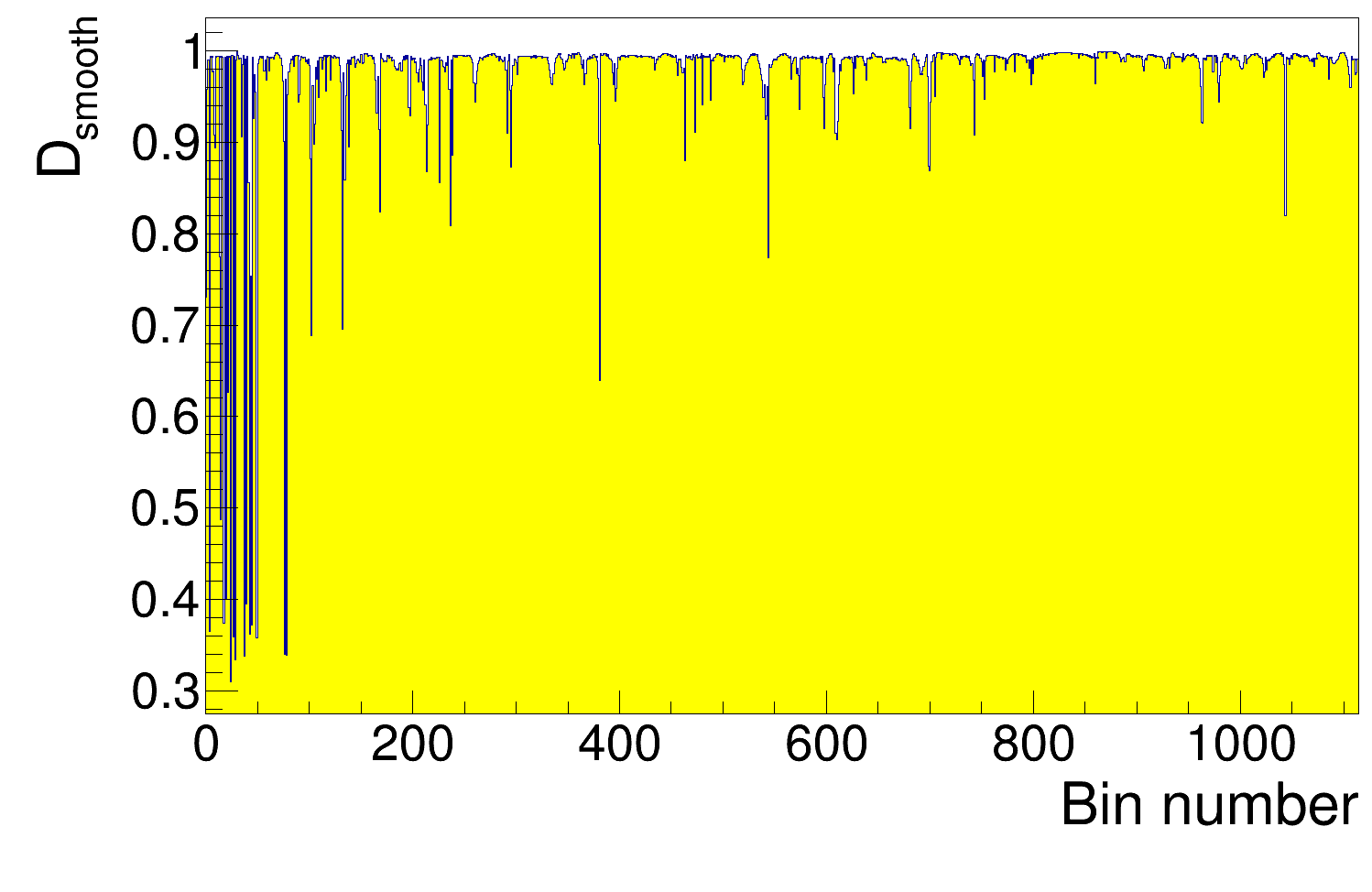}
	\caption{Distribution of $D_{smooth}$ values for all interpolated data points using the adaptive interpolation.}
	\label{fig:dsmooth}
\end{figure}

Since the $\chi^2$ and $\chi^2_{red}$ values do not explain the very high $\mathcal{F}$ values in the adaptive interpolation approach, the second parameter needed for the construction of that criterion will be investigated. The $D_{smooth}$ parameter per interpolated data point is shown in Fig.~\ref{fig:dsmooth}. This graph shows that especially the first data points have a very low smoothness. Due to the non-linear usage of this parameter for the $\mathcal{F}$ calculation, its impact seems higher than the $\chi^2_{red}$ values. Another way to express this observation is that the optimization process was actually judged based on the smoothness and only slightly on the distances.

\begin{figure}[htbp]
\makebox[\textwidth][c]{
	\begin{adjustbox}{max width=1.\textwidth}
	\includegraphics[width=\textwidth]{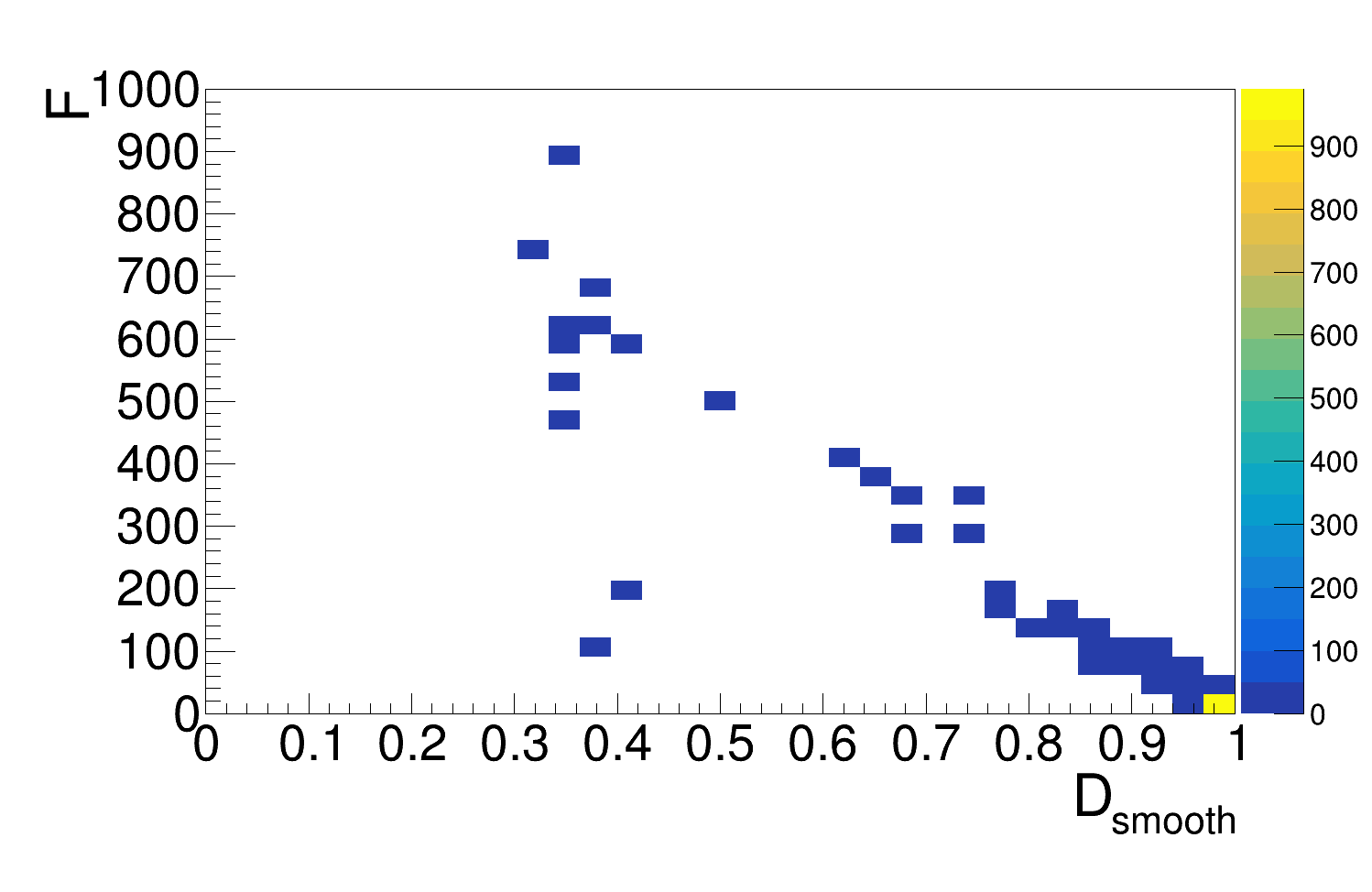}
	\includegraphics[width=\textwidth]{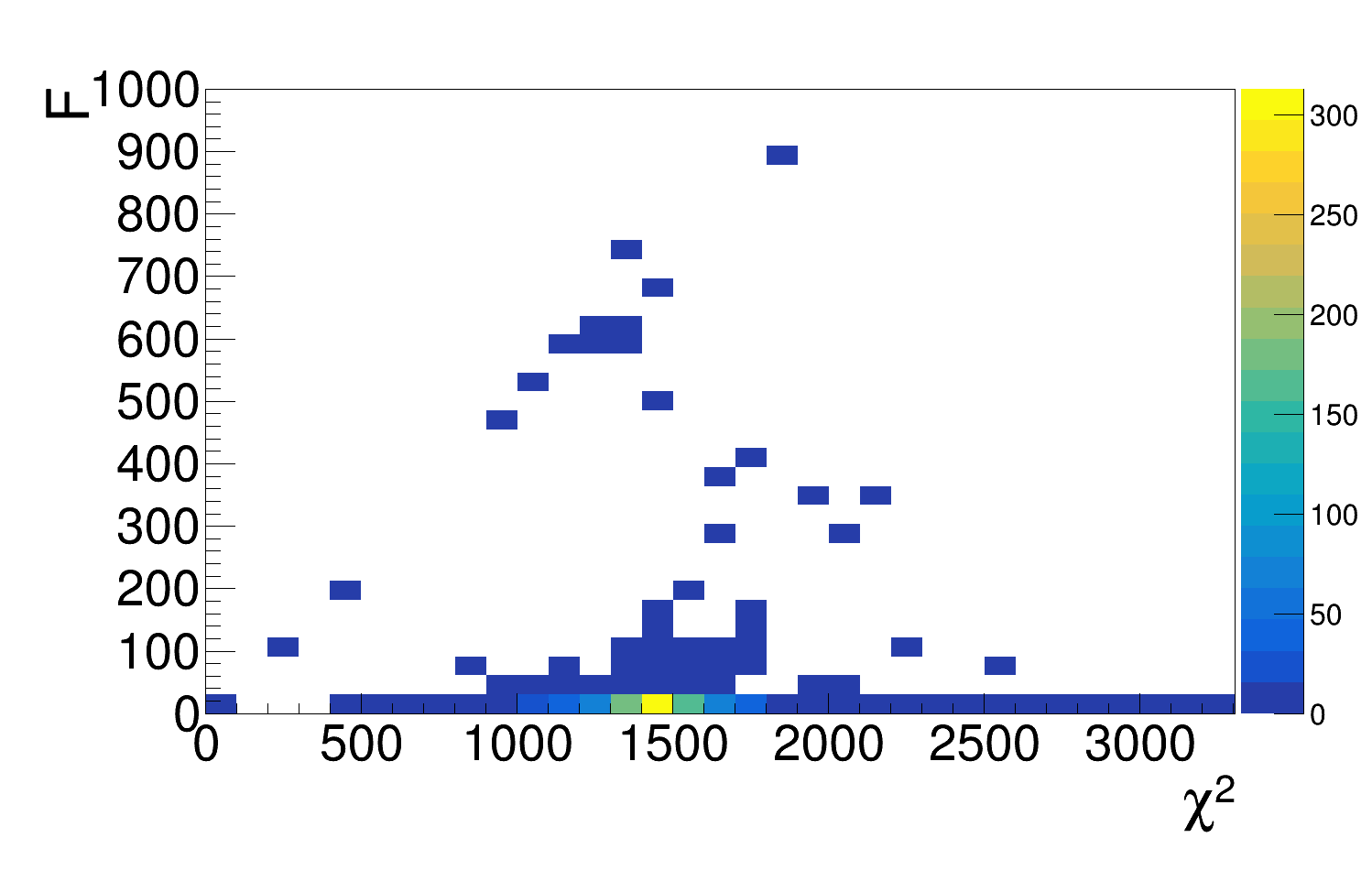} 
	\end{adjustbox}
}
\caption{Impact of $D_{smooth}$ (left) and of $\chi^2$ (right) on the resulting $\mathcal{F}$ values using the adaptive interpolation.}
\label{fig:impact_f}
\end{figure}

In order to investigate the actual impact of both parameters on the $\mathcal{F}$ values, those were shown individually as a function of the $\mathcal{F}$ value in Fig.~\ref{fig:impact_f}. As assumed, the $D_{smooth}$ parameter is strongly anti-correlated with the $\mathcal{F}$ value. On the other hand, $\mathcal{F}$ and $\chi^2$ do not seem to have a strong correlation.


\begin{figure}[htbp]
	\centering
	\includegraphics[width=0.8\textwidth]{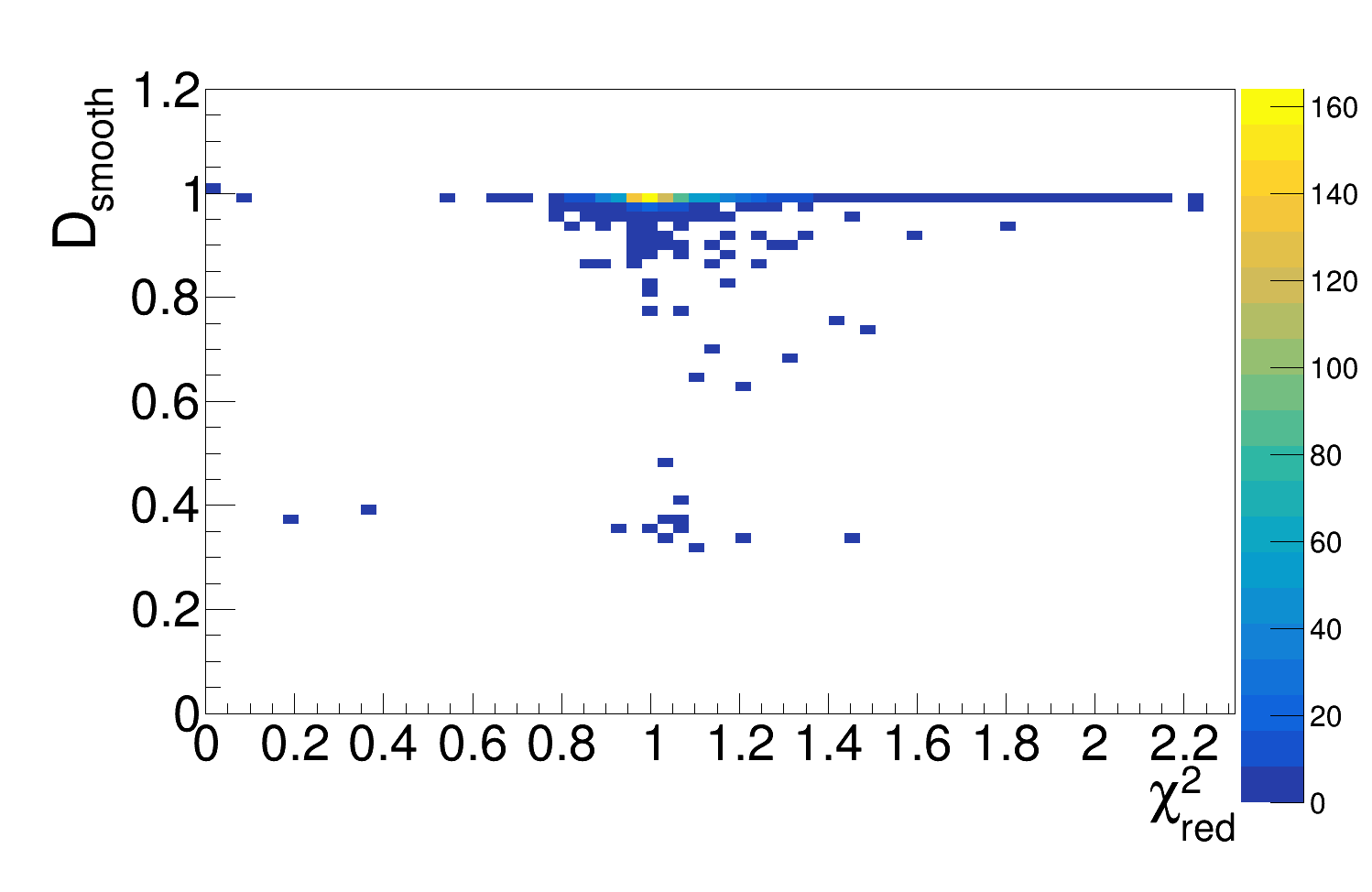}
	\caption{$D_{smooth}$ and $\chi^2$ shown for every interpolated data point using the adaptive interpolation.}
	\label{fig:dsmooth_chi2}
\end{figure}

As a last investigation of the fitting quality, the question remains if $\chi^2$ and $D_{smooth}$ are correlated. If so, then the introduction of $\mathcal{F}$ would be meaningless and could be expressed as a function of one of these parameters. The result of such a possible parametrization would be that the shape could be ignored. In order to investigate this dependency, both parameters were plotted against each other. This is shown in Fig.~\ref{fig:dsmooth_chi2}. As the image shows, both parameters do not show a strong correlation, since decreasing $D_{smooth}$ values are mostly centered around constant values of the reduced $\chi^2$.

Until now, only the possible benefit of $\mathcal{F}$ values upon the fitting procedure was presented. In order to compare between the adaptive and the fixed order polynomial functions, the means of several parameters are shown in Tab.~\ref{tab:compare_means}. Both interpolations represent the fitting of the MC generator tuning problem. Looking only at the $\overline{\chi^2}$ values would lead to the assumption that the fixed order polynomial would perform much better, but since the expectation value of a single $\chi^2$ value is the number of degrees of freedom ($n_{dof}$), numerically given by subtracting the number of fit parameters from the number of parameter vectors, the adaptive polynomial functions have multiple expectation values. This leads to a non-comparability of the parameters. For that purpose, the $\overline{\chi^2_{red}}$ values are shown. Those are independent of the individual $n_{dof}$. The comparison between those parameters shows that both deliver similar results.

The parameter $\overline{D_{smooth}}$ is independent of the $n_{dof}$ and therefore delivers a shape dependent judgment. A direct comparison between both algorithms shows that the results are quite similar with a slightly better value in the case of the new algorithm. This result is not surprising, since the new algorithm also searched for a maximum in this parameter, especially with respect to the impact of that parameter as shown in Fig.~\ref{fig:impact_f} upon $\mathcal{F}$.

In Tab.~\ref{tab:compare_means}, two values of $\mathcal{F}$ are compared. While $\overline{\mathcal{F}}$ takes into account the simple $\chi^2$ definition, $\overline{\mathcal{F}_{red}}$ is based on $\chi^2_{red}$, and is therefore independent of the $n_{dof}$. Since the impact of the smoothness on that parameter was already shown to be strong, the resulting $\overline{\mathcal{F}_{red}}$ parameter leads in the case of the adaptive algorithm to a much better result.

\begin{table}[htbp]
 	\centering
  	\begin{tabular}{ l c c }
    		\hline
    		Parameter & Professor & Adaptive polynomial \\ \hline
		$\overline{\chi^2}$ & 466.84 & 1472.40 \rule{0pt}{2.6ex} \\
		$\overline{\chi^2_{red}}$ & 1.059 & 1.077 \\  
		$\overline{D_{smooth}}$ & 0.945 & 0.977 \\            
		$\overline{\mathcal{F}}$ & 14.32 & 18.72 \\
		$\overline{\mathcal{F}_{red}}$ & 0.032 & 0.014 \\
    		\hline
  	\end{tabular}
\caption{Summary of interpolation quality parameters for comparison between the \textsc{Professor} and the adaptive interpolation.}
\label{tab:compare_means}
\end{table}  

The most important test of a possible improvement of the adaptive interpolation algorithm in contrast to the fixed order polynomial function algorithm of \textsc{Professor} is the reproduction of the run-combinations as described in Chapter \ref{ch:alter_ansatz}. In order to investigate the tuning behavior of the new algorithm, the same model parameters, the same configuration of \textsc{Pythia 8} and the same observables and data points are used. Therefore, the only change in the tuning approach becomes the interpolation itself. Since the number of used parameter vectors was increased, 600 subsets with 1000 out of the overall 1500 vectors are used for the adaptive interpolation.

\begin{figure}[htbp]
\makebox[\textwidth][c]{
	\begin{adjustbox}{max width=1.\textwidth}
	\includegraphics[width=\textwidth]{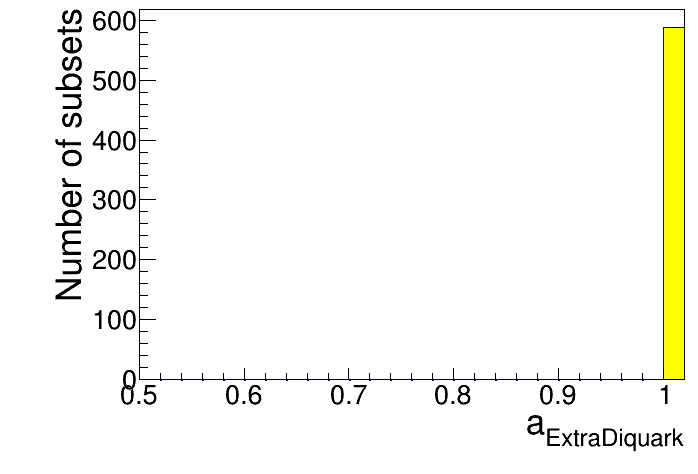}
	\includegraphics[width=\textwidth]{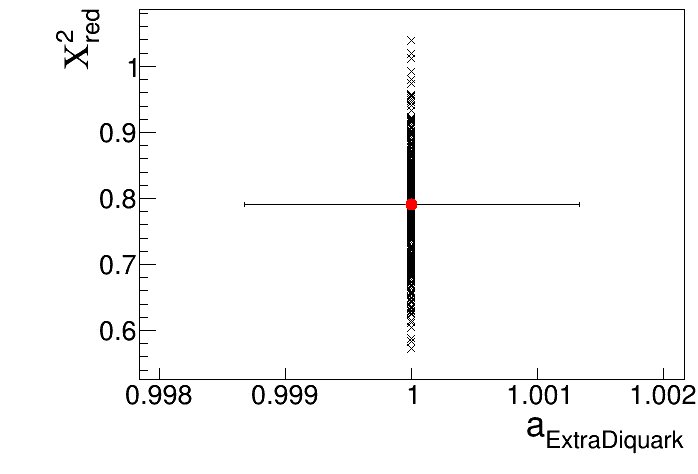} 
	\end{adjustbox}
}
	\caption{Distribution of the run-combinations obtained with the adaptive interpolation algorithm using the neutral weights from \cref{tab:particle_spectrum,tab:pdg_multiplicities1,tab:pdg_multiplicities2,tab:b_fragmentation,tab:event_shape,tab:diff_jet_rate} in App.~\ref{asec:observablesandweights}. The left-hand side graph shows the histogram with number of tuned parameter values of $a_{\textrm ExtraDiquark}$. The right-hand side shows the corresponding $\chi^2_{red}$-value of the tuned parameter values of $a_{\textrm ExtraDiquark}$. The red dot represents the tuning result of a single tuning using all 1500 samples.}
	\label{fig:aed_stable}
\end{figure}

Firstly, Fig.~\ref{fig:aed_stable} shows the distribution of the run-combinations of the model parameter $a_{\textrm ExtraDiquark}$. This parameter was chosen in order to compare it to the result from Sec.~\ref{sec:double_peak}. As already shown there, the distribution itself is at the upper limit of the given parameter range.


\begin{figure}[htbp]
\makebox[\textwidth][c]{
	\begin{adjustbox}{max width=1.\textwidth}
	\includegraphics[width=\textwidth]{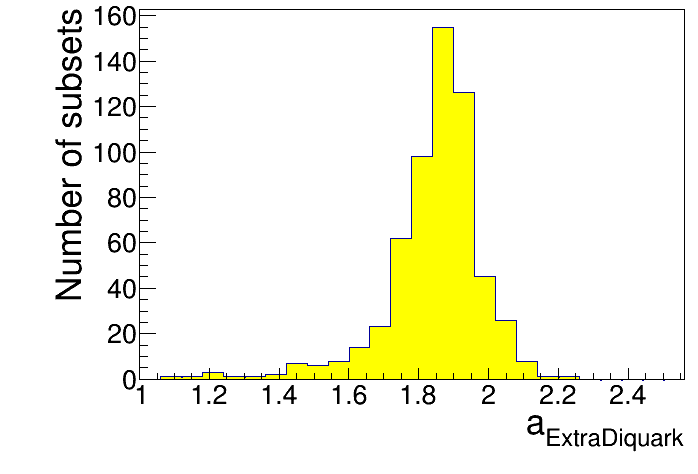}
	\includegraphics[width=\textwidth]{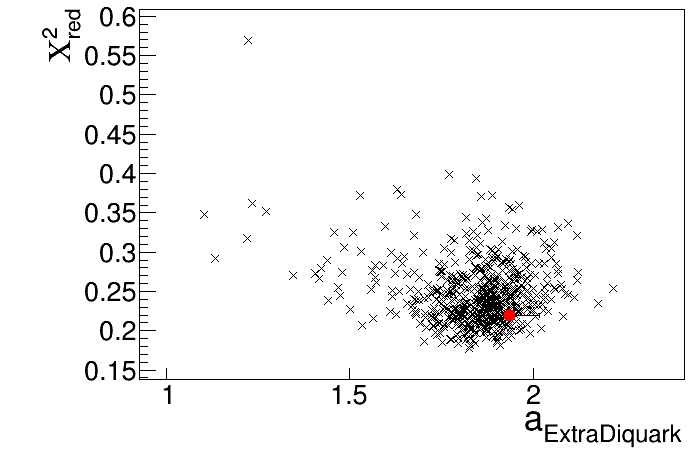} 
	\end{adjustbox}
}
	\caption{Distribution of the run-combinations obtained with the adaptive interpolation algorithm using the neutral weights from \cref{tab:particle_spectrum,tab:pdg_multiplicities1,tab:pdg_multiplicities2,tab:b_fragmentation,tab:event_shape,tab:diff_jet_rate} in App.~\ref{asec:observablesandweights} without parameter limits. The left-hand side graph shows the histogram with number of tuned parameter values of $a_{\textrm ExtraDiquark}$. The right-hand side shows the corresponding $\chi^2_{red}$-value of the tuned parameter values of $a_{\textrm ExtraDiquark}$. The red dot represents the tuning result of a single tuning using all 1500 samples.}
	\label{fig:aed_stable_nolimits}
\end{figure}

Again, in order to investigate the behavior of the function itself, the parameter limits are dropped and a re-minimization of the unchanged interpolation function is performed. The result of this change is shown in Fig.~\ref{fig:aed_stable_nolimits}. In comparison to the previous approach shown in Fig.~\ref{fig:aed_prof_nolimits}, the result shows a distribution around a central value. The $\chi^2_{red}$ distribution is similar to a normal distribution around a single central value on both axes. A complete overview over all model parameters for the limited and unlimited case is shown in App.~\ref{asec:adaptive_runcombs}.\\


The observation that the values are distributed around a single center allow the conclusion that the run-combinations approach can be seen as perturbation of a best tuning. Since only a single peak exists and the tuned parameter values of the interpolation function using all 1500 parameter vectors deliver a result close to or at the best tuned parameter values, the run-combinations are not necessarily needed by this interpolation algorithm. Furthermore, since only a single peak is observed, reasonable tuned parameter values can be extracted out of the run-combinations.

\section{Comparison between \textsc{Minuit} and BAT}
\label{sec:adaptive_comparison}
The minimization was yet performed by using \textsc{Minuit}. As shown in Sec.~\ref{sec:result_bat}, BAT is able to reveal further information about the function behavior. For that purpose, the minimization of the adaptive interpolation function will be performed using this toolkit. Like in the case of the \textsc{Professor} interpolation in Sec.\ref{sec:result_bat}, all parameter vectors will be used for the calculation of a single function. In this case, those are 1500. The minimization is performed with \textsc{Minuit}. The corresponding results are shown in Tab.~\ref{tab:compare_stable_tunes}.

\begin{table}[htbp]
 	\centering
  	\begin{tabular}{ l c c c }
    		\hline
				&&& BAT with \\
    		Parameter & \textsc{Minuit} & BAT & $-\chi^2$-log-likelihood \\ \hline
		$a_{\textrm Lund}$ & 0.70 & 0.70 & 0.70 \\
		$b_{\textrm Lund}~[GeV^{-2}]$ & 1.09 & 0.84 & 1.50 \\  
		$a_{\textrm ExtraDiquark}$ & 1.00 & 1.00 & 1.00 \\            
		$\sigma~[GeV]$ & 0.331 & 0.297 & 0.315 \\
		$\alpha_s(M_Z)$ & 0.139 & 0.139 & 0.139 \\
		$p_{T, min}~[GeV]$ & 1.00 & 1.00 & 1.00 \\
    		\hline
  	\end{tabular}
\caption{Tuning results of the interpolation function using the adaptive algorithm and all 1500 parameter vectors. The \textsc{Minuit} minimization is performed by the \textsc{Professor} framework. BAT is used once with a normal distribution as likelihood and once with the $-\chi^2$-function as log-likelihood. The stated values represent the mode of the posterior likelihood.}
\label{tab:compare_stable_tunes}
\end{table}

\begin{figure}[htbp]
\makebox[\textwidth][c]{
	\begin{adjustbox}{max width=1.\textwidth}
	\includegraphics[width=\textwidth]{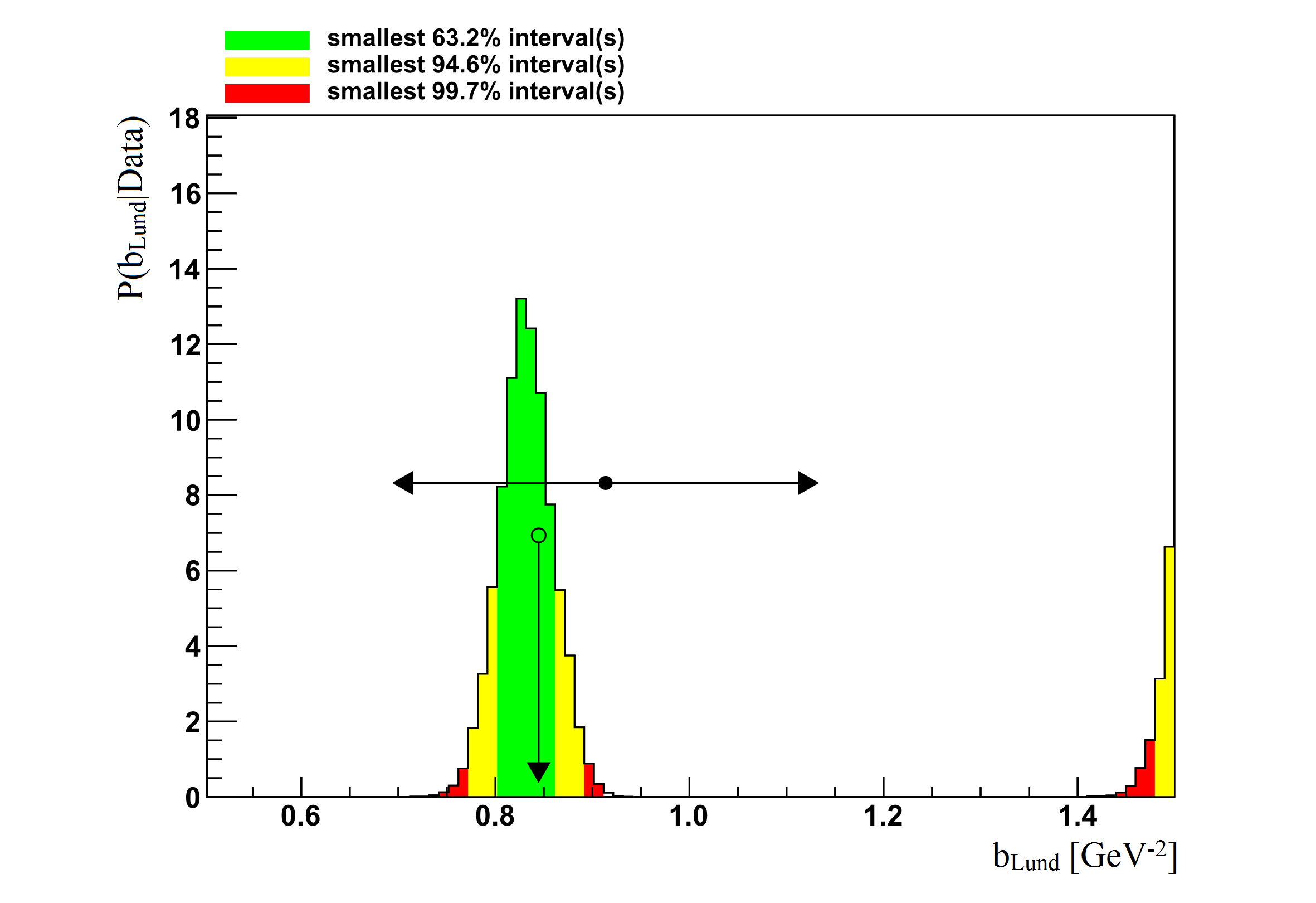}
	\includegraphics[width=\textwidth]{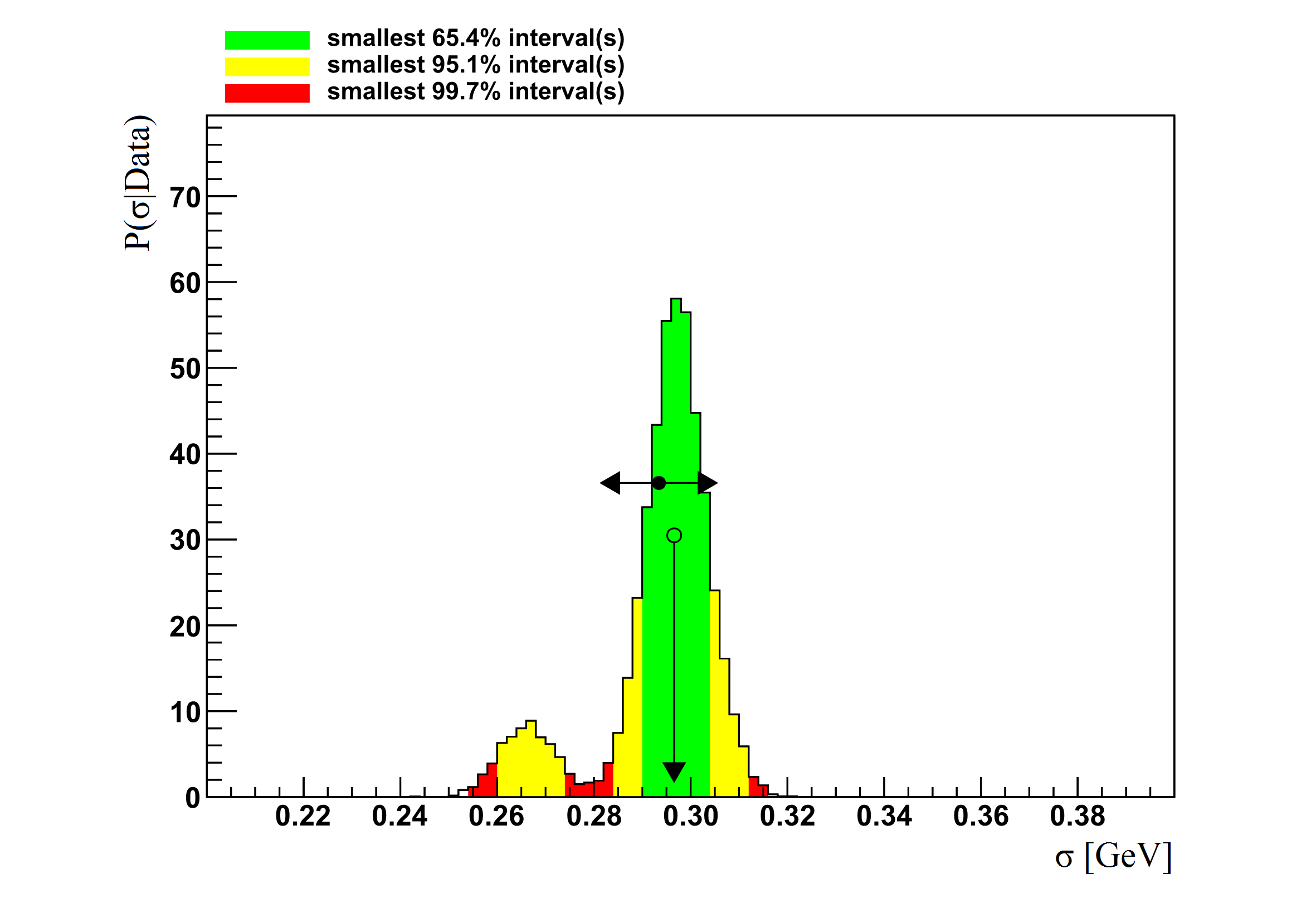}
	\end{adjustbox}
}
	\caption{Marginalized posterior likelihood of the interpolation function using the new algorithm and 1500 parameter vectors. Left: Distribution of $b_{\textrm Lund}$ in $[GeV^{-2}]$. Right: Distribution of $\sigma$ in $[GeV]$. The height depends on the number of steps which were performed in each data point by all Markov Chains combined.}
	\label{fig:stable_bat}
\end{figure}

Minimizing the interpolation function in the given ranges from Tab.~\ref{tab:parameter_ranges_nadine} with BAT leads to the results shown in Tab.~\ref{tab:compare_stable_tunes}. The results are almost identical to the result of \textsc{Minuit} with the exception of $b_{\textrm Lund}$ and $\sigma$. Since the results differ in those two parameters, the corresponding marginalized posterior likelihood is shown in Fig.~\ref{fig:stable_bat}. As the graphs show, the likelihood function has at least two regions of a higher likelihood. On the other hand, none of the regions is around the values stated by \textsc{Minuit}. Furthermore, the other four parameters are at the limit of the given parameter range. This allow the conclusion that the difference is not only a matter of the scaling factor as described in Sec.~\ref{sec:result_bat}.

A possible explanation of the difference could be within the likelihood function. Since the used likelihood function in BAT is a normal distribution as given by Eq.~\ref{eq:log-likelihood} and \textsc{Minuit} minimizes a $\chi^2$-function, the result differs by $\mathcal{O}(\ln(\sigma))$ per data point. The definition of the width $\sigma$ of the normal distribution is given by $\sqrt{\sigma_f^2 + \sigma_R^2}$ and thus related to both the variance of the function $\sigma_f^2$ evaluated at a certain point in the parameter space and the variance of the simulated data $\sigma_R^2$. As a result of this difference, the non-constant offset could lead to the difference in the marginalized posterior likelihood. Furthermore, the uncertainty of the interpolation function is given by the diagonal terms of Eq.~\ref{eq:covmatchi2}. As a result of this, the uncertainty tends to result in an underestimation of the actual uncertainty of the interpolation function, since the correlation was neglected. The \textsc{Professor} interpolation overestimated it. Therefore, the contribution of $\mathcal{O}(\ln(\sigma))$ per data point resulted in a smaller difference in the previous chapter than in this case and could be neglected in the previous application. In order to compare between the minimization procedures, the likelihood function in BAT will be replaced by the $-\chi^2$-function as used by \textsc{Minuit}. The corresponding modes of the likelihood are shown in the right most column of Tab.~\ref{tab:compare_stable_tunes}. The result shows a difference between the \textsc{Minuit} minimization in the parameters $b_{\textrm Lund}$ and $\sigma$, too. Compared to the previous BAT minimization, both parameters lead to higher values, but $\sigma$ still has a smaller value than in the \textsc{Minuit} minimization. Using the $-\chi^2$ likelihood leads to a bigger $b_{\textrm Lund}$ value than in both previously performed minimizations.

\begin{figure}[htbp]
\makebox[\textwidth][c]{
	\begin{adjustbox}{max width=1.\textwidth}
	\includegraphics[page=2,width=\textwidth]{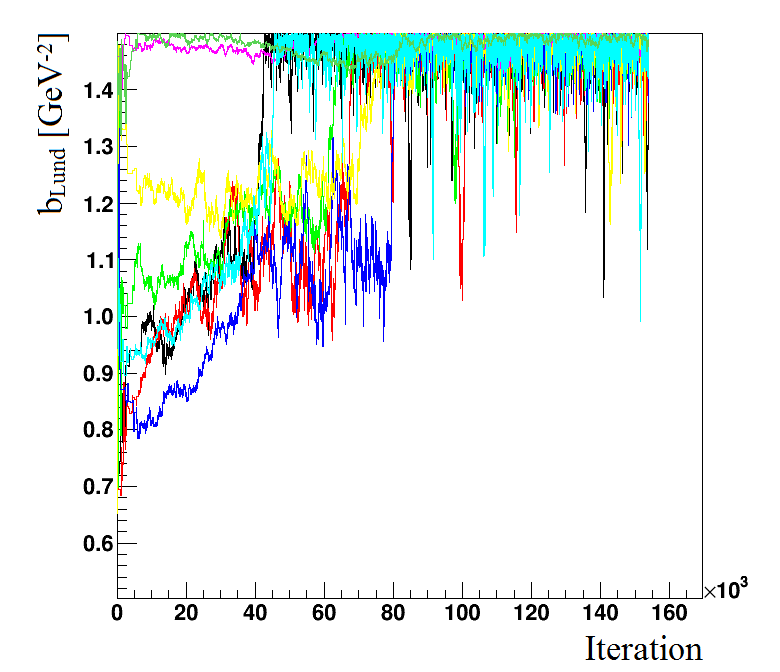}
	\includegraphics[page=4,width=\textwidth]{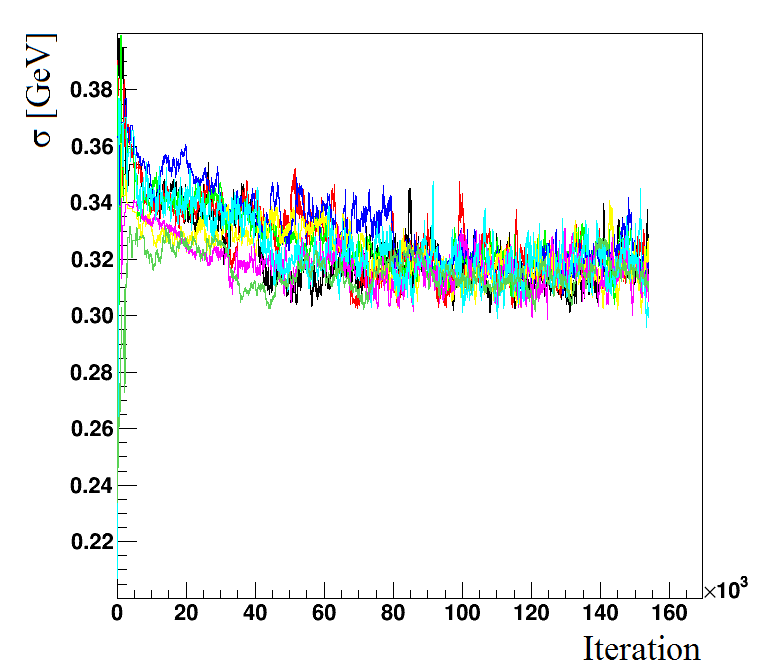}
	\end{adjustbox}
}
	\caption{Evolution of the Markov Chains for $b_{\textrm Lund}$ in $[GeV^{-2}]$ (left) and $\sigma$ in $[GeV]$ (right) with a $-\chi^2$-likelihood function during the pre-run.}
	\label{fig:stable_bat_chi2}
\end{figure}

Besides the resulting mode given by BAT, the toolkit provides additional information about the topology during the iterations of the Markov Chains as shown in Fig.~\ref{fig:stable_bat_chi2}. These graphs show the evolution of the Markov Chains for the parameters $b_{\textrm Lund}$ and $\sigma$ while using the $-\chi^2$ likelihood during the pre-run. The interesting area is between $40\cdot10^3$ and $80\cdot10^3$ iterations. Within this interval, multiple Markov Chains are close to the optimal values as they are stated by \textsc{Minuit}. The problem that occurs in this case is very similar to the problem during the re-run of BAT in Sec.~\ref{sec:result_bat}. Because of the limited parameter space, the efficiencies are reduced and the Markov Chains get caught around the parameter limits, leading to misleading final statements of BAT itself.\\

BAT is able to reproduce the results of \textsc{Minuit} but they need to be extracted in a more complicated way. On the other hand, BAT delivers additional information about the likelihood behavior itself. As a result of these observations, a tuned parameter set can be extracted using BAT only in the case that the regions of interest in the likelihood function are inside the parameters space and not at the border. 

\section{Uncertainty calculations using covariance matrices}
\label{sec:covmat}
As it was shown, BAT is able to reproduce the results given by \textsc{Minuit}. On the other hand, the maximum likelihood needs to be inside the parameter space in order to obtain a reasonable result. In this section, a convergence of the Markov Chains inside this parameter space will be searched. In order to receive a reasonable result, the uncertainty calculation of the interpolation function will be used as given by Eq.~\ref{eq:sigmaf}. The whole covariance matrix of the interpolation parameters will be used from now on. This uncertainty propagation leads to a much more complicated likelihood function. Since BAT handles the parameters as independent, which can be correlated by the interpolation function itself. The usage of the covariance matrix of the interpolation parameters leads to further correlation boundaries of the model parameters. Furthermore, a data point-wise sum of normal distributions will be used again as log-likelihood function.

\begin{figure}[htbp]
\makebox[\textwidth][c]{
	\begin{adjustbox}{max width=1.\textwidth}
	\includegraphics[width=\textwidth]{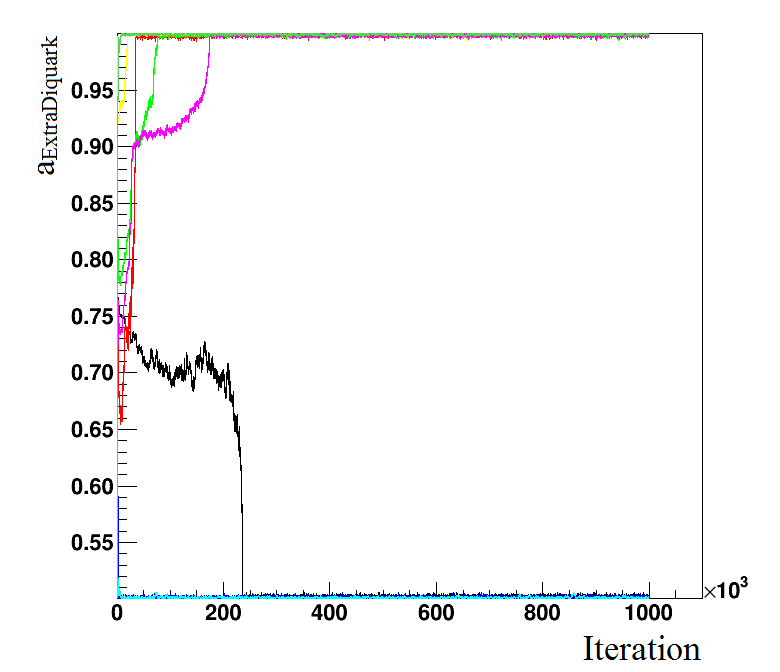}
	\includegraphics[width=\textwidth]{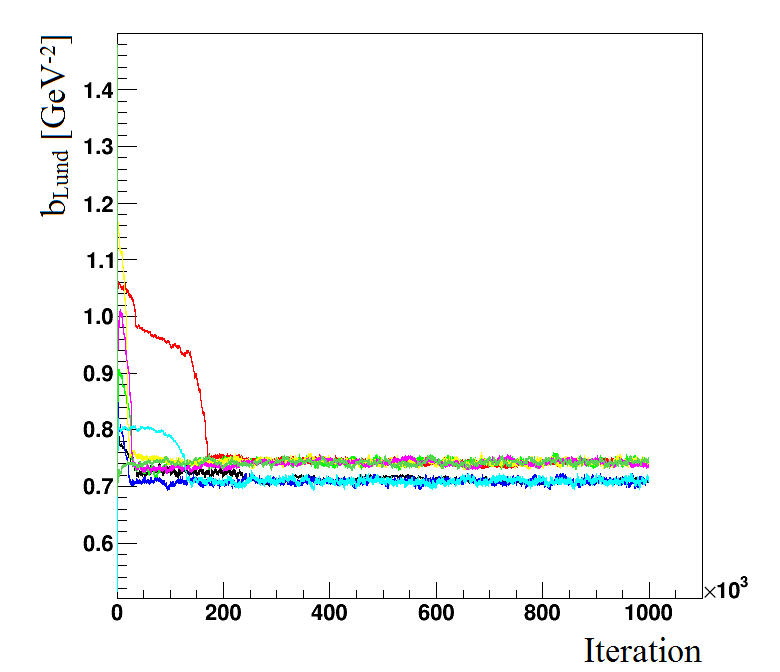} 
	\end{adjustbox}
}
	\caption{Evolution of the Markov Chains for $a_{\textrm ExtraDiquark}$ (left) and $b_{\textrm Lund}$ in $[GeV^{-2}]$ (right) with neutral weights during the pre-run. The uncertainty propagation is based upon the covariance matrix of the interpolation parameters.}
	\label{fig:covmat_1500}
\end{figure}

A first run of BAT using the same interpolation function as in the previous section with neutral weights as given in \cref{tab:particle_spectrum,tab:pdg_multiplicities1,tab:pdg_multiplicities2,tab:b_fragmentation,tab:event_shape,tab:diff_jet_rate} in App.~\ref{asec:observablesandweights} is performed. The result of the pre-run is shown in Fig.~\ref{fig:covmat_1500}.

As the left-hand side evolution of the Markov Chains shows, at least two attractors for different $a_{\textrm ExtraDiquark}$ values exist. One is around 0.5, the other one around 1.0. Like in the previous BAT runs, the problem concerning the scaling factors of the individual Markov Chains does also occur at this point and thus the Markov Chains remain at the upper and lower parameter range limit of that parameter. The problem is even worse, since a jump of a Markov Chain over the whole parameter range of the parameter $a_{\textrm ExtraDiquark}$ would be necessary in order to receive a converged result. This behavior further affects the behavior of the parameter $b_{\textrm Lund}$ as shown on the right-hand side of Fig.~\ref{fig:covmat_1500}. The two attractors of $a_{\textrm ExtraDiquark}$ lead to two different attractors of $b_{\textrm Lund}$ in such a way that they do not jump between each other due to the small scaling factor of the Markov Chains. Additionally to the two shown Markov Chain evolutions, the chains in other parameters are at the limit of their given parameter range. As a result of this, the scaling factors are rather small and do almost not allow any jumps. This observation can be made during the main-run, too.\\
\medskip

In order to encounter the problem of the attractors at the limit of the given parameter ranges, new parameter ranges are defined. After the definition, new parameter vectors need to be sampled, events simulated, observables extracted and the interpolation function calculated. Then, extrapolation should not be needed. In order to estimate a larger parameter space to receive a convergence inside the parameter space while still keeping it small enough to receive an almost equal parameter vector density inside the space, it is based upon the result from the first run. Since no convergence was reached during the first run, the ``tuning'' is based on the Markov Chains attractors by eye. An overview over the extracted tuned values, the old and the estimated new parameter ranges are given in Tab.~\ref{tab:3500}.

\begin{table}[htbp]
\makebox[\textwidth][c]{
  	\begin{tabular}{ l | c | c c | c c }
    		\hline
    		& Initial run & \multicolumn{2}{c}{Initial ranges} & \multicolumn{2}{| c}{Updated ranges} \\
    		Parameter & Attractors & Minimum & Maximum & Minimum & Maximum \\ \hline
		$a_{\textrm Lund}$ & 0.20 & 0.2 & 0.7 & 0.1 & 1.0 \\
		$b_{\textrm Lund}~[GeV^{-2}]$ & 0.72 \& 0.74 & 0.5 & 1.5 & 0.5 & 1.5 \\  
		$a_{\textrm ExtraDiquark}$ & 0.50 \& 1.00 & 0.5 & 1.0 & 0.2 & 1.5 \\            
		$\sigma~[GeV]$ & 0.283 & 0.2 & 0.4 & 0.24 & 0.34 \\
		$\alpha_s(M_Z)$ & 0.139 & 0.120 & 0.139 & 0.134 & 0.150 \\
		$p_{T, min}~[GeV]$ & 0.40 & 0.4 & 1.0 & 0.1 & 0.6 \\
    		\hline
  	\end{tabular}
}
\caption{Extraction of the attractors of the minimization with BAT using the adaptive interpolation method in the old ranges. Based upon this information, the new parameter ranges are defined.}
\label{tab:3500}
\end{table}

Comparing the results from the initial run to the chosen parameter ranges shows that both $a_{\textrm Lund}$ and $p_{T,min}$ need an extension of the parameter range towards lower values. Furthermore $\alpha_s$ needs an extension towards higher values and $a_{\textrm ExtraDiquark}$ a two-sided extension due to the two attractors found in the initial run. If appropriate as in the case of $\sigma$, the range is reduced.

Since the parameter vector density is an important feature of the adaptive interpolation algorithm, an extension of the parameter space while keeping the number of vectors constant could lead to a more problematic behavior in order to estimate $D_{smooth}$. For that purpose, the volume of the parameter spaces needs to be calculated and the number of needed parameter vectors based on this volume. In the old parameter space, the volume was $5.7\cdot10^{-4}$. The new parameter space has a volume of $9.4\cdot10^{-4}$. Since the volume almost doubled, the number of parameter vectors is also increased. While the old tuning used 1500 vectors, the new one uses 3500. All other settings are unchanged.

\begin{figure}[htbp]
\makebox[\textwidth][c]{
	\begin{adjustbox}{max width=1.\textwidth}
	\includegraphics[width=\textwidth]{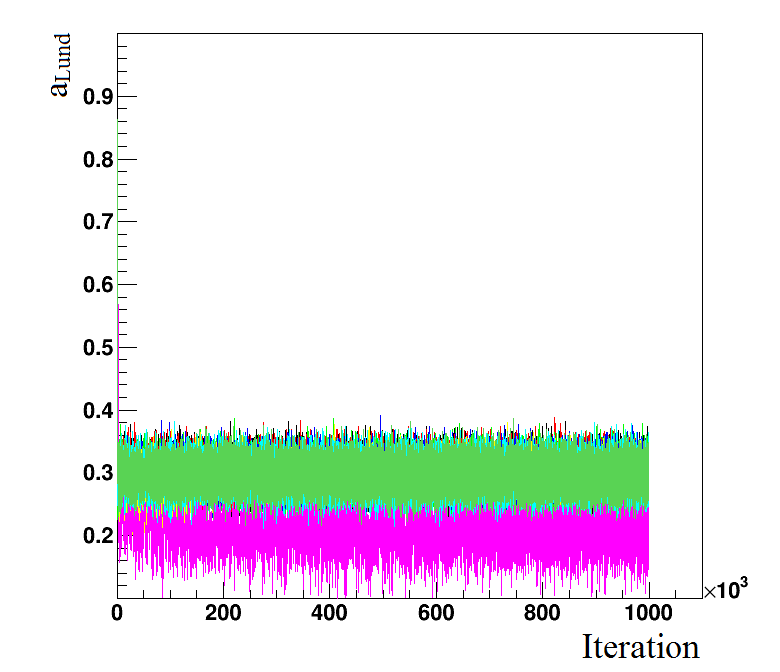}
	\includegraphics[width=\textwidth]{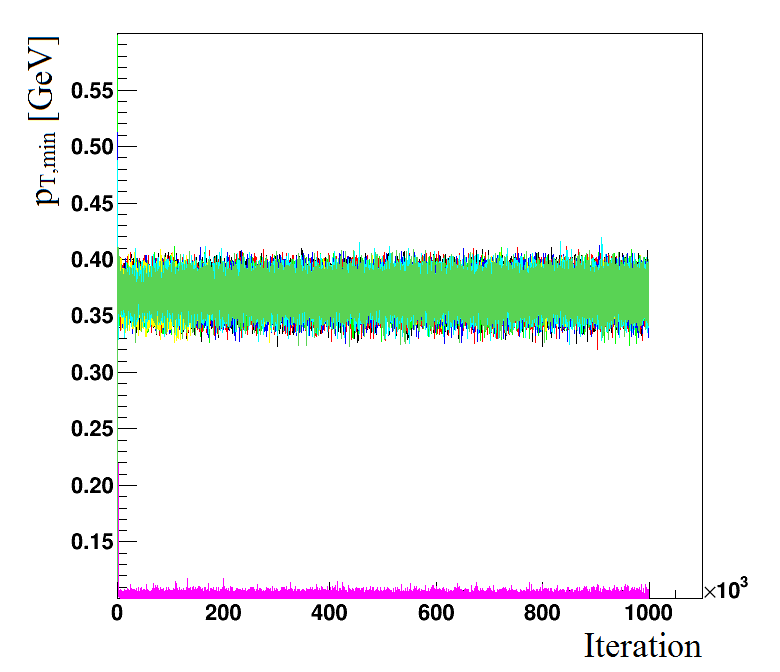} 
	\end{adjustbox}
}
	\caption{Evolution of the Markov Chains for $a_{\textrm Lund}$ (left) and $p_{T, min}$ in $[GeV]$ (right) with neutral weights during the pre-run in the updated parameter ranges from Tab.~\ref{tab:3500}. The uncertainty propagation is based on the covariance matrix of the interpolation parameters.}
	\label{fig:covmat_3500}
\end{figure}

Performing the minimization in this way leads to a similar behavior of $a_{\textrm ExtraDiquark}$. This parameter shows two attractors again, but this time around 0.2 and 1.5. Those values represent again the limits of the parameter range. Also, the value of $a_{\textrm ExtraDiquark}$ affects the other parameters as shown in Fig.~\ref{fig:covmat_3500}.

The purple Markov Chain corresponds to $a_{\textrm ExtraDiquark}$ values around 1.5, the other chains correspond to $a_{\textrm ExtraDiquark}$ around 0.2. With this updated parameter range, only $a_{\textrm ExtraDiquark}$ and $p_{T,min}$ are at the limit of their parameter ranges. Therefore, the parameter ranges need to be bigger for $a_{\textrm ExtraDiquark}$ and $p_{T, min}$. Further investigations with various parameter spaces lead to an even more modified parameter space. The resulting parameter ranges are given in Tab.~\ref{tab:finalspace} and are partly limited on the software-side by the \textsc{Pythia 8} framework. This is the case for every parameter but $\sigma$ and $\alpha_s(M_Z)$.

\begin{table}[htbp]
 	\centering
  	\begin{tabular}{ l c c }
    		\hline
    		Parameter & Minimum & Maximum \\ \hline
		$a_{\textrm Lund}$ & 0.0 & 1.0 \\
		$b_{\textrm Lund}~[GeV^{-2}]$ & 0.5 & 2.0 \\
		$a_{\textrm ExtraDiquark}$ & 0.0 & 2.0 \\           
		$\sigma~[GeV]$ & 0.24 & 0.34 \\
		$\alpha_s(M_Z)$ & 0.135 & 0.145 \\
		$p_{T, min}~[GeV]$ & 0.10 & 0.75 \\
    		\hline
  	\end{tabular}
\caption{Definition of the parameter ranges in their final configuration.}
\label{tab:finalspace}
\end{table} 

\begin{figure}[htbp]
\makebox[\textwidth][c]{
	\begin{adjustbox}{max width=1.\textwidth}
	\includegraphics[width=\textwidth]{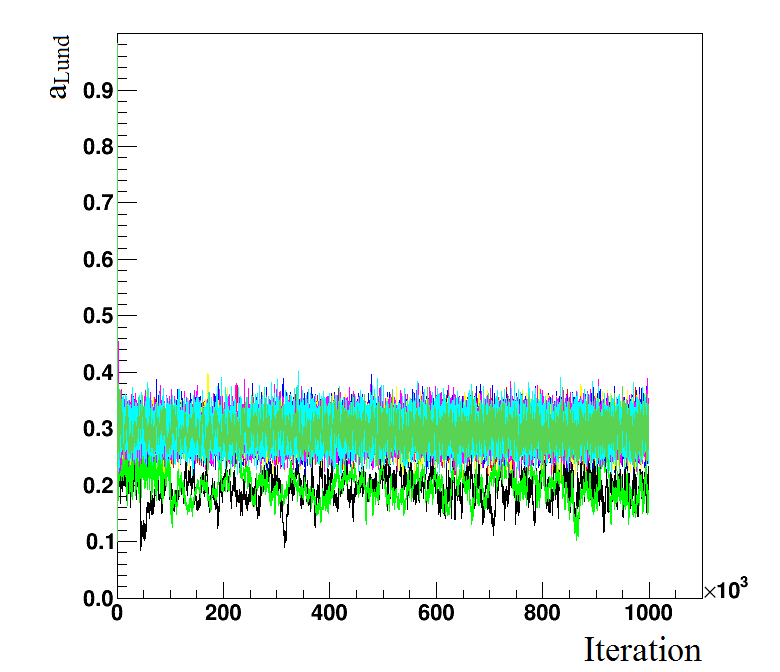}
	\includegraphics[width=\textwidth]{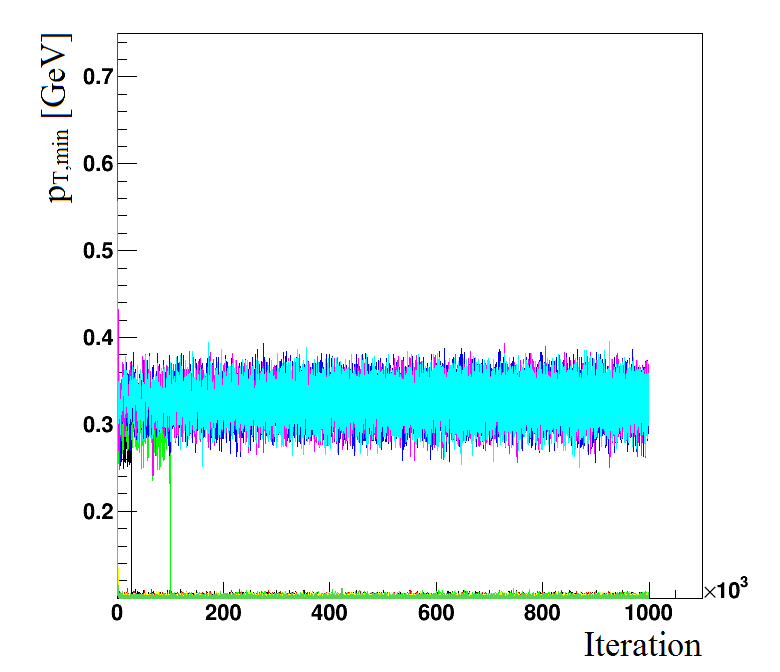} 
	\end{adjustbox}
}
	\caption{Evolution of the Markov Chains for $a_{\textrm Lund}$ (left) and $p_{T, min}$ in $[GeV]$ (right) with neutral weights during the pre-run in the parameter ranges of Tab.~\ref{tab:finalspace}. The uncertainty propagation is based upon the covariance matrix of the interpolation parameters.}
	\label{fig:covmat_050417}
\end{figure}

Re-performing the tuning process with 3000 samples, the resulting $a_{\textrm ExtraDiquark}$ values are again at the lower and upper limit of the given parameter range. As a result of this behavior, the parameter $a_{\textrm Lund}$ and $p_{T,min}$ tend to two different attractors again as shown in Fig.~\ref{fig:covmat_050417}.

This observation is quite similar to Fig.~\ref{fig:covmat_3500}. A more detailed observation of the results even shows more close attractors in the parameter space (cf. App.~\ref{asec:adaptivetune}). On the other hand, the parameter spaces cannot be stretched further in order to receive a convergence inside the parameter space. The consequence of that situation is that the problem cannot be solved that way and that a reasonable result can only be obtained by another method.

\subsection{Exchanging the observables}
Since an extension of the parameter ranges does not lead to a convergence of the Markov Chains inside the parameter space, the sensitivity of the selected analyzes and observables on the model parameters to tune is investigated. For that purpose, the envelopes of the simulated data points are investigated. The result is that some mean multiplicities are insensitive to the used model parameters or are not modeled correctly by the MC generator. Therefore, those observables can be neglected. On the other hand, the list of observables is extended by further analyzes provided in the \textsc{Rivet} framework but not used yet. The reason for the adding of new observables is driven by the problem of $a_{\textrm ExtraDiquark}$. If the used analyzes are not sensitive (enough) to this parameter, then others could be. In the context of correlations between the model parameters, this would affect other parameters, too. The changes in the set of used observables is shown in Tab.~\ref{tab:exchangeobservables} in App.~\ref{asec:observablesandweights}.

Adding observables related to baryons can lead to additional constraints on the tuned $a_{\textrm ExtraDiquark}$ value and thus to a reasonable solution. Additionally, since several momentum spectra of mesons and baryons are added, this does not affect only a single model parameter. The parameter ranges for this tune with the modified list of observables is given in Tab.~\ref{tab:finalspace}. The remainder of the tuning setups stays unchanged.

Performing the minimization leads to a single value of $a_{\textrm ExtraDiquark}$ after the convergence, close to/at the lower limit of the parameter range. Every other parameter converges inside the given parameter space. The corresponding evolution of the Markov Chains is shown in App.~\ref{asec:adaptivemodified}. The convergence leads to a (more) reasonable result and tuned parameter values can be stated. The corresponding modes are given in Tab.~\ref{tab:moreanalyzes}.\\

\begin{table}[htbp]
 	\centering
  	\begin{tabular}{ l c c c c }
    		\hline
    		& \multicolumn{2}{c}{All observables} & \multicolumn{2}{c}{Re-selected observables} \\
    		Parameter & Value & Uncertainty & Value & Uncertainty\\ \hline
		$a_{\textrm Lund}$ & 0.34 & 0.02 & 0.34 & 0.02 \\
		$b_{\textrm Lund}~[GeV^{-2}]$ & 0.96 & 0.03 & 0.93 & 0.03 \\
		$a_{\textrm ExtraDiquark}$ & $0.20\cdot 10^{-3}$ & $4.81\cdot 10^{-3}$ & $0.20\cdot 10^{-3}$ & $4.99\cdot 10^{-3}$ \\           
		$\sigma~[GeV]$ & 0.280 & 0.001 & 0.284 & 0.002 \\
		$\alpha_s(M_Z)$ & 0.140 & 0.001 & 0.140 & 0.001 \\
		$p_{T, min}~[GeV]$ & 0.31 & 0.02 & 0.32 & 0.02 \\
    		\hline
  	\end{tabular}
\caption{Overview of the tuned model parameters. The middle columns show the values and uncertainties using the modified list from Tab.~\ref{tab:exchangeobservables}. The right columns show the tuned values and uncertainties of the modified list using additional outlier rejection.}
\label{tab:moreanalyzes}
\end{table}

\subsection{Outlier rejection}
Replacing the list of observables leads to a first convergence using the adaptive interpolation with the full covariance matrix of the interpolation parameters. The modification was performed on the envelope of the simulations. Only data points that are completely insensitive (e.g. a constant MC response for every parameter vector) were neglected until now. This involves the assumption that the remaining observables are well modeled in the MC generator, even in LO calculations. In order to control this, every data point of the modified list will be controlled for the quality of the MC generator. Under the assumption that the MC generator is able to reproduce the measurements with the given model and a certain parameter vector, then each data point should have at least one parameter vector that has an intersection interval with respect to the MC statistical uncertainty with the measurement and its uncertainty. If that is not the case, then the conclusion is that the MC generator is not able to model a certain data point. This argumentation differs from the previously performed pure insensitivity check such that the MC response can be sensitive upon the model parameters but it does not reproduce the measurement well. Performing this outlier rejection leads to a modification of the set of observables as given in Tab.~\ref{tab:outlier}. Most of the removed observables are related to hadrons with $s$-, $c$- or $b$-content. This leads to the question, if heavy quarks are described well by the used tune from Tab.~\ref{tab:parameters_settings_pythia} or modeled well in the \textsc{Pythia 8} framework. Within this tuning procedure, enhancement or suppression of individual quark flavors were not tuned at all. Such model parameters would need to be added and the complete tuning process must be repeated in order to (possibly) reproduce the measurements. That involves the definition of parameter ranges for those parameters and an adaption of the number of needed samples in order to receive an appropriate parameter vector density.

In addition, observables related to scalar mesons like $a_0(980)$ are removed due to outlier rejection. Since similar observables involving those were already removed in the first modification (cf. Tab.~\ref{tab:exchangeobservables}), the conclusion is that the modeling in the used configuration of \textsc{Pythia 8} is not sufficient in order to use those observables.\\
	
The tune with the outlier rejected set of the modified list of observables is performed by using BAT. This is the only modification compared to the previous BAT run. The corresponding result is shown in the right column of Tab.~\ref{tab:moreanalyzes}. The result is very similar to the previous one with only minor changes. A possible explanation of the changes is given by the fact that the removed data points are sensitive to the model parameters and therefore can contribute to the tuning, too. The removing of the data points/observables on the other hand re-weighted the importance of the remaining data points.\\

As a result of the changes of the tuned model parameters by performing the outlier rejection, two different tuned parameter vectors exist. This leads to the question, which tune is better in order to describe the actual measurements. Since both tunes rely on interpolation functions, a calculation of the distance of the function using the tuned vector and the measurements involves the behavior of the interpolation functions itself. In order to get rid of that and to perform the fairest comparison, new event simulations are performed. The configuration of App.~\ref{asec:pythia_configuration} is used with the values of the model parameters given in Tab.~\ref{tab:moreanalyzes}. Furthermore, events are generated using the tunes performed in Ref.~\cite{ref:nadine_fischer}. In order to state the quality of the tune and to compare them, the $\chi^2$-values of the simulations and the measurements are computed. The $\chi^2$-function is given by

\begin{equation}
\chi^2(\vec{p}) = \sum_\mathcal{O}\sum_{b\in \mathcal{O}}\frac{(\mathcal{S}_b-\mathcal{R}_b)^2}{\Delta^2_b}
\label{eq:chi2_compare}
\end{equation}

with the simulated data point value $\mathcal{S}_b$ and the measured data point value $\mathcal{R}_b$ of a certain data point $b$ of an observable $\mathcal{O}$. $\Delta^2_b$ describes the sum of the variances of the uncertainty of the simulations and the measurements. This calculation is performed for every set of observables used in this thesis yet. The corresponding result is shown in Tab.~\ref{tab:on_tune}.

\begin{table}[htbp]
\makebox[\textwidth][c]{
  	\begin{tabular}{ l c c c c }
    		\hline
				& $\chi^2$ of & $\chi^2$ of & $\chi^2$ of & $\chi^2$ of \\
    		Tune & default set & modified set & outlier rejected set & set of ``Tune 2'' \\ \hline
    		Tune 1 + default & 6512.69 & 7404.34 & 5502.13 & 6403.42 \\
		Tune 2 + default & 6692.86 & 7559.72 & 5645.77 & 6583.77 \\
		Adaptive + modified & 6363.82 & 7159.45 & 5213.83 & 6255.84 \\
		Adaptive + outlier & 6433.72 & 7235.76 & 5288.63 & 6324.53 \\ \hline
		$n_{dof}$ & 1109 & 1227 & 1159 & 1058 \\
    		\hline
  	\end{tabular}
}
\caption{Overview of $\chi^2$ values of the different tunes and the different sets of observables. The default set refers to \cref{tab:particle_spectrum,tab:pdg_multiplicities1,tab:pdg_multiplicities2,tab:b_fragmentation,tab:event_shape,tab:diff_jet_rate}, the modified set refers to the changes in Tab.~\ref{tab:exchangeobservables}, the outlier rejected set refers to the changes in Tab.~\ref{tab:outlier} and the rightmost column refers to the set of observables used for the calculation of ``Tune 2''. The tunes ``Tune 1'' and ``Tune 2'' refer to the tunes of Ref.~\cite{ref:nadine_fischer} using the weights given by \cref{tab:particle_spectrum,tab:pdg_multiplicities1,tab:pdg_multiplicities2,tab:b_fragmentation,tab:event_shape,tab:diff_jet_rate}. The term ``Adaptive'' refers to the tuning using the adaptive interpolation on the different observable sets as given in Tab.~\ref{tab:moreanalyzes}. ``default'', ``modified'' and ``outlier'' refer to the used observable set for the tuning. The bottom row shows the $n_{dof}$ of the set of observables.}
\label{tab:on_tune}
\end{table}  

Since the number of data points is different for every set of observables, a direct comparison within a row is not possible but within a column. A comparison between the previously performed tunes shows, that the ``Tune 2'' produces a worse $\chi^2$-value with every set of observables compared to the ``Tune 1''. The difference is above 100 in every case. Noticeable is that ``Tune 1'' was designed as general purpose tune, while ``Tune 2'' was performed ``with emphasis on event shapes and jet rates''\cite[p.~50]{ref:nadine_fischer} and is therefore more specialized than ``Tune 1''. This could be the reason for the difference in the $\chi^2$-values. Remarkable is that ``Tune 2'' seems to be worse than ``Tune 1'' by using the set of observables that were used for the calculation of ``Tune 2''. Furthermore, both tunes produce a worse $\chi^2$-value on every set observables than the tunes using the adaptive interpolation. Regarding both new tunes, the tune using the full modified set of observables seems to produce a better $\chi^2$-value than the tune using the outlier rejected. This remains true for the outlier rejected set of observables, although the latter was tuned by using this set.

Based on this parameter it can be stated that the adaptive interpolation using the full modified set of observables seems to produce the best results in order the reproduce the measurements. Remarkable is also that the tunes using the adaptive interpolation produce better results than the fixed order interpolation without relying on predefined sets of weights.

\chapter{Summary and conclusions}
\label{ch:summary}
\setcounter{page}{89}

MC generators are a useful tool in order to test models in HEP processes, reproduce measurements and to make predictions. For that purpose MC generators need to be tuned. This thesis considered a possible tuning process that is based on a parametrization approach. This approach considers multiple random parameter value sampling that is used as input for the MC generator in order to calculate events. From those, observables are extracted and interpolated for each data point as function of the model parameters. The interpolation functions are compared to reference data and optimal values of the MC parameters are extracted. Within this thesis it was shown that an interpolation using a fixed order polynomial function does not guarantee reasonable tuning results. This result was obtained by a reproduction of a previously performed tune.

The first investigative step was to control the minimization algorithm that is provided by \textsc{Minuit}. This was done by implementing the problem into BAT. By using this toolkit, another optimization algorithm was used that is based upon Bayes' theorem. This theorem was already used for similar tuning problems\cite{ref:scheiss_tune} in a non-parametrization based approach. It was shown within this thesis that the algorithm provided the same results as the \textsc{Professor} minimization such that the conclusion that the potential instability does lie within the interpolation was drawn. Besides, BAT enabled an insight into the function behavior but caused difficulties because of the algorithmic implementation of BAT.

That conclusion led to an investigation of the interpolation algorithm. This was performed by the implementation of an adaptive algorithm that estimates the number of terms in a polynomial function depending on the data point that needed to be interpolated. The adaption itself is implemented in an iterative way such that after each iteration a term is added to the function until the quality of the function is optimal. For that purpose a new stopping criterion was constructed, used and tested. The criterion combined a distance related parameter and a function shape dependent parameter. Using this algorithm led to a stable first result. It was further shown that the result does not necessarily rely on the construction of run-combinations.

Using the new interpolation algorithm and combining it with the information gathered by using it with a BAT-based optimization led to a deeper insight into the previously performed tuning approach. As a result of the investigation, the ranges of the model parameters were adapted and observables removed or added in two steps. This was done firstly by removing insensitive observables and those which are not simulated by the MC generator. On the other hand, multiple observables were added. As second modification of the set of observables, data points were rejected in which no set of model parameters was able to reconstruct the measurements. Such a data point-wise outlier rejection revealed an insight into poorly described observables such as multiplicities and momentum spectra of hadrons involving $s$-, $c$- or $b$-quarks. The result of those modifications led to two sets of tuned model parameters. These results were compared with the previously performed tunes and it was shown that the modifications of the tuning approach that were introduced in this thesis led to a result that has a much better agreement with data than previous tunes. Obtaining such a result is remarkable if one considers that data point-wise defined weights were not used within this thesis. Thus it does not seem to be necessary to define a set of weights in order to obtain a result that is able to reproduce the measurements of certain observables sufficiently well.

These new applications in parametrization based tuning deliver results on a current regular desktop PC (e.g. with a processor frequency of 8x3.40$~GHz$) in 3-4 days, if the simulation and observable extraction is already done. Since the tradeoff by using these algorithms is given by a higher CPU and RAM consumption, a more complex approach within this parametrization framework is currently difficult to realize. Other interpolations like polyharmonic splines, multiple local interpolations inside the parameter space or couplings between the individual parts of the tuning approach in order to further improve a parametrization of the MC generator response are imaginable in the future.

The tune performed in this thesis was based on six parameters. Other \textsc{Pythia~8} tunes like the Monash tune\cite{ref:monasch}, which involves 19 model parameters, could lead to even better results by re-tuning it with the new method. On the other hand, performing a parametrization based tune with that many parameters would create a massive computing and memory resources consumption. With respect to the hardware of a currently regular desktop PC, the amount of model parameters that can be tuned simultaneously is limited. Also an alternating tune of multiple disjunctive sets of model parameters can just be performed with limited quality, since this approach assumes independent sets. Therefore in order to obtain possibly better results a (much) higher dimensional tuning needs to be performed either on a more powerful system (e.g. computing cluster) or on a future system.

\appendix
\chapter{Tuning setup}
\setcounter{page}{91}
\section{\textsc{Pythia 8} configuration}
\label{asec:pythia_configuration}
\begin{table}[htbp]
 	\centering
  	\begin{tabular}{ l c }
    		\hline
    		Parameter & Value \\ \hline
		Beams:eCM & 91.188 \\              
		Beams:idA & 11 \\   
		Beams:idB & -11 \\                  
		Main:numberOfEvents & 500000 \\      
		Next:numberCount & 50000 \\    
		Next:numberShowEvent & 0 \\        
		WeakSingleBoson:ffbar2gmZ & on \\    
		23:onMode & off \\
		23:onIfAny & 1 2 3 4 5 \\
		PDF:lepton & off \\
		SpaceShower:QEDshowerByL & off \\
		HadronLevel:all & on \\
    		\hline
  	\end{tabular}
\caption{Basic \textsc{Pythia 8} configurations used for the simulations. An explanation of the used parameters can be found in Ref.~\cite{ref:pythia_online_manual}.}
\label{tab:basic_settings_pythia}
\end{table}  

\begin{table}[htbp]
 	\centering
  	\begin{tabular}{ l c c }
    		\hline
    		Parameter & Value & Kept fixed? \\ \hline
 		StringFlav:probStoUD & 0.19 & $\surd$ \\
   		StringFlav:probQQtoQ & 0.09 & $\surd$ \\
  		StringFlav:probSQtoQQ & 1.00 & $\surd$ \\
    		StringFlav:probQQ1toQQ0 & 0.027 & $\surd$ \\
   		StringFlav:mesonUDvector & 0.62 & $\surd$ \\
    		StringFlav:mesonSvector & 0.725 & $\surd$ \\
    		StringFlav:mesonCvector & 1.06 & $\surd$ \\
    		StringFlav:mesonBvector & 3.0 & $\surd$ \\
    		StringFlav:etaSup & 0.63 & $\surd$ \\
    		StringFlav:etaPrimeSup & 0.12 & $\surd$ \\
    		StringFlav:popcornSpair & 0.5 & $\surd$ \\
    		StringFlav:popcornSmeson & 0.5 & $\surd$ \\
    		StringFlav:suppressLeadingB & false & $\surd$ \\
    		StringZ:aLund & 0.386 or 0.351 & X \\
    		StringZ:bLund & 0.977 or 0.942 & X \\
    		StringZ:aExtraSquark & 0.00 & $\surd$ \\
    		StringZ:aExtraDiquark & 0.940 or 0.547 & X \\
    		StringZ:rFactC & 1.00 & $\surd$ \\
    		StringZ:rFactB & 0.67 & $\surd$ \\
    		StringPT:sigma & 0.286 or 0.283 & X \\
    		StringPT:enhancedFraction & 0.01 & $\surd$ \\
    		StringPT:enhancedWidth & 2.0 & $\surd$ \\
    		TimeShower:alphaSvalue & 0.139 (both) & X \\
    		TimeShower:alphaSorder & 1 & $\surd$ \\
    		TimeShower:alphaSuseCMW & false & $\surd$ \\
    		TimeShower:pTmin & 0.409 or 0.406 & X \\
    		TimeShower:pTminChgQ & 0.409 or 0.406 & X \\
    		\hline
  	\end{tabular}
\caption{\textsc{Pythia 8} model parameters and their default values in the configuration set eeTune = 5 or eeTune = 6. The columns that contain two parameters show the only difference between both predefined parameter sets. The first value always refers to the setting 5, the second to the setting 6. The last column indicates if a parameter is tuned in this thesis or held constant. The parameters are taken from the \textsc{src/Settings.cc}-file from Ref.~\cite{ref:pythia_dl}. An explanation of the used parameters can be found in Ref.~\cite{ref:pythia_online_manual}.}
\label{tab:parameters_settings_pythia}
\end{table}   
  
\FloatBarrier  
\section{\textsc{Rivet} analyzes}
\label{asec:rivet_analyzes}
\begin{table}[htbp]
\begin{adjustbox}{max width=\textwidth}
  	\begin{tabular}{ l l | l }
    		\hline
    		\multicolumn{2}{c |}{Rivet name-giving}&\\
    		Analysis & Observable & Observable\\ \hline
		\verb|ALEPH_1996_S3486095| & d01-x01-y01 & Sphericity, $S$ (charged) \\
		\verb|ALEPH_1996_S3486095| & d02-x01-y01 & Aplanarity, $A$ (charged) \\	
		\verb|ALEPH_1996_S3486095| & d03-x01-y01	& 1-Thrust, $1-T$ (charged) \\
		\verb|ALEPH_1996_S3486095| & d04-x01-y01	& Thrust minor, $m$ (charged) \\
		\verb|ALEPH_1996_S3486095| & d07-x01-y01	& $C$ parameter (charged) \\
		\verb|ALEPH_1996_S3486095| & d08-x01-y01	& Oblateness, $M - m$ (charged) \\
    		\verb|ALEPH_1996_S3486095| & d09-x01-y01	& Scaled momentum, $x_p = |p|/|p_\text{beam}|$ (charged) \\
		\verb|ALEPH_1996_S3486095| & d11-x01-y01	& In-plane $p_T$ in GeV w.r.t. sphericity axes (charged) \\
		\verb|ALEPH_1996_S3486095| & d12-x01-y01	& Out-of-plane $p_T$ in GeV w.r.t. sphericity axes (charged) \\
		\verb|ALEPH_1996_S3486095| & d17-x01-y01	& Log of scaled momentum, $\log(1/x_p)$ (charged) \\
		\verb|ALEPH_1996_S3486095| & d18-x01-y01	& Charged multiplicity distribution \\
		\verb|ALEPH_1996_S3486095| & d19-x01-y01	& Mean charged multiplicity \\
		\verb|ALEPH_1996_S3486095| & d25-x01-y01	& $\pi^\pm$ spectrum \\
		\verb|ALEPH_1996_S3486095| & d26-x01-y01	& $K^\pm$ spectrum \\
		\verb|ALEPH_1996_S3486095| & d29-x01-y01	& $\pi^0$ spectrum \\
		\verb|ALEPH_1996_S3486095| & d30-x01-y01	& $\eta$ spectrum \\
		\verb|ALEPH_1996_S3486095| & d31-x01-y01	& $\eta'$ spectrum \\
		\verb|ALEPH_1996_S3486095| & d32-x01-y01	& $K^0$ spectrum \\
		\verb|ALEPH_1996_S3486095| & d33-x01-y01	& $\Lambda^0$ spectrum \\
		\verb|ALEPH_1996_S3486095| & d34-x01-y01	& $\Xi^-$ spectrum \\
		\verb|ALEPH_1996_S3486095| & d35-x01-y01	& $\Sigma^\pm(1385)$ spectrum \\
		\verb|ALEPH_1996_S3486095| & d36-x01-y01	& $\Xi^0(1530)$ spectrum \\
		\verb|ALEPH_1996_S3486095| & d37-x01-y01	& $\rho$ spectrum \\
		\verb|ALEPH_1996_S3486095| & d38-x01-y01	& $\omega(782)$ spectrum \\
		\verb|ALEPH_1996_S3486095| & d39-x01-y01	& $K^{*0}(892)$ spectrum \\
		\verb|ALEPH_1996_S3486095| & d40-x01-y01	& $\phi$ spectrum \\
		\verb|ALEPH_1996_S3486095| & d43-x01-y01	& $K^{*\pm}(892)$ spectrum \\ \hline
		\verb|ALEPH_2001_S4656318| & d01-x01-y01	& $b$ quark fragmentation function $f(x_B^\text{weak})$ \\
		\verb|ALEPH_2001_S4656318| & d07-x01-y01	& Mean of $b$ quark fragmentation function $f(x_B^\text{weak})$ \\ \hline
  	\end{tabular}
\end{adjustbox}
\caption{Translation between the \textsc{Rivet} name-giving and the observables based on Ref.~\cite{ref:rivet_analysis_link}.}
\end{table}

\begin{table}[htbp]
\begin{adjustbox}{max width=\textwidth}
  	\begin{tabular}{ l l | l }
    		\hline
    		\multicolumn{2}{c |}{Rivet name-giving}&\\
    		Analysis & Observable & Observable\\ \hline
    		\verb|DELPHI_1996_S3430090| & d01-x01-y01 & In-plane $p_\perp$ in GeV w.r.t. thrust axes \\
		\verb|DELPHI_1996_S3430090| & d02-x01-y01 & Out-of-plane $p_\perp$ in GeV w.r.t. thrust axes \\
		\verb|DELPHI_1996_S3430090| & d03-x01-y01 & In-plane $p_\perp$ in GeV w.r.t. sphericity axes \\
		\verb|DELPHI_1996_S3430090| & d04-x01-y01 & Out-of-plane $p_\perp$ in GeV w.r.t. sphericity axes \\
		\verb|DELPHI_1996_S3430090| & d07-x01-y01 & Scaled momentum, $x_p = |p|/|p_\text{beam}|$ \\
		\verb|DELPHI_1996_S3430090| & d08-x01-y01 & Log of scaled momentum, $\log(1/x_p)$ \\
		\verb|DELPHI_1996_S3430090| & d09-x01-y01 & Mean out-of-plane $p_\perp$ in GeV w.r.t. thrust axes vs. $x_p$ \\
		\verb|DELPHI_1996_S3430090| & d10-x01-y01 & Mean $p_\perp$ in GeV vs. $x_p$ \\
		\verb|DELPHI_1996_S3430090| & d11-x01-y01 & $1-\text{Thrust}$ \\
		\verb|DELPHI_1996_S3430090| & d12-x01-y01 & Thrust major, $M$ \\
		\verb|DELPHI_1996_S3430090| & d13-x01-y01 & Thrust minor, $m$ \\
		\verb|DELPHI_1996_S3430090| & d14-x01-y01 & Oblateness = $M - m$ \\
		\verb|DELPHI_1996_S3430090| & d15-x01-y01 & Sphericity, $S$ \\
		\verb|DELPHI_1996_S3430090| & d16-x01-y01 & Aplanarity, $A$ \\
		\verb|DELPHI_1996_S3430090| & d17-x01-y01 & Planarity, $P$ \\
		\verb|DELPHI_1996_S3430090| & d18-x01-y01 & $C$ parameter \\
		\verb|DELPHI_1996_S3430090| & d19-x01-y01 & $D$ parameter \\
		\verb|DELPHI_1996_S3430090| & d33-x01-y01 & Energy-energy correlation, EEC \\
		\verb|DELPHI_1996_S3430090| & d35-x01-y01 & Mean charged multiplicity \\ \hline
		\verb|JADE_OPAL_2000_S4300807| & d26-x01-y01	& Differential 2-jet rate with Durham algorithm (91.2 GeV) \\
		\verb|JADE_OPAL_2000_S4300807| & d26-x01-y02	& Differential 3-jet rate with Durham algorithm (91.2 GeV) \\
		\verb|JADE_OPAL_2000_S4300807| & d26-x01-y03	& Differential 4-jet rate with Durham algorithm (91.2 GeV) \\
		\verb|JADE_OPAL_2000_S4300807| & d26-x01-y04	& Differential 5-jet rate with Durham algorithm (91.2 GeV) \\ \hline
		\verb|PDG_HADRON_MULTIPLICITIES| & d01-x01-y03 & Mean $\pi^+$ multiplicity \\
		\verb|PDG_HADRON_MULTIPLICITIES| & d02-x01-y03 & Mean $\pi^0$ multiplicity \\
		\verb|PDG_HADRON_MULTIPLICITIES| & d03-x01-y03 & Mean $K^+$ multiplicity \\
		\verb|PDG_HADRON_MULTIPLICITIES| & d04-x01-y03 & Mean $K^0$ multiplicity \\
		\verb|PDG_HADRON_MULTIPLICITIES| & d05-x01-y03 & Mean $\eta$ multiplicity \\
		\verb|PDG_HADRON_MULTIPLICITIES| & d06-x01-y03 & Mean $\eta'(958)$ multiplicity \\
		\verb|PDG_HADRON_MULTIPLICITIES| & d07-x01-y03 & Mean $D^+$ multiplicity \\
		\verb|PDG_HADRON_MULTIPLICITIES| & d08-x01-y03 & Mean $D^0$ multiplicity 	\\
		\verb|PDG_HADRON_MULTIPLICITIES| & d09-x01-y03 & Mean $D^+_s$ multiplicity \\
		\verb|PDG_HADRON_MULTIPLICITIES| & d10-x01-y01 & Mean $B^+, B^0_d$ multiplicity \\
		\verb|PDG_HADRON_MULTIPLICITIES| & d11-x01-y01 & Mean $B^+_u$ multiplicity \\
		\verb|PDG_HADRON_MULTIPLICITIES| & d12-x01-y01 & Mean $B^0_s$ multiplicity \\
		\verb|PDG_HADRON_MULTIPLICITIES| & d13-x01-y03 & Mean $f_0(980)$ multiplicity \\
		\verb|PDG_HADRON_MULTIPLICITIES| & d14-x01-y01 & Mean $a_0^+(980)$ multiplicity \\
		\verb|PDG_HADRON_MULTIPLICITIES| & d15-x01-y03 & Mean $\rho^0(770)$ multiplicity \\
		\verb|PDG_HADRON_MULTIPLICITIES| & d16-x01-y01 & Mean $\rho^+(770)$ multiplicity \\
		\verb|PDG_HADRON_MULTIPLICITIES| & d17-x01-y02 & Mean $\omega(782)$ multiplicity \\
		\verb|PDG_HADRON_MULTIPLICITIES| & d18-x01-y03 & Mean $K^{*+}(892)$ multiplicity \\
		\verb|PDG_HADRON_MULTIPLICITIES| & d19-x01-y03 & Mean $K^{*0}(892)$ multiplicity \\
		\verb|PDG_HADRON_MULTIPLICITIES| & d20-x01-y03 & Mean $\phi(1020)$ multiplicity \\
		\verb|PDG_HADRON_MULTIPLICITIES| & d21-x01-y03 & Mean $D^{*+}(2010)$ multiplicity \\
		\verb|PDG_HADRON_MULTIPLICITIES| & d23-x01-y02 & Mean $D^{*+}_s(2112)$ multiplicity \\
		\verb|PDG_HADRON_MULTIPLICITIES| & d24-x01-y01 & Mean $B^*$ multiplicity \\
 		\hline
  	\end{tabular}
\end{adjustbox}
\caption{Translation between the \textsc{Rivet} name-giving and the observables based on Ref.~\cite{ref:rivet_analysis_link}.}
\end{table}

\begin{table}[htbp]
\begin{adjustbox}{max width=\textwidth}
  	\begin{tabular}{ l l | l }
    		\hline
    		\multicolumn{2}{c |}{Rivet name-giving}&\\
    		Analysis & Observable & Observable\\ \hline
    		\verb|PDG_HADRON_MULTIPLICITIES| & d25-x01-y02 & Mean $J/\psi(1S)$ multiplicity \\
		\verb|PDG_HADRON_MULTIPLICITIES| & d26-x01-y01 & Mean $\psi(2S)$ multiplicity \\
		\verb|PDG_HADRON_MULTIPLICITIES| & d27-x01-y01 & Mean $\Upsilon(1S)$ multiplicity \\
		\verb|PDG_HADRON_MULTIPLICITIES| & d28-x01-y01 & Mean $f_1(1285)$ multiplicity \\
		\verb|PDG_HADRON_MULTIPLICITIES| & d29-x01-y01 & Mean $f_1(1420)$ multiplicity \\
		\verb|PDG_HADRON_MULTIPLICITIES| & d30-x01-y01 & Mean $\chi_{c1}(3510)$ multiplicity \\
    		\verb|PDG_HADRON_MULTIPLICITIES| & d31-x01-y03 & Mean $f_2(1270)$ multiplicity \\
		\verb|PDG_HADRON_MULTIPLICITIES| & d32-x01-y01 & Mean $f_2'(1525)$ multiplicity \\
		\verb|PDG_HADRON_MULTIPLICITIES| & d34-x01-y02 & Mean $K_2^{*0}(1430)$ multiplicity \\
		\verb|PDG_HADRON_MULTIPLICITIES| & d35-x01-y01 & Mean $B^{**}$ multiplicity \\
		\verb|PDG_HADRON_MULTIPLICITIES| & d36-x01-y01 & Mean $D_{s1}^+$ multiplicity \\
		\verb|PDG_HADRON_MULTIPLICITIES| & d37-x01-y01 & Mean $D_{s2}^+$ multiplicity \\
		\verb|PDG_HADRON_MULTIPLICITIES| & d38-x01-y03 & Mean $p$ multiplicity \\
		\verb|PDG_HADRON_MULTIPLICITIES| & d39-x01-y03 & Mean $\Lambda$ multiplicity \\
		\verb|PDG_HADRON_MULTIPLICITIES| & d40-x01-y02 & Mean $\Sigma^0$ multiplicity \\
		\verb|PDG_HADRON_MULTIPLICITIES| & d41-x01-y01 & Mean $\Sigma^-$ multiplicity \\
		\verb|PDG_HADRON_MULTIPLICITIES| & d42-x01-y01 & Mean $\Sigma^+$ multiplicity \\
		\verb|PDG_HADRON_MULTIPLICITIES| & d43-x01-y01 & Mean $\Sigma^\pm$ multiplicity \\
		\verb|PDG_HADRON_MULTIPLICITIES| & d44-x01-y03 & Mean $\Xi^-$ multiplicity \\
		\verb|PDG_HADRON_MULTIPLICITIES| & d45-x01-y02 & Mean $\Delta^{++}(1232)$ multiplicity \\
		\verb|PDG_HADRON_MULTIPLICITIES| & d46-x01-y03 & Mean $\Sigma^-(1385)$ multiplicity \\
		\verb|PDG_HADRON_MULTIPLICITIES| & d47-x01-y03 & Mean $\Sigma^+(1385)$ multiplicity \\
		\verb|PDG_HADRON_MULTIPLICITIES| & d48-x01-y03 & Mean $\Sigma^\pm(1385)$ multiplicity \\
		\verb|PDG_HADRON_MULTIPLICITIES| & d49-x01-y02 & Mean $\Xi^0(1530)$ multiplicity \\
		\verb|PDG_HADRON_MULTIPLICITIES| & d50-x01-y03 & Mean $\Omega^-$ multiplicity \\
		\verb|PDG_HADRON_MULTIPLICITIES| & d51-x01-y03 & Mean $\Lambda_c^+$ multiplicity \\
		\verb|PDG_HADRON_MULTIPLICITIES| & d52-x01-y01 & Mean $\Lambda_b^0$ multiplicity \\
		\verb|PDG_HADRON_MULTIPLICITIES| & d54-x01-y02 & Mean $\Lambda(1520)$ multiplicity \\
    		\hline
  	\end{tabular}
\end{adjustbox}
\caption{Translation between the \textsc{Rivet} name-giving and the observables based on Ref.~\cite{ref:rivet_analysis_link}.}
\end{table}
  
\FloatBarrier  
\section{Observables and weights}
\label{asec:observablesandweights}
\begin{table}[htbp]
 	\centering
  	\begin{tabular}{ l c c c }
    		\hline
    		Observable & \multicolumn{3}{c}{Weights}\\ 
    		& Tune 1 & Tune 2 & Tune neutral\\ \hline
    		$K^{*\pm}(892)$ spectrum & 1.0 & 1.0 & 1.0\\
    		$\rho$ spectrum & 1.0 & 1.0 & 1.0\\
    		$\omega (782)$ spectrum & 1.0 & 1.0 & 1.0\\
		$\Xi^-$ spectrum & 1.0 & 1.0 & 1.0\\
		$K^{*0}$ spectrum & 1.0 & 1.0 & 1.0\\
		$\phi$ spectrum & 1.0 & 1.0 & 1.0\\
		$\Sigma^\pm$ spectrum & 1.0 & 1.0 & 1.0\\
		$\gamma$ spectrum & 1.0 & 1.0 & 1.0\\
		$K^\pm$ spectrum & 1.0 & 1.0 & 1.0\\
		$\Lambda^0$ spectrum & 1.0 & 1.0 & 1.0\\
		$\pi^0$ spectrum & 1.0 & 1.0 & 1.0\\
		$p$ spectrum & 1.0 & 1.0 & 1.0\\
		$\eta'$ spectrum & 1.0 & 1.0 & 1.0\\
		$\Xi^0 (1530)$ spectrum & 1.0 & 1.0 & 1.0\\
		$\pi^\pm$ spectrum & 1.0 & 1.0 & 1.0\\
		$\eta$ spectrum & 1.0 & 1.0 & 1.0\\
		$K^0$ spectrum & 1.0 & 1.0 & 1.0\\
    		\hline
  	\end{tabular}
\caption{Weights of momentum spectra of identified particles from Ref.~\cite{ref:nadine_fischer}. A neutral tune using weights of 1 for every observable was added. The data was originally published in Ref.~\cite{ref:aleph_1996}.}
\label{tab:particle_spectrum}
\end{table}

\begin{table}[htbp]
 	\centering
  	\begin{tabular}{ l c c c }
    	\hline
    		Observable & \multicolumn{3}{c}{Weights}\\ 
    		& Tune 1 & Tune 2 & Tune neutral\\ \hline
		Mean $\rho^0 (770)$ multiplicity & 10.0 & 0.0 & 1.0 \\
		Mean $\Delta^{++} (1232)$ multiplicity & 10.0 & 0.0 & 1.0 \\
		Mean $K^{*+} (892)$ multiplicity & 10.0 & 0.0 & 1.0 \\
		Mean $\Sigma^0$ multiplicity & 10.0 & 0.0 & 1.0 \\
		Mean $\Lambda^0_b$ multiplicity & 10.0 & 0.0 & 1.0 \\
		Mean $K^+$ multiplicity & 10.0 & 0.0 & 1.0 \\
		Mean $\Xi^0 (1530)$ multiplicity & 10.0 & 0.0 & 1.0 \\
		Mean $\Lambda (1520)$ multiplicity & 10.0 & 0.0 & 1.0 \\
		Mean $D^{*+}_2 (2112)$ multiplicity & 10.0 & 0.0 & 1.0 \\
		Mean $\Sigma^- (1385)$ multiplicity & 10.0 & 0.0 & 1.0 \\
		Mean $f_1 (1420)$ multiplicity & 10.0 & 0.0 & 1.0 \\
		Mean $\phi (1020)$ multiplicity & 10.0 & 0.0 & 1.0 \\
		Mean $K^{*0}_2$ multiplicity & 10.0 & 0.0 & 1.0 \\
		Mean $\Omega^-$ multiplicity & 10.0 & 0.0 & 1.0 \\
		Mean $\Sigma^\pm (1385)$ multiplicity & 10.0 & 0.0 & 1.0 \\
		Mean $\psi(2S)$ multiplicity & 10.0 & 0.0 & 1.0 \\
		Mean $D^{*+}$ multiplicity & 10.0 & 0.0 & 1.0 \\
		Mean $B^*$ multiplicity & 10.0 & 0.0 & 1.0 \\
		Mean $\pi^0$ multiplicity & 10.0 & 0.0 & 1.0 \\
		Mean $\eta$ multiplicity & 10.0 & 0.0 & 1.0 \\
		Mean $a^+_0 (980)$ multiplicity & 10.0 & 0.0 & 1.0 \\
		Mean $D^+_{s1}$ multiplicity & 10.0 & 0.0 & 1.0 \\
		Mean $\rho^+ (770)$ multiplicity & 10.0 & 0.0 & 1.0 \\
		Mean $\Xi^-$ multiplicity & 10.0 & 0.0 & 1.0 \\
		Mean $\omega (782)$ multiplicity & 10.0 & 0.0 & 1.0 \\
		Mean $\Upsilon (1S)$ multiplicity & 10.0 & 0.0 & 1.0 \\
		\hline
  	\end{tabular}
\caption{Weights of multiplicities from Ref.~\cite{ref:nadine_fischer}. A neutral tune using weights of 1 for every observable was added. The data was originally published in Ref.~\cite{ref:pdg_multiplicities}.}
\label{tab:pdg_multiplicities1}
\end{table}
		
\begin{table}[htbp]
 	\centering
  	\begin{tabular}{ l c c c }
    		\hline
    		Observable & \multicolumn{3}{c}{Weights}\\ 
    		& Tune 1 & Tune 2 & Tune neutral\\ \hline
		Mean $\chi_{c1} (3510)$ multiplicity & 10.0 & 0.0 & 1.0 \\
		Mean $D^+$ multiplicity & 10.0 & 0.0 & 1.0 \\
		Mean $\Sigma^+$ multiplicity & 10.0 & 0.0 & 1.0 \\
		Mean $f_1 (1285)$ multiplicity & 10.0 & 0.0 & 1.0 \\
		Mean $f_2(1270)$ multiplicity & 10.0 & 0.0 & 1.0 \\
		Mean $J/\psi (1S)$ multiplicity & 10.0 & 0.0 & 1.0 \\
		Mean $B^+_u$ multiplicity & 10.0 & 0.0 & 1.0 \\
		Mean $B^{**}$ multiplicity & 10.0 & 0.0 & 1.0 \\
		Mean $\Lambda^+_c$ multiplicity & 10.0 & 0.0 & 1.0 \\
		Mean $D^0$ multiplicity & 10.0 & 0.0 & 1.0 \\
		Mean $f_2' (1525)$ multiplicity & 10.0 & 0.0 & 1.0 \\
		Mean $\Sigma^\pm$ multiplicity & 10.0 & 0.0 & 1.0 \\
		Mean $D^+_{s2}$ multiplicity & 10.0 & 0.0 & 1.0 \\
		Mean $K^{*0} (892)$ multiplicity & 10.0 & 0.0 & 1.0 \\
		Mean $\Sigma^-$ multiplicity & 10.0 & 0.0 & 1.0 \\
		Mean $\pi^+$ multiplicity & 10.0 & 0.0 & 1.0 \\
		Mean $f_0(980)$ multiplicity & 10.0 & 0.0 & 1.0 \\
		Mean $\Sigma^+ (1385)$ multiplicity & 10.0 & 0.0 & 1.0 \\
		Mean $D^+_s$ multiplicity & 10.0 & 0.0 & 1.0 \\
		Mean $p$ multiplicity & 10.0 & 0.0 & 1.0 \\
		Mean $B^0_s$ multiplicity & 10.0 & 0.0 & 1.0 \\
		Mean $K^0$ multiplicity & 10.0 & 0.0 & 1.0 \\
		Mean $B^+, B^0_d$ multiplicity & 10.0 & 0.0 & 1.0 \\
		Mean $\Lambda$ multiplicity & 10.0 & 0.0 & 1.0 \\
		Mean $\eta' (958)$ multiplicity & 10.0 & 0.0 & 1.0 \\
    		\hline
  	\end{tabular}
\caption{Weights of multiplicities from Ref.~\cite{ref:nadine_fischer}. A neutral tune using weights of 1 for every observable was added. The data was originally published in Ref.~\cite{ref:pdg_multiplicities}.}
\label{tab:pdg_multiplicities2}
\end{table}

\begin{table}[htbp]
 	\centering
  	\begin{tabular}{ l c c c }
    		\hline
    		Observable & \multicolumn{3}{c}{Weights}\\ 
    		& Tune 1 & Tune 2 & Tune neutral\\ \hline
		$b$ quark fragmentation function $f(x^{weak}_B)$ & 7.0 & 35.0 & 1.0 \\
		Mean of $b$ quark fragmentation function $f(x^{weak}_B)$ & 3.0 & 15.0 & 1.0 \\
    		\hline
  	\end{tabular}
\caption{Weights of observables related to the fragmentation of $b$ quarks from Ref.~\cite{ref:nadine_fischer}. A neutral tune using weights of 1 for every observable was added. The data was originally published in Ref.~\cite{ref:aleph_2001}.}
\label{tab:b_fragmentation}
\end{table}

\begin{table}[htbp]
\begin{adjustbox}{max width=\textwidth}
  	\begin{tabular}{ l c c c }
    		\hline
    		Observable & \multicolumn{3}{c}{Weights}\\ 
    		& Tune 1 & Tune 2 & Tune neutral\\ \hline
		In-plane $p_\perp$ in $GeV$ w.r.t. sphericity axes & 1.0 & 5.0 & 1.0 \\
		In-plane $p_\perp$ in $GeV$ w.r.t. thrust axes & 1.0 & 5.0 & 1.0 \\
		Out-of-plane $p_\perp$ in $GeV$ w.r.t. sphericity axes & 1.0 & 5.0 & 1.0 \\
		Out-of-plane $p_\perp$ in $GeV$ w.r.t. thrust axes & 1.0 & 5.0 & 1.0 \\
		Mean out-of-plane $p_\perp$ in $GeV$ w.r.t. thrust axis vs. $x_p = |p|/|p_{beam}|$ & 1.0 & 5.0 & 1.0 \\
		Mean $p_\perp$ in $GeV$ vs. $x_p$ & 1.0 & 5.0 & 1.0 \\
		Scaled momentum $x_p = |p|/|p_{beam}|$ & 1.0 & 5.0 & 1.0 \\
		Log of scaled momentum, $log(1/x_p)$ & 1.0 & 5.0 & 1.0 \\
		Energy-energy correlation, EEC & 1.0 & 5.0 & 1.0 \\
		Sphericity, $S$ & 1.0 & 5.0 & 1.0 \\
		Aplanarity, $A$ & 2.0 & 10.0 & 1.0 \\
		Planarity, $P$ & 1.0 & 5.0 & 1.0 \\
		$D$ Parameter & 1.0 & 5.0 & 1.0 \\
		$C$ Parameter & 1.0 & 5.0 & 1.0 \\
		1-Thrust & 1.0 & 5.0 & 1.0 \\
		Thrust major, $M$ & 1.0 & 5.0 & 1.0 \\
		Thrust minor, $m$ & 2.0 & 10.0 & 1.0 \\
		Oblatness $O = M - m$ & 1.0 & 5.0 & 1.0 \\
		Charged multiplicity distribution & 2.0 & 10.0 & 1.0 \\
		Mean charged multiplicity & 150.0 & 750.0 & 1.0 \\
    		\hline
  	\end{tabular}
\end{adjustbox}
\caption{Weights of event shape observables from Ref.~\cite{ref:nadine_fischer}. A neutral tune using weights of 1 for every observable was added. The data was originally published in Refs.~\cite{ref:aleph_1996,ref:delphi_1996}.}
\label{tab:event_shape}
\end{table}

\begin{table}[htbp]
 	\centering
  	\begin{tabular}{ l c c c }
    	\hline
    		Observable & \multicolumn{3}{c}{Weights}\\ 
    		& Tune 1 & Tune 2 & Tune neutral\\ \hline
    		Differential 2-jet rate & 2.0 & 10.0 & 1.0\\
    		Differential 3-jet rate & 2.0 & 10.0 & 1.0\\
    		Differential 4-jet rate & 2.0 & 10.0 & 1.0\\
    		Differential 5-jet rate & 2.0 & 10.0 & 1.0\\
    		\hline
  	\end{tabular}
\caption{Weights of differential jet rates from Ref.~\cite{ref:nadine_fischer}. A neutral tune using weights of 1 for every observable was added. The data was originally published in Ref.~\cite{ref:jade_opal}.}
\label{tab:diff_jet_rate}
\end{table}

\begin{table}[htbp]
 	\centering
  	\begin{tabular}{ c l l }
  		\hline
    		& Analysis & Observable \\ \hline
		\parbox[t]{2mm}{\multirow{9}{*}{\rotatebox[origin=c]{90}{Removed}}} & \verb|PDG_HADRON_MULTIPLICITIES| & Mean $f_0(980)$ multiplicity \\
		& \verb|PDG_HADRON_MULTIPLICITIES| & Mean $a_0^+(980)$ multiplicity \\
		& \verb|PDG_HADRON_MULTIPLICITIES| & Mean $f_1(1285)$ multiplicity \\
		& \verb|PDG_HADRON_MULTIPLICITIES| & Mean $f_1(1420)$ multiplicity \\
    		& \verb|PDG_HADRON_MULTIPLICITIES| & Mean $f_2(1270)$ multiplicity \\
		& \verb|PDG_HADRON_MULTIPLICITIES| & Mean $f_2'(1525)$ multiplicity \\
		& \verb|PDG_HADRON_MULTIPLICITIES| & Mean $K_2^{*0}(1430)$ multiplicity \\
		& \verb|PDG_HADRON_MULTIPLICITIES| & Mean $B^{**}$ multiplicity \\
		& \verb|PDG_HADRON_MULTIPLICITIES| & Mean $\Lambda(1520)$ multiplicity \\
    		\hline
    		\parbox[t]{2mm}{\multirow{12}{*}{\rotatebox[origin=c]{90}{Added}}} & \verb|DELPHI_1999_S3960137| & $\rho^0$ scaled momentum \\
		& \verb|OPAL_1994_S2927284| & $p,\bar{p}$ momentum \\
		& \verb|OPAL_1995_S3198391| & $\Delta^{++}$ scaled momentum \\
		& \verb|OPAL_1996_S3257789| & $J/\psi$ scaled momentum \\
    		& \verb|OPAL_1997_S3396100| & $\Sigma^+(1385)$ scaled momentum \\
		& \verb|OPAL_1997_S3396100| & $\Sigma^-(1385)$ scaled momentum \\
		& \verb|OPAL_1997_S3396100| & $\Lambda^0(1520)$ scaled momentum \\
		& \verb|OPAL_1998_S3702294| & $f_0(980)$ scaled momentum \\
		& \verb|OPAL_1998_S3702294| & $f_2(1270)$ scaled momentum \\
		& \verb|OPAL_1998_S3749908| & $\rho^\pm$ scaled momentum \\
		& \verb|OPAL_1998_S3749908| & $\eta'$ scaled momentum \\
		& \verb|OPAL_1998_S3749908| & $a_0^\pm$ scaled momentum \\
    		\hline
  	\end{tabular}
\caption{Removing of model parameters insensitive observables from \cref{tab:particle_spectrum,tab:pdg_multiplicities1,tab:pdg_multiplicities2,tab:b_fragmentation,tab:event_shape,tab:diff_jet_rate} and adding of new observables.}
\label{tab:exchangeobservables}
\end{table} 

\begin{table}[htbp]
 	\centering
  	\begin{tabular}{ l c }
    		\hline
    		Observable & Modification \\ \hline
		Mean out-of-plane $p_\perp$ in GeV w.r.t. thrust axes vs. $x_p$ & Truncation of the edges \\
		Mean $p_\perp$ in GeV vs. $x_p$ & Truncation of the edges \\
		$K^\pm$ spectrum & Removed data points in the center \\
		$\eta$ spectrum & Removed data points in the center \\
		$K^0$ spectrum & Removed data points in the center \\
		$\Sigma^\pm(1385)$ spectrum & Truncation of the edges \\
		$\Xi^0(1530)$ spectrum & Removed completely \\
		$K^{*0}(892)$ spectrum & Removed data point in the center \\
		Mean $D^+$ multiplicity & Removed completely \\
		Mean $D^0$ multiplicity & Removed completely \\
		Mean $B^+_u$ multiplicity & Removed completely \\
		Mean $B^0_s$ multiplicity & Removed completely \\
		Mean $D^{*+}(2010)$ multiplicity & Removed completely \\
		Mean $\psi(2S)$ multiplicity & Removed completely \\
		Mean $\Upsilon(1S)$ multiplicity & Removed completely \\
		Mean $\chi_{c1}(3510)$ multiplicity & Removed completely \\
		Mean $D_{s2}^+$ multiplicity & Removed completely \\
		Mean $\Delta^{++}(1232)$ multiplicity & Removed completely \\
		Mean $\Omega^-$ multiplicity & Removed completely \\
		Mean $\Lambda_c^+$ multiplicity & Removed completely \\
		$p,\bar{p}$ momentum & Removed data points in the center \\
		$\Delta^{++}$ scaled momentum & Truncation of the edges \\
		$\Sigma^+(1385)$ scaled momentum & Truncation of the edges \\
		$\Sigma^-(1385)$ scaled momentum & Truncation of the edges \\
		$\Lambda^0(1520)$ scaled momentum & Removed completely \\
		$f_0(980)$ scaled momentum & Removed completely \\
		$f_2(1270)$ scaled momentum & Removed completely \\
		$a_0^\pm$ scaled momentum & Removed completely \\
    		\hline
  	\end{tabular}
\caption{Changes due to outlier rejection performed on the modified list of observables in Tab.~\ref{tab:exchangeobservables}.}
\label{tab:outlier}
\end{table}

\FloatBarrier
\chapter{\textsc{Professor} based tuning}
\setcounter{page}{102}

\begin{table}[htbp]
  \centering
  \begin{tabular}{ l l l }
    \hline
    Parameter & Value (left distribution) & Value (right distribution) \\ \hline
    $a_{\textrm Lund}$ & 0.42 & 0.56 \\
    $b_{\textrm Lund}~[GeV^{-2}]$ & 0.82 & 1.00 \\
    $a_{\textrm ExtraDiquark}$ & 0.99 & 0.93 \\
    $\sigma~[GeV]$ & 0.259 & 0.399 \\
    $\alpha_s(M_Z)$ & 0.120 & 0.126 \\
    $p_{T,min}~[GeV]$ & 0.42 & 0.95 \\
    \hline
  \end{tabular}
  \caption{Parameter values used for the MC simulation in Fig.~\ref{fig:example_param_variation}.}
  \label{tab:parameter_values_example}
\end{table}

\begin{table}[htbp]
  \centering
  \begin{tabular}{ l l l }
    \hline
    Parameter & Tune 1 & Tune 2 \\ \hline
    $a_{\textrm Lund}$ & 0.35 & 0.42 \\
    $b_{\textrm Lund}~[GeV^{-2}]$ & 0.87 & 0.98 \\
    $a_{\textrm ExtraDiquark}$ & 0.61 & 0.62 \\
    $\sigma~[GeV]$ & 0.284 & 0.289 \\
    $\alpha_s(M_Z)$ & 0.139 & 0.139 \\
    $p_{T,min}~[GeV]$ & 0.42 & 0.41 \\
    \hline
  \end{tabular}
  \caption{Tuned parameter values using the run-combinations in \textsc{Professor} with the weights of \cref{tab:particle_spectrum,tab:pdg_multiplicities1,tab:pdg_multiplicities2,tab:b_fragmentation,tab:event_shape,tab:diff_jet_rate} in App.~\ref{asec:observablesandweights}. The values of $a_{\textrm Lund}$ are fixed to the tuning results of \cite{ref:nadine_fischer}.}
  \label{tab:fixed_alund}
\end{table}

\newpage
\section{Distribution of the run-combinations}
\label{asec:distribution_reproduction}

\begin{figure}[htbp]
	\centering
	\includegraphics[width=0.445\textwidth]{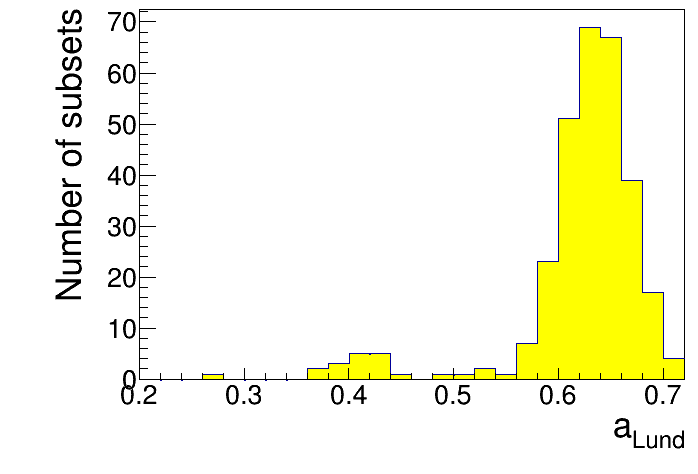}
	\includegraphics[width=0.445\textwidth]{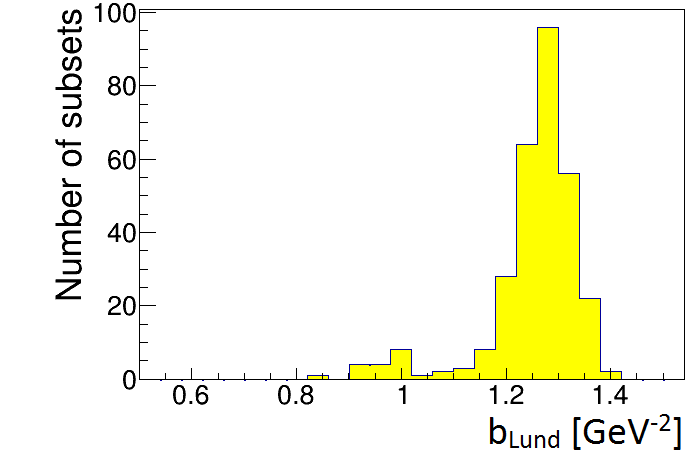}
	\includegraphics[width=0.445\textwidth]{Abbildungen/tune_nadine/weights0/aextradiquark_distr.png}
	\includegraphics[width=0.445\textwidth]{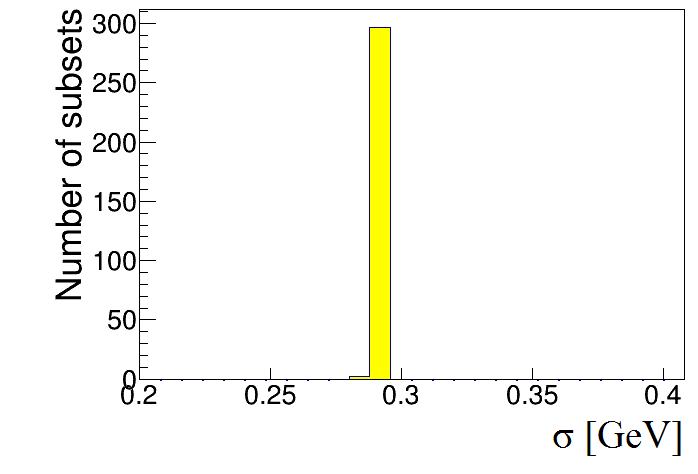}
	\includegraphics[width=0.445\textwidth]{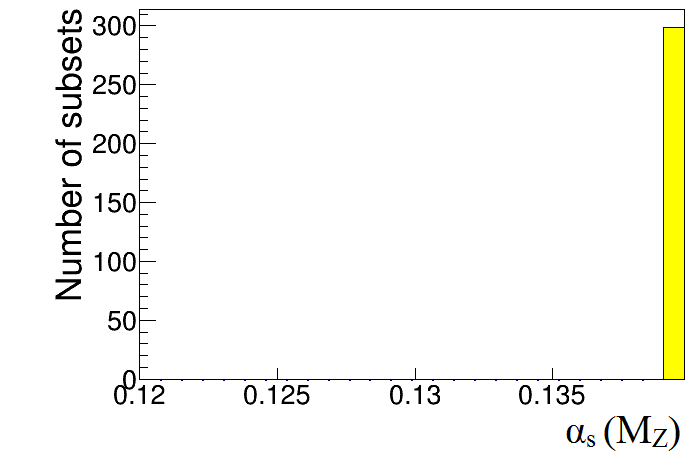}
	\includegraphics[width=0.445\textwidth]{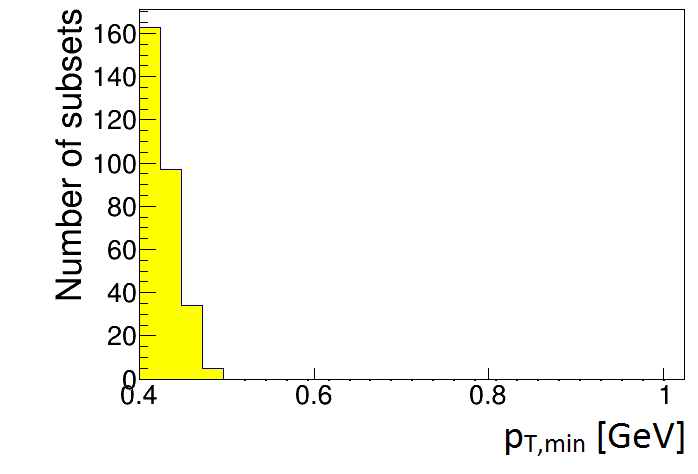}
	\caption{Distribution of the tuned model parameters using the run-combinations from Subsection \ref{subsec:tuning_setup} and the weights ``Tune neutral'' from \cref{tab:particle_spectrum,tab:pdg_multiplicities1,tab:pdg_multiplicities2,tab:b_fragmentation,tab:event_shape,tab:diff_jet_rate} in App.~\ref{asec:observablesandweights}.}
	\label{fig:tune_nadine_weights0}
\end{figure}

\begin{figure}
	\centering
	\includegraphics[width=0.45\textwidth]{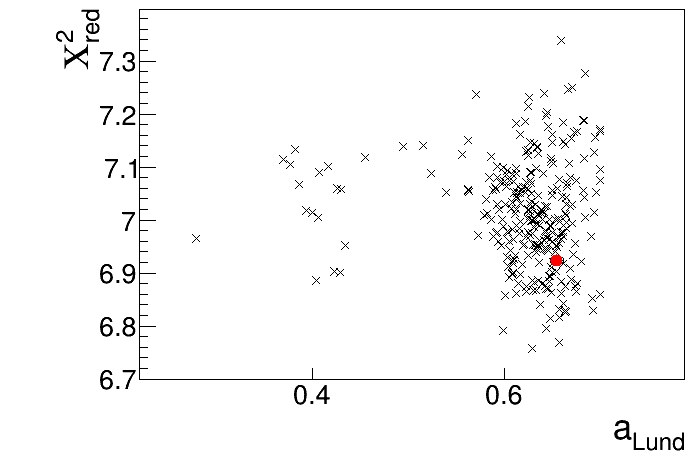}
	\includegraphics[width=0.45\textwidth]{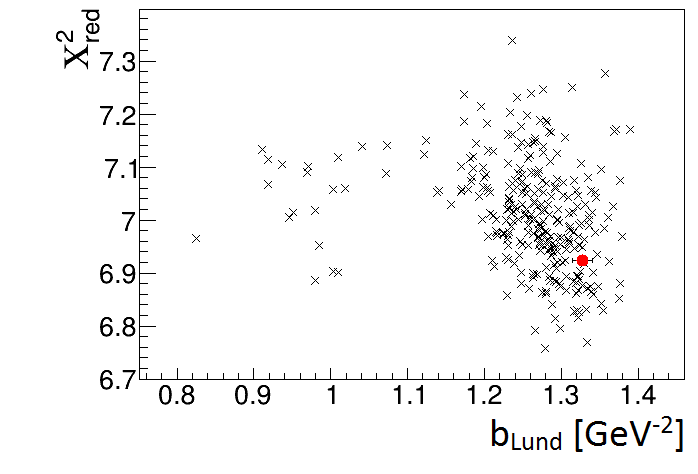}
	\includegraphics[width=0.45\textwidth]{Abbildungen/tune_nadine/weights0/aextradiquark_chi2.png}
	\includegraphics[width=0.45\textwidth]{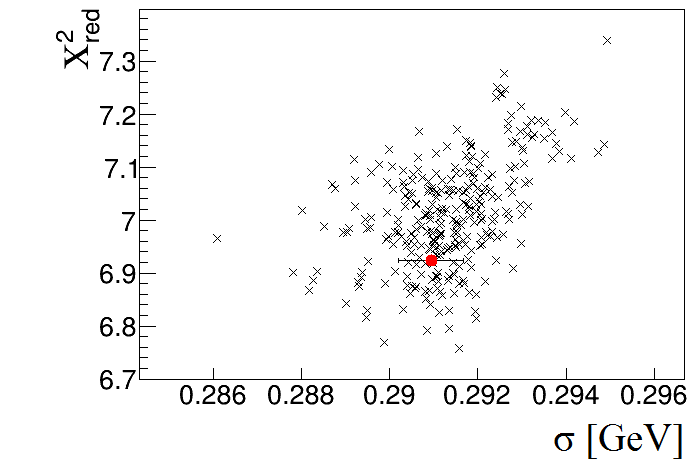}
	\includegraphics[width=0.45\textwidth]{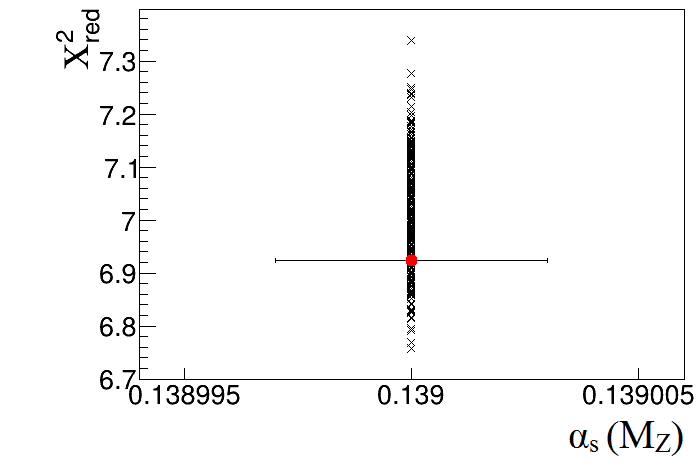}
	\includegraphics[width=0.45\textwidth]{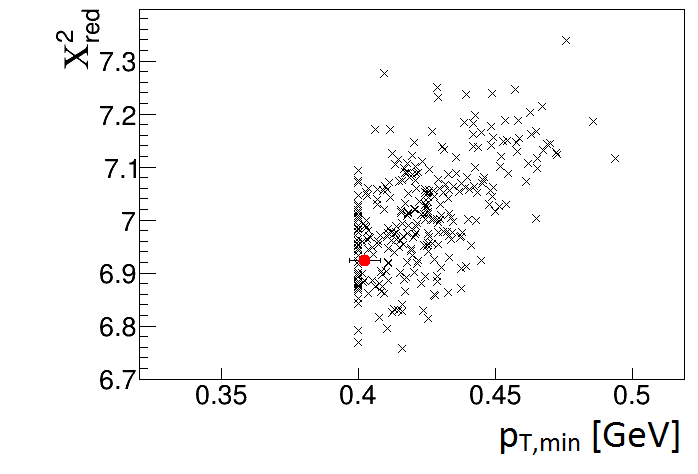}
	\caption{Distribution of the $\chi^2_{red}$ values of the tuned model parameters using the run-combinations from Subsection \ref{subsec:tuning_setup} and the weights ``Tune neutral'' from \cref{tab:particle_spectrum,tab:pdg_multiplicities1,tab:pdg_multiplicities2,tab:b_fragmentation,tab:event_shape,tab:diff_jet_rate} in App.~\ref{asec:observablesandweights}.}
	\label{fig:tune_nadine_weights0_chi2}
\end{figure}

\begin{figure}
	\centering
	\includegraphics[width=0.45\textwidth]{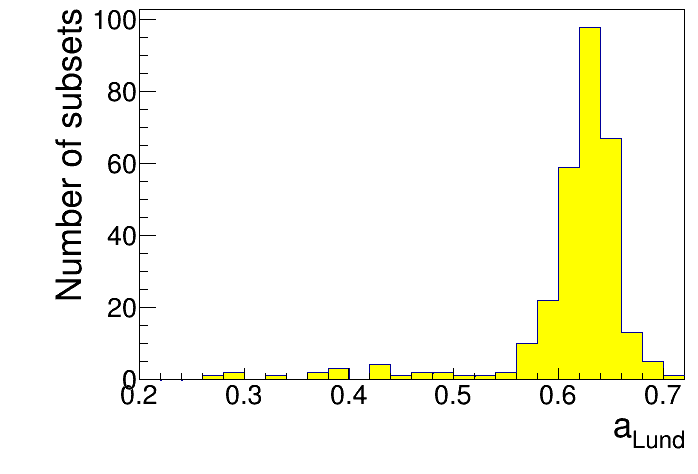}
	\includegraphics[width=0.45\textwidth]{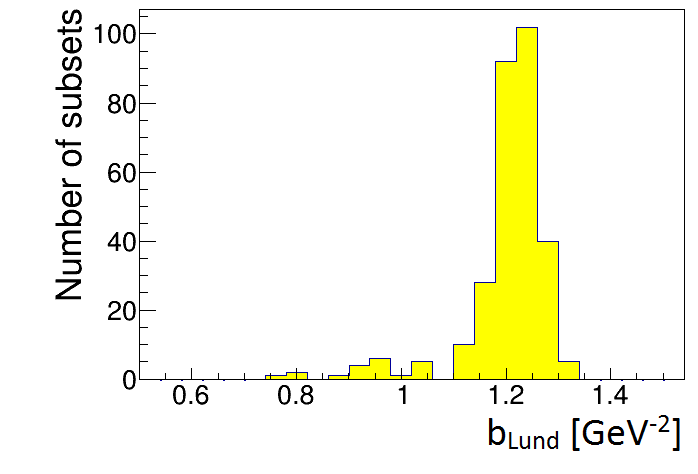}
	\includegraphics[width=0.45\textwidth]{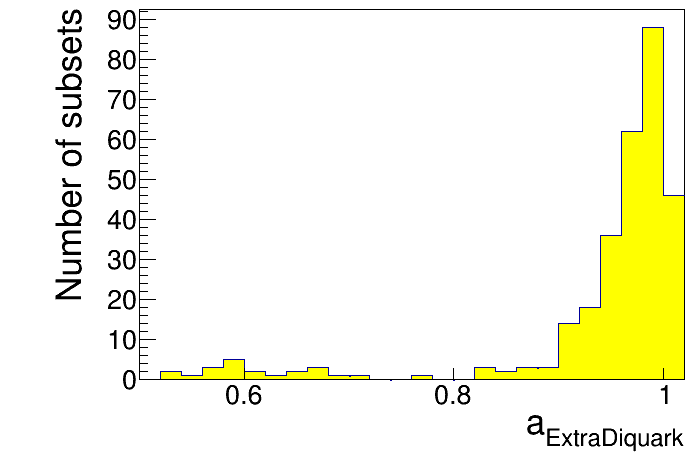}
	\includegraphics[width=0.45\textwidth]{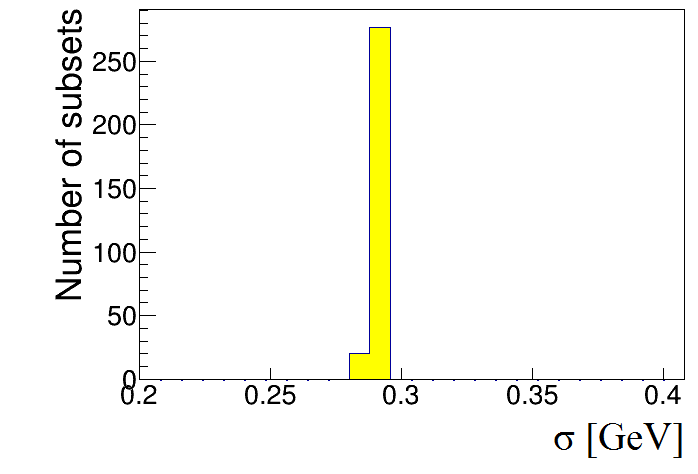}
	\includegraphics[width=0.45\textwidth]{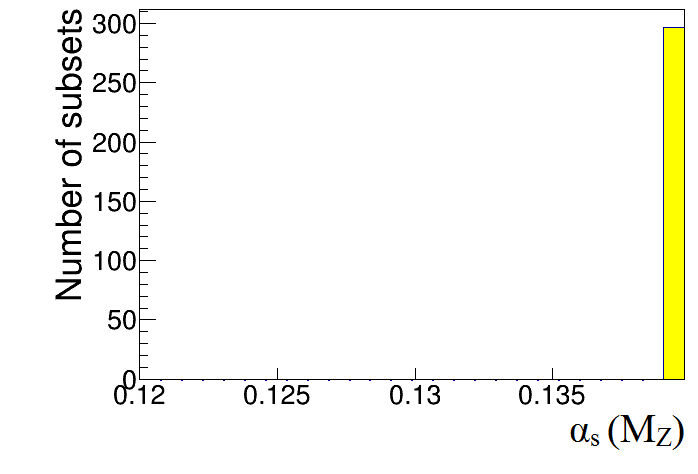}
	\includegraphics[width=0.45\textwidth]{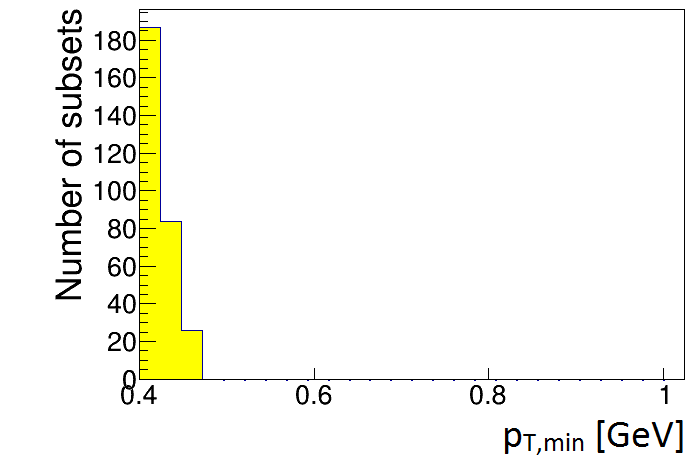}
	\caption{Distribution of the tuned model parameters using the run-combinations from Subsection \ref{subsec:tuning_setup} and the weights ``Tune 1'' from \cref{tab:particle_spectrum,tab:pdg_multiplicities1,tab:pdg_multiplicities2,tab:b_fragmentation,tab:event_shape,tab:diff_jet_rate} in App.~\ref{asec:observablesandweights}.}
	\label{fig:tune_nadine_weights1}
\end{figure}

\begin{figure}
	\centering
	\includegraphics[width=0.45\textwidth]{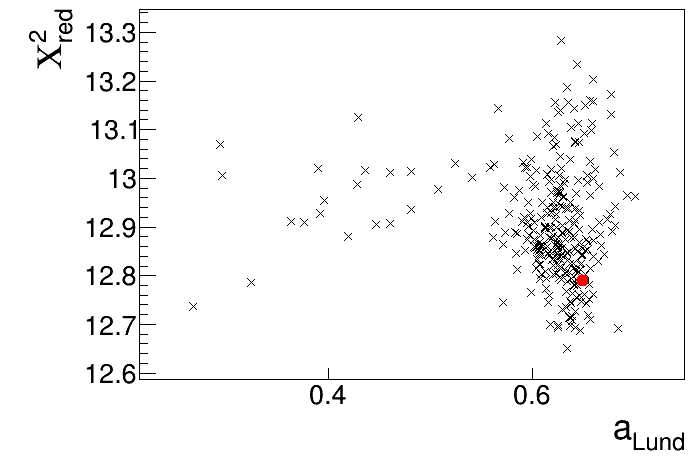}
	\includegraphics[width=0.45\textwidth]{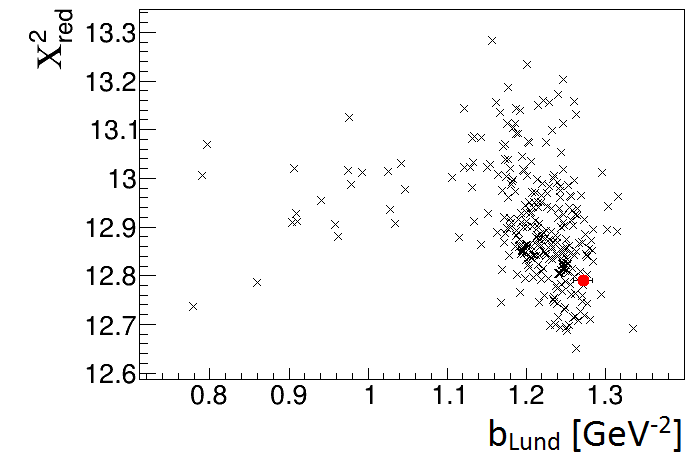}
	\includegraphics[width=0.45\textwidth]{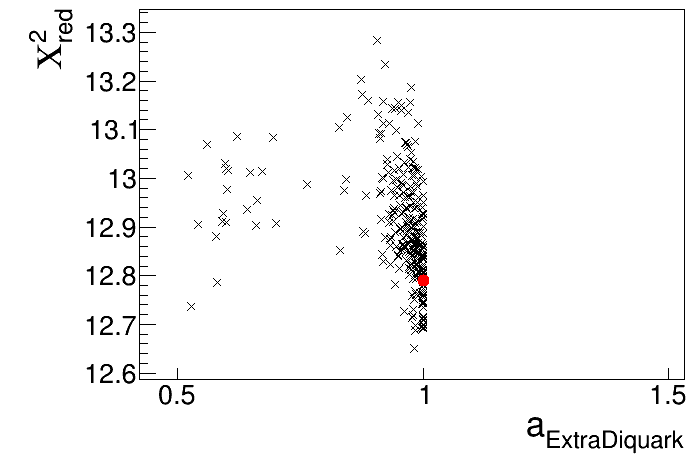}
	\includegraphics[width=0.45\textwidth]{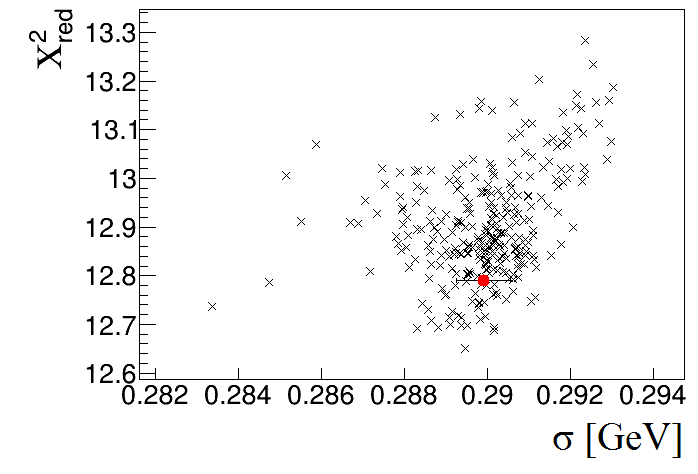}
	\includegraphics[width=0.45\textwidth]{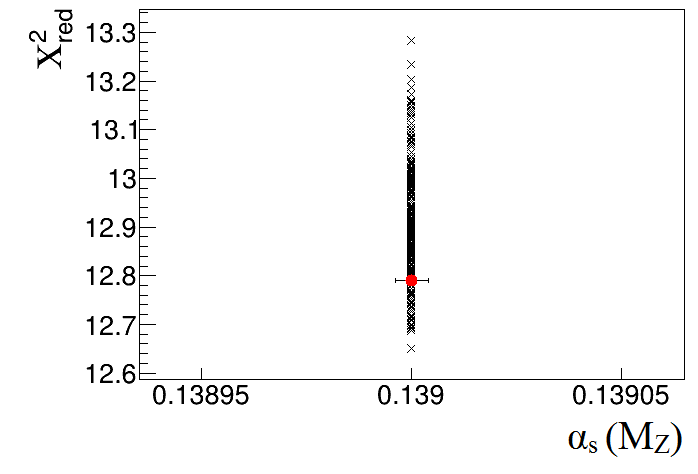}
	\includegraphics[width=0.45\textwidth]{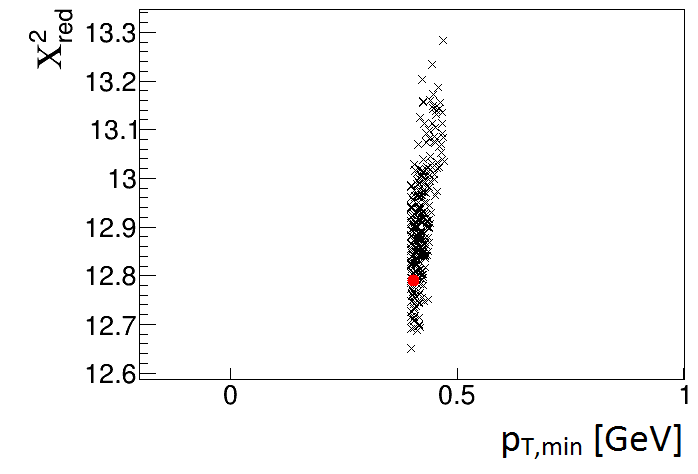}
	\caption{Distribution of the $\chi^2_{red}$ values of the tuned model parameters using the run-combinations from Subsection \ref{subsec:tuning_setup} and the weights ``Tune 1'' from \cref{tab:particle_spectrum,tab:pdg_multiplicities1,tab:pdg_multiplicities2,tab:b_fragmentation,tab:event_shape,tab:diff_jet_rate} in App.~\ref{asec:observablesandweights}.}
	\label{fig:tune_nadine_weights1_chi2}
\end{figure}

\begin{figure}
	\centering
	\includegraphics[width=0.45\textwidth]{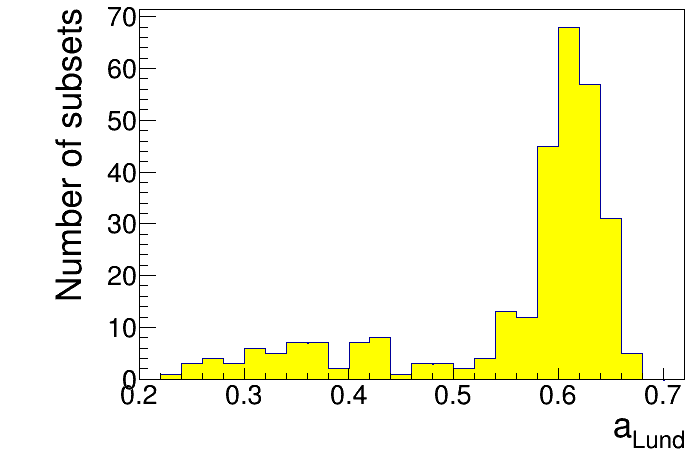}
	\includegraphics[width=0.45\textwidth]{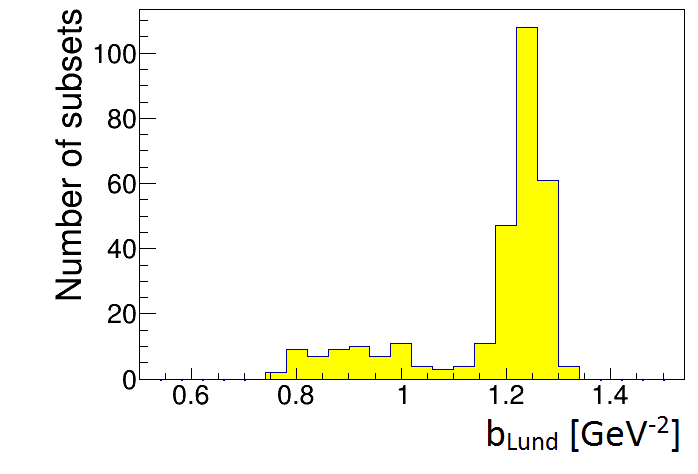}
	\includegraphics[width=0.45\textwidth]{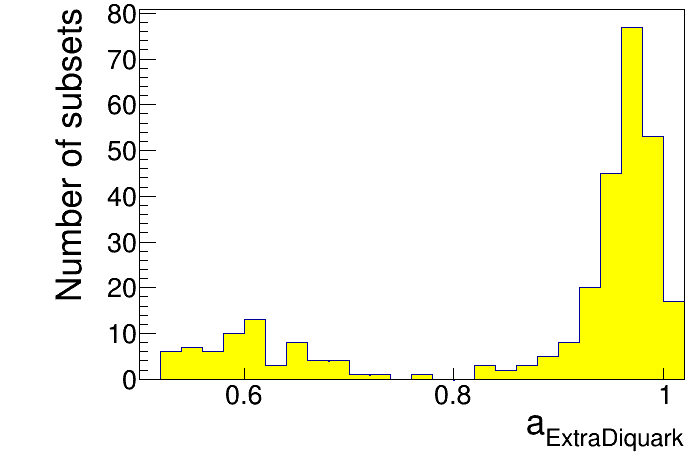}
	\includegraphics[width=0.45\textwidth]{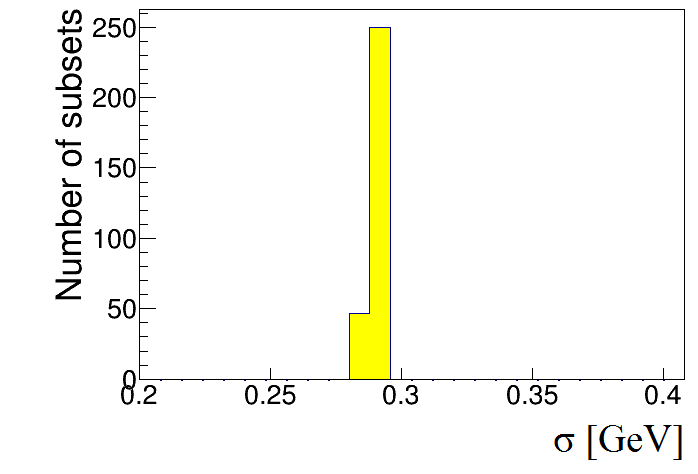}
	\includegraphics[width=0.45\textwidth]{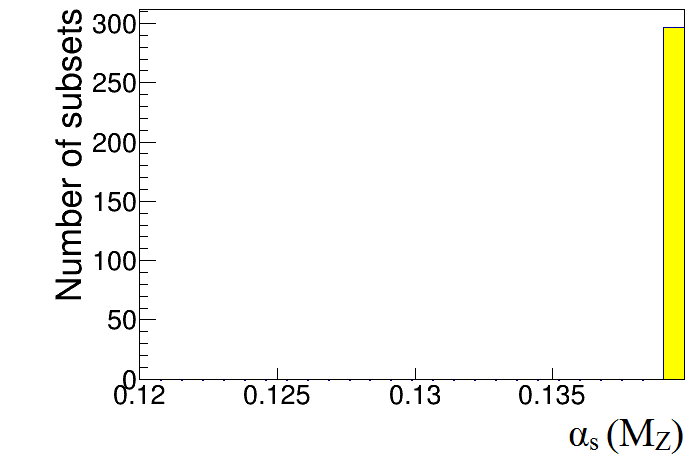}
	\includegraphics[width=0.45\textwidth]{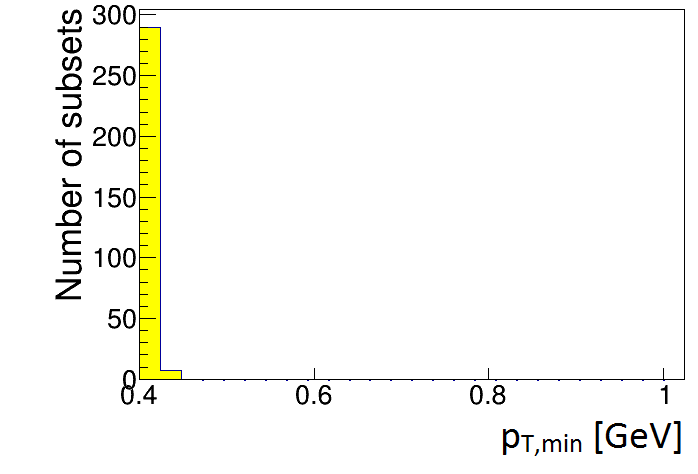}
	\caption{Distribution of the tuned model parameters using the run-combinations from Subsection \ref{subsec:tuning_setup} and the weights ``Tune 2'' from \cref{tab:particle_spectrum,tab:pdg_multiplicities1,tab:pdg_multiplicities2,tab:b_fragmentation,tab:event_shape,tab:diff_jet_rate} in App.~\ref{asec:observablesandweights}.}
	\label{fig:tune_nadine_weights2}
\end{figure}

\begin{figure}
	\centering
	\includegraphics[width=0.45\textwidth]{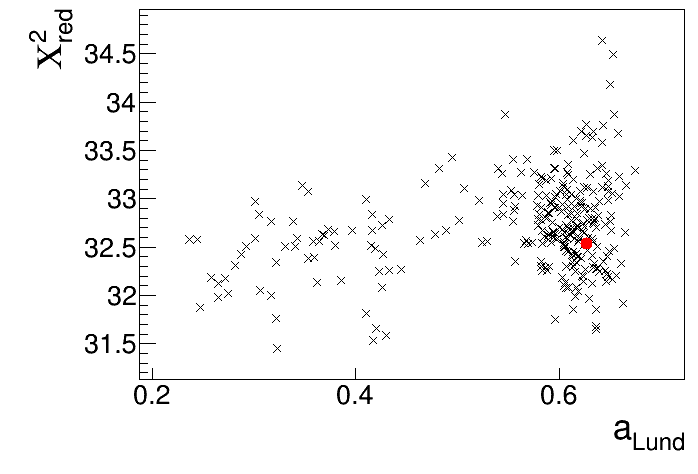}
	\includegraphics[width=0.45\textwidth]{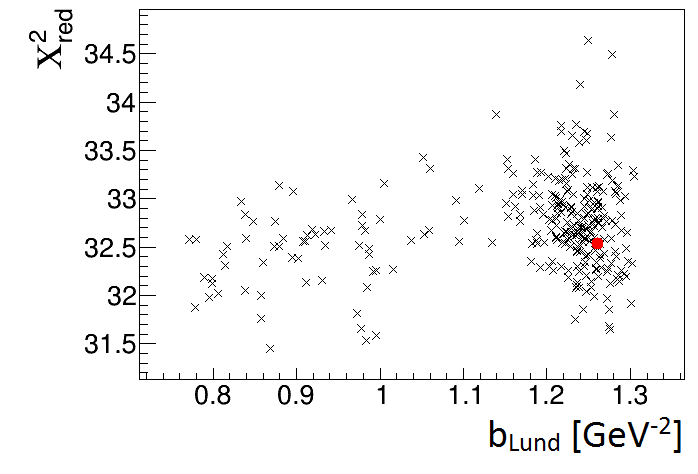}
	\includegraphics[width=0.45\textwidth]{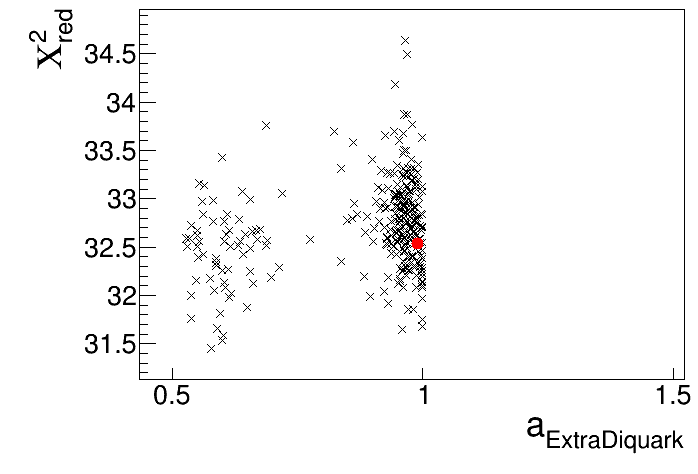}
	\includegraphics[width=0.45\textwidth]{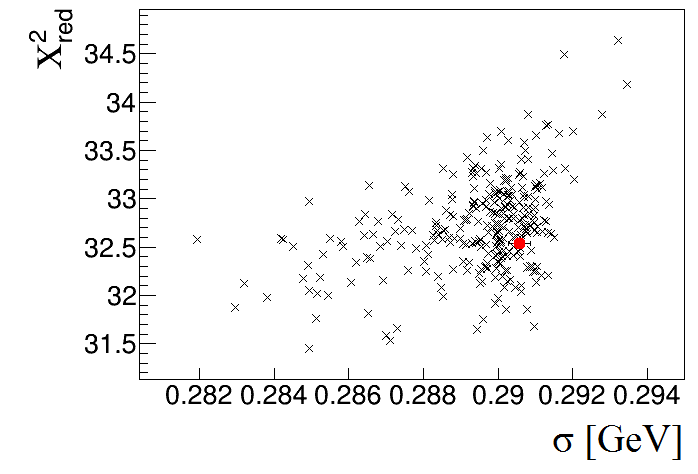}
	\includegraphics[width=0.45\textwidth]{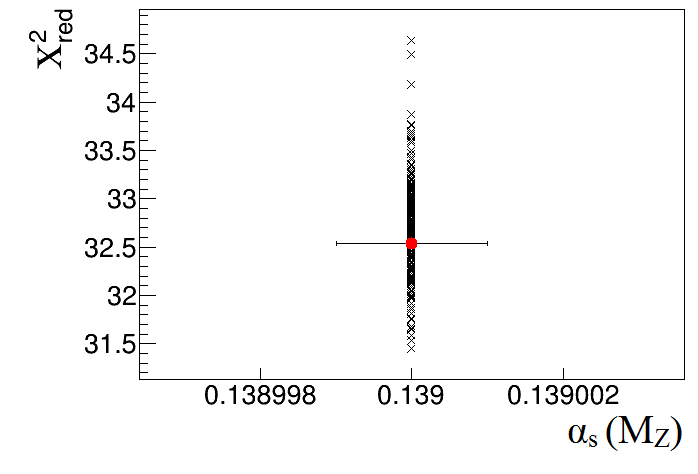}
	\includegraphics[width=0.45\textwidth]{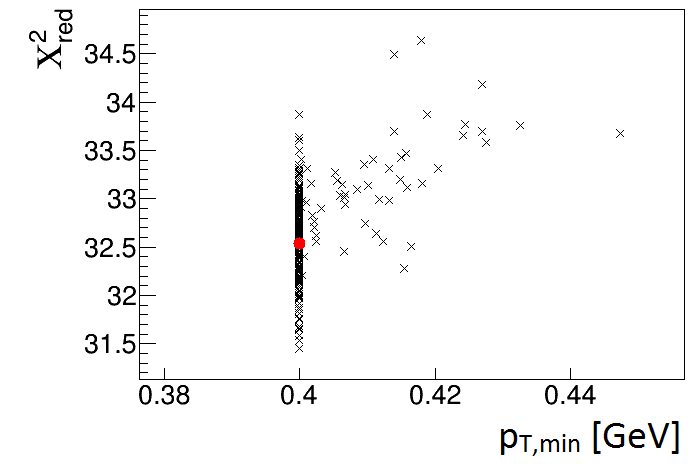}
	\caption{Distribution of the $\chi^2_{red}$ values of the tuned model parameters using the run-combinations from Subsection \ref{subsec:tuning_setup} and the weights ``Tune 2'' from \cref{tab:particle_spectrum,tab:pdg_multiplicities1,tab:pdg_multiplicities2,tab:b_fragmentation,tab:event_shape,tab:diff_jet_rate} in App.~\ref{asec:observablesandweights}.}
	\label{fig:tune_nadine_weights2_chi2}
\end{figure}

\FloatBarrier
\section{Distribution of the run-combinations without parameter ranges}
\label{asec:distribution_reproduction_nolimits}

\begin{figure}[h]
	\centering
	\includegraphics[width=0.41\textwidth]{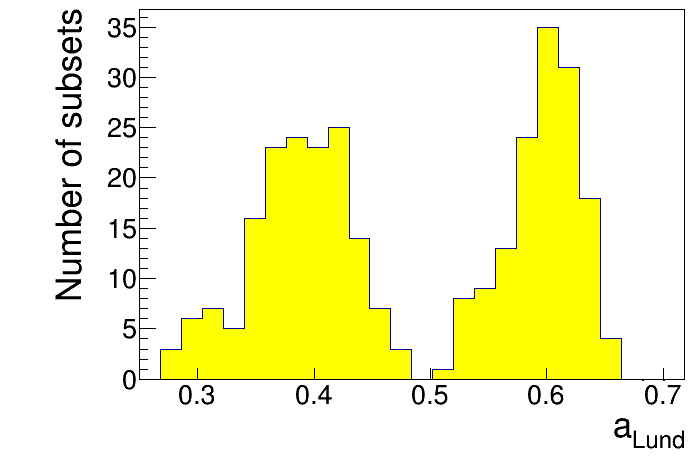}
	\includegraphics[width=0.41\textwidth]{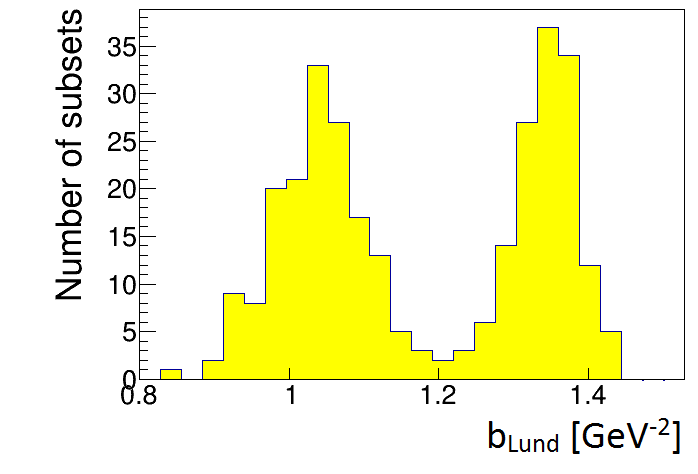}
	\includegraphics[width=0.41\textwidth]{Abbildungen/tune_nadine/no_limits/weights0/aextradiquark_distr.png}
	\includegraphics[width=0.41\textwidth]{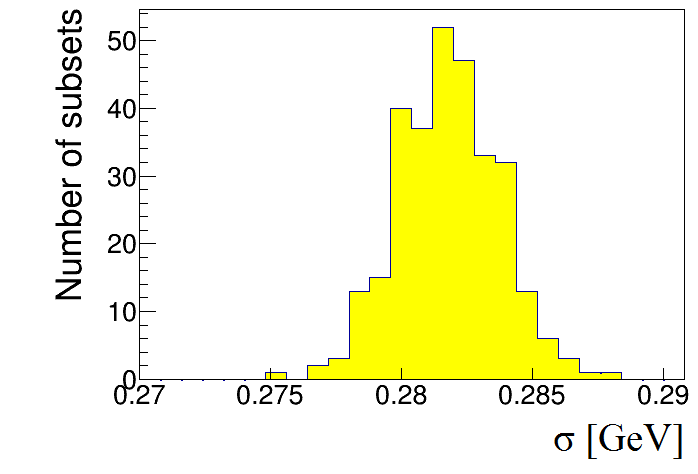}
	\includegraphics[width=0.41\textwidth]{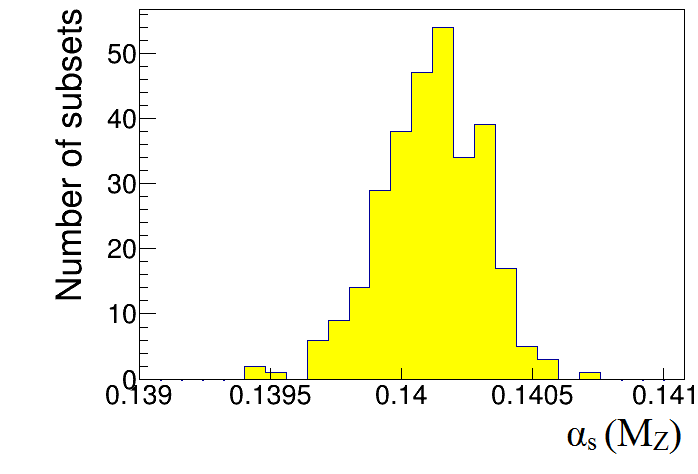}
	\includegraphics[width=0.41\textwidth]{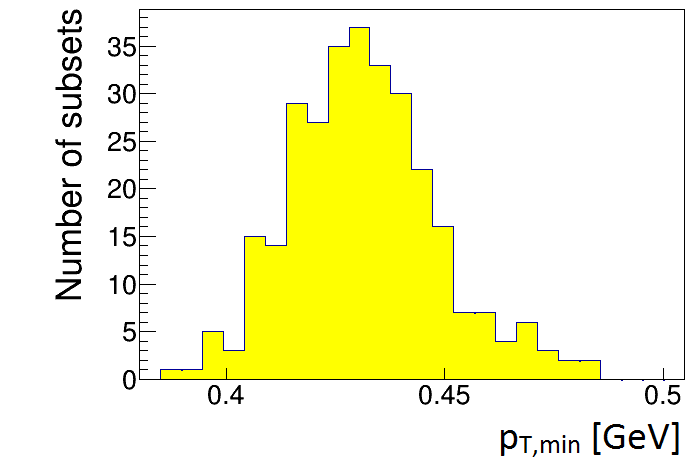}
	\caption{Distribution of the tuned model parameters using the run-combinations from Subsection \ref{subsec:tuning_setup} without range limitations and the weights ``Tune neutral'' from \cref{tab:particle_spectrum,tab:pdg_multiplicities1,tab:pdg_multiplicities2,tab:b_fragmentation,tab:event_shape,tab:diff_jet_rate} in App.~\ref{asec:observablesandweights}.}
	\label{fig:tune_nadine_weights0_nolimits}
\end{figure}

\begin{figure}
	\centering
	\includegraphics[width=0.45\textwidth]{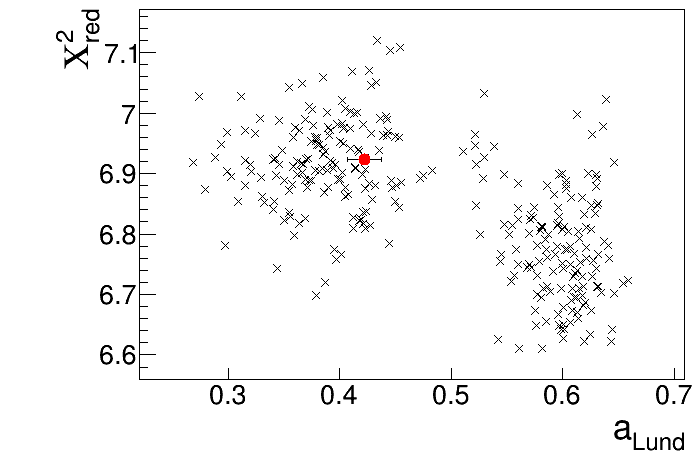}
	\includegraphics[width=0.45\textwidth]{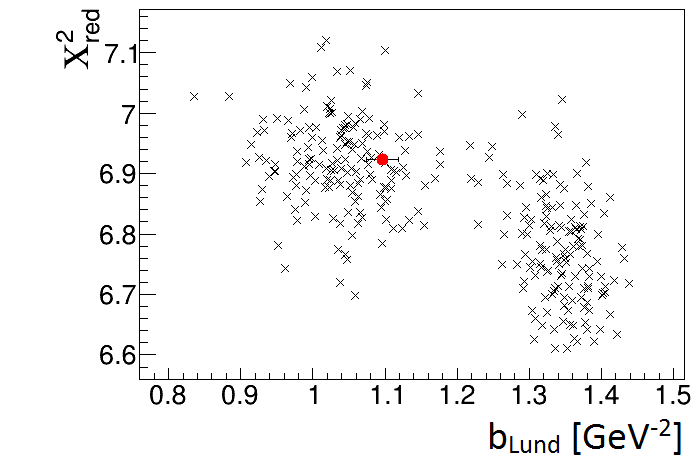}
	\includegraphics[width=0.45\textwidth]{Abbildungen/tune_nadine/no_limits/weights0/aextradiquark_chi2.png}
	\includegraphics[width=0.45\textwidth]{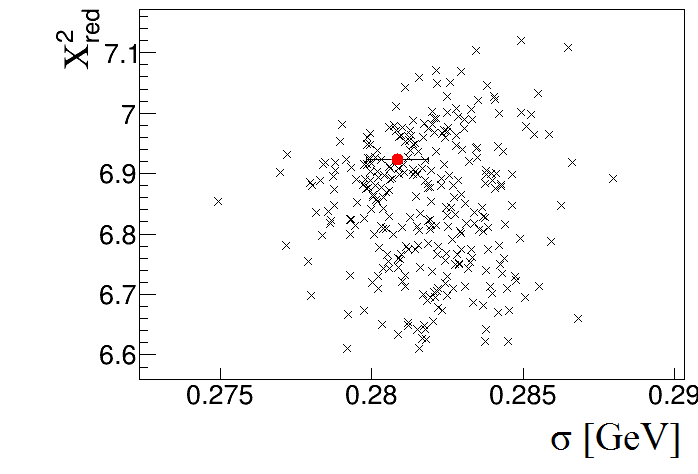}
	\includegraphics[width=0.45\textwidth]{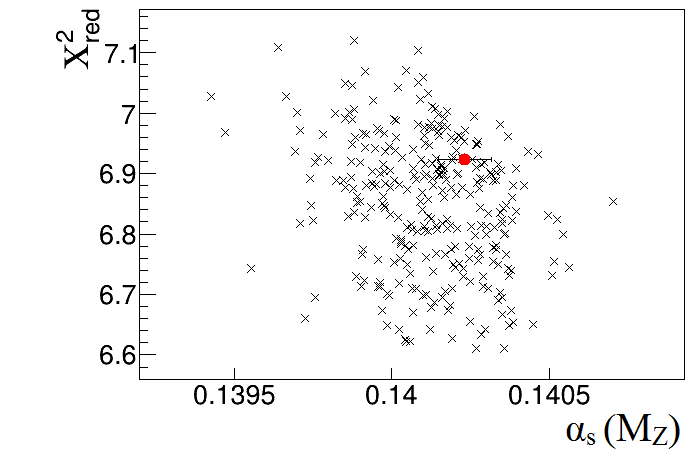}
	\includegraphics[width=0.45\textwidth]{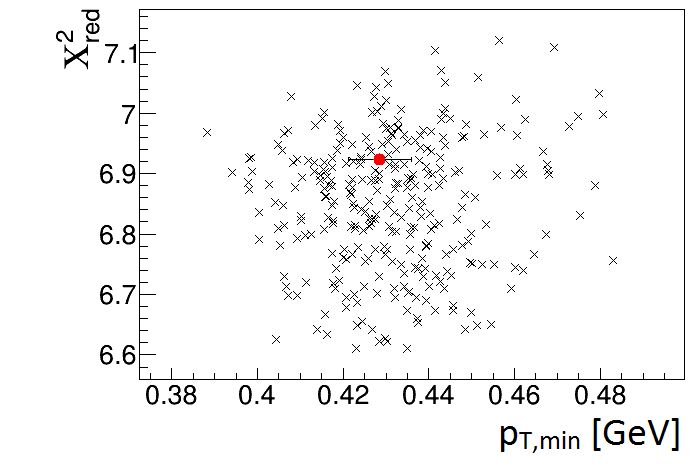}
	\caption{Distribution of the $\chi^2_{red}$ values of the tuned model parameters using the run-combinations from Subsection \ref{subsec:tuning_setup} without range limitations and the weights ``Tune neutral'' from \cref{tab:particle_spectrum,tab:pdg_multiplicities1,tab:pdg_multiplicities2,tab:b_fragmentation,tab:event_shape,tab:diff_jet_rate} in App.~\ref{asec:observablesandweights}.}
	\label{fig:tune_nadine_weights0_nolimits_chi2}
\end{figure}

\begin{figure}
	\centering
	\includegraphics[width=0.45\textwidth]{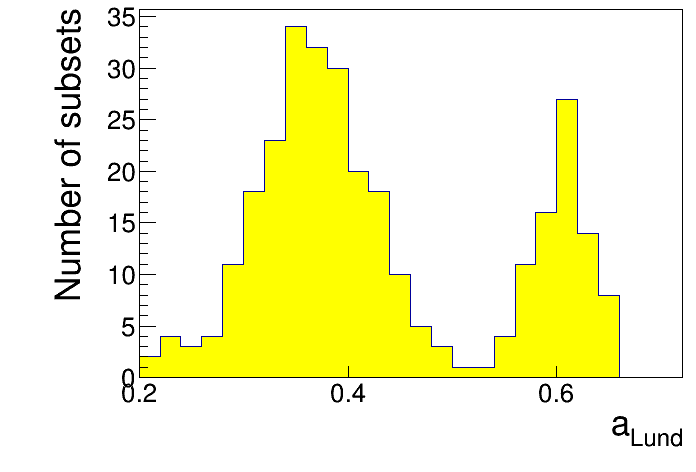}
	\includegraphics[width=0.45\textwidth]{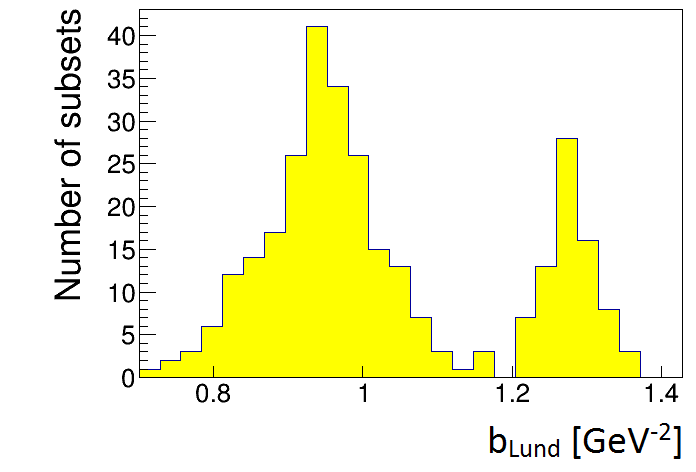}
	\includegraphics[width=0.45\textwidth]{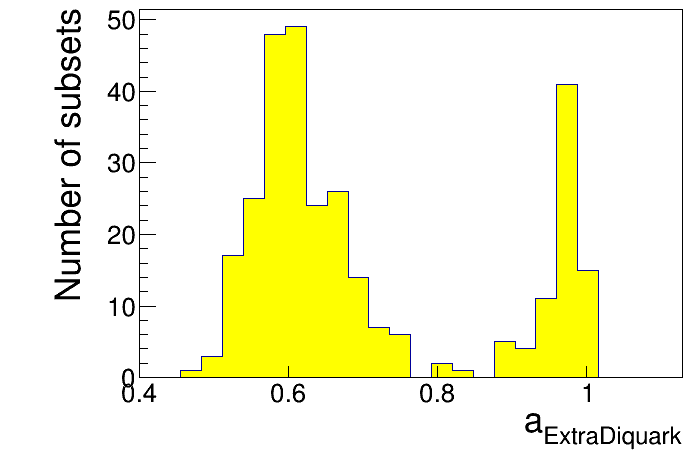}
	\includegraphics[width=0.45\textwidth]{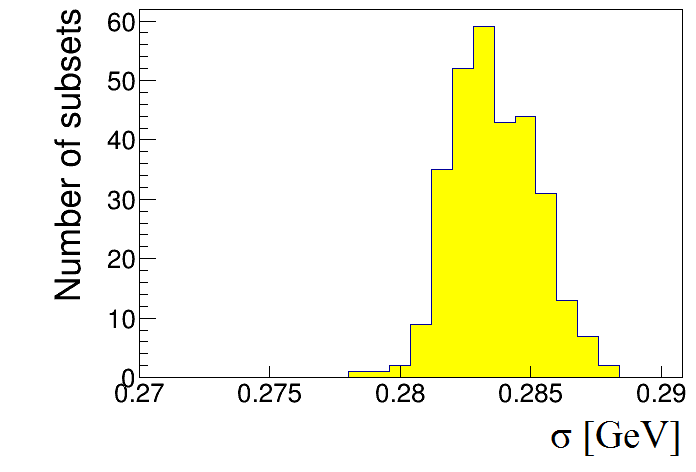}
	\includegraphics[width=0.45\textwidth]{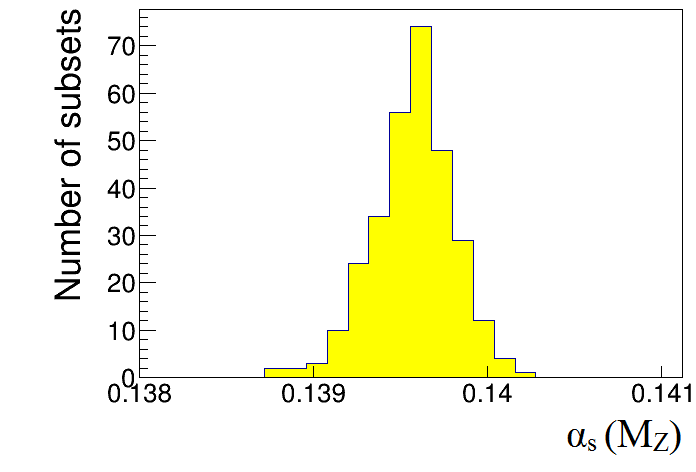}
	\includegraphics[width=0.45\textwidth]{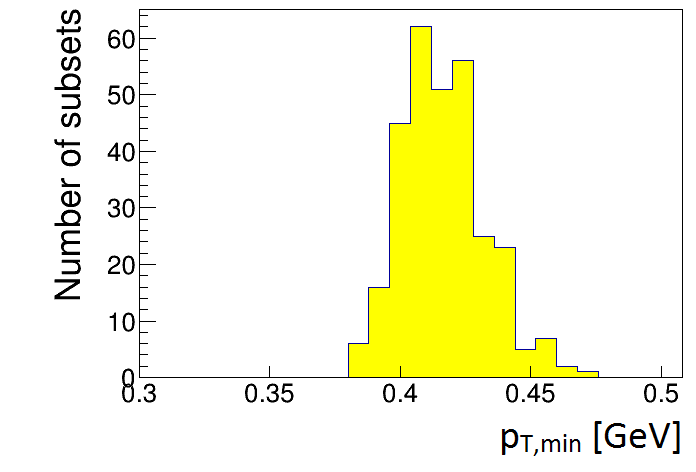}
	\caption{Distribution of the tuned model parameters using the run-combinations from Subsection \ref{subsec:tuning_setup} without range limitations and the weights ``Tune 1'' from \cref{tab:particle_spectrum,tab:pdg_multiplicities1,tab:pdg_multiplicities2,tab:b_fragmentation,tab:event_shape,tab:diff_jet_rate} in App.~\ref{asec:observablesandweights}.}
	\label{fig:tune_nadine_weights1_nolimits}
\end{figure}

\begin{figure}
	\centering
	\includegraphics[width=0.45\textwidth]{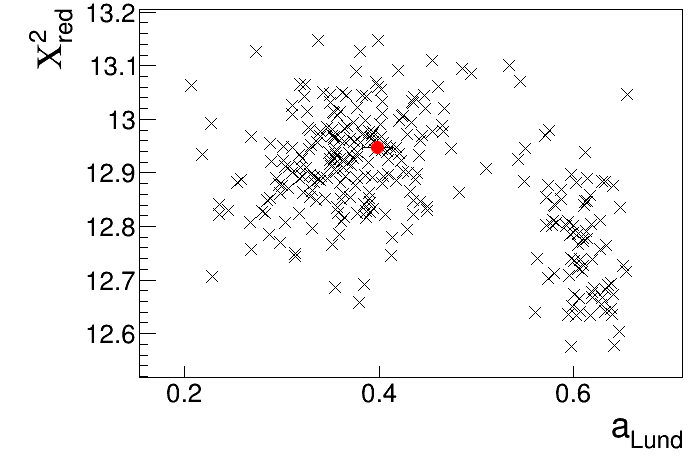}
	\includegraphics[width=0.45\textwidth]{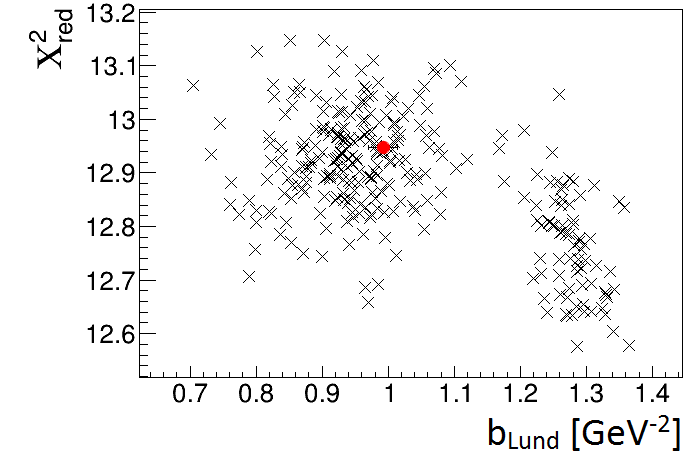}
	\includegraphics[width=0.45\textwidth]{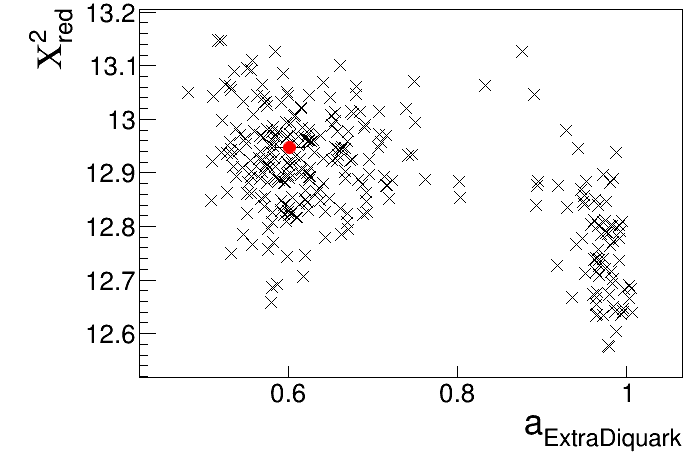}
	\includegraphics[width=0.45\textwidth]{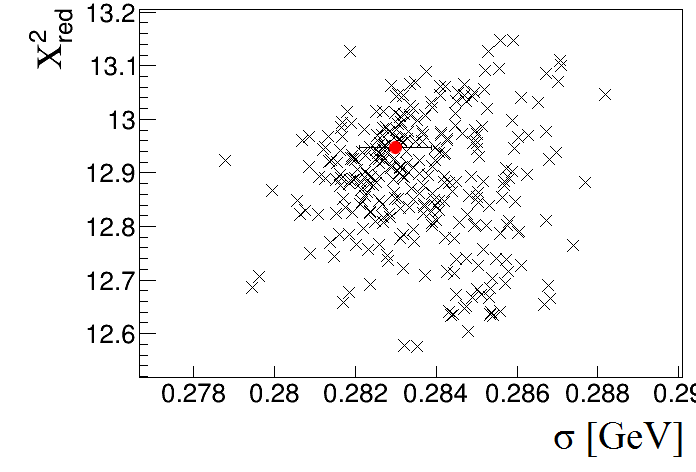}
	\includegraphics[width=0.45\textwidth]{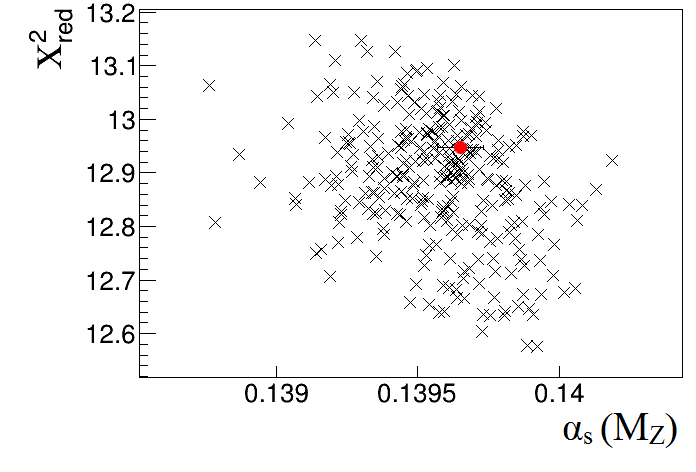}
	\includegraphics[width=0.45\textwidth]{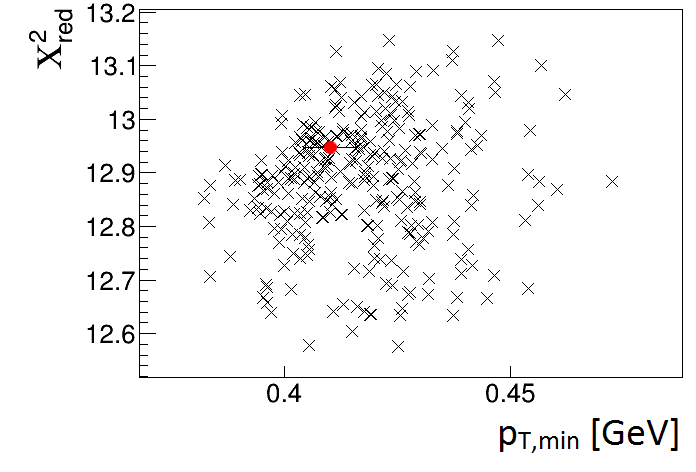}
	\caption{Distribution of the $\chi^2_{red}$ values of the tuned model parameters using the run-combinations from Subsection \ref{subsec:tuning_setup} without range limitations and the weights ``Tune 1'' from \cref{tab:particle_spectrum,tab:pdg_multiplicities1,tab:pdg_multiplicities2,tab:b_fragmentation,tab:event_shape,tab:diff_jet_rate} in App.~\ref{asec:observablesandweights}.}
	\label{fig:tune_nadine_weights1_nolimits_chi2}
\end{figure}

\begin{figure}
	\centering
	\includegraphics[width=0.45\textwidth]{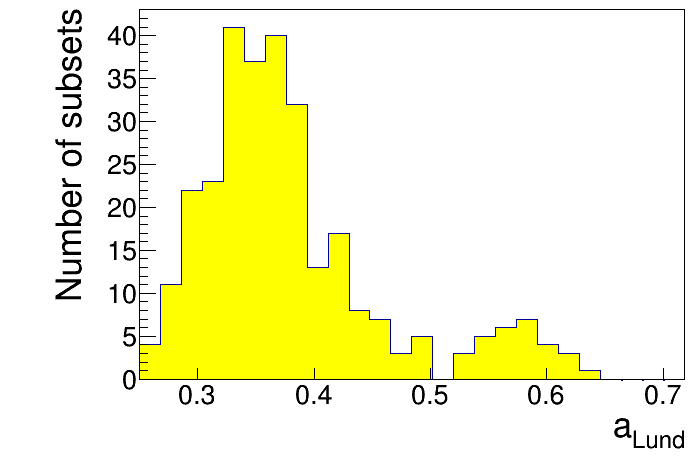}
	\includegraphics[width=0.45\textwidth]{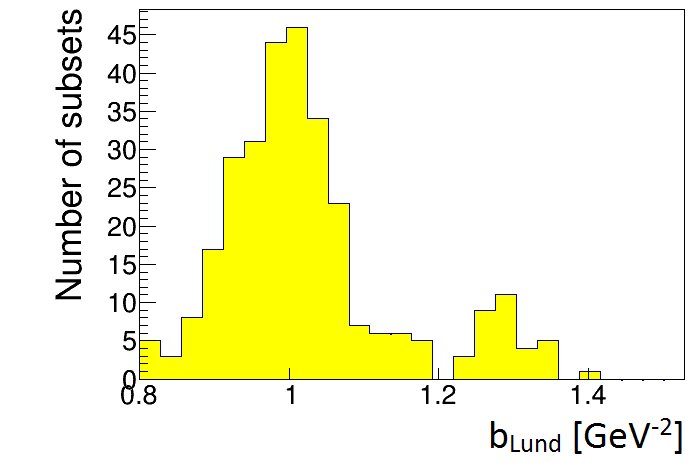}
	\includegraphics[width=0.45\textwidth]{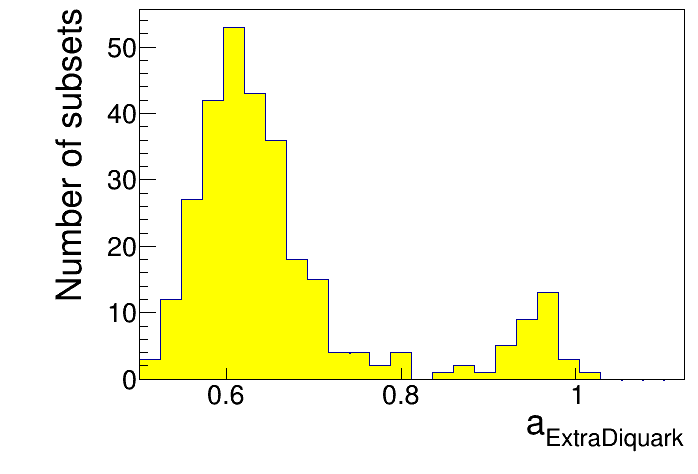}
	\includegraphics[width=0.45\textwidth]{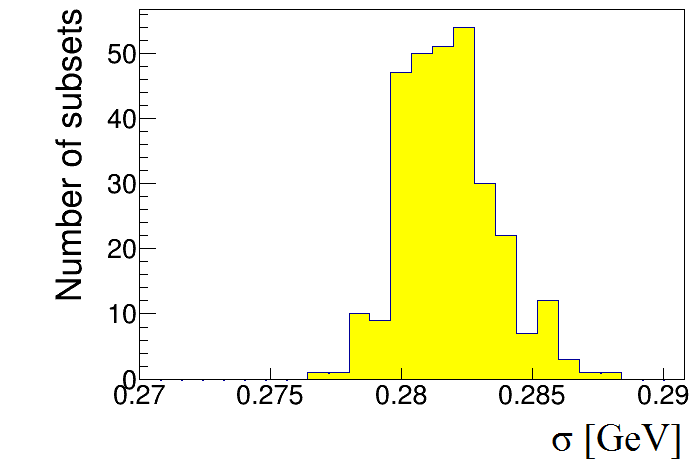}
	\includegraphics[width=0.45\textwidth]{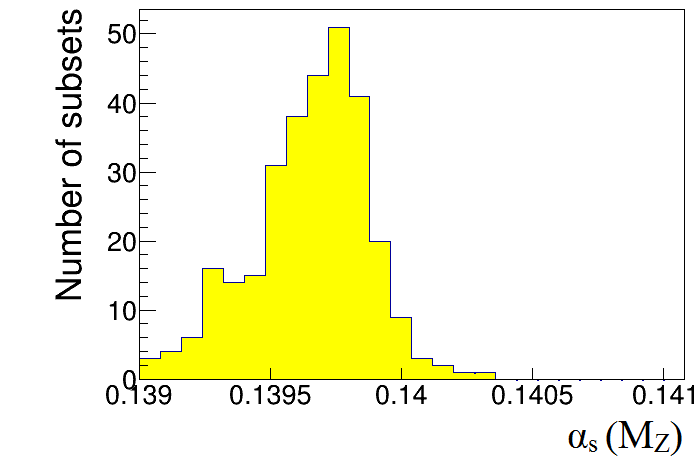}
	\includegraphics[width=0.45\textwidth]{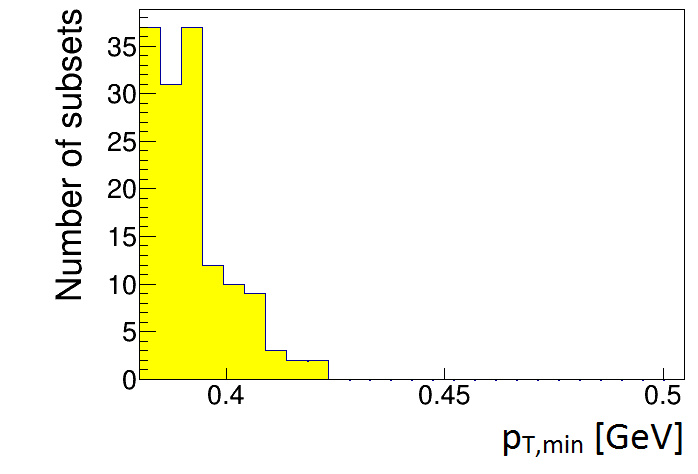}
	\caption{Distribution of the tuned model parameters using the run-combinations from Subsection \ref{subsec:tuning_setup} without range limitations and the weights ``Tune 2'' from \cref{tab:particle_spectrum,tab:pdg_multiplicities1,tab:pdg_multiplicities2,tab:b_fragmentation,tab:event_shape,tab:diff_jet_rate} in App.~\ref{asec:observablesandweights}.}
	\label{fig:tune_nadine_weights2_nolimits}
\end{figure}

\begin{figure}
	\centering
	\includegraphics[width=0.45\textwidth]{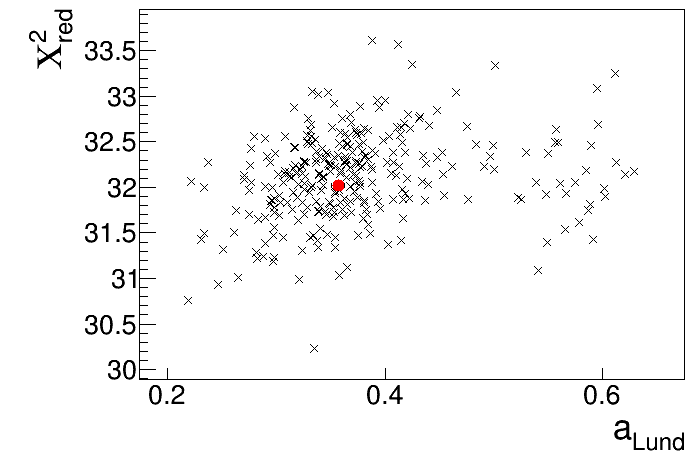}
	\includegraphics[width=0.45\textwidth]{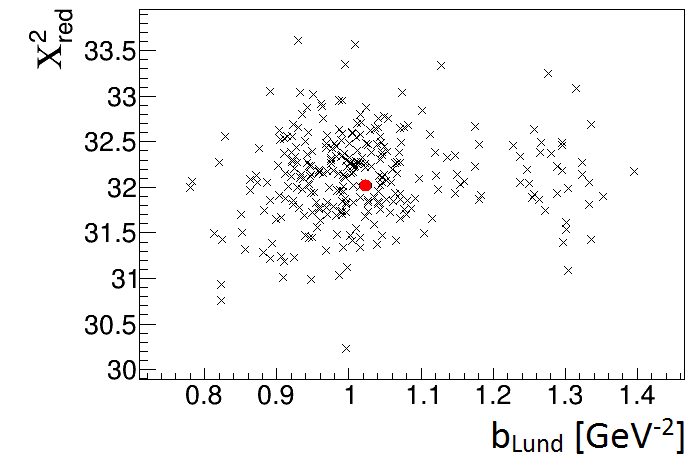}
	\includegraphics[width=0.45\textwidth]{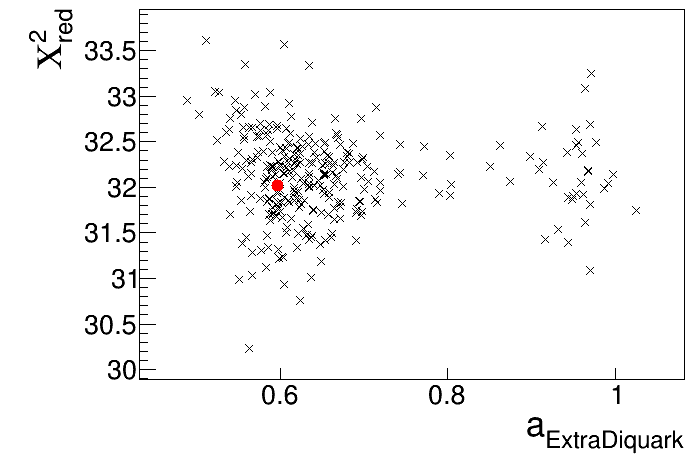}
	\includegraphics[width=0.45\textwidth]{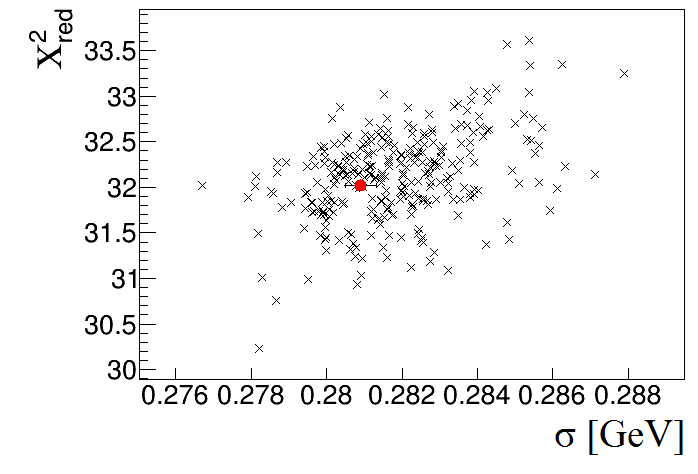}
	\includegraphics[width=0.45\textwidth]{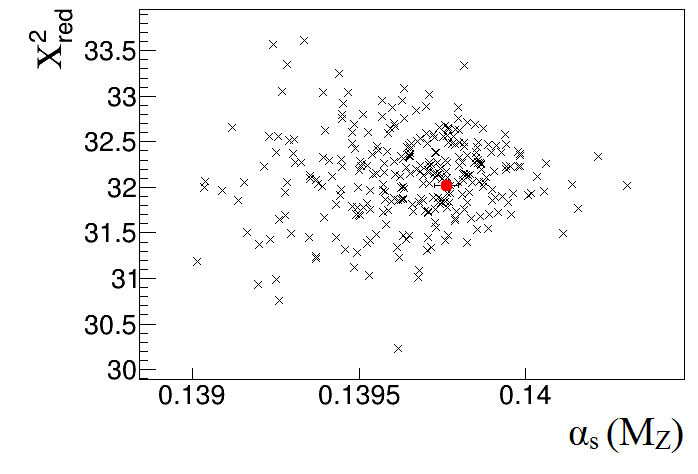}
	\includegraphics[width=0.45\textwidth]{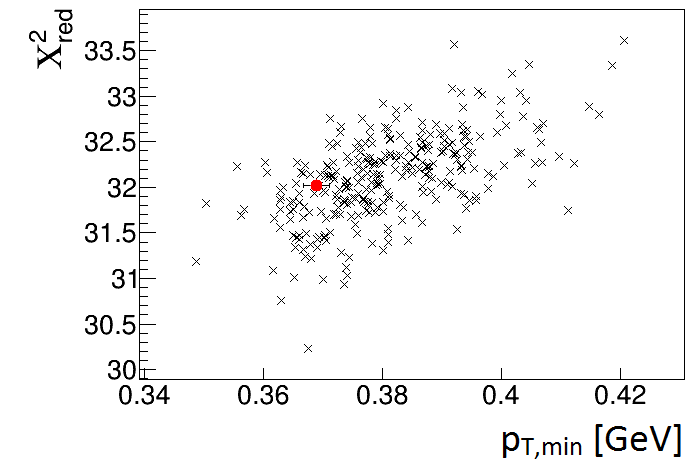}
	\caption{Distribution of the $\chi^2_{red}$ values of the tuned model parameters using the run-combinations from Subsection \ref{subsec:tuning_setup} without range limitations and the weights ``Tune 2'' from \cref{tab:particle_spectrum,tab:pdg_multiplicities1,tab:pdg_multiplicities2,tab:b_fragmentation,tab:event_shape,tab:diff_jet_rate} in App.~\ref{asec:observablesandweights}.}
	\label{fig:tune_nadine_weights2_nolimits_chi2}
\end{figure}

\FloatBarrier
\chapter{Minimization using BAT}
\label{asec:bat_prof}
\setcounter{page}{115}

\section{Program structure and Implementation}
\label{sec:implementation}
This Section will introduce the implemented program structure of BAT in the version 1.0.0-RC1\cite{ref:batdl}. Since BAT does neither have an I/O-interface for the \textsc{Professor} and \textsc{Rivet} data implemented nor a log-likelihood function for adjusting the model parameter of a MC generator to fit a response function to a measurement, those need to implemented. Therefore the inheritance and the implemented functions will be presented here. The introduction into the structure of the toolkit will also show an interface which will be later used for the implementation of the optimization process. In order to load the data from a file that contains the weights (cf. Eq.~\ref{eq:chi2_prof}), the interpolation file and from the measurements provided by \textsc{Rivet}, a couple of helper-classes are created. At first, a configuration file is read. This file contains the paths to the needed files, the number of Markov Chains that are used and a boolean flag for plotting the chain behavior during the pre-run. For the loading of the data itself, at first an object is created that loads the information of the interpolation file. For the purpose of extracting this data, the file is read line-wise and the corresponding information is stored by using string-manipulation. Since the interpolation file is loaded first, the analyzes and observables which were used for the interpolations are known and can be used to only load the corresponding data from the measurements by searching after the explicit name of a histogram. This is performed immediately afterwards. The last loading process is performed for the weight file. The information from that file is simply stored in the same order as they are extracted from the file. In order to create a matching order with the interpolation file, a re-ordering is performed afterwards. That way, analyzes and observables that are mentioned in the weight file but which do not have a corresponding interpolation function in the interpolation file can be deleted, too. After all information is read, it is passed to the BAT related object that will be explained in the following.\\
\medskip

Since the toolkit provides many more possibilities than are used or needed for the purpose of MC tuning, only the most important parts for the log-likelihood calculation will be introduced in the following. Since BAT is written in \textsc{C++}, the following parts will be presented in that programming language.

In order to get an interface for the implementation of a custom log-likelihood function, a class needs to be created that inherits from the class \textsc{BCModel}\cite{ref:batblack}. This custom class will be called \textsc{CC}. The inheritance will from \textsc{BCModel} will be \textit{public} and no further inheritances will be used. In order to use \textsc{CC}, and therewith BAT, the model parameters need to be added via the function \mbox{\textsc{void BCEngineMCMC::AddParameter(``name'', minValue, maxValue)}}. The string \textit{``name''} represents a custom name that can be given for the parameter. The two other parameters define the range [\textit{minValue}, \textit{maxValue}], in which the parameter can be varied.

Since BAT is based upon Bayes' theorem, a prior needs to be defined. In order to be as conservative as possible, the prior will be assumed as flat in all parameters via the function \textsc{void BCEngineMCMC::SetPriorConstantAll()}.

As apparent, both functions are not part of \textsc{BCModel} but from \mbox{\textsc{BCEngineMCMC}}. This super class provides at first the derived class \mbox{\textit{BCIntegrate}} its member variables and functions. The class \textit{BCModel} further inherits from \mbox{\textit{BCIntegrate}} its functionality. The class \textsc{BCEngineMCMC} is the class that provides functions, stores all needed parameters, tracks the behavior of them and drives the MCMC itself.

The last function of importance is\\ \mbox{\textsc{virtual double BCModel::log-likelihood(std::vector<double>\& params)}}. As the name already suggests, this function calculates the log-likelihood. Unlike the two other functions mentioned before, this function is purely virtual and needs to be implemented by the user. The basic construction of the implementation for the purpose of MC generator tuning as it is used is given by:\\

\begin{adjustbox}{max width=1.2\textwidth}
\begin{lstlisting}[language=C++, directivestyle={\color{black}},
                   emph={int,char,double,float,unsigned},
                   emphstyle={\color{blue}}]
double CC::loglikelihood(const std::vector<double>& parameters)
{
double logprob = 0;
std::vector<double> tmp_parameters;
tmp_parameters.resize(parameters.size());
std::vector<double> anchors;
anchors.assign(_power.size(), 1);
	
//Mapping parameters into [0,1]-interval.
for(size_t i = 0; i < parameters.size(); i++)
	tmp_parameters[i] = (parameters[i] - var_minvalue[i]) 
					/ (var_maxvalue[i] - var_minvalue[i]);

//Calculate fit-parameter independent part of
//every polynomial term.
for(size_t i = 0; i < _power.size(); i++)
	for(size_t j = 0; j < _power[0].size(); j++)
		anchors[i] *= pow(tmp_parameters[j], _power[i][j]);
		
//Walk over every data point and put information together.			
#pragma omp parallel for reduction(+:logprob) schedule(guided)
for(size_t i = 0; i < ref_values.size(); i++)
	for(size_t j = 0; j < ref_values[i].size(); j++)
	{
		logprob += weights[i] * BCMath::LogGaus(
				fit_value(anchors, num_bins_cumulative[i] + j), 
				ref_values[i][j], 
				sqrt(ref_values_errors[i][j] * ref_values_errors[i][j] 
				+ fit_var(anchors, num_bins_cumulative[i] + j)));	
	}
	
//Return the log of the conditional probability p(data|pars).
return logprob;
}
\end{lstlisting}
\end{adjustbox}

The parameter \textit{parameters} is a vector that contains the current position of the Markov chain. The local variable \textit{logprob} represents the value of the log-likelihood that will be calculated within this function. \textit{tmp\_parameters} is used as the storage of a [0,1]-mapping of the parameters. This is necessary since the fit-parameters calculated by \textsc{Professor} are based upon this rescaling of the parameter ranges. The last local variable \textit{anchors} serves as speed-up of the calculation. Its length is determined by the member variable \textit{\_power}. This member variable carries the information about the power of every parameter in every term of the polynomial function. Therefore, the length of this list is equal to the number of terms in the polynomial function. The variable \textit{anchors} is used as a storage of the fit-parameter independent part of the polynomial function. Since this part is equal for every data point in every histogram, this just needs to be calculated once for a certain value of \textit{parameters}. Additionally, in order to speed-up the code, the \textsc{pow}-function needs a long time to calculate the value of the problem $a^b$. Therefore, it is necessary to reduce the function calls.

As a first step to receive the log-likelihood, the mapping is performed. Secondly the entries in \textit{anchors} need to be set. After every helper-variable is calculated, the log-likelihood itself can be calculated. For that purpose, two \textit{for}-loops iterate over every data point and every histogram available in the given interpolation file. For every single data point, the weight defined by the variable \textit{weight} will be multiplied with the natural logarithm of a normal distribution. This is a built-in function represented by \textsc{double BCMath::LogGaus($x$, $\mu$ = 0, $\sigma$ = 1, norm = false)}. The arguments of the function represent the value ($x$), the mean ($\mu$) and the standard deviation ($\sigma$) of a normal distribution. As a last parameter, the parameter \textit{norm} can be passed to the function. This parameter is a boolean flag for a normalized function. Since the log-likelihood will not be normalized, the single normal distributions do not need to be normalized either and the parameter can be neglected. The value of $x$ is given by the value of the polynomial function with the given \textit{parameters}. The mean is given by the measured data and the standard deviation is given by square root of the squared sum of the uncertainties of the measurement and the interpolation. This form is equal to the uncertainty used in Eq.~\ref{eq:chi2_prof}.

A difference to the $\chi^2$-based approach performed by \textsc{Professor} in order to receive the tuning parameters is based in the \textsc{BCMath::LogGaus($x$, $\mu$, $\sigma$)}-function. Since the $\chi^2$-function as defined in Eq.~\ref{eq:chi2_prof} represents an analogous representation of the natural logarithm of a normal distribution as shown in Eq.~\ref{eq:loggauschi2} just in the case of $\sigma$ being independent of the model parameters. Therefore, using the function \textsc{BCMath::LogGaus($x$, $\mu$, $\sigma$)} and therewith the non-constant and computing-intensive offset shown in the first term on the right hand side of Eq.~\ref{eq:log-likelihood} could produce a different result than the \textsc{Professor}-tune based on the impact.

One last point in the log-likelihood function is the parallelized computing using \textsc{OpenMP}. This is represented by \textit{\#pragma omp parallel for reduction(+:logprob) schedule(guided)}. At that point a parallel environment with several threads will be created for a shared-memory system. This parallel environment wraps \mbox{one/several} iteration(s) from the \textit{for}-loop and passes them on a core that will calculate it. Therefore, using multiple cores reduces the run-time significantly. Since multiple threads could write on \textit{logprob} simultaneously, this could produce wrong results. This is a typical problem in parallel environments. Therefore, the \textit{reduction}-clause is used to protect this variable against multiple simultaneously accesses. The \textit{schedule}-clause represents the splitting-tactic of \textsc{OpenMP} of iterations in the top \textit{for}-loop. The parameter \textit{guided} represents the fact that the first thread encountering this loop takes the biggest amount of iterations and starts to calculate them. The next thread will get a smaller amount and so on. The reason for this tactic is based on the implementation of the multiple Markov chain approach implemented in \mbox{\textsc{BCEngineMCMC}}. Using multiple chains creates in this upper level class a parallel environment with a Markov chain per thread that will be evaluated. The thread creation in \mbox{\textsc{CC::loglikelihood(const std::vector<double>\& parameters)}} creates therefore a nested team of threads. Under the assumption that more threads than chains are available, the overflow of threads is used to assist the designated threads that are bound to a certain Markov chain in the calculation of the log-likelihood. Since the assisting threads will start later with the log-likelihood-calculation of a certain chain, they should take less work. If the chain itself is finished, the assisting threads can perform calculations in another chain, but since this chain had more time in order to compute the log-likelihood, the threads only need to help with the remaining parts and so on.

The calculations performed for the problem are too simple and therefore do not provide a huge speed-up of the program, but this will become a more important aspect in Sec.~\ref{sec:covmat}.

\newpage
\section{Marginalized posterior pdfs}
\begin{figure}[h]
\makebox[\textwidth][c]{
	\begin{adjustbox}{max width=1.\textwidth}
	\includegraphics[width=\textwidth]{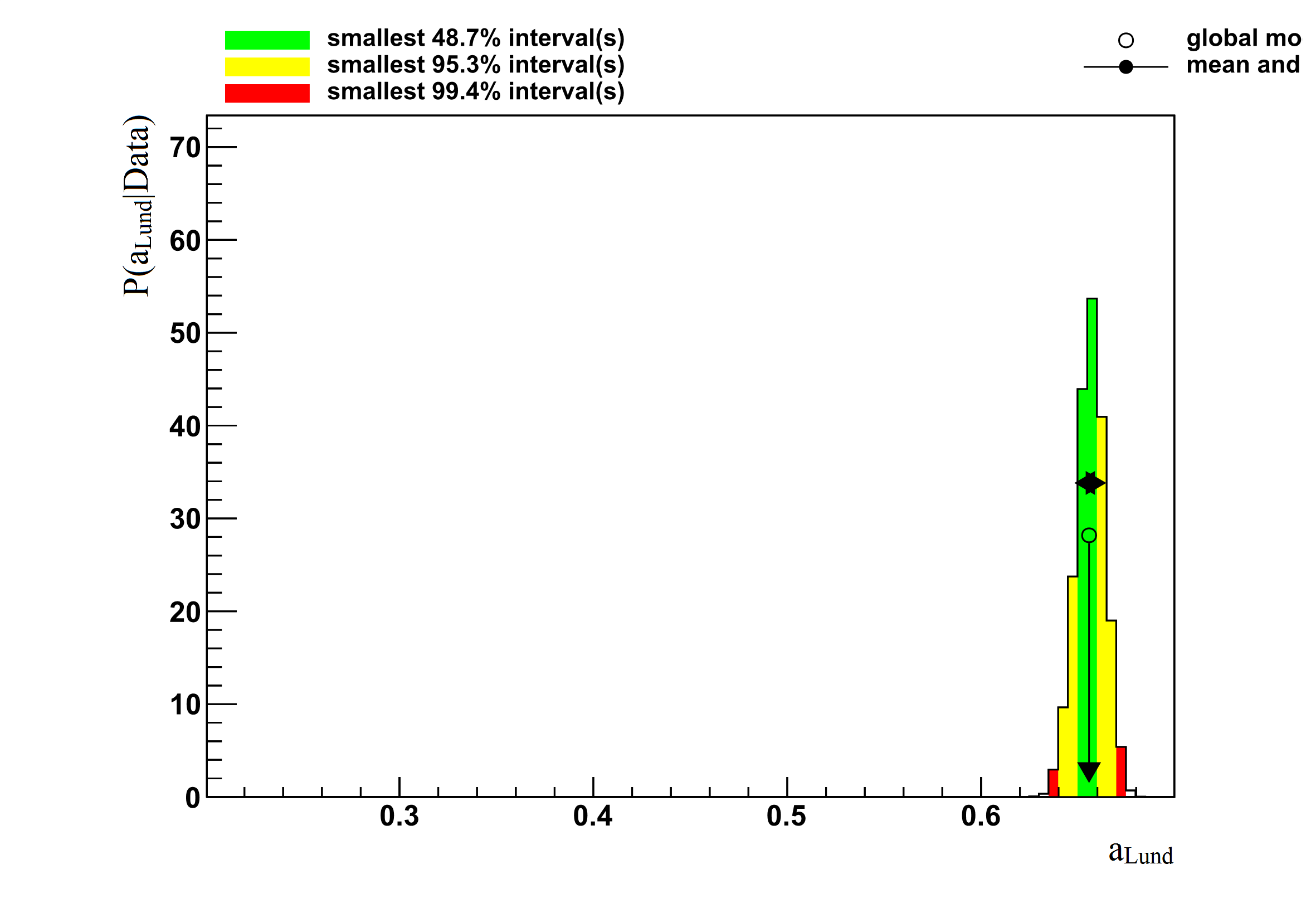}
	\includegraphics[width=\textwidth]{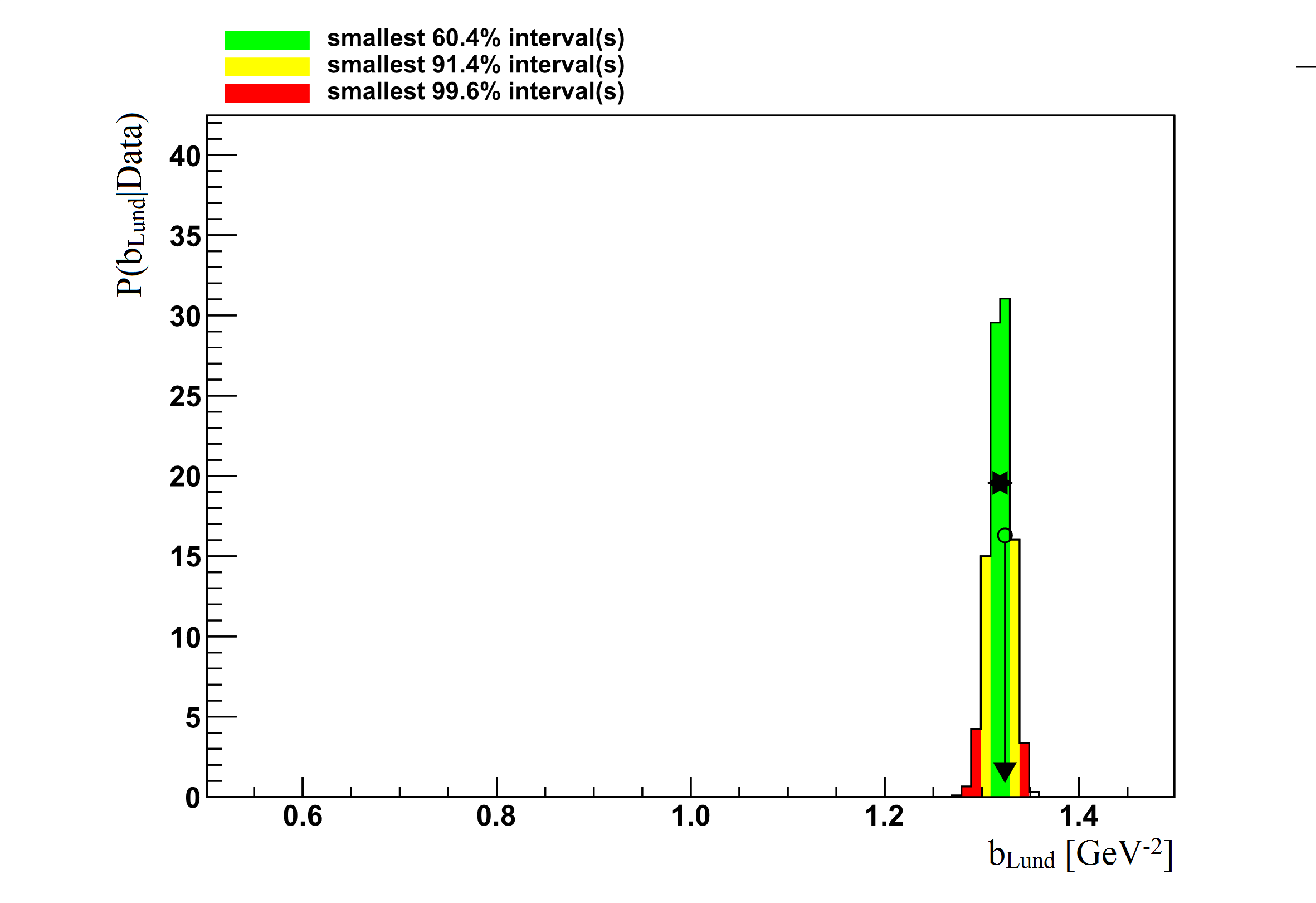} 
	\end{adjustbox}
}
\makebox[\textwidth][c]{
	\begin{adjustbox}{max width=1.\textwidth} 
	\includegraphics[width=\textwidth]{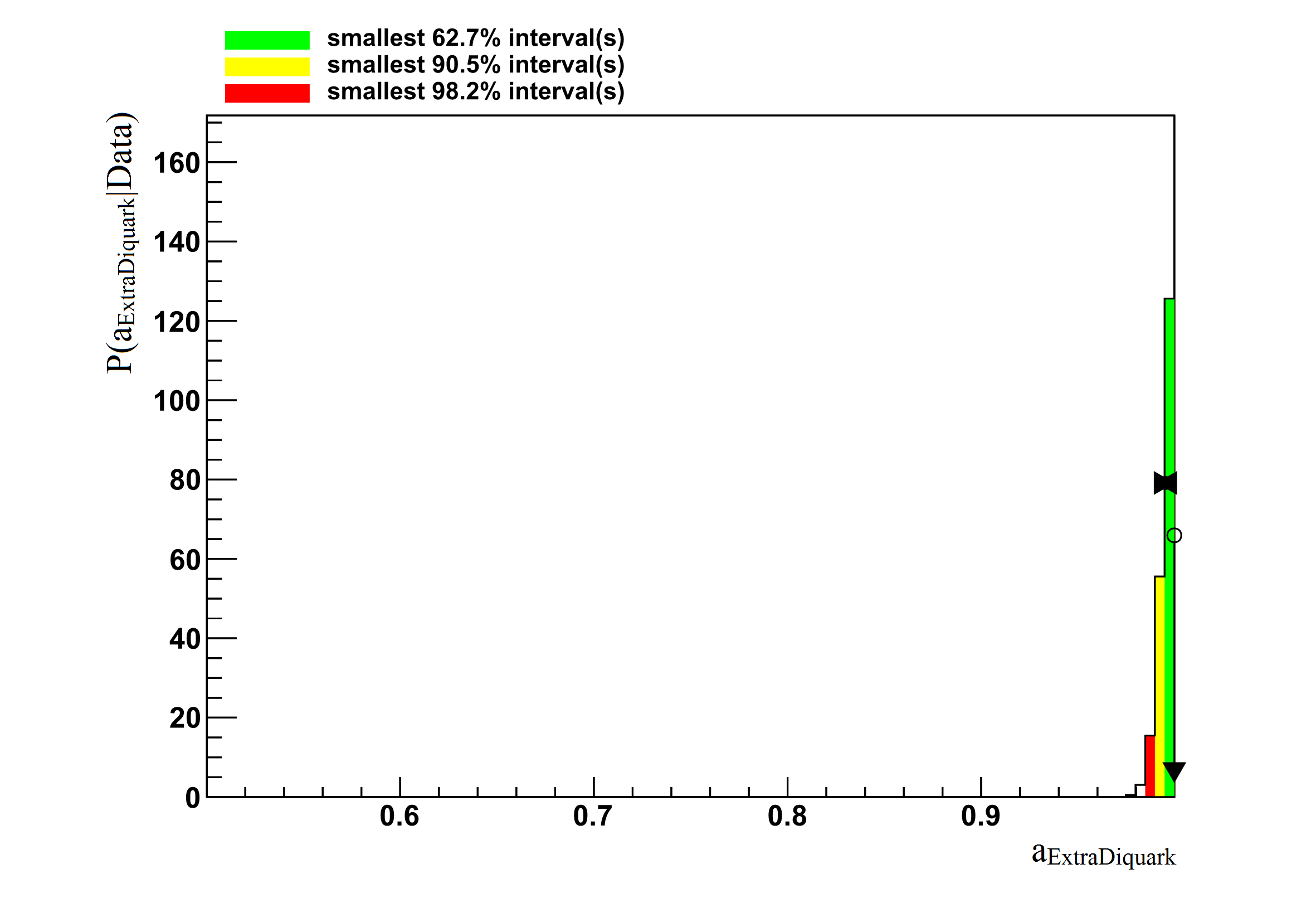}
	\includegraphics[width=\textwidth]{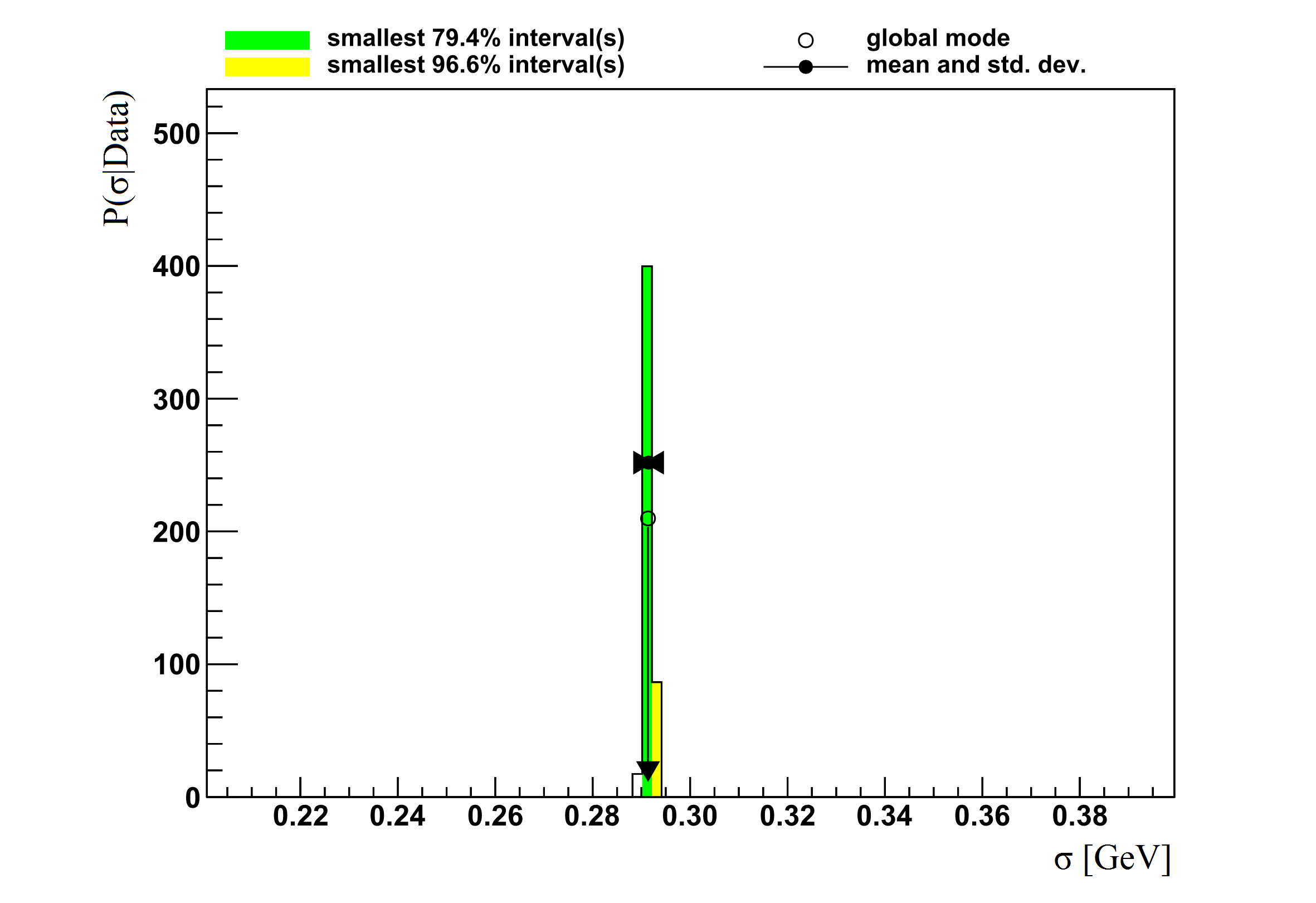} 
	\end{adjustbox}
}
	\caption{Posterior pdf of the re-run of the \textsc{Professor} based interpolation file using BAT. The parameters refer to $a_{\textrm Lund}$ (top left), $b_{\textrm Lund}$ in $[GeV^{-2}]$ (top right), $a_{\textrm ExtraDiquark}$ (bottom left) and $\sigma$ in $[GeV]$ (bottom right).}
	\label{fig:bat_prof_tune1}
\end{figure}

\begin{figure}[htbp]
	\centering
	\includegraphics[width=0.75\textwidth]{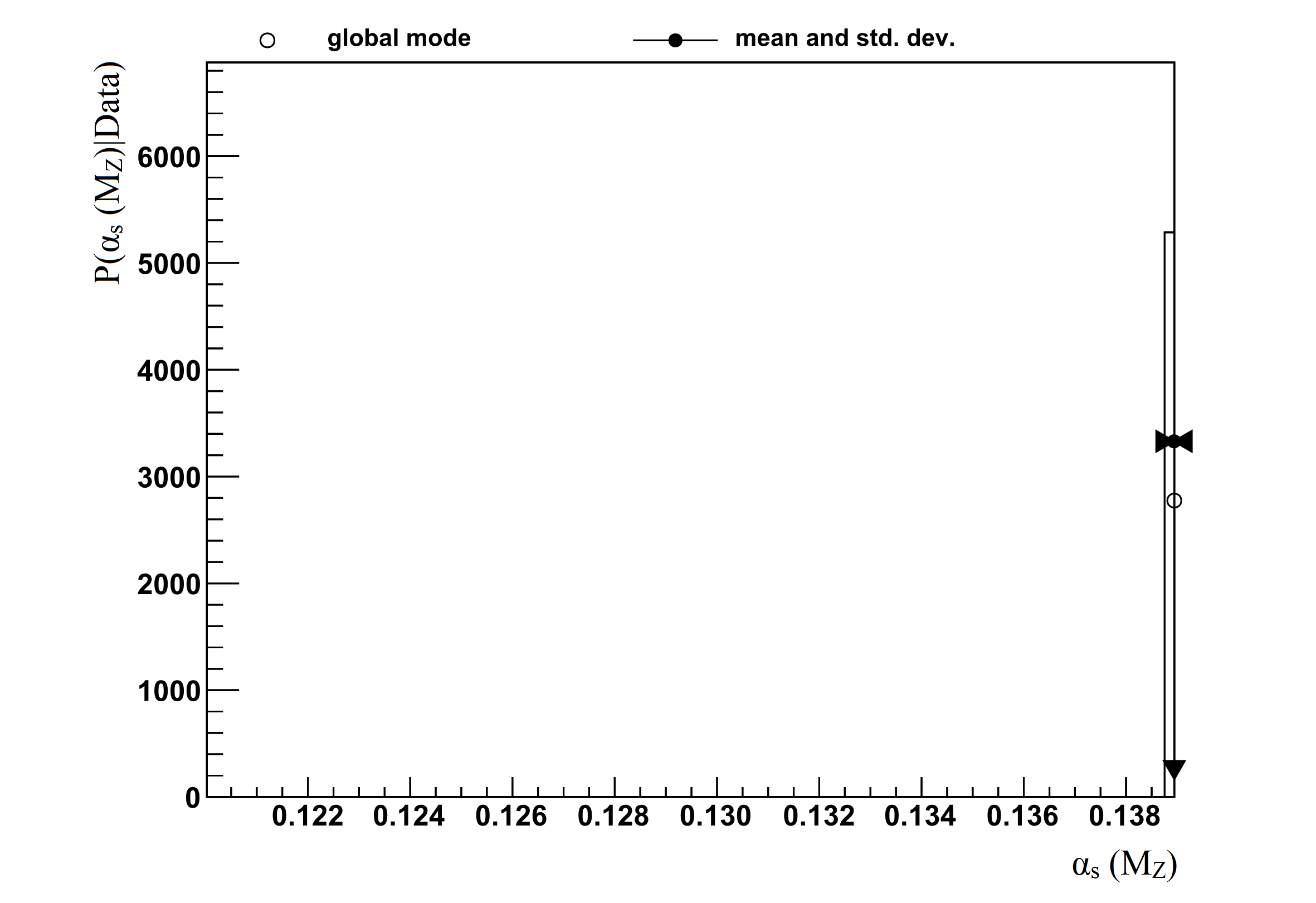}
	\includegraphics[width=0.75\textwidth]{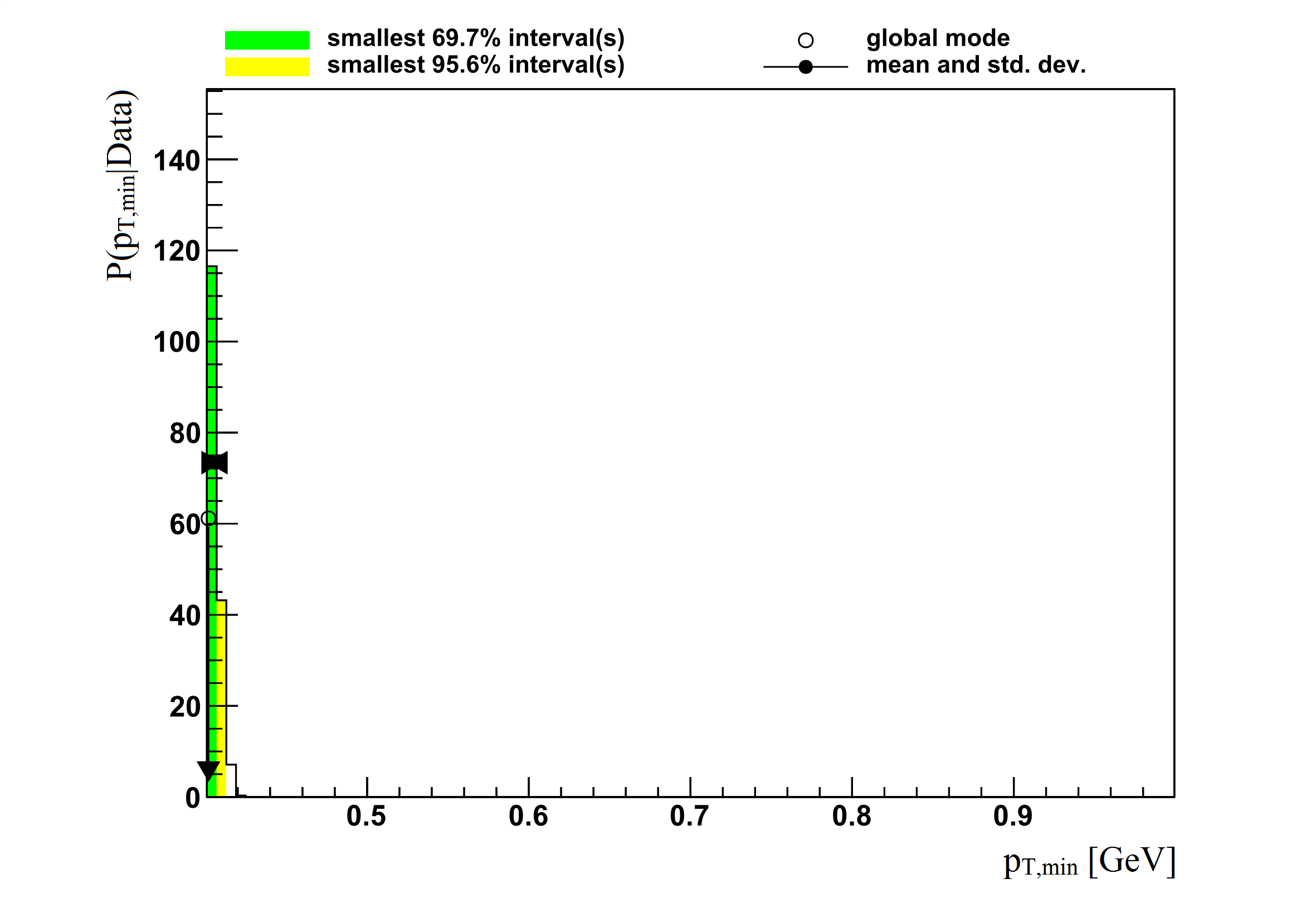}
	\caption{Posterior pdf of the re-run of the \textsc{Professor} based interpolation file using BAT. The parameters refer from top to down to $\alpha_s(M_Z)$ and $p_{T,min}$ in $[GeV]$.}
	\label{fig:bat_prof_tune2}
\end{figure}

\FloatBarrier
\section{Re-run}
\label{asec:rerun}

\begin{figure}[h]
\makebox[\textwidth][c]{
	\begin{adjustbox}{max width=1.\textwidth}
	\includegraphics[width=\textwidth]{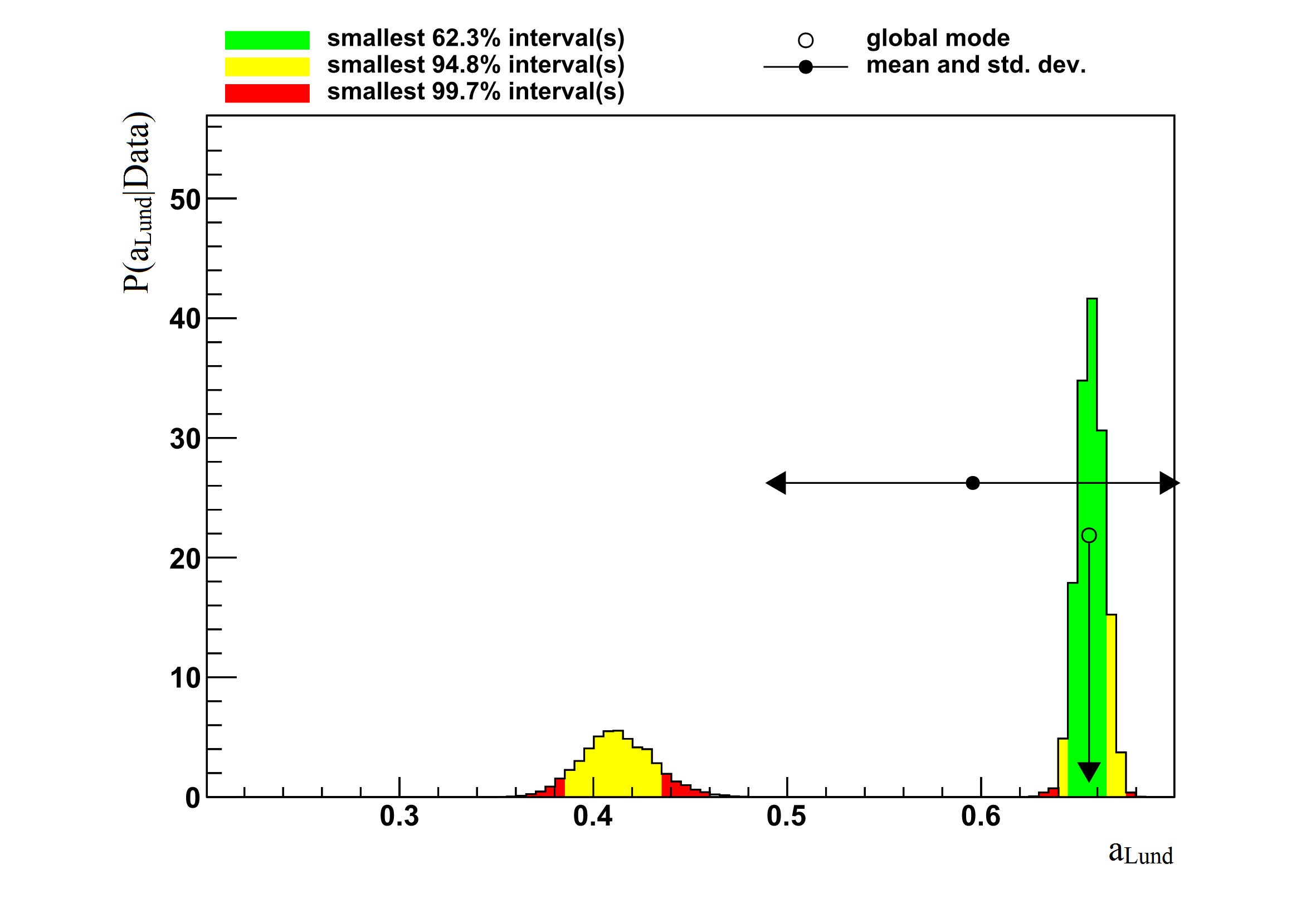}
	\includegraphics[width=\textwidth]{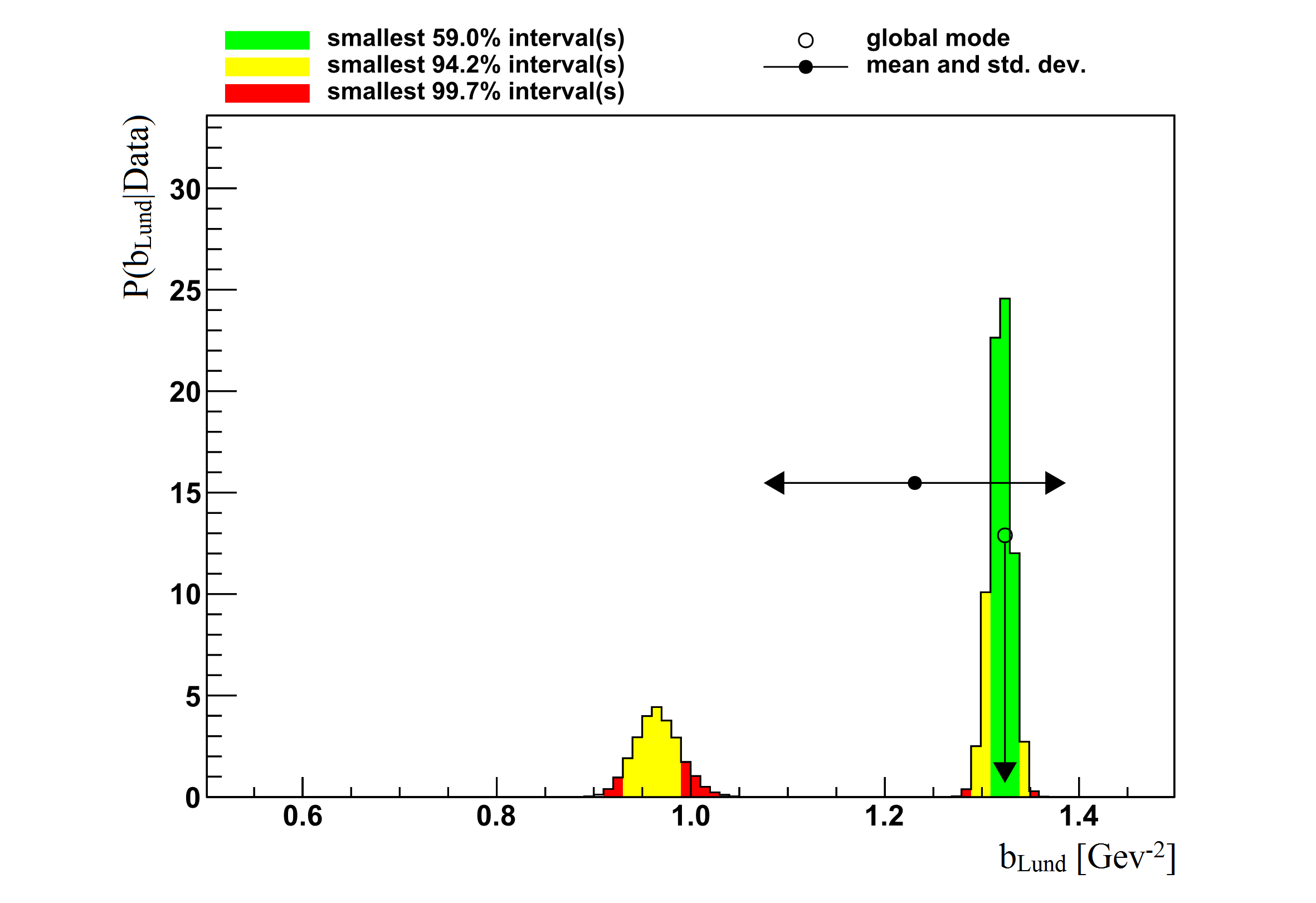} 
	\end{adjustbox}
}
\makebox[\textwidth][c]{
	\begin{adjustbox}{max width=1.\textwidth} 
	\includegraphics[width=\textwidth]{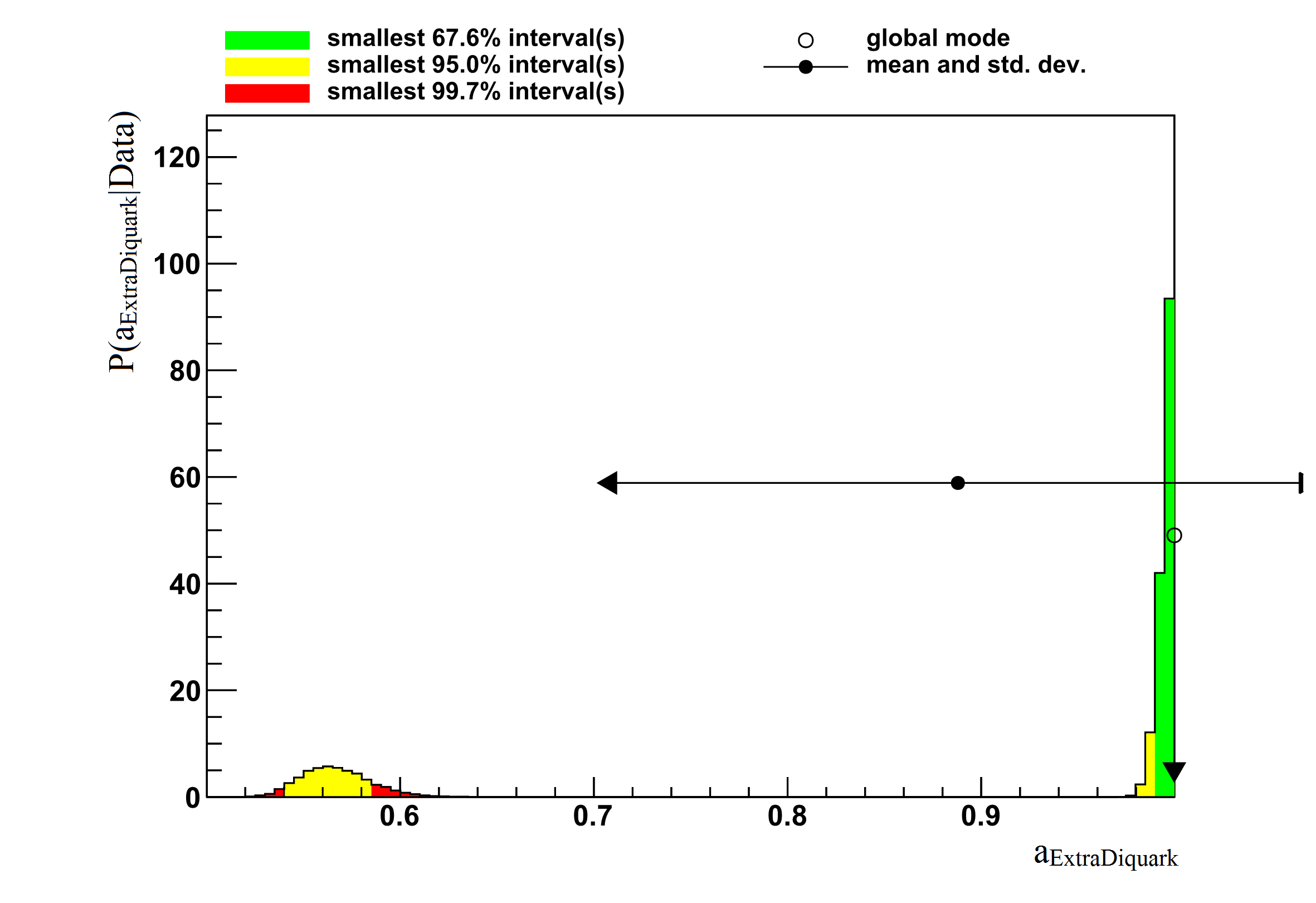}
	\includegraphics[width=\textwidth]{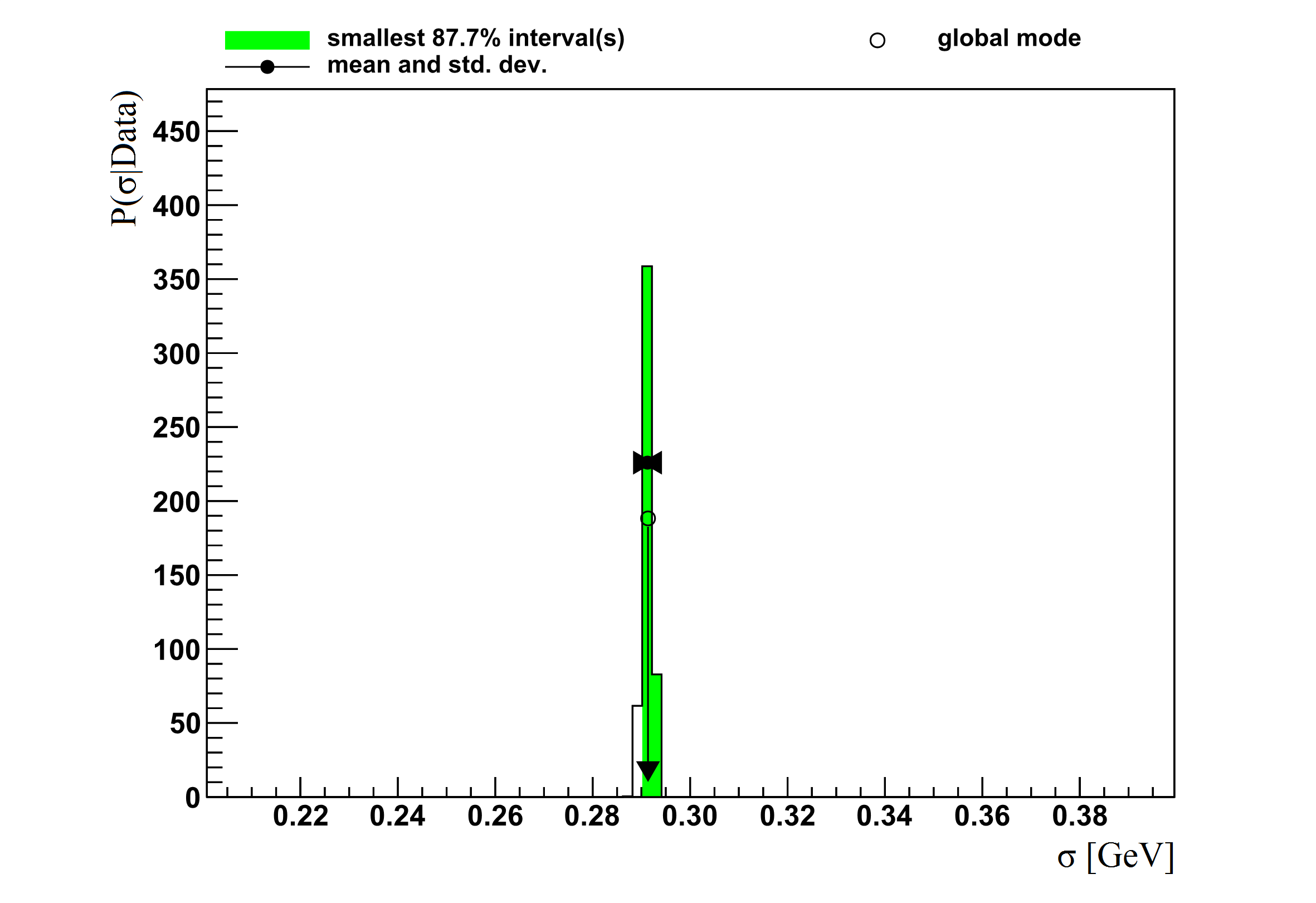} 
	\end{adjustbox}
}
	\caption{Posterior pdf of the re-run of the \textsc{Professor} based interpolation file using BAT. The parameters refer to $a_{\textrm Lund}$ (top left), $b_{\textrm Lund}$ in $[GeV^{-2}]$ (top right), $a_{\textrm ExtraDiquark}$ (bottom left) and $\sigma$ in $[GeV]$ (bottom right).}
	\label{fig:bat_prof_tune1_rerun}
\end{figure}

\begin{figure}[htbp]
	\centering
	\includegraphics[width=0.75\textwidth]{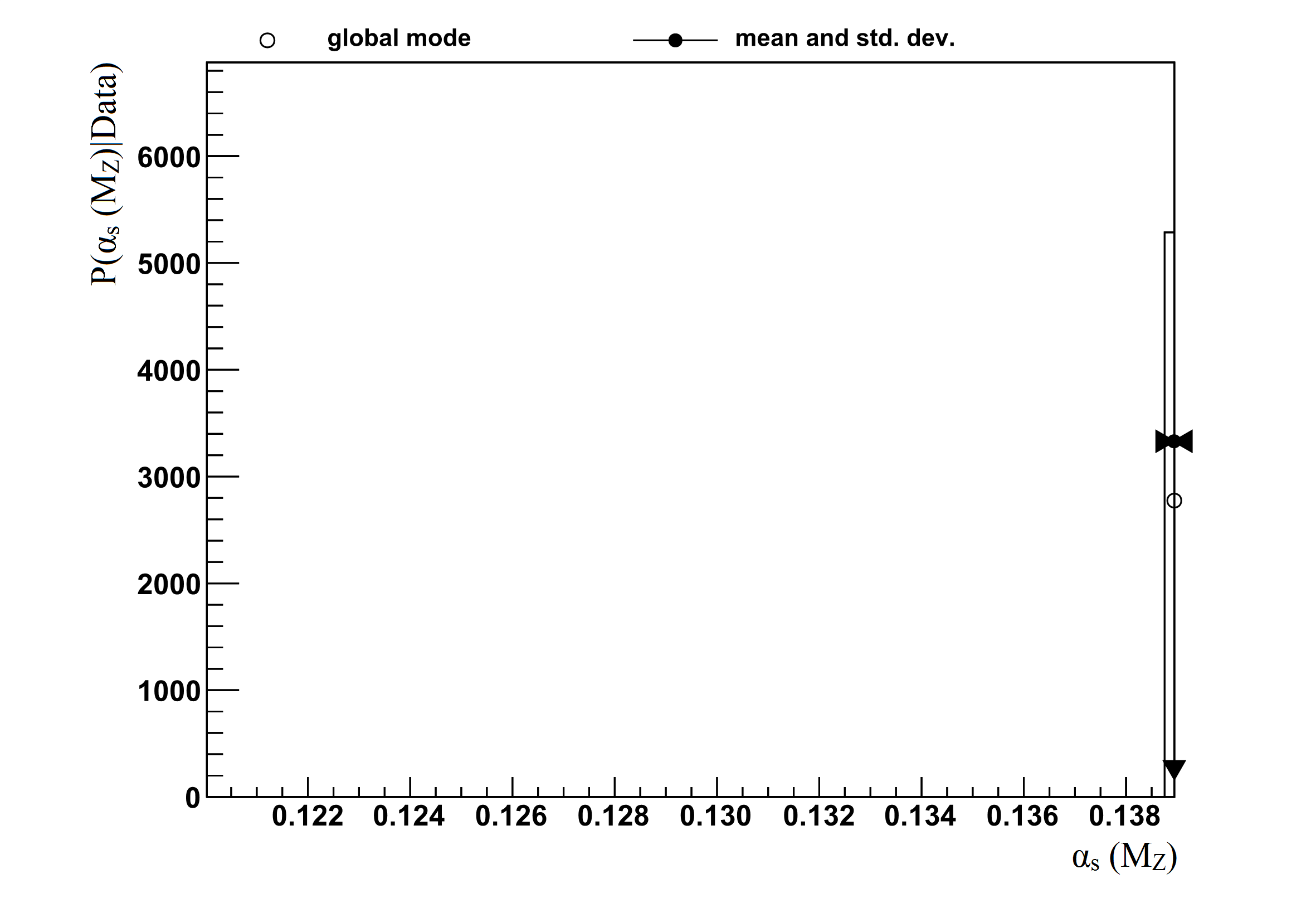}
	\includegraphics[width=0.75\textwidth]{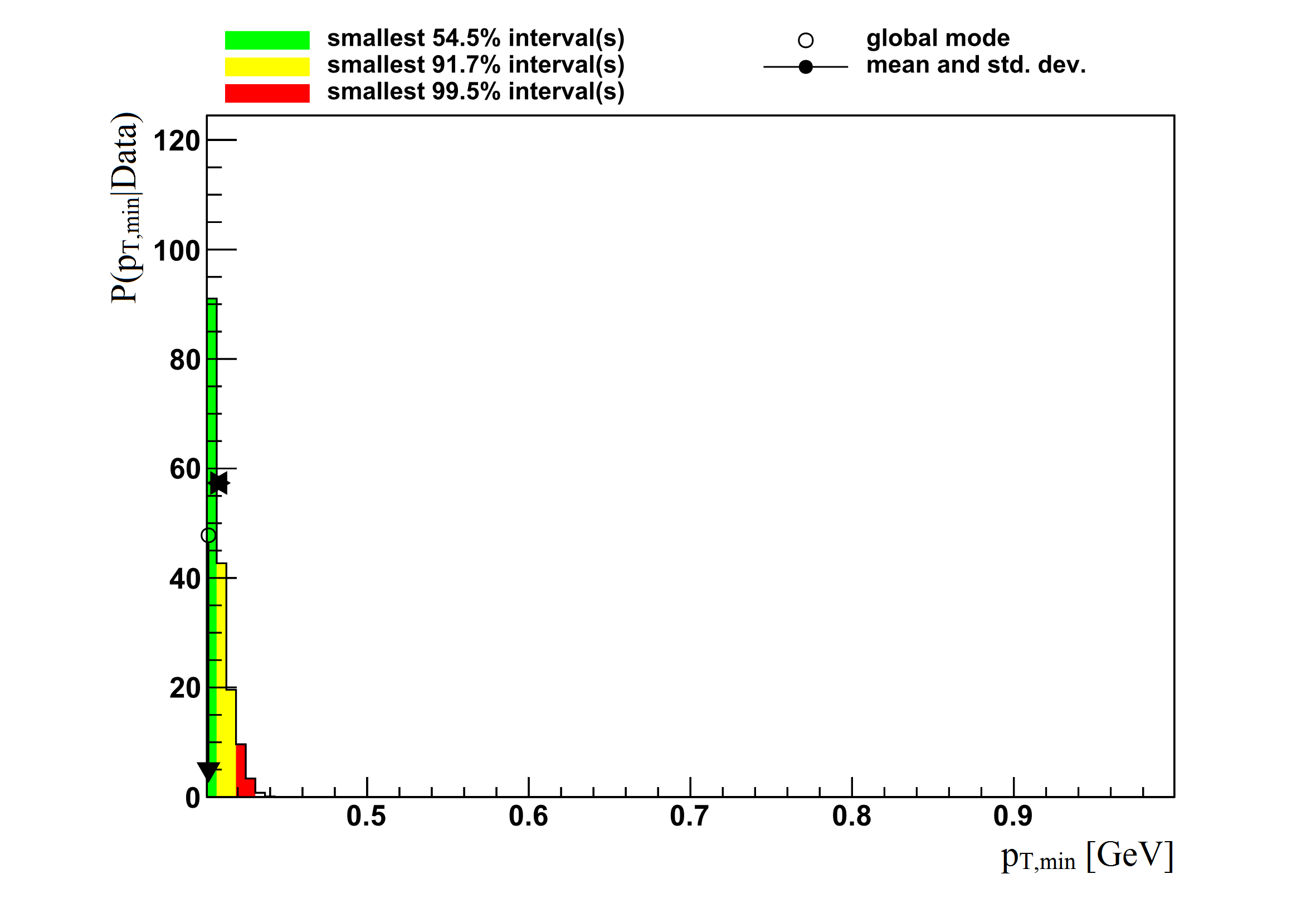}
	\caption{Posterior pdf of the re-run of the \textsc{Professor} based interpolation file using BAT. The parameters refer from top to down to $\alpha_s(M_Z)$ and $p_{T,min}$ in $[GeV]$.}
	\label{fig:bat_prof_tune2_rerun}
\end{figure}

\FloatBarrier
\chapter{Algorithmic implementation of the adaptive interpolation}
\setcounter{page}{124}

\section{Implementation structure}
\label{sec:implementationipol}

This Section considers the implementation of the algorithmic approach mentioned in the previous Section. In general, the program is written in \textsc{C++} and is optimized for multiple CPUs in a shared memory architecture using \textsc{OpenMP}\cite{ref:openmp}.

The first part of the program code is meant to read the necessary data. First of all a configuration file is read. This file is used to perform simple changes in the program behavior without having to recompile the code. The content of the file itself are the paths of the data, which are needed for the interpolation like the parameter vectors and the corresponding values of the data points of each observable. Furthermore, behavior related parameters can be set like the threshold of a RR-constraint or the $\kappa$ value in Eq.~\ref{eq:kappa}. Another information handed over when calling the program is the weights file. The actual values are not important in this file but the naming of analyzes and their observables. Those are the only ones that are interpolated during the execution of the program.

After the given files are read, the corresponding reference data are loaded. Every parameter vector, data point value and corresponding uncertainty is stored together in a single object called \textsc{RefHandler}. Since this object contains every parameter vector used for the simulation, the minimum and maximum value of each model parameter can be extracted. If a rescaling is necessary, every parameter vector is component-wise rescaled on a [0,1]-interval. Furthermore, in order to initialize the actual interpolation process, an object for writing the full output of the program and an object containing information about the powers of every model parameter in every term of the polynomial function are declared. Both objects are created in a static-like way. The first can be called in a thread-safe way and therefore ensures a non-corrupted output. The second calculates the needed numbers once and stores them for future usage. Since the used algorithm that is taken from \textsc{Professor 2.1.4} has a long run-time, the storage of pre-computed values allows faster access to the numbers even for different data points.\\
\medskip

After the ETL is performed, the main part of the interpolation algorithm is given by a \textit{for}-loop, which iterates over every data point of every observable. This loop is parallelized by \textsc{OpenMP} such that every thread gets another data point to interpolate. A simplified version of the actual code is shown in App.~\ref{asec:adaptivecode} together with further explanation.

In each iteration of the loop, an object, called \textsc{FitHandler}, is created, that obtains the necessary information e.g. from \textsc{RefHandler} in order to calculate the actual fit function for a certain data point. The object \textsc{FitHandler} starts the interpolation with a polynomial function of zeroth order. The corresponding normalized gradient vector of the simulated data points, as mentioned in the previous Section, is calculated in \textsc{RefHandler}. After both are calculated, the first $\mathcal{F}$ value is calculated. Since the application of a sliding mean is used in order to find the best iteration, the number of terms in the polynomial function of \textsc{FitHandler} is at least increased iteratively to the number of terms specified by the configuration file. Within each iteration in a certain \textit{for}-loop iteration, the currently smallest $\mathcal{F}$ value and the corresponding iteration is tracked. Once a single $\mathcal{F}$ value is bigger than the sliding mean, the tracked best iteration is re-calculated and the corresponding uncertainty of the interpolation parameters is calculated.\\

Until now, the objects \textsc{RefHandler} and \textsc{FitHandler} were, beside the theoretical explanation in the Sec. \ref{sec:mathematicalmodel}, treated as black-boxes. The \textsc{FitHandler} serves already as a good description with some minor adaptions and optimizations in order to treat the specific problem. The first considerable function is the calculation of the $\chi^2$ value. Since the simulated data is based upon MC simulations with limited number of events, the statistical uncertainty in a certain data point can be zero for a certain parameter vector. In order to treat those problems, the simulated data gets an arbitrary but small uncertainty assigned. Furthermore, the limited number of events leads to the observation that for some parameter vectors the value of a data point is zero with a very small uncertainty. This can occur spottily in the parameter space. Those points are surrounded by data point values unequal zero with an uncertainty that is bigger by some orders of magnitude. This behavior is treated as problem in the simulation and/or observable extraction and the parameter vectors that are responsible for data point values of zero are neglected.

The second run-time improving part in \textsc{FitHandler} occurs during the calculation of the normalized gradient vectors of the interpolation function. Since the used parameter vectors remain the same during the execution of the program, the parameter vector dependent part of the gradient remains the same, too. As a consequence of this, a list containing this information needs to be calculated only once and can be stored in a static list. The static property allows the usage of the list for every instance of \textsc{FitHandler}. Furthermore, an expansion of the list in the case of additional needed terms in a polynomial function is performed in a thread safe construction. An important definition of the gradient calculating function is the definition of the norm of a zero-vector. In the implementation, a normalized zero-vector is defined as a zero-vector.

The normalized gradient vectors of the interpolation function and those from the simulated data are used inside this object in order to calculate the $D_{smooth}$ parameter. Since the norm of a zero-vector is defined as zero-vector, the corresponding dot product between two zero-vectors is defined as one. The reason for this result is given by the construction of $D_{smooth}$. This value states a value close to one as good coincidence between the function and the data behavior.

The essential feature of \textsc{FitHandler} is the uncertainty calculation of the interpolation parameters. As mentioned in Eq.~\ref{eq:covmatchi2}, the inverse of the precision matrix needs to be calculated. Since such a calculation is numerically complicated, an approximation of the true result is obtained. In order to use the symmetry property of the precision matrix, the \textsc{ROOT} object \textsc{TMatrixDSym} is used. This implementation ensures that a matrix is at least symmetric, even after the inversion.\\
\medskip

A fundamental component of \textsc{RefHandler} is referred to the calculation of the normalized gradient vectors of the simulated data. Using the property of the constant parameter vectors leads in this object to the property that the surrounding points around a central point remain the same and therefore need to be calculated only once and can be stored for the whole run-time. This implementation speeds the code up, especially in the case of huge sets of parameter vectors. The selection of the points is performed in list structure that contains $2^{dimension}$ elements. This list is initialized with \textit{nan} values. In case that a point cannot be assigned in a certain area, the value is used as an indicator for that situation. Like the encoding in Fig.~\ref{fig:nasa}, the list entries are filled in the same way. In order to find the points with the smallest distance to the central point, every point but the central point is tested component-wise. If the $i$-th component of the test point is bigger than the $i$-th component of the central point then the $i$-th component of the binary representation as shown in Fig.~\ref{fig:nasa} is set to 1. The translation of the so resulting binary number to a decimal number lead to the entry of the list. If a point is currently stored in this element of the list, the point with the smaller distance to the central point is set as the element in the list. The resulting list is used together with the central point in order to calculate the polynomial fit function of first order and to extract the normalized gradient vector out of it.

\section{Implementation of the \textit{for}-loop}
\label{asec:adaptivecode}

\begin{adjustbox}{max width=1.2\textwidth}
\begin{lstlisting}[language=C++, directivestyle={\color{black}},
                   emph={int,char,double,float,unsigned},
                   emphstyle={\color{blue}}]
#pragma omp parallel for firstprivate(rh) shared(pow, ch, oh)
for(size_t num_bin = 0; num_bin < num_bins_total; num_bins++)
{
	bool quitflag = true;
	
	//Initialize FitHandler
	FitHandler fh(rh, num_bin, ch, pow);
	fh.interpolate();
			
	//Calculate gradient vectors of the data
	rh.calculate_normalvectors(num_bin, ch);
		
	//Track the best result and store the history
	track_result(fh);
		
	//Increase the number of terms
	while(quitflag)
	{					
		//Calculate the next iteration
		fh.next_iteration();
		
		//Track the best result, the history and the cutoff condition
		track_result(fh);
		check_cutoff(quitflag);
	}	
	//Calculate the best iteration and the uncertainty
	fh.set_best_iteration(best_iteration);			
	fh.set_fit_uncertainties(oh);
	
	//Write the output
	oh.write_output();
}
\end{lstlisting}
\end{adjustbox}

This code snippet shows a simplified description of the main work performed during the interpolation. The used variables in the code are listed and described in Tab.~\ref{tab:used_variables}.

\begin{table}[htbp]
 	\centering
  	\begin{tabular}{ l | l }
    	\hline
    	Variable & Application \\ \hline
    	rh & \textsc{RefHandler} object that stores every loaded data \\
    	pow & Object that stores the list of powers for every parameter in every term \\
    	ch & Object that stores the configuration \\
    	oh & Object that handles the output of the interpolation results \\
    	num\_bin & Variable that counts from zero up to num\_bins\_total \\
    	num\_bins\_total & Total number of data points in every used observable \\	 
    	quitflag & Boolean expression that can stop the iteration \\
    	fh & \textsc{FitHandler} object that calculates the interpolation \\
    	best\_iteration & Storage of the iteration with the best interpolation \\	
    	\hline
  	\end{tabular}
\caption{Explanation of the used variables in the \textsc{C++}-code in App.~\ref{asec:adaptivecode}.}
\label{tab:used_variables}
\end{table}

The first line of the code represent the creation of a team of threads in order to enable the execution of the code in parallel. Given in the \textit{firstprivate} clause, the \textsc{RefHandler} object is handed over to every thread in the initialization at the beginning. The following \textit{shared} clause defines objects, that are used by every thread, mostly due to decreasing the run-time of the program.

In the loop itself, in line seven, a \textsc{FitHandler} object is created and initialized with the necessary information. The following line is used to create calculate the interpolation function of zeroth order. In order to receive the normalized gradient vectors of the simulated data, that only needs to be calculated once, this is performed in line eleven. The next code line is used to track the $\mathcal{F}$ value of the current result and set it as the current best iteration for later comparison.

The \textit{while}-loop in line 17 is performed as long as the interpolation functions do not get worse. Inside the loop itself, the object iteratively increases the number of terms. After the new interpolation function is calculated, the current $\mathcal{F}$ value and the best results is further tracked. In line 24, a check is performed, if the interpolation function is getting worse. In that case, the value of \textit{quitflag} is set to false. Beginning in line 27, the best iteration is re-calculated based on the tracked best iteration in \textit{best\_iteration}. Afterwards, the uncertainty of the interpolation parameters is calculated and the results are getting printed either onto the terminal or into files.

\FloatBarrier
\chapter{Distribution of the run-combinations with adaptive interpolation}
\label{asec:adaptive_runcombs}
\setcounter{page}{129}

\begin{figure}[h]
	\centering
	\includegraphics[width=0.41\textwidth]{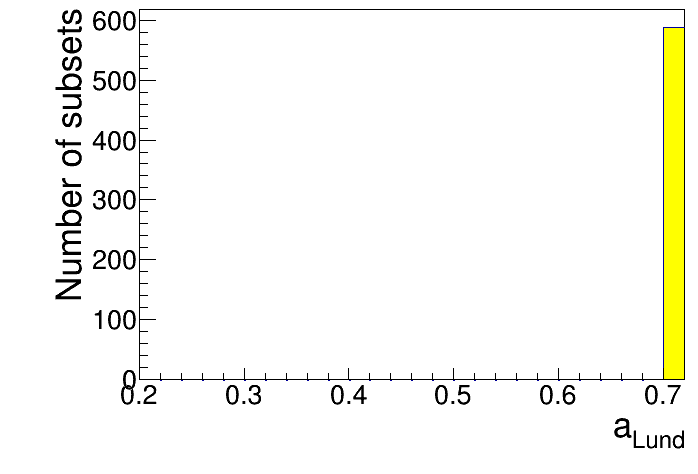}
	\includegraphics[width=0.41\textwidth]{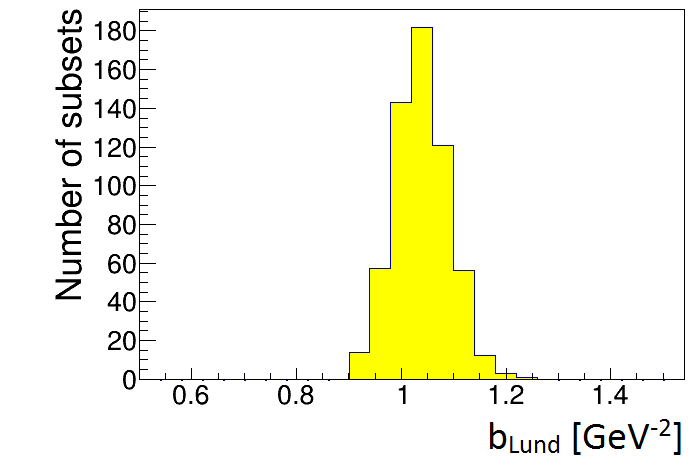}
	\includegraphics[width=0.41\textwidth]{Abbildungen/stable/aextradiquark_distr.png}
	\includegraphics[width=0.41\textwidth]{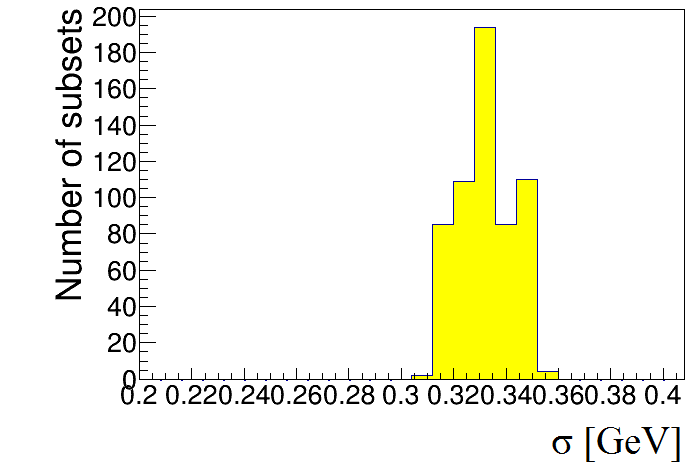}
	\includegraphics[width=0.41\textwidth]{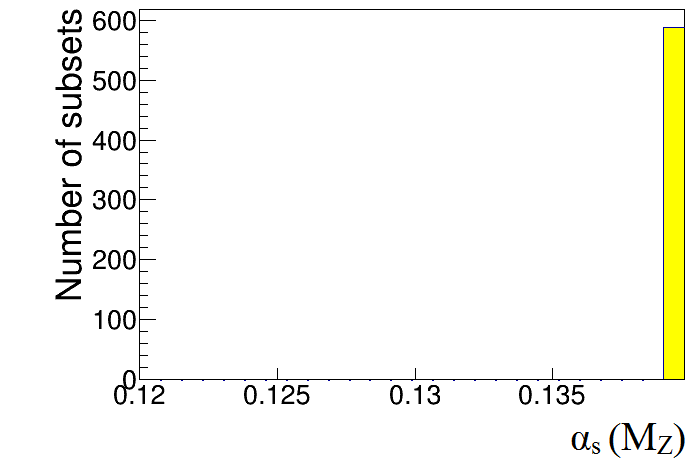}
	\includegraphics[width=0.41\textwidth]{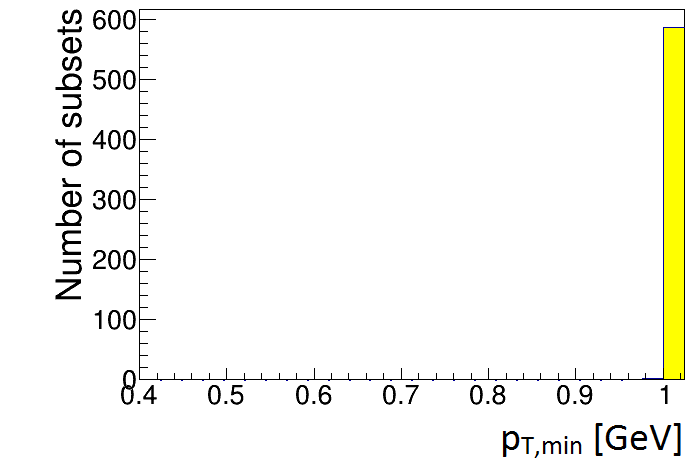}
	\caption{Distribution of the tuned model parameters using the run-combinations from Subsection \ref{subsec:tuning_setup} with the interpolation algorithm from Chapter \ref{ch:adaptive} and the weights ``Tune neutral'' from \cref{tab:particle_spectrum,tab:pdg_multiplicities1,tab:pdg_multiplicities2,tab:b_fragmentation,tab:event_shape,tab:diff_jet_rate} in App.~\ref{asec:observablesandweights}.}
	\label{fig:tune_adaptive_weights0}
\end{figure}

\begin{figure}
	\centering
	\includegraphics[width=0.45\textwidth]{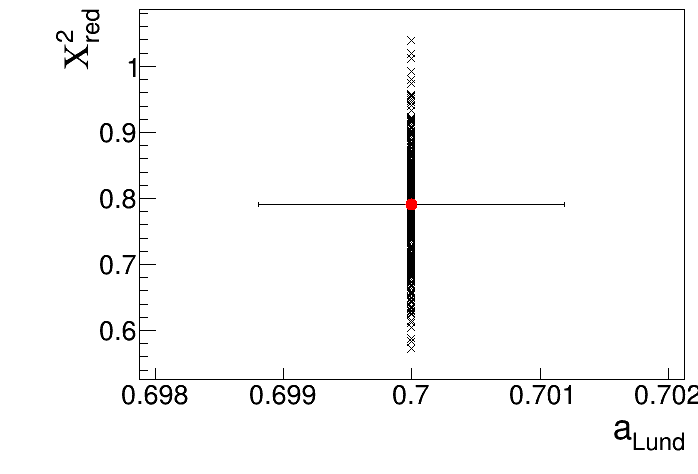}
	\includegraphics[width=0.45\textwidth]{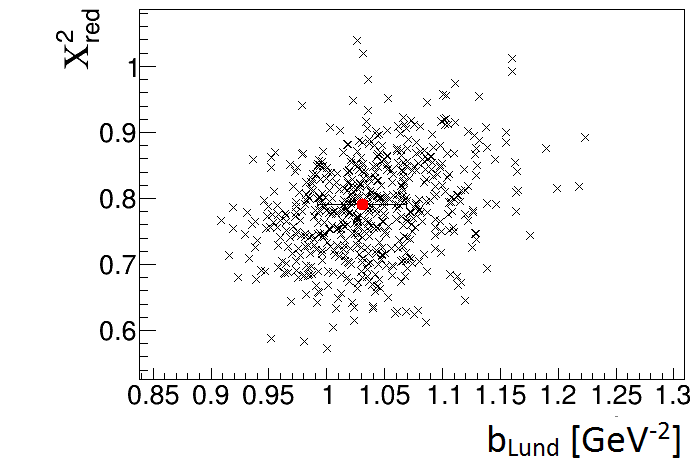}
	\includegraphics[width=0.45\textwidth]{Abbildungen/stable/aextradiquark_chi2.png}
	\includegraphics[width=0.45\textwidth]{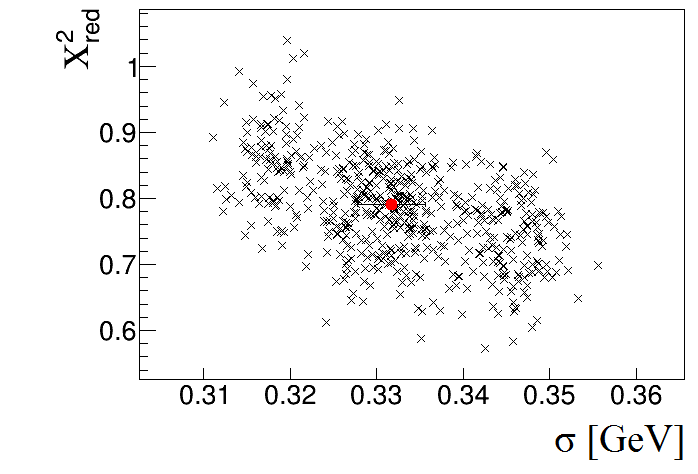}
	\includegraphics[width=0.45\textwidth]{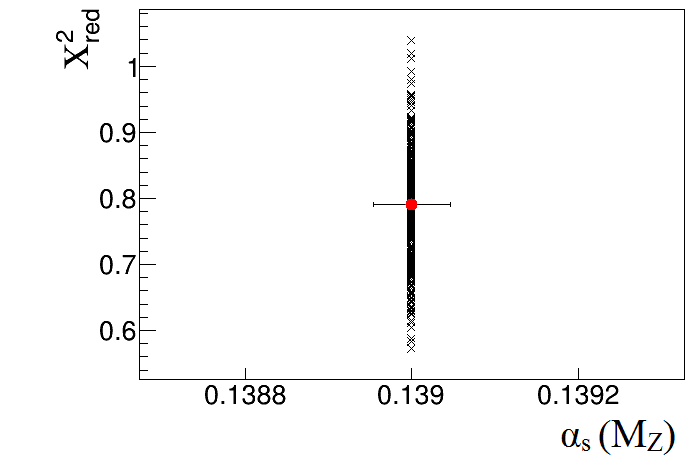}
	\includegraphics[width=0.45\textwidth]{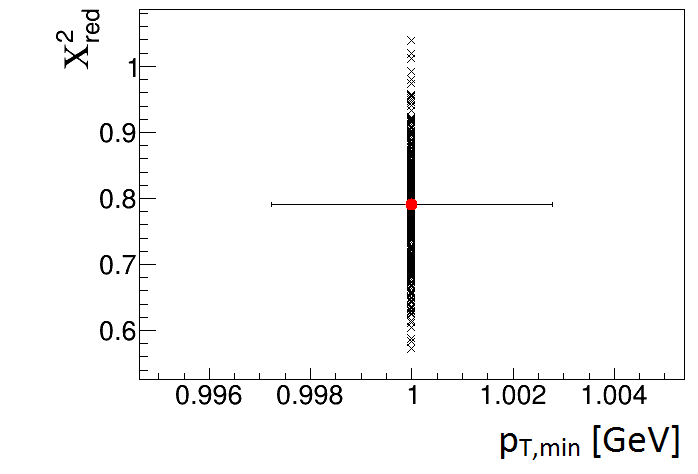}
	\caption{Distribution of the $\chi^2_{red}$ values of the tuned model parameters using the run-combinations from Subsection \ref{subsec:tuning_setup} with the interpolation algorithm from Chapter \ref{ch:adaptive} and the weights ``Tune neutral'' from \cref{tab:particle_spectrum,tab:pdg_multiplicities1,tab:pdg_multiplicities2,tab:b_fragmentation,tab:event_shape,tab:diff_jet_rate} in App.~\ref{asec:observablesandweights}.}
	\label{fig:tune_adaptive_weights0_chi2}
\end{figure}

\begin{figure}[h]
	\centering
	\includegraphics[width=0.45\textwidth]{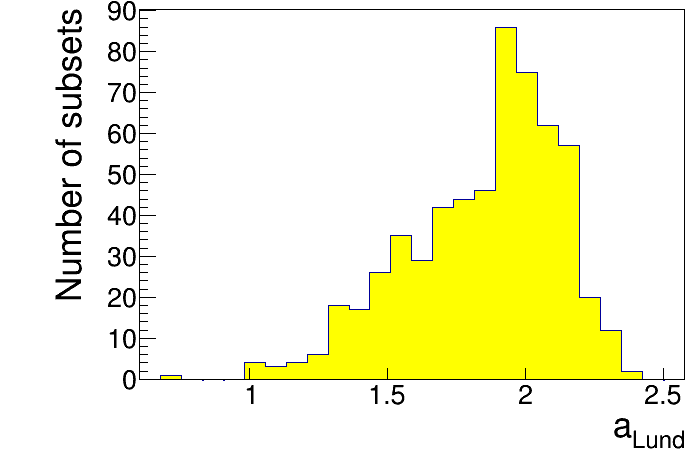}
	\includegraphics[width=0.45\textwidth]{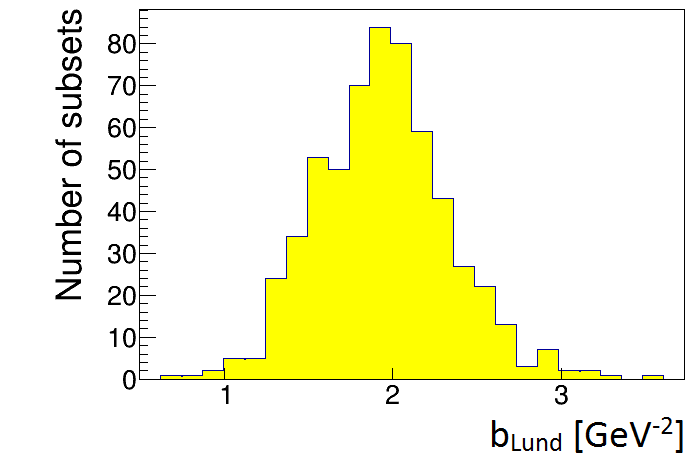}
	\includegraphics[width=0.45\textwidth]{Abbildungen/stable/no_limits/aextradiquark_distr.png}
	\includegraphics[width=0.45\textwidth]{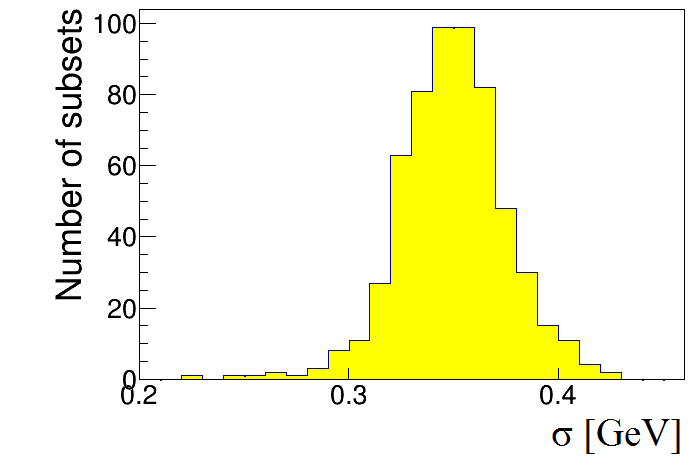}
	\includegraphics[width=0.45\textwidth]{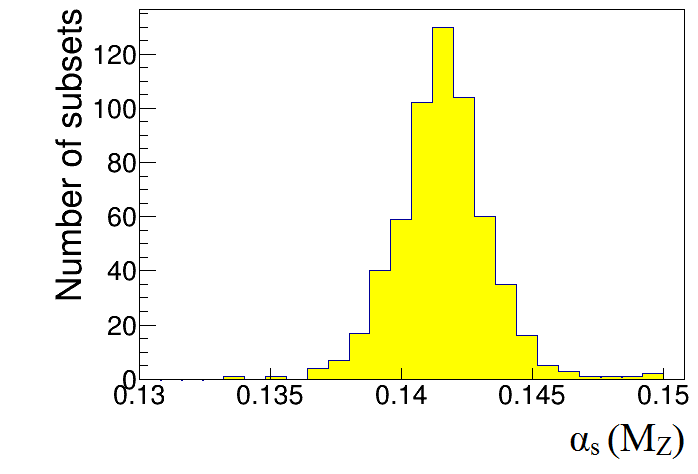}
	\includegraphics[width=0.45\textwidth]{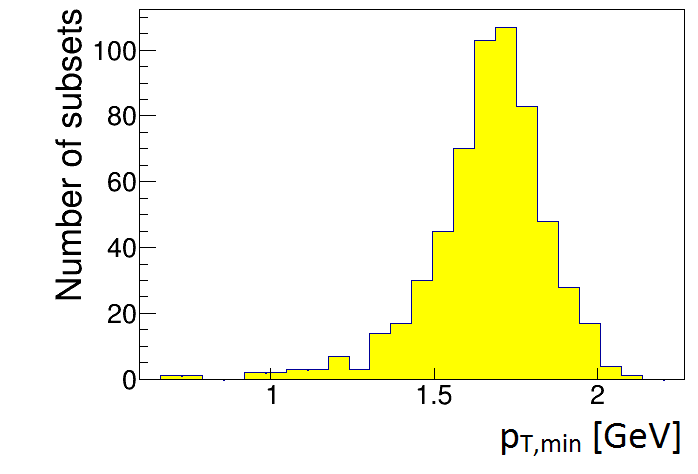}
	\caption{Distribution of the tuned model parameters using the run-combinations from Subsection \ref{subsec:tuning_setup} with the interpolation algorithm from Chapter \ref{ch:adaptive} and the weights ``Tune neutral'' from \cref{tab:particle_spectrum,tab:pdg_multiplicities1,tab:pdg_multiplicities2,tab:b_fragmentation,tab:event_shape,tab:diff_jet_rate} in App.~\ref{asec:observablesandweights} without parameter limits.}
	\label{fig:tune_adaptive_weights0_nolimits}
\end{figure}

\begin{figure}
	\centering
	\includegraphics[width=0.45\textwidth]{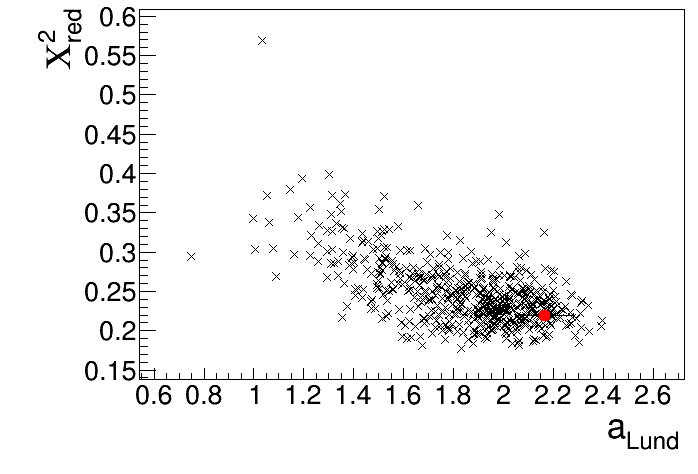}
	\includegraphics[width=0.45\textwidth]{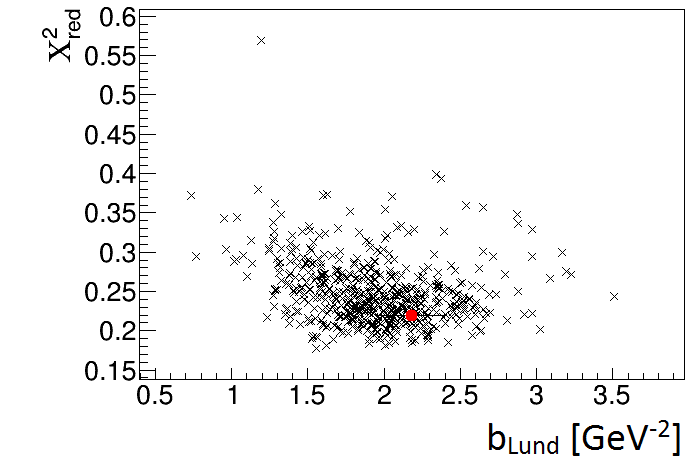}
	\includegraphics[width=0.45\textwidth]{Abbildungen/stable/no_limits/aextradiquark_chi2.png}
	\includegraphics[width=0.45\textwidth]{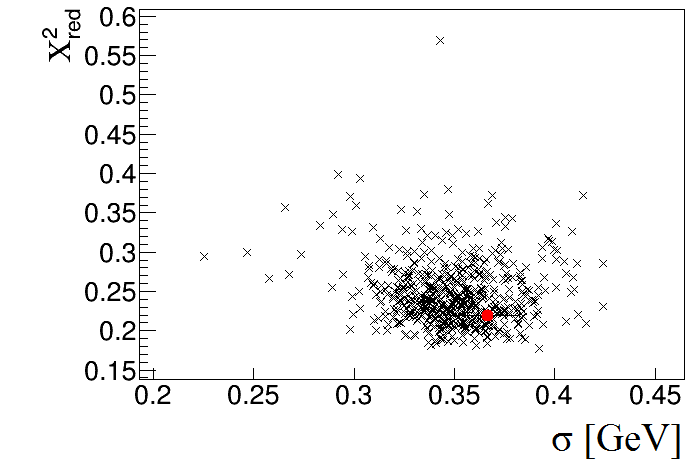}
	\includegraphics[width=0.45\textwidth]{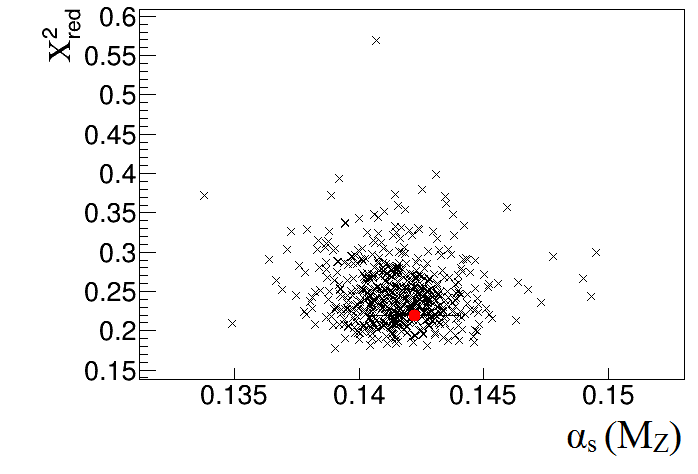}
	\includegraphics[width=0.45\textwidth]{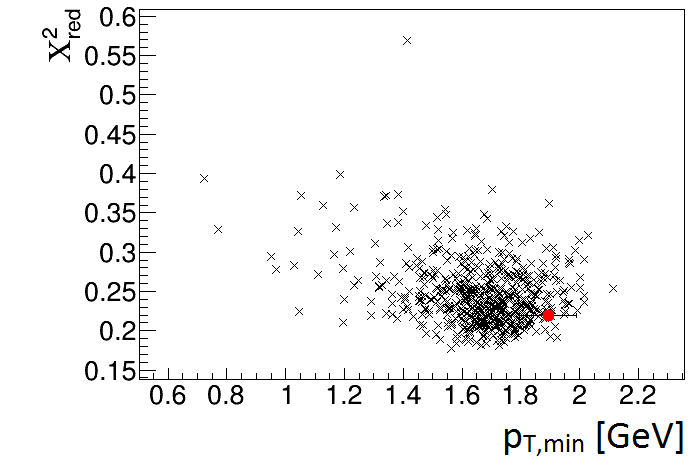}
	\caption{Distribution of the $\chi^2_{red}$ values of the tuned model parameters using the run-combinations from Subsection \ref{subsec:tuning_setup} with the interpolation algorithm from Chapter \ref{ch:adaptive} and the weights ``Tune neutral'' from \cref{tab:particle_spectrum,tab:pdg_multiplicities1,tab:pdg_multiplicities2,tab:b_fragmentation,tab:event_shape,tab:diff_jet_rate} in App.~\ref{asec:observablesandweights} without parameter limits.}
	\label{fig:tune_adaptive_weights0_chi2_nolimits}
\end{figure}

\chapter{Tuning the adaptive interpolation with BAT}
\setcounter{page}{133}

\section{Default set of observables}
\label{asec:adaptivetune}
\begin{figure}[h]
\makebox[\textwidth][c]{
	\begin{adjustbox}{max width=1.\textwidth}
	\includegraphics[width=\textwidth]{Abbildungen/covmat/050417/alund_history_prerun.png}
	\includegraphics[width=\textwidth]{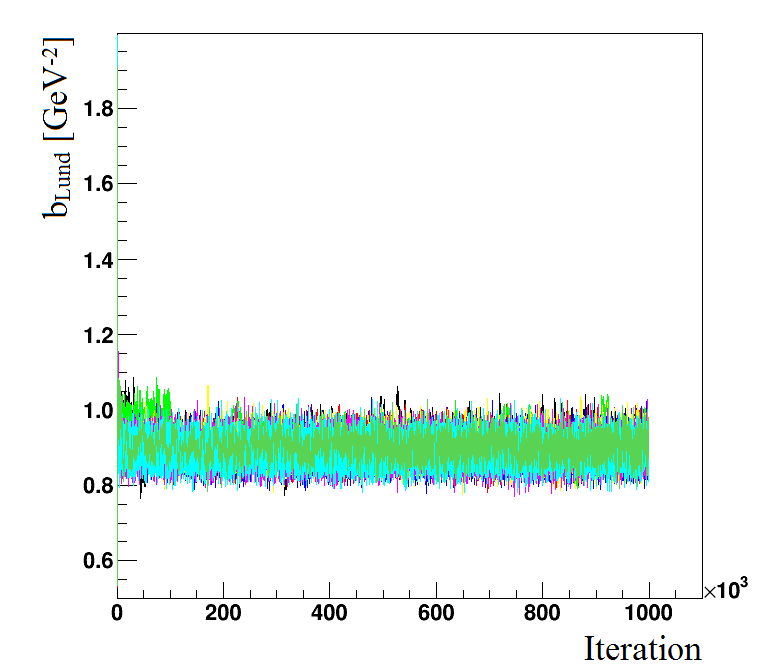}
	\end{adjustbox}
}
	\caption{Evolution of the Markov Chains during the pre-run of the adaptive interpolation in the parameter space of Tab.~\ref{tab:finalspace} using neutral weights. The left image corresponds to $a_{\textrm Lund}$, the one to $b_{\textrm Lund}$ in $[GeV^{-2}]$.}
	\label{fig:adaptive_covmat1}
\end{figure}

\begin{figure}[h]
\makebox[\textwidth][c]{
	\begin{adjustbox}{max width=1.\textwidth}
	\includegraphics[width=\textwidth]{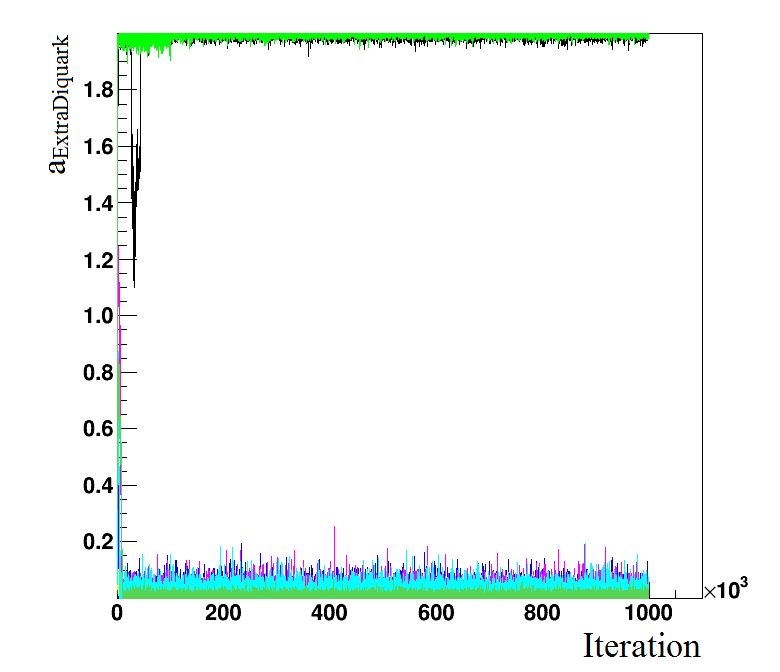}
	\includegraphics[width=\textwidth]{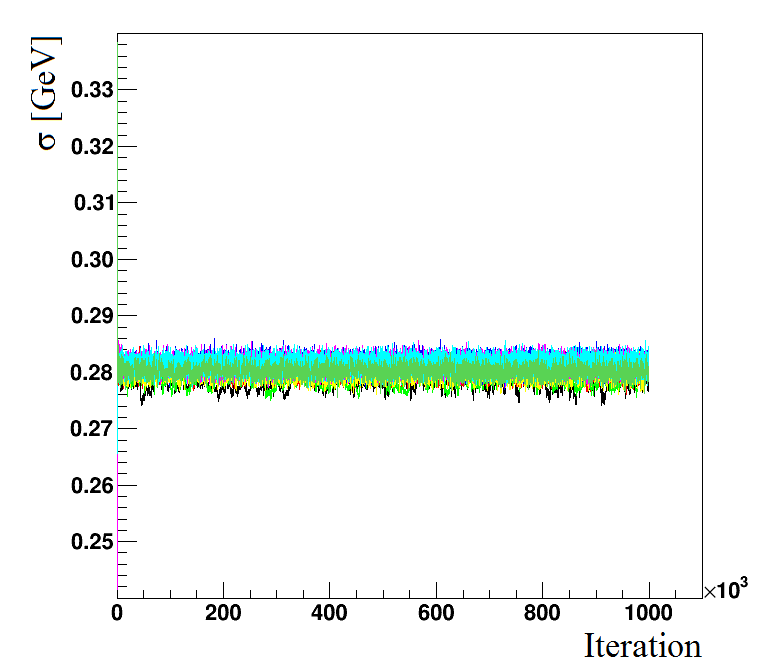}
	\end{adjustbox}
}
\makebox[\textwidth][c]{
	\begin{adjustbox}{max width=1.\textwidth} 
	\includegraphics[width=\textwidth]{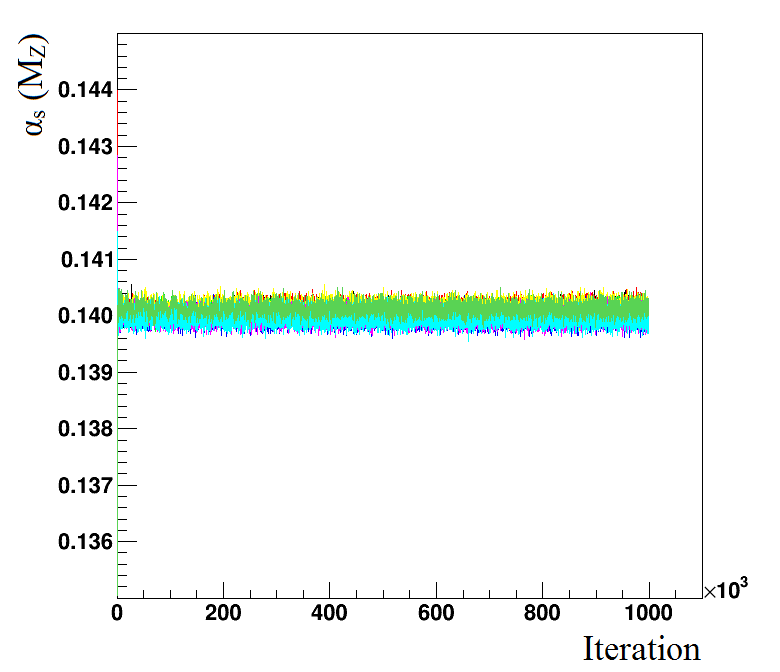}
	\includegraphics[width=\textwidth]{Abbildungen/covmat/050417/ptmin_history_prerun.png}
	\end{adjustbox}
}
	\caption{Evolution of the Markov Chains during the pre-run of the adaptive interpolation in the parameter space of Tab.~\ref{tab:finalspace} using neutral weights. In top left right image corresponds to $a_{\textrm ExtraDiquark}$, the top right one to $\sigma$ in $[GeV]$, the bottom left one to $\alpha_s(M_Z)$ and the bottom right one to $p_{T,min}$ in $[GeV]$.}
	\label{fig:adaptive_covmat2}
\end{figure}

\FloatBarrier
\section{Modified set of observables}
\label{asec:adaptivemodified}

\begin{figure}[h]
\makebox[\textwidth][c]{
	\begin{adjustbox}{max width=1.\textwidth}
	\includegraphics[width=\textwidth]{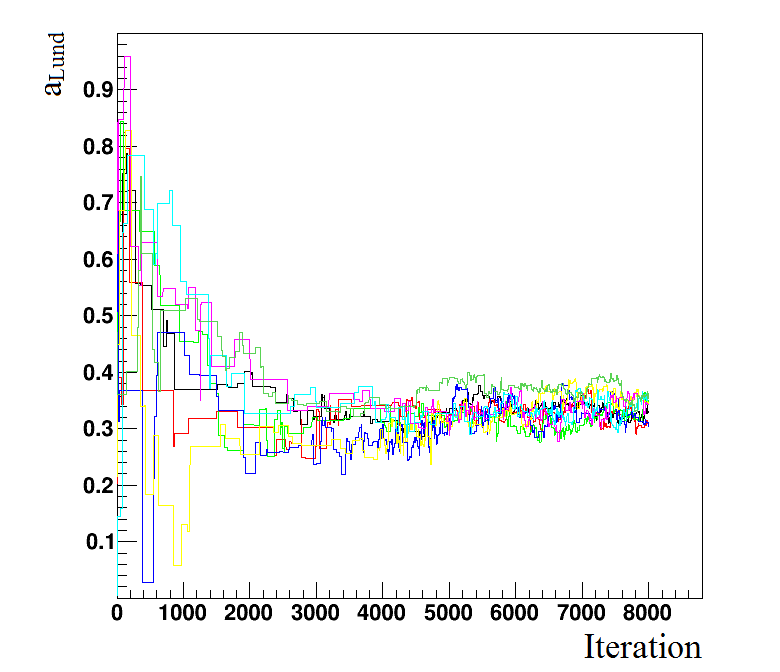}
	\includegraphics[width=\textwidth]{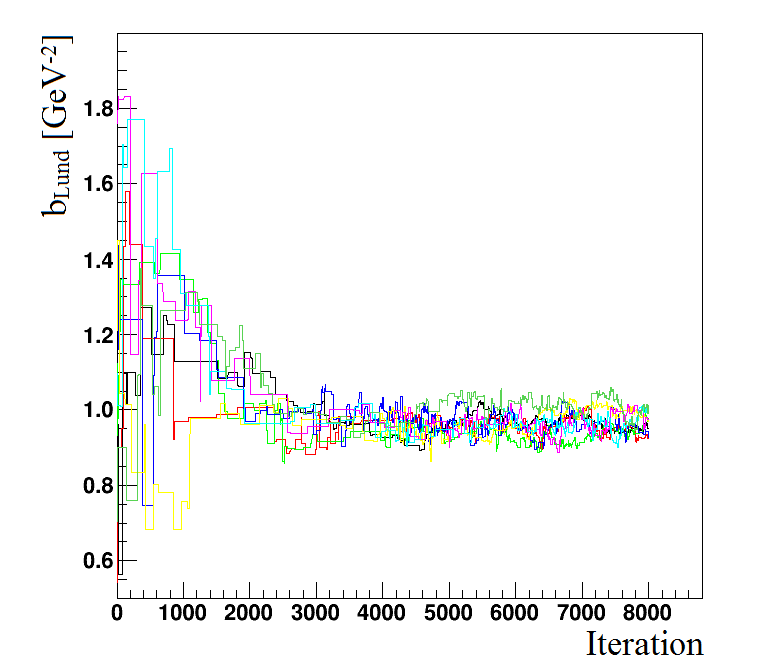}
	\end{adjustbox}
}
\makebox[\textwidth][c]{
	\begin{adjustbox}{max width=1.\textwidth}
	\includegraphics[width=\textwidth]{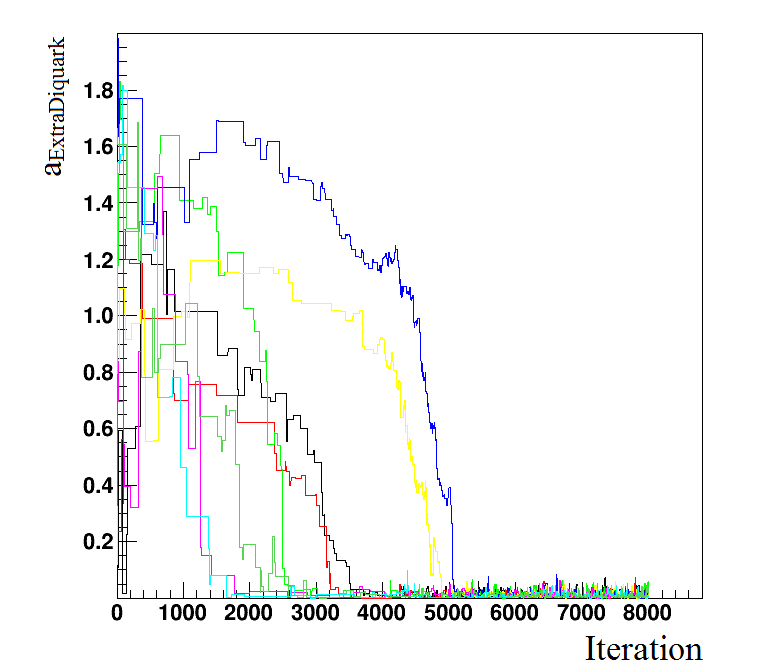}
	\includegraphics[width=\textwidth]{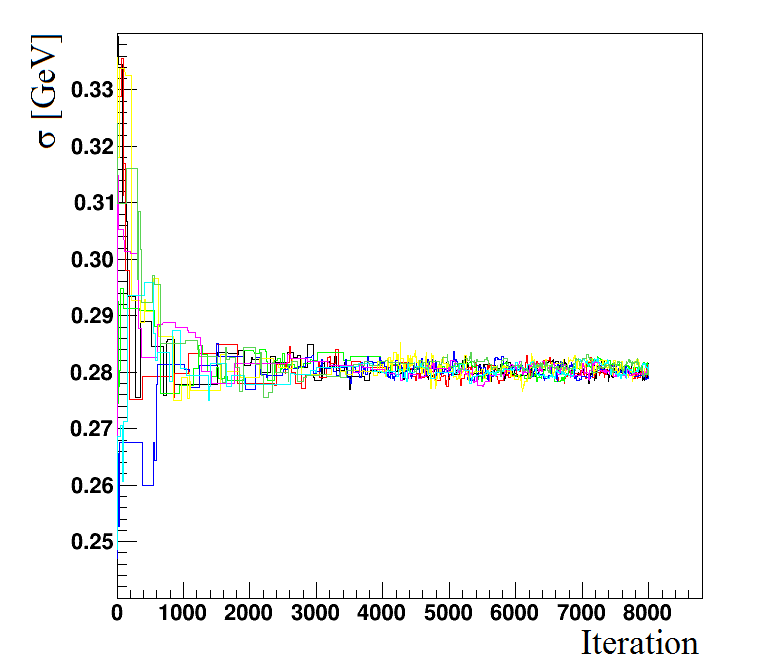}
	\end{adjustbox}
}
	\caption{Evolution of the Markov Chains during the pre-run of the adaptive interpolation in the parameter space of Tab.~\ref{tab:finalspace} using neutral weights and the modification from Tab.~\ref{tab:exchangeobservables}. The top left image corresponds to $a_{\textrm Lund}$, the top right one to $b_{\textrm Lund}$ in $[GeV^{-2}]$, bottom left one to $a_{\textrm ExtraDiquark}$ and the bottom right one to $\sigma$ in $[GeV]$.}
	\label{fig:modified1}
\end{figure}

\begin{figure}[h]

\makebox[\textwidth][c]{
	\begin{adjustbox}{max width=1.\textwidth} 
	\includegraphics[width=\textwidth]{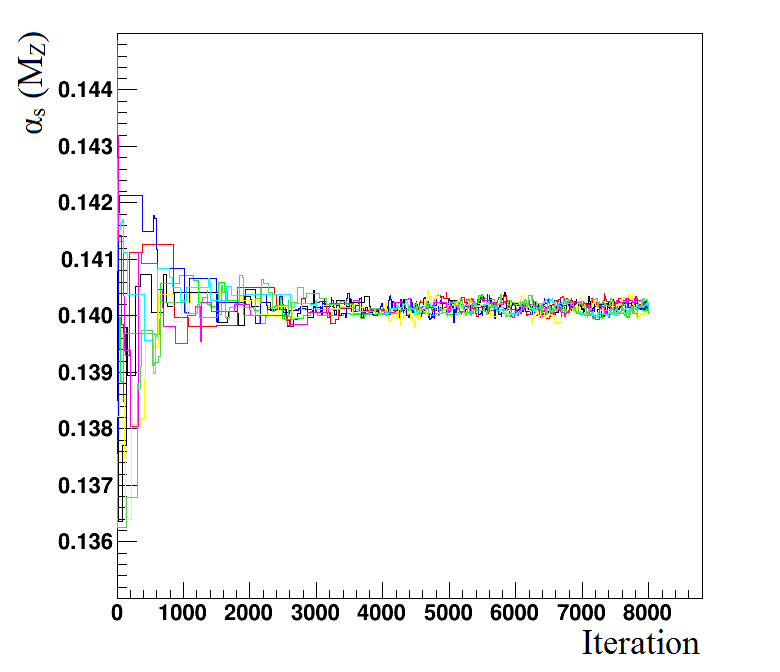}
	\includegraphics[width=\textwidth]{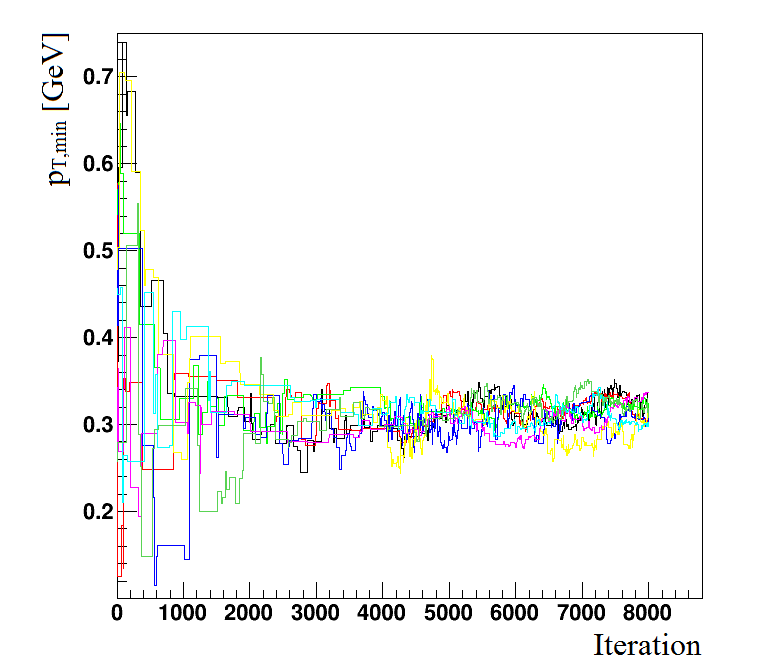}
	\end{adjustbox}
}
	\caption{Evolution of the Markov Chains during the pre-run of the adaptive interpolation in the parameter space of Tab.~\ref{tab:finalspace} using neutral weights and the modification from Tab.~\ref{tab:exchangeobservables}. The left image corresponds to $\alpha_s(M_Z)$ and the right one to $p_{T,min}$ in $[GeV]$.}
	\label{fig:modified2}
\end{figure}

\backmatter

\chapter*{Acknowledgment}
\setcounter{page}{142}
Zunächst möchte ich meiner Familie dafür danken, dass sie mir während der Zeit meines Studiums stets zur Seite standen und in der gesamten Zeit unterstützt haben. Dank dieser Unterstützung konnte ich die Studienzeit so angenehm wie möglich gestalten.\\
Weiter gilt mein Dank der gesamten Arbeitsgruppe, in welcher ich das vergangene Jahr verbringen durfte. In dieser Gruppe habe ich nicht nur in einem überaus freundlichen Arbeitsumfeld meine Masterarbeit anfertigen können, sondern konnte auch jederzeit Fragen und Anregungen einholen. Ganz besonderer Dank gilt dabei Andrea, welche sich immer unverzüglich meinen Problemen annahm und sich immer Zeit für mich genommen hat. Auch gilt mein besonderer Dank Stefan, der für meine Anliegen immer eine offene Tür hatte. Weiter möchte ich Andrii und Ludo dafür danken, dass diese mir zur Seite standen und mich in dem vergangenen Jahr unterstützt haben.
\end{document}